\documentclass[a4paper,fleqn,usenatbib]{mnras}

\usepackage{newtxtext,newtxmath}
\usepackage[T1]{fontenc}
\usepackage{ae,aecompl}

\usepackage{graphicx}
\usepackage{longtable}
\usepackage[labelfont=bf]{caption}
\usepackage{lscape,supertabular}
\usepackage{natbib}
\usepackage{hyperref}
\usepackage{subfig}
\usepackage{enumitem}
\usepackage{inputenc}
\usepackage[range-units = brackets,tophrase={-},seperr,repeatunits=false]{siunitx}

\title[HICOSMO I]{HICOSMO - Cosmology with a complete sample of galaxy clusters. \\\Large I. Data analysis, sample selection and luminosity-mass scaling-relation}

\author[Schellenberger et al.]{
G. Schellenberger,$^{1, 2}$\thanks{E-mail: gerrit.schellenberger@cfa.harvard.edu}
and T. H. Reiprich$^{1}$
\\
$^{1}$Argelander-Institut f\"ur Astronomie, Universit\"at Bonn, Auf dem H\"ugel 71, 53121 Bonn, Germany\\
$^{2}$Harvard-Smithsonian Center for Astrophysics, 60 Garden Street, Cambridge, MA 02138, USA
}

\date{Accepted by MNRAS}

\pubyear{2016}

\begin{document}
\label{firstpage}
\pagerange{\pageref{firstpage}--\pageref{lastpage}}
\maketitle

\begin{abstract}
The X-ray regime, where the most massive visible component of galaxy clusters, the intra cluster medium (ICM), is visible, offers directly measured quantities, like the luminosity, and derived quantities, like the total mass, to characterize these objects.
The aim of this project is to analyze a complete sample of galaxy clusters in detail and constrain cosmological parameters, like the matter density, $\Omega_\mathrm{m}$, or the amplitude of initial density fluctuations, $\sigma_8$. The purely X-ray flux-limited sample (\textit{HIFLUGCS}) consists of the 64 X-ray brightest galaxy clusters, which are excellent targets to study the systematic effects, that can bias results. 
We analyzed in total 196 Chandra observations of the 64 HIFLUGCS clusters, with a total exposure time of 7.7 Ms. 
Here we present our data analysis procedure (including an automated substructure detection and an energy band optimization for surface brightness profile analysis) which gives individually determined, robust total mass estimates. 
These masses are tested against dynamical and Planck Sunyaev-Zeldovich (SZ) derived masses of the same clusters, where good overall agreement is found with the dynamical masses. 
The Planck SZ masses seem to show a mass dependent bias to our hydrostatic masses; possible biases in this mass-mass comparison are discussed including the Planck selection function.
Furthermore, we show the results for the $\SIrange{0.1}{2.4}{keV}$-luminosity vs. mass scaling-relation. The overall slope of the sample (1.34) is in agreement with expectations and values from literature. Splitting the sample into galaxy groups and clusters reveals, even after a selection bias correction, that galaxy groups exhibit a significantly steeper slope (1.88) compared to clusters (1.06).
\end{abstract}


%%%%%%%%%%%%%%%%%%%%%%%%%%%%%%%%%%%%%%%%%%%%%%%%%%

%%%%%%%%%%%%%%%%% BODY OF PAPER %%%%%%%%%%%%%%%%%%

\section{Introduction}
\label{sec:intro}
Galaxy clusters are thought to resemble the intersections of the filamentary structure of the Dark Matter. Therefore, these largest gravitationally bound systems witness the growth of structure in the Universe and are excellent objects for cosmological studies. With cosmological parameters it is possible to predict the cluster mass function, i.e. the number density of Dark Matter halos, reflected by the observed galaxy clusters. Important parameters are the normalized matter density, $\Omega_{\rm M}$, and the amplitude of initial density fluctuations, $\sigma_8$. Even with local galaxy clusters these quantities can be constrained.

Unfortunately, X-ray observations of galaxy clusters as a cosmological probe require assumptions on how the total gravitating mass can be obtained. Either one can assume that the intra cluster medium (ICM) is in hydrostatic equilibrium or use tracers like the luminosity or temperature for the total mass. Calibrating these scaling relations of observables and the total mass, e.g., using weak lensing observations might provide a way to reliably estimate masses for big samples of galaxy clusters. Unfortunately other (maybe unknown) biases are connected with weak lensing studies, such as noise bias (e.g., \citealp{2013MNRAS.429..661M}), mass sheet degeneracy (e.g., \citealp{1995A&A...294..411S,2004A&A...424...13B}), asymmetry of the point spread function (e.g., \citealp{2003MNRAS.343..459H}), false photometric redshifts and miscentering (\citealp{2015arXiv150805308K}). Also selection effects, that enter by the composition of a galaxy cluster sample (e.g., selecting only massive or intrinsically brighter objects) can bias cosmological results, if they are not accounted properly.

Any observed sample of galaxy clusters has been subject to a selection process, which properties are essential for any cosmological study. For the application of a halo mass function, a clear relation between any quantity (like X-ray flux) used to select objects, and the total mass of galaxy clusters is essential. 
X-ray flux selected samples tend to include more relaxed objects, while, for example, galaxy clusters selected via the Sunyaev-Zeldovich (SZ) effect are close to be limited in mass and have slightly more disturbed clusters (\citealp{2016MNRAS.457.4515R}). For X-ray flux selected samples the luminosity mass relation takes a key role for the analysis. 
Especially for future surveys (like eROSITA), the X-ray luminosity will be the most important quantity for cosmology, since it is very easy to obtain with already tens of photons. For a plasma in complete hydrostatic equilibrium, the luminosity of the ICM is only dependent on the total gravitating mass of the cluster. First correlations have been found observationally by \cite{1977MNRAS.181P..25M}. But the X-ray luminosity is also strongly dependent on cluster physics close to the core, which introduces an intrinsic scatter for the luminosity mass relation. This intrinsic scatter together with selection effects of clusters is responsible for biased constraints on the slope and normalization of an observed distribution of luminosities and total cluster masses.
Ideally the luminosity mass relation is calibrated simultaneously with the cosmological analysis, in this way it is also possible to correct for biases which affect the observed luminosity mass relation.

For a cosmological application an X-ray flux limited sample like HIFLUGCS is of special interest: It provides high quality data of nearby galaxy clusters, which can be studied in detail including a treatment of possible substructure and contaminating point sources to get precise temperature and surface brightness profiles. It has been shown in \cite{reiprich_hiflugcs} that with such a sample, $\Omega_{\rm m}$ and $\sigma_8$ can be quantified, so one has an independent probe for cosmological parameters in hand.

This analysis will enable us to put constraints on at least two cosmological parameters and also to gain knowledge about the physical processes in the X-ray brightest galaxy cluster sample.
For the first time individually X-ray derived hydrostatic mass estimates of a complete sample of galaxy clusters are used to constrain cosmological parameters. The sample of interest here consists of the X-ray brightest galaxy clusters, with very high data quality available. Not only will we study the halo mass function and the cosmological implications, but also evaluate and quantify many sources of systematic biases.

The HIFLUGCS Cosmology Study (HICOSMO) is separated into two papers: This work (paper I) is focused on the Chandra data analysis to obtain total masses for all HIFLUGCS clusters individually, including crucial steps, e.g., the extrapolation technique. The cluster mass estimates are shown to be robust by a comparison to dynamical mass estimates (from optical velocity dispersion) and SZ derived masses. Furthermore, here we investigate the observed $L_x-M$ relation and compare it to the biased corrected one, as well as to other scaling relations from literature. 
The results and discussion of the cosmological analysis will be presented in \cite{2017arXiv170505843S} (Paper II).

Unless stated otherwise, we assume a flat $\Lambda$CDM cosmology with the following parameters: $\Omega_\mathrm{M} = 0.27$, $\Omega_\Lambda = 0.73$, $H_0 = h \cdot \SI{100}{km\,s^{-1}\,Mpc^{-1}}$ and $h = 0.71$.

\section{Galaxy cluster sample}
\label{ch:hiflugcs_sample}
\begin{figure*}
	\centering
	\includegraphics[trim=0px 100px 0px 100px,clip,width=0.79\textwidth]{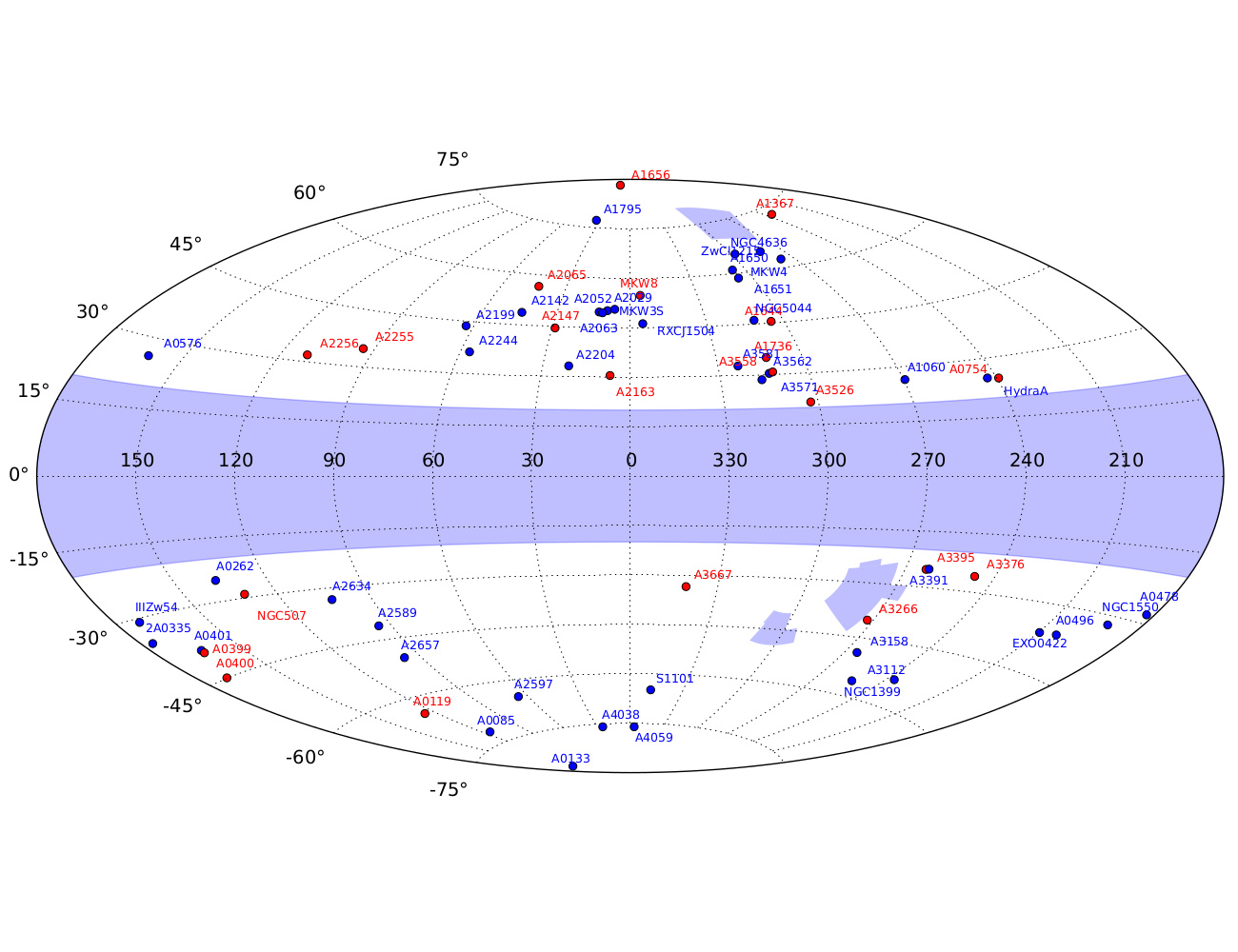}
	\caption{The 64 HIFLUGCS galaxy clusters (red are clusters marked as merging in \citealp{2009ApJ...692.1033V}) in Galactic coordinates. The blue shaded regions denote the excluded regions (Milky Way plane, Magellanic Clouds and Virgo cluster).}
	\label{fig:hiflugcs_aitoff}
\end{figure*}
The \textbf{HI}ghest X-ray \textbf{FLU}x \textbf{G}alaxy \textbf{C}luster \textbf{S}ample (HIFLUGCS, \citealp{reiprich_hiflugcs}) was selected from the ROSAT All-Sky Survey (RASS, \citealp{1993Sci...260.1769T,1999A&A...349..389V}). It consists of 64 galaxy cluster above a flux limit of $\SI{2e-11}{erg\,s^{-1}\,cm^{-2}}$ in the $\SIrange{0.1}{2.4}{keV}$ band (ROSAT band) and within a region outside the Milky Way disk ($|b|\geq 20\degree$), the Magellanic Clouds and the Virgo cluster, which sums up to $\SI{64.78}{\percent}$ of the sky (see Fig. \ref{fig:hiflugcs_aitoff}). This results in a sample of very bright and local galaxy clusters. Looking at it in more detail highlights the effort spent on creating a complete, flux limited sample. 

First candidates were selected from 4 different ROSAT catalogs:
\begin{itemize}
	\item REFLEX, \cite{2001A&A...369..826B}: This catalog covers the southern part of the sky and optical follow-up observations have been made within the ESO programme. It comprises 452 galaxy clusters above a flux limit of $\SI{3e-12}{erg\,\s^{-1}\,cm^{-2}}$ in the ROSAT band. At the flux limit it is supposed to be complete at the 90\% level. At redshifts above 0.2 the fraction of non-extended REFLEX sources rises above 30\%, which makes it also crucial to select a local sample. Otherwise the number of AGNs falsely identified as clusters becomes significant. REFLEX was constructed with special attention on cosmological application.
	\item NORAS, \cite{2000ApJS..129..435B}: The northern sky of RASS is covered by NORAS, which consists of 378 clusters. The completeness is stated with 50\% at the same flux level as REFLEX. 
	\item NORAS II, see \cite{reiprich_hiflugcs} for reference: NORAS II has a similar structure as NORAS but with a lower flux limit of $\SI{2e-12}{erg\,s^{-1}\,cm^{-2}}$.
	\item BCS, \cite{1998MNRAS.301..881E}: BCS consists of 201 bright galaxy clusters in the northern hemisphere above a flux of $\SI{4.4e-12}{erg\,s^{-1}\,cm^{-2}}$ and 90\% completeness. 
\end{itemize}

Clusters were selected from these catalogs with a slightly lower flux limit of $\SI{1.7e-11}{erg\,s^{-1}\,cm^{-2}}$. This extended sample was then reanalyzed in \cite{reiprich_hiflugcs} to obtain a homogeneous flux limited sample. It should be noted that RXCJ1504 was originally not included in HIFLUGCS, but follow-up observations later revealed that the contribution of AGN emission to the total flux is not as high as initially thought. The median redshift of the final sample is $\num{0.05}$. 
Detailed studies of several aspects of this sample have been carried out in the past: \cite{2007A&A...466..805C} (cooling flow), \cite{hudson_what_2009} (cool core definition), \cite{2009A&A...501..835M} (radio AGNs), \cite{zhang_hiflugcs:_2010} (scaling relations), \cite{2011A&A...535A..78Z} (star formation), \cite{2011A&A...532A.133M} ($L_x - T_{\rm vir}$ relation), \cite{2011A&A...526A..79E} (selection biases).

Despite this interesting opportunities for the future by the construction of even bigger samples, it is essential to have a reliable data analysis structure, which takes care of many sources of biases like substructure or extrapolation techniques. For this purpose, HIFLUGCS is the sample of choice to create a baseline for cluster analysis and cosmological interpretation. 

\section{Data analysis}
\label{sec:data}
All galaxy clusters of the HIFLUGCS sample have been observed at least once with the Chandra X-ray observatory. For reliable hydrostatic masses one has to not only follow the standard tasks for data reduction, but also account for the following aspects:
\begin{itemize}
	\item Contamination by point sources and substructure
	\item Instrumental and astrophysical background components
	\item Density profile from surface brightness in appropriate energy band
	\item Temperature profile parametrization
	\item Mass extrapolation
\end{itemize}

After giving an overview on the basic steps of the Chandra data reduction, each of the above mentioned points will be discussed in this section. Many Chandra observations are available for the 64 HIFLUGCS clusters. The basic selection criteria are:
\begin{itemize}
	\item Observation was public by Dec 2015,
	\item Advanced CCD Imaging Spectrometer (ACIS) detector was used,
	\item no grating observations,
	\item and exposure time at least $\SI{9}{ks}$.
\end{itemize}
This selects 336 observations with a total exposure time of $\SI{12.2}{Ms}$. A subset of these observations was actually analyzed, which usually excludes short observations with a high off-axis angle (cluster outskirts), observations that were not publicly available at the time of analysis and observations that would not add much more to an already large summed exposure time (if $\Delta t_\mathrm{exp} < 20\,\%$). 
In total 196 observations ($\SI{7.6}{Ms}$) were analyzed.
Note that some recent observations were added later\footnote{OBSID 15186, 17168, 17492, 16129, 16514, 16515, 16516, 13442, 13448, 13452, 14338, 13450, 2941, 7232, 7217, 4189\label{note1}}  (see below).

\subsection{Chandra standard data reduction tasks}
The procedure for the raw Chandra data reduction follows the description in \cite{Schellenberger2015}. The important parts are briefly mentioned here. The analysis software that was used is
\begin{itemize}
	\item CIAO software version 4.6, with CALDB (calibration database) version 4.6.5 (released in December 2014 and compatible for observations until October 2014), only for very recent observations (see footnote \ref{note1}) CALDB 4.7.0 was used.
	\item Heasoft 6.16 including \verb|Xspec| 12.8.2e, see also Section \ref{ch:gasmass} for the use of an updated \verb|Xspec| version.
\end{itemize}
The \verb|chandra_repro| task applies a default treatment to the raw events files, for example badpixels and afterglows are detected and marked, and also the latest calibration is applied to update time, coordinate and pulse-height, grade and status information of the events. 
A lightcurve cleaning is performed in order to remove flared periods from the events file: The events file is binned in the time domain to have 1000 counts in the $\SIrange{0.7}{7}{keV}$ band in each time bin (but at least $\SI{200}{s}$ per bin). The \verb|deflare| task using the \verb|lc_clean| algorithm with a $3\sigma$ clipping is the default method, but for some very flared observations we switched to the iterative \verb|lc_sigma_clip| algorithm, which is more stable, but sometimes cuts out more clean time. Many Chandra observations are not affected by flares.
Any of the following region selection, either for the temperature or surface brightness profile, is centered on the emission weighted cluster center. Since Chandra does not provide a large field of view that is appropriate for this quantity, we take the emission weighted centers determined by \cite{zhang_hiflugcs:_2010} using XMM-Newton (see details in Tab. \ref{tab:hiflugcs_details}).

\subsection{Point source treatment}
\begin{figure*}
	\centering
	\includegraphics[width=0.475\textwidth]{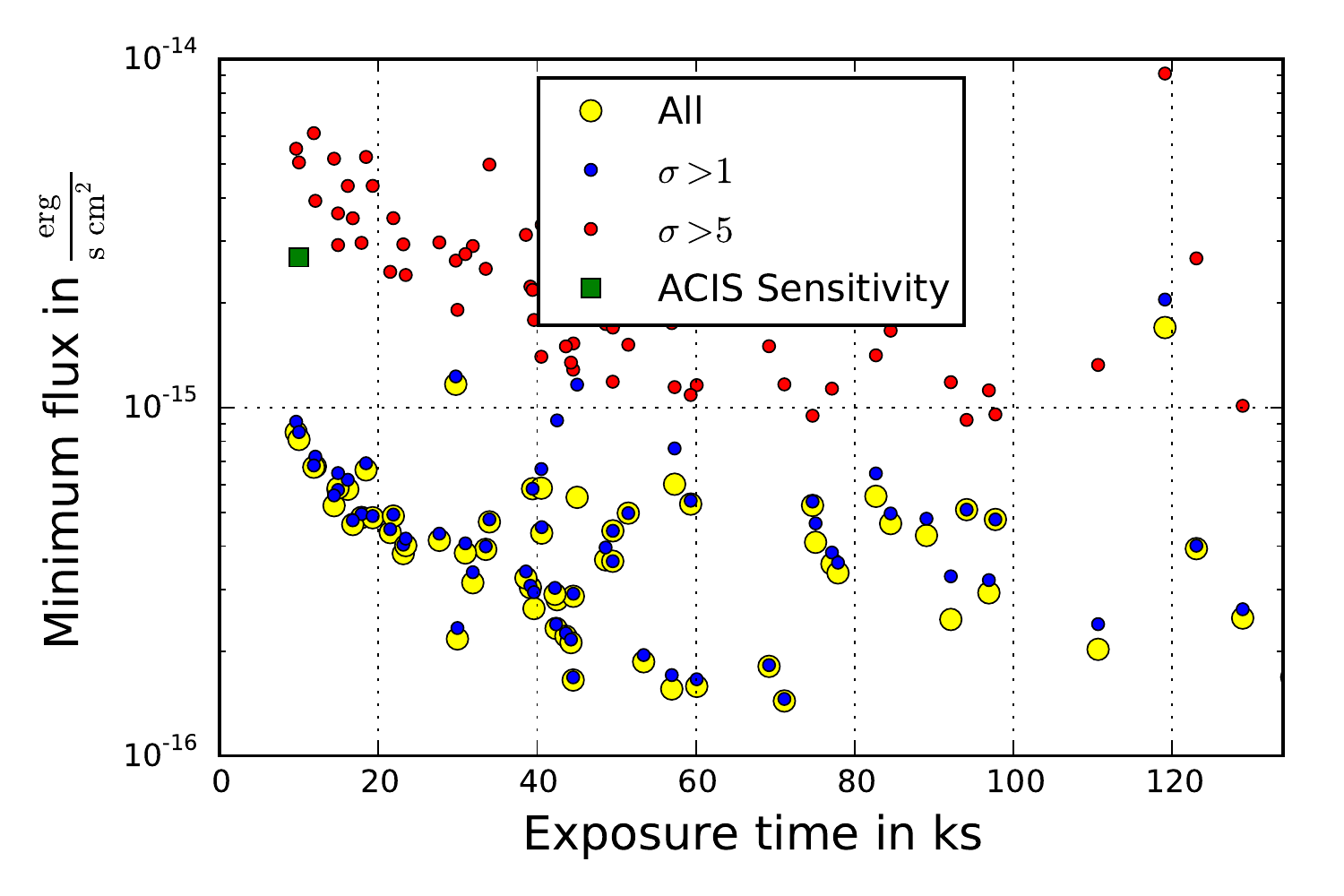}
	\includegraphics[width=0.475\textwidth]{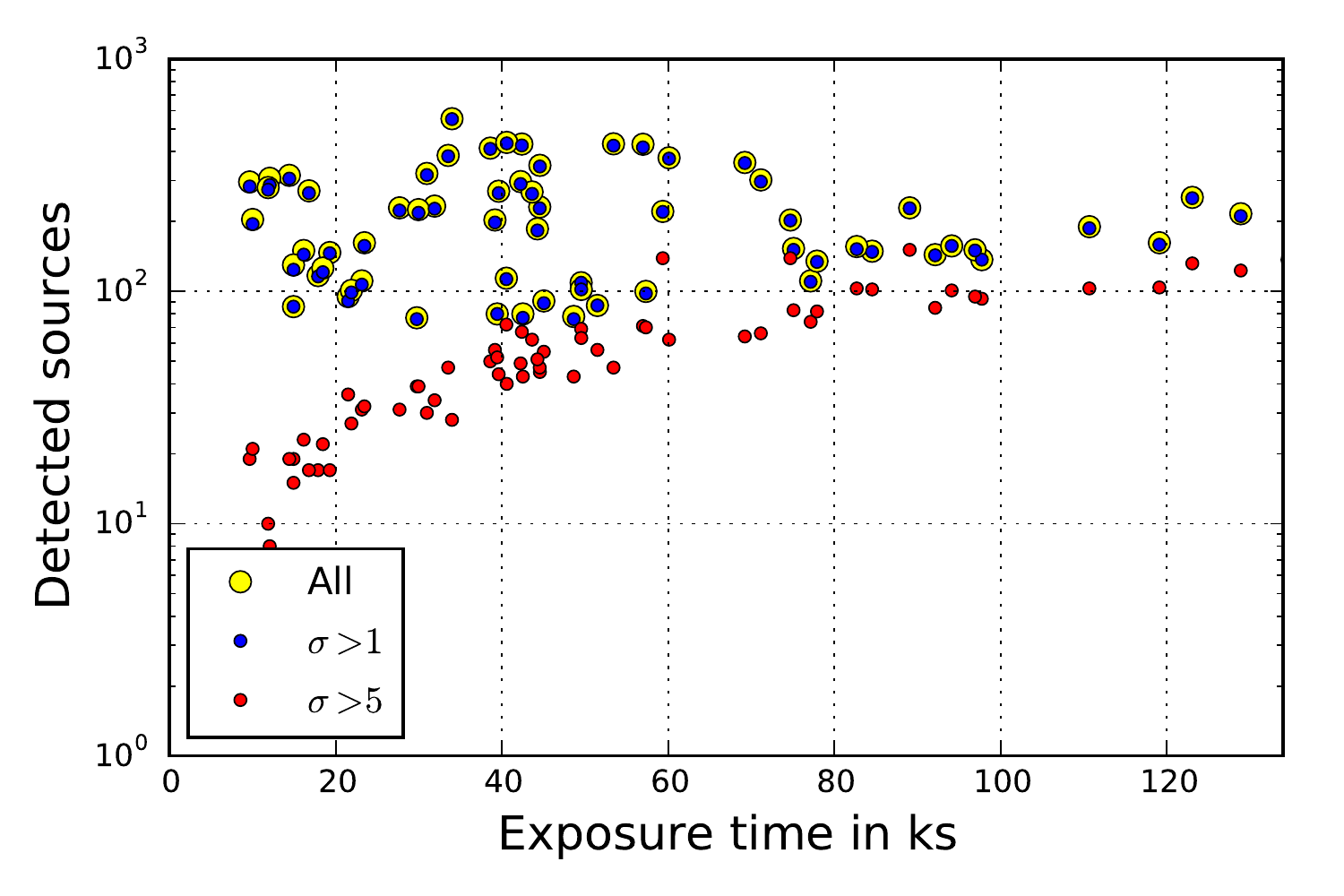}
	\caption{\textit{Left:} Minimum "flux" of detected point sources as a function of the exposure time for different selection criteria: Blue and red points are point sources with a \textit{wavdetect} detection significance above 1 or 5, respectively. Yellow points mark the minimum flux among all detected sources. \textit{Right:} Number of detected sources for each selection criterion as a function of the exposure time.}
	\label{fig:minflux}
\end{figure*}

Point sources are sources which have an angular size smaller than the point spread function (PSF) of the instrument and are broadened to the PSF scale. Chandra has arcsecond spatial resolution on axis, so one can assume that all the detected point sources are AGNs. The \verb|wavdetect| task, which uses a Mexican-Hat Wavelet transformation for the automated detection, is applied. Different scales (\verb|1 2 4 8 16| pixel, which corresponds to roughly the same number in arcseconds due to the binned image) are used for the correlation of the image with the wavelet. Each source detection is characterized by a significance value, which we use to clip spurious detections.
The significance is calculated from the net source counts and the (Gehrels) error of background counts, which can create misleading values for the low counts regime.
Figure \ref{fig:minflux} (left) shows the correlation between the minimum detected flux and the exposure time. 
The flux here describes only a rough estimate of the source flux and has not been recomputed properly by, e.g., the ciao-task \textit{srcflux}, because we don't use it later in the analysis.
Note that in Fig. \ref{fig:minflux} only the main observation for each cluster (see Tab. \ref{tab:observations}) is shown. The three different colors correspond to a significance threshold of 0, 1 or 5. Above $\sim\SI{50}{ks}$ the flux limit does not decrease any more. 
The fluxes were computed in the $\SIrange{0.5}{2}{keV}$ band corrected for vignetting effects, dead area and detector quantum efficiency and assuming a powerlaw with 2 as spectral index. In the Chandra Proposers' Observator Guide\footnote{\url{http://cxc.harvard.edu/proposer/POG/arch_pdfs/POG_cyc13.pdf}} a sensitivity of Chandra/ACIS of $\SI{4e-15}{erg\,s^{-1}\,cm^{-2}}$ is mentioned in the $\SIrange{0.4}{6}{keV}$ band for a $\SI{10}{ks}$ observation. This value roughly corresponds to $\SI{2.7e-15}{erg\,s^{-1}\,cm^{-2}}$ in the energy band for Fig. \ref{fig:minflux}. Applying the $5\sigma$ threshold means that some faint point sources may not be cut out (since the ACIS sensitivity limit is lower) but one gets rid of most of the spurious detections. 
Although the ACIS sensitivity is well above the $1\sigma$ significance limit in Fig. \ref{fig:minflux}, we do not trust the determined fluxes for these faint sources given the reasons described above.
Also the expected increasing behavior of the detected number of sources as a function of exposure time strongly favors the $5\sigma$ threshold (Fig. \ref{fig:minflux} right). For galaxy clusters with more than one observation, the point source detection was run for all observations and the complete point source list was excluded for each observation. Only multiple detected sources (in different observations) were only removed once before taking spectra or surface brightness profiles. So all observations of one cluster use the same point source catalog.

\subsection{Substructure Selection}
\label{sec:substructure}
By assuming hydrostatic equilibrium one requires idealized objects that undergo no interaction with the surrounding environment. But a complete sample of galaxy clusters also contains merging systems. In some cases, these interactions are spatially localized in one part of the cluster. By excluding the substructure area from the extraction region of the profiles it is possible to minimize the bias that would arise from disturbed regions. 

\begin{figure*}
	\centering
	\includegraphics[width=500px]{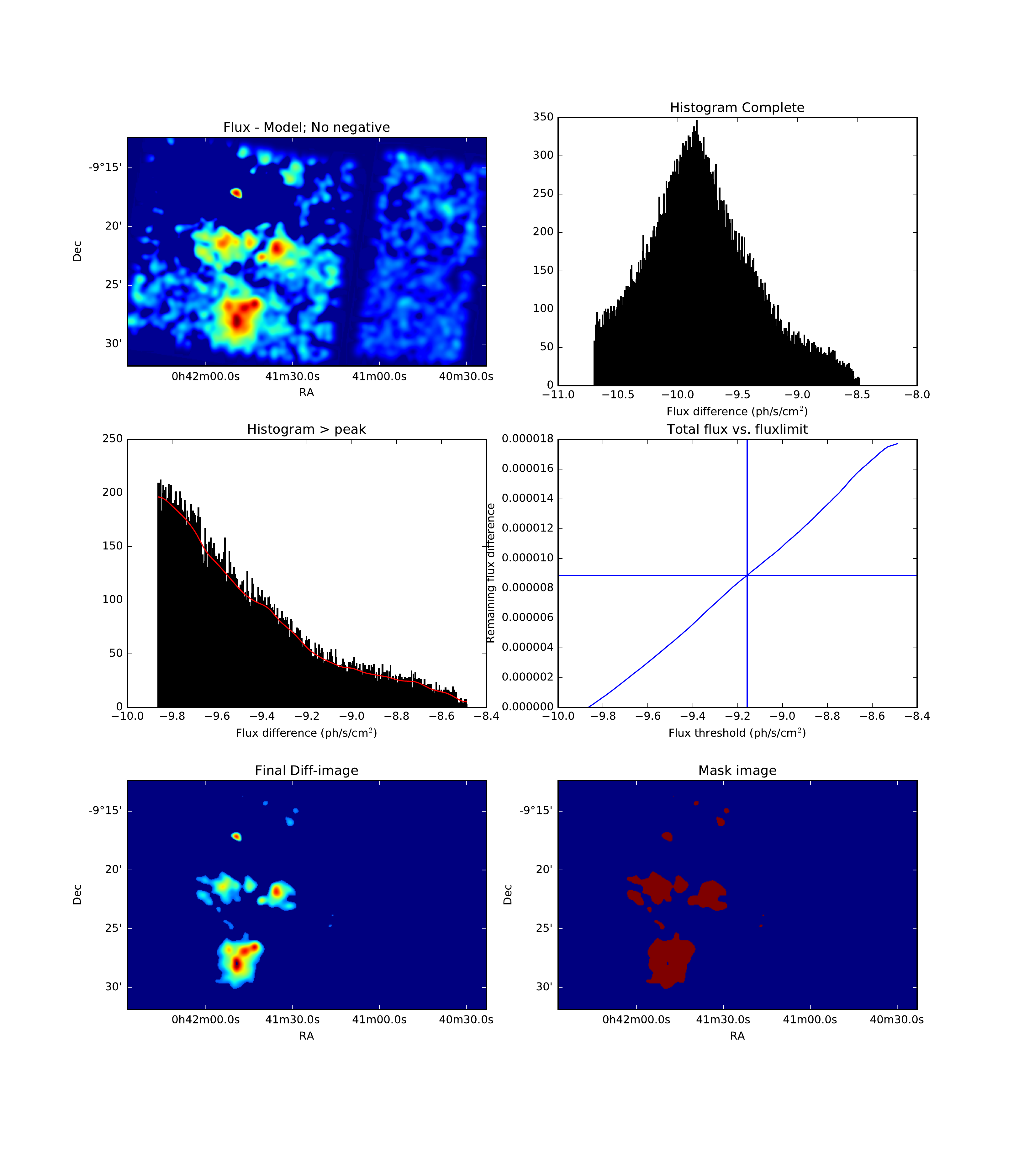}
	\caption{Substructure selection procedure for A85: Substructure is selected by filtering the pixels based on the flux difference to the best-fit surface brightness model. The algorithm selects only regions with a larger flux difference than half the total cumulative flux excess. Note flux values are all negative because the $\log_{10}$ of the flux with respect to the given unit is plotted. (details see text).}
	\label{fig:contours}
\end{figure*}

The dense substructure is usually visible in X-rays as excess emission to the normal ICM radiation, but often not detected by the \verb|wavdetect| task, which is more efficient for point sources. We assume that in general the surface brightness profile of the ICM emission follows a double $\beta$-model,
\begin{equation}
S(r) = S_{01} \left( 1 +  \frac{r^2}{r_{\rm c1}^2}  \right)^{-3\beta_1 + 0.5} + S_{02} \left( 1 + \frac{r^2}{r_{\rm c2}^2}  \right)^{-3\beta_2 + 0.5}~,
\label{eq:double_beta}
\end{equation}
where for each component $i = 1,2$, $S_{0i}$ denotes the central surface brightness value, $r_{{\rm c}i}$ the core radius and $\beta_i$ characterizes the decrease of the surface brightness to the outskirts.
Starting from a least square fit to the surface brightness distribution using this model (centered on the emission weighted cluster center as defined in \citealp{zhang_hiflugcs:_2010}), one can detect local excess emission. To reduce the noise we smooth both, the model and the photon flux image (corrected for vignetting, exposure time variations due to Chandra dither motion, quantum efficiency, bad pixels) of the observation with a Gaussian ($\SI{16}{arcsec}$ width), before subtracting the model from the observation to get the flux difference image (see Fig. \ref{fig:contours}, top left). 

It is now important tune the threshold of excess emission to be cut out, that not too much emission gets excluded from very small fluctuations with respect to the model. In a histogram of the excess flux per pixel (Fig. \ref{fig:contours}, top right) one can recognize a peak which represents the abundant flux difference between measurement and model. Left of this peak are smaller differences and to the right the larger ones. As default we chose the threshold to cut at a flux-difference limit, where the cumulative flux-excess is half the total flux difference summed over all pixels (note that only positive flux differences are included in this analysis):
\begin{equation}
\sum \limits_i \left( f_\mathrm{OBS} - f_\mathrm{MODEL} \right) > 0.5 \sum \limits_{\rm all\,pixel} \left( f_\mathrm{OBS} - f_\mathrm{MODEL} \right)~,
\label{eq:flux_thresh} 
\end{equation}
where $i$ are all pixels with a flux difference above a threshold (see also Fig. \ref{fig:contours}, middle right). 
This has been computed by smoothing with histogram distribution as shown in Fig. \ref{fig:contours}, middle left.
The factor of $0.5$ is slightly adjusted for a few observations, where still too much noise was included. This means that half of the excess flux is removed  (Fig. \ref{fig:contours}, bottom left). A mask file  (Fig. \ref{fig:contours}, bottom right) containing all the selected pixels (as polygons) is created and saved in a standard regions file. 
Since this procedure is very time consuming it is only done once per cluster for the longest exposure of a large part of the cluster (main observation ID).

\begin{figure*}
	\centering
	\includegraphics[width=300px]{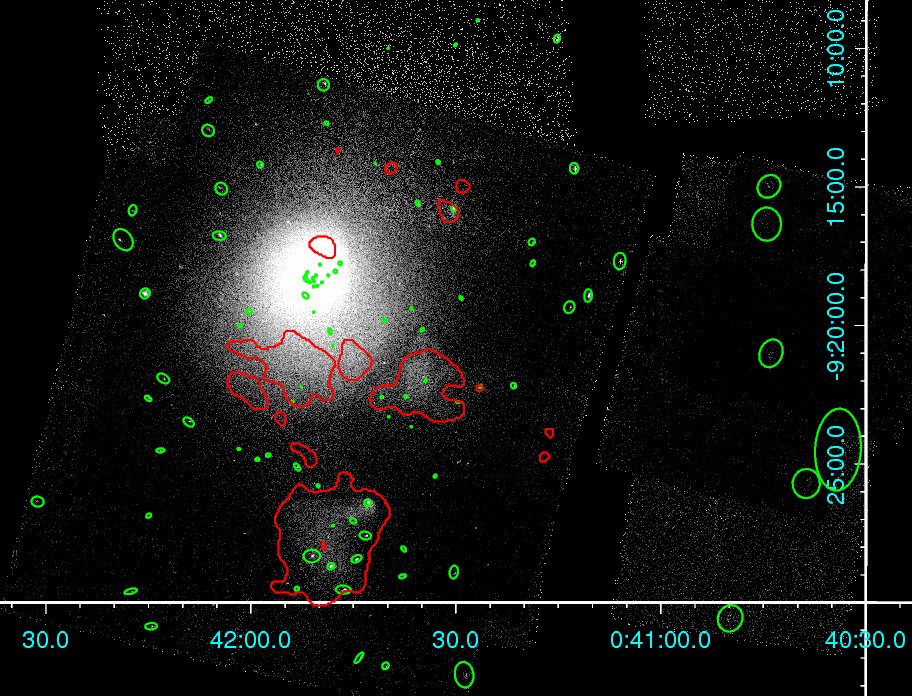}
	\caption{Mosaic image of all exposure corrected A85 observations that were used. Green ellipses mark the wavelet-detected sources, red polygons the substructure that has been marked.}
	\label{fig:A85mosaic}
\end{figure*}

Together with the combined point source region file, these are the regions that are being excluded for the spectral and surface brightness analysis in the following. One can see the final result, for Abell 85 in Fig. \ref{fig:A85mosaic}, which has some bright in-falling structure to the south.

\subsection{Surface brightness and gas density profile}
For an optically thin plasma the X-ray emissivity $\epsilon$ is a function of the luminosity $L_x$ and the emitting volume $V$,
\begin{equation}
\epsilon = \frac{\mathrm{d}L_x}{\mathrm{d}V}~,
\end{equation}
and can be written as the product of the number densities of electrons $n_\mathrm{e}$ and Hydrogen atoms $n_\mathrm{H}$ and the cooling function $\Lambda$. The cooling function describes how much energy is radiated by a plasma with a density of unity and depends on the electron temperature $T_\mathrm{e}$ and the abundance of heavy elements $Z$ in the plasma. The surface brightness $S_X$ can be described as the flux of photons within a certain energy band and per solid angle. Combining these definitions one can write
\begin{equation}
\label{eq:sbr1}
S_X^\mathrm{phot} = \frac{L}{E_\mathrm{mean} \Omega 4\pi D_\mathrm{L}^2} = \frac{\int \epsilon \, \mathrm{d}V}{E_\mathrm{mean} \Omega 4 \pi D_\mathrm{L}^2}~,
\end{equation}
where $\Omega = A/D_\mathrm{A}^2$ is the solid angle and $E_\mathrm{mean}$ is the emission weighted mean photon energy within an energy band,
\begin{equation}
E_\mathrm{mean} = h_\mathrm{p} \cdot \frac{\int \limits_{\nu_\mathrm{min}}^{\nu_\mathrm{max}} \nu \,\epsilon(\nu) \,\mathrm{d}\nu}{\int \limits_{\nu_\mathrm{min}}^{\nu_\mathrm{max}} \epsilon(\nu)\, \mathrm{d}\nu}~,
\label{eq:emean}
\end{equation}
where $h_\mathrm{p}$ is the Planck constant.

Inserting in Eq. \ref{eq:sbr1} the definition of the emissivity gives
\begin{equation}
S_X^\mathrm{phot} =  \frac{1}{E_\mathrm{mean}\,4\pi (1+z)^4} \int n_\mathrm{e}\, n_\mathrm{H}\,\Lambda(T_\mathrm{e},Z) \mathrm{d}l ~.
\label{eq:flux_density}
\end{equation}

Since the cooling function depends on the temperature which is variable across the cluster one cannot easily separate the $\Lambda$ from the integral. Usually one chooses an energy band where the dependence of the cooling function on the temperature is very small, so one can calculate the gas mass from the surface brightness. 

An analytic description derived from a King galaxy density model with isotropic velocity dispersion can be given for the gas density distribution,
\begin{eqnarray}
\rho_\mathrm{gas}(r) &=& \rho_\mathrm{gas}(0) \left( 1 + \frac{r^2}{r_c^2}  \right)^{-\frac{3}{2}\beta}~,\\
n_\mathrm{e}(r) &=& n_\mathrm{e}(0) \left( 1 + \frac{r^2}{r_c^2}  \right)^{-\frac{3}{2}\beta} 
\label{eq:gas_density}
\end{eqnarray}
The assumption entering here is an isothermal, ideal gas in hydrostatic equilibrium.
By inserting \ref{eq:gas_density} in \ref{eq:flux_density} one can simplify this to 
\begin{eqnarray}
S_X^\mathrm{phot}(r) &=& \frac{1}{E_\mathrm{mean} 4 \pi (1+z)^{4}} \int \zeta  n_\mathrm{e}^2(0) \left( 1 + \frac{r^2}{r_c^2}  \right)^{-3\beta} \Lambda \, \mathrm{d}l \label{eq:sx} \nonumber \\
&=& S_0 \left( 1 + \frac{r^2}{r_c^2} \right)^{-3\beta + 0.5}~, \label{eq:sxfinal}
\end{eqnarray}
where $\zeta$ is the ratio of Hydrogen and electron number densities.

For the double $\beta$ model the procedure is more complex and has been presented in \cite{hudson_what_2009}. While the two components of the surface brightness distribution are simply added up, for the density model the components are added in quadrature,
\begin{equation}
n_e(r) = \sqrt{  n_{\mathrm{e}1}^2 \left( 1 + \frac{r^2}{r_{c1}^2} \right)^{-3\beta_1} + n_{\mathrm{e}2}^2 \left( 1 + \frac{r^2}{r_{c2}^2} \right)^{-3\beta_2} }~,
\label{eq:double_density}
\end{equation}
where $n_{\mathrm{e}i}$ are the central electron densities of the two components.

\begin{figure*}
	\centering
	\resizebox{1.\hsize}{!}{\includegraphics[width=0.98\textwidth]{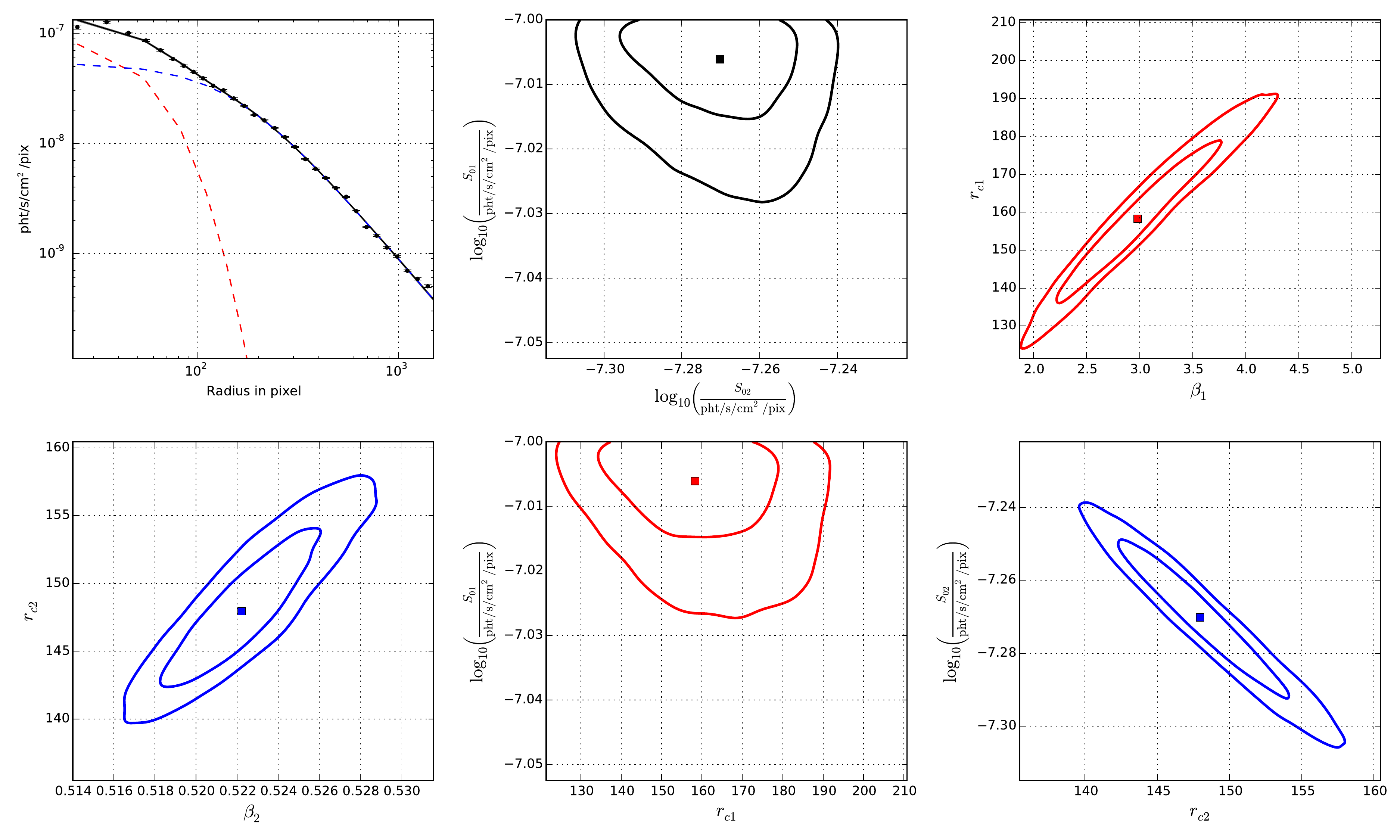}}
	\caption[Surface brightness profile of A85]{Surface brightness profile (excluding substructure) and constraints on double $\beta$ model parameters for A85. Red corresponds to the inner component, blue to the outer one.}
	\label{fig:sbr_constraints}
\end{figure*}
\begin{figure*}
	\centering
	\resizebox{0.49\hsize}{!}{\includegraphics[height=149px]{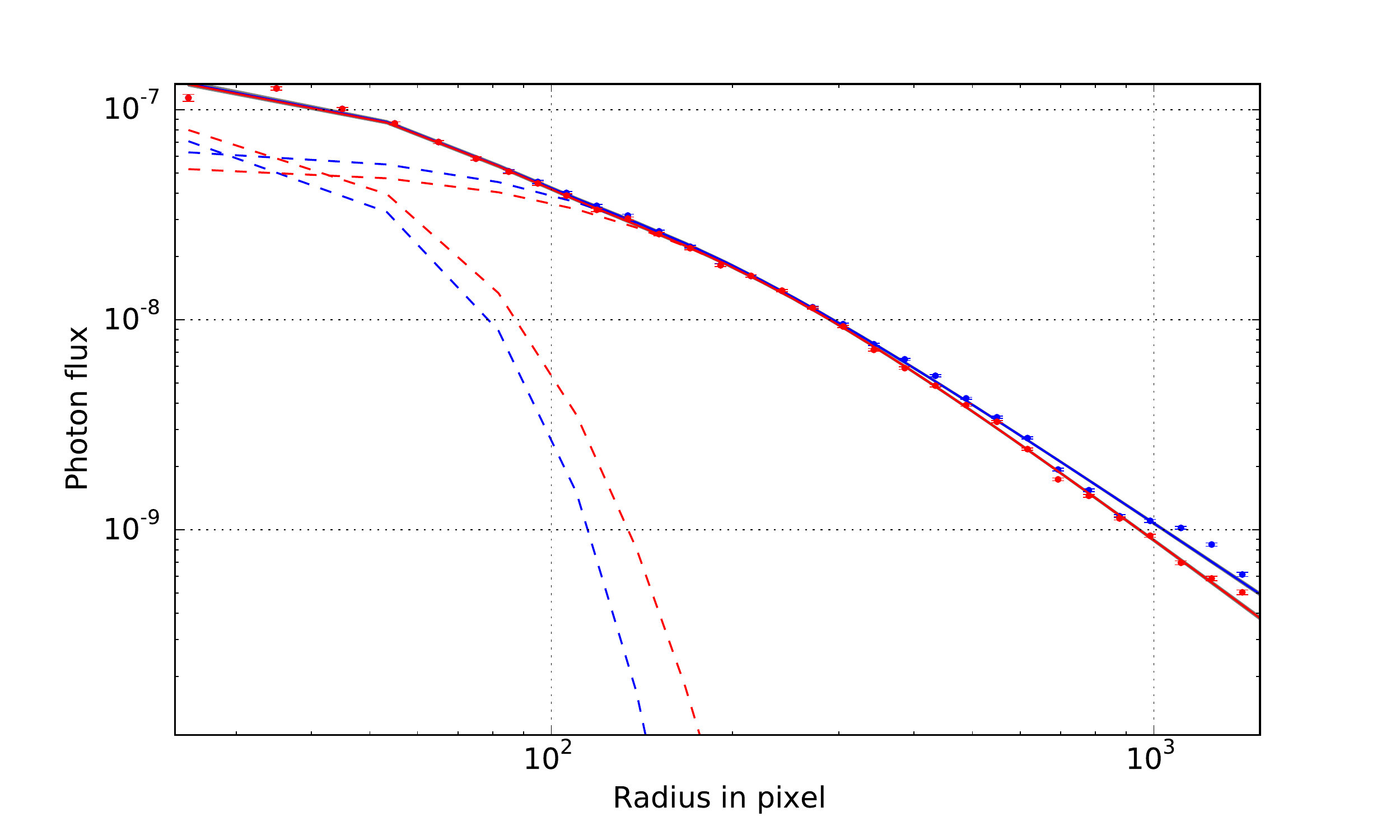}}
	\resizebox{0.49\hsize}{!}{\includegraphics[height=149px]{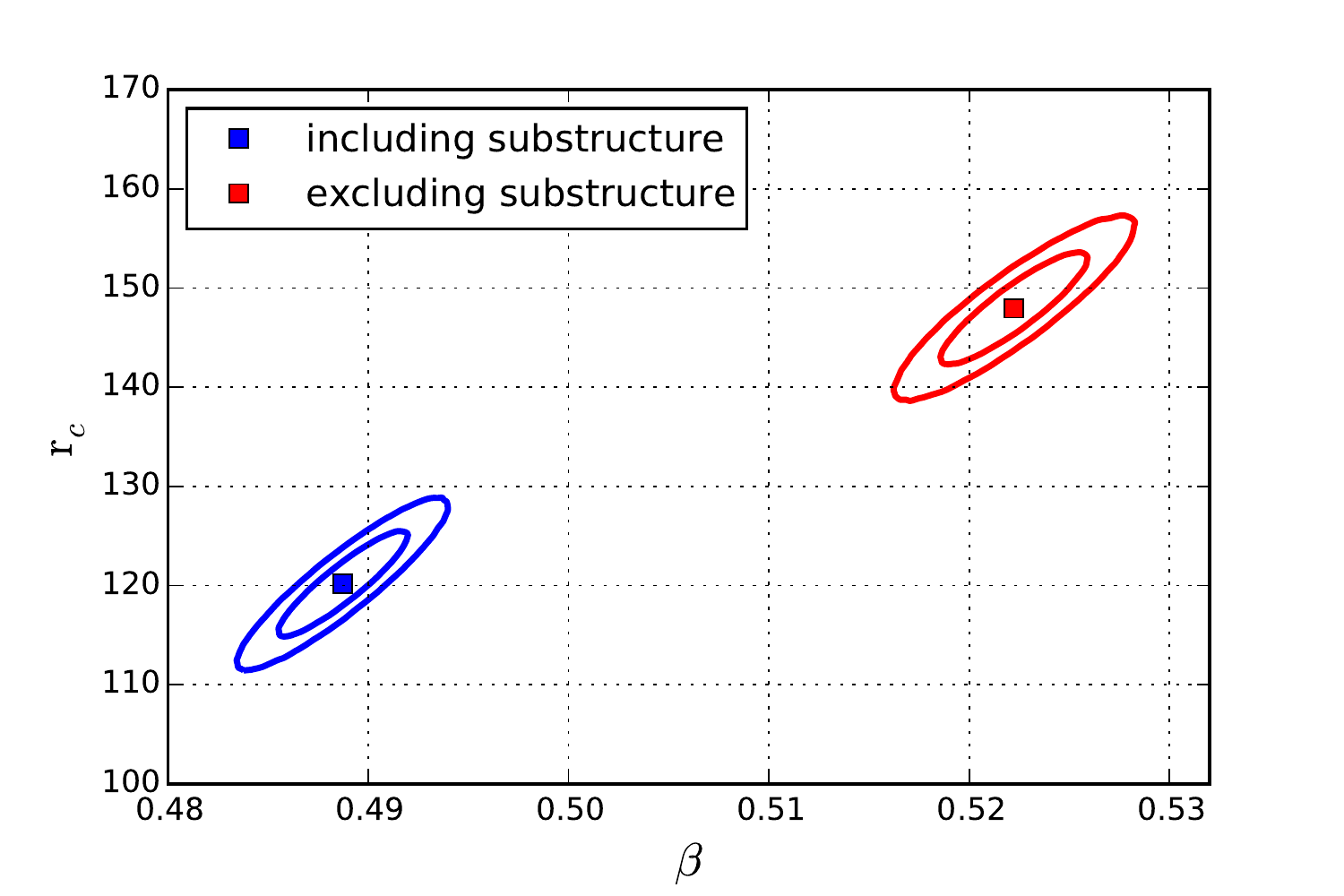}}
	\caption{Impact on the surface brightness profile for A85 when including (blue) or excluding (red) the substructure. Point sources are excluded in both cases. 1 pixel corresponds to $\SI{0.492}{\arcsec}$. \textit{Left:} Surface brightness profile in $\si{pht\,s^{-1}\,cm^{-2}\,pix^{-1}}$, with the two components of the double $\beta$ model indicated (dashed). \textit{Right:} Constraints (68\% and 95\%) on $\beta$ and the core radius (in pixel) for the outermost component of the double $\beta$ model when including (blue) or excluding (red) the substructure in the fit.}
	\label{fig:sbr_compare}
\end{figure*}

As it can be seen in Fig. \ref{fig:sbr_constraints} (top left) the surface brightness profile is well fit by a double $\beta$ model. Every radial bin in the surface brightness profile is usually calculated from several 1000 counts, only in very short observations of faint clusters from at least 70 counts, so the Gaussian probability distribution is a good approximation. The other graphs in Fig. \ref{fig:sbr_constraints} show that there exists a strong degeneracy between some double $\beta$ model parameters. It is very important to take this degeneracy into account when calculating the total or gas mass from the surface brightness model, otherwise the uncertainties will be overestimated.

The impact of the substructure removal procedure is illustrated in Fig. \ref{fig:sbr_compare} for the case of A85. Especially the outer component of the model is affected by the substructure. If one would include substructure in the analysis, the total mass will be underestimated (if the temperature is not be affected).

\subsubsection{Cooling Function}
\label{ch:coolingfunction}
\begin{figure*}
	\centering
	\includegraphics[width=225px]{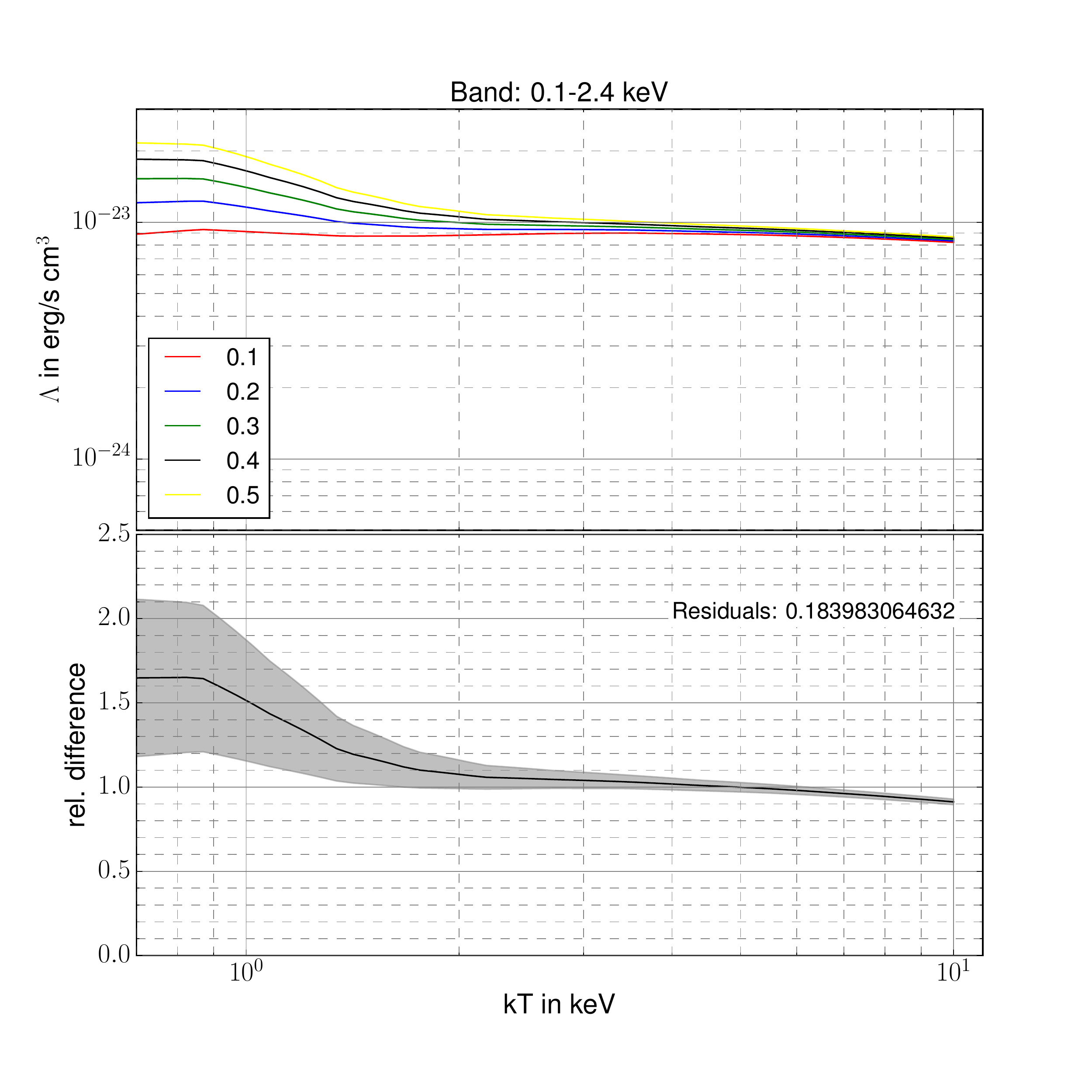}
	\includegraphics[width=225px]{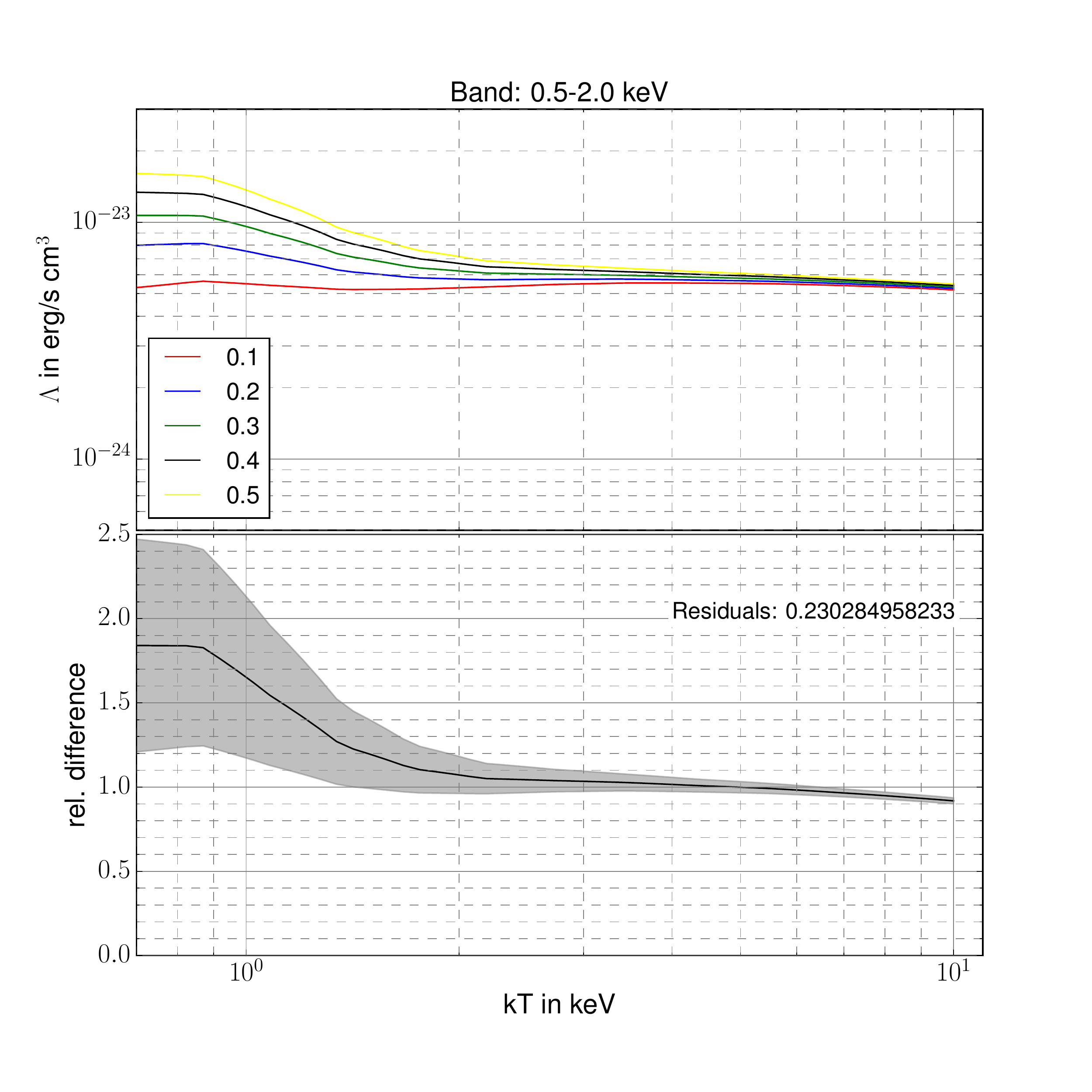}
	\includegraphics[width=225px]{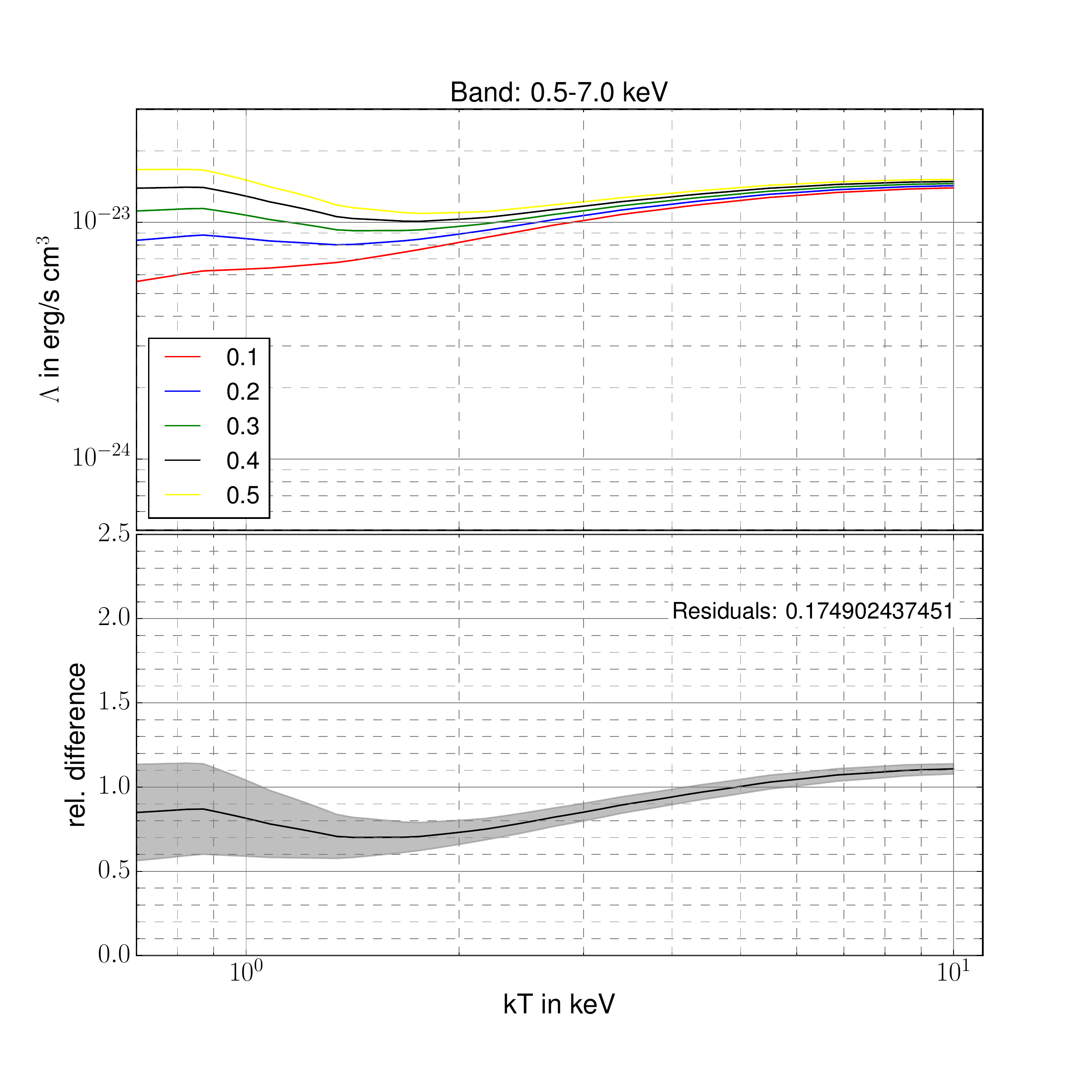}
	\includegraphics[width=225px]{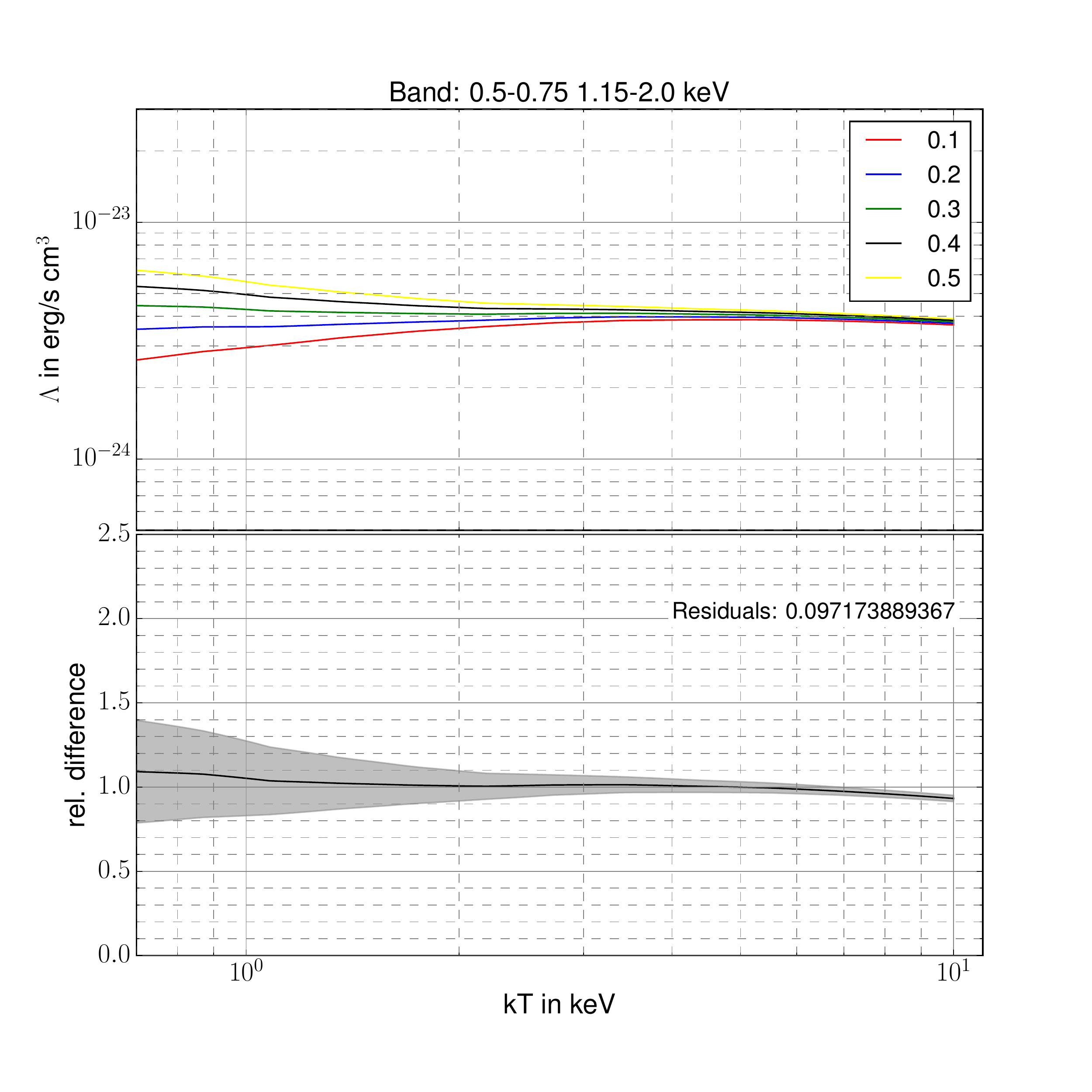}
	\caption[Cooling function in different energy bands]{Cooling function in different energy bands for five relative abundances. Each lower panel shows the relative difference with respect to the value at $\SI{5}{keV}$. In the lower right panel the cooling function in the optimal panel is plotted (see text for details).}
	\label{fig:coolingfunc}
\end{figure*}

As it has been indicated in the previous Section, one assumption to convert a surface brightness profile into a density profile (from Eq. \ref{eq:sx} to Eq. \ref{eq:sxfinal}) is that the cooling function does not depend on the radius, i.e. the cluster temperature or relative abundance of heavy elements is also constant with radius.

An appropriate energy band is defined by the detector sensitivity and the emission characteristics of the source. Very common for X-ray instruments are the following bands: 
\begin{itemize}
	\item $\SIrange{0.5}{2.0}{keV}$ -- Chandra and XMM-Newton have a very high effective area in this band so it is often used for these instruments.  
	\item $\SIrange{0.1}{2.4}{keV}$ -- This band was used for the ROSAT satellite, which was sensitive to softer energies. It is still used because many galaxy cluster catalogs are selected from the ROSAT All-Sky Survey.
\end{itemize}
It turns out that for plasma temperatures $> \SI{2}{keV}$ the cooling function in the energy bands quoted above is almost constant with temperature/abundance, which can also be seen in the upper panels of Fig. \ref{fig:coolingfunc}. 

The Fe-L line complex around $\SI{1}{keV}$ is very temperature and abundance sensitive. A combination of small bands below and above this feature turns out to be very insensitive to the plasma temperature even at temperatures below $\SI{1}{keV}$. As it can be seen in the bottom right panel of Fig. \ref{fig:coolingfunc} for a typical heavy element abundance of 0.3 the relative change in the cooling function is below 10\% over the range of temperatures from $\SI{0.7}{keV}$ to $\SI{10}{keV}$. In the same temperature range the relative change of $\Lambda$ in the $\SIrange{0.1}{2.4}{keV}$ band is 4 times larger. 
The optimal energy band was determined by calculating the residuals of the cooling function to a flat distribution over $\SIrange{0.7}{10}{keV}$ temperature range and a typical abundance of $Z=0.3 Z_\odot$. 
The variable parameters are the lower and upper boundaries of the energy band, while also a gap in the middle was allowed (as it happens for the optimal band).
These residuals are minimized and at the same time the average value of the cooling function maximized (to get more counts and better statistics later on for the surface brightness). The resulting energy band was found to be the combination of the 
$\SIrange{0.5}{0.75}{keV}$ band plus the $\SIrange{1.15}{2.0}{keV}$ band. 

\subsubsection{Gas mass}
\label{ch:gasmass}
The gas mass $M_\mathrm{gas}$ of the galaxy cluster is obtained by simply integrating the electron density profile Eq. \ref{eq:double_density} until $r_{500}$ (which is iteratively calculated when deriving the total mass $M_\mathrm{tot}$ in Section \ref{ch:he}),
\begin{equation}
M_\mathrm{gas} = 4\pi \xi \int \limits_0^{r_{500}}  n_e(r) \,r^2 \,\mathrm{d}r,
\label{eq:mgas}
\end{equation}
where $\xi$ is the ratio of the gas density $\rho_\mathrm{gas}$ and electron density $n_e$ assuming a constant abundance of heavy elements:
\begin{equation}
\xi = m_\mathrm{p} \frac{1 + 2 \left( \frac{n_\mathrm{e}}{n_\mathrm{H}} -1 \right)}{\frac{n_\mathrm{e}}{n_\mathrm{H}}} = m_\mathrm{p} \left(2 - \frac{n_\mathrm{e}}{n_\mathrm{H}} \right)~,
\end{equation}  
where $m_\mathrm{p}$ is the proton mass and assuming a neutron to proton ratio of 1 in atoms (except Hydrogen), which is correct for most elements that are considered here (i.e. included in the abundance tables). Note that for heavier elements the number of neutrons increases. The ratio of electrons per Hydrogen atom is directly calculated from the abundance of the individual elements $A_i$,
\begin{equation}
\frac{n_\mathrm{e}}{n_\mathrm{H}} = 1 + 2\cdot 0.083 + \sum \limits_{i=3} A_i \cdot i~.
\end{equation}
$\num{0.083}$ is the Helium atom abundance. For the analysis we use the \cite{2009ARA&A..47..481A} solar abundance table and assume that all elements heavier than Helium have 0.3 solar abundances, which corresponds to $\xi \approx 1.144 \cdot m_\mathrm{p}$. The difference in $\xi$ for 0.2 and 0.4 solar abundances is around 0.1\%. Another quantity derived from the abundance of heavy elements which is of special interest for the hydrostatic mass, is the mean molecular weight $\mu$, which gives the average mass of a particle in $m_\mathrm{p}$ mass units,
\begin{equation}
\mu = \frac{\rho}{ (n_\mathrm{e} + n_\mathrm{ion}) m_\mathrm{p} } = \frac{2 \frac{n_\mathrm{e}}{n_\mathrm{H}}-1}{\frac{n_\mathrm{e}}{n_\mathrm{H}} + \sum \limits_i A_i } \approx 0.59 ~.	
\label{eq:mu}
\end{equation}

The parameters of the surface brightness profile are calculated by a Markov Chain Monte Carlo simulation with $\num{200000}$ samples (out of which the first 50\% are ignored), which also determine the uncertainty for the gas mass. The central value of the electron density is obtained from the spectrum of the innermost region outside $\SI{50}{kpc}$ in order to be clearly outside the possible emission of the BCG. This region was never distributed on more than one ACIS chip. 

The gas mass is of particular interest for cosmology e.g., for $f_\mathrm{gas}$ test or as a calibrator for the total mass. 
The gas mass fraction $f_\mathrm{gas} = \frac{M_\mathrm{gas}}{M_\mathrm{tot}}$ can constrain the matter density of the Universe $\Omega_\mathrm{m}$ assuming a Baryon density $\Omega_\mathrm{b}$ (e.g., from CMB measurements or Big Bang Nucleosynthesis data). This has been shown initially by \cite{1993Natur.366..429W} and more recently, e.g., by \cite{2014MNRAS.440.2077M}. 
Note that the uncertainties of the total and gas mass, which both enter in $f_\mathrm{gas}$, are both strongly dependent on the uncertainty of the radius. We take this into account when calculating $f_\mathrm{gas}$.
The gas mass can also be used as a tracer for the total mass which experiences low scatter (\citealp{2008A&A...482..451Z,2009ApJ...692.1033V}). 

Note that with a recent patch version 12.9.0o for \verb|Xspec|\footnote{\url{https://heasarc.gsfc.nasa.gov/docs/xanadu/xspec/issues/issues.html}} a problem concerning the calculation of the normalization of the \verb|apec| model was solved. This will affect all our gas masses, so we recalculated the normalizations of the innermost annuli using the updated \verb|Xspec| version to calibrate the density profile. Since the total mass depends only on the density gradient, this normalization bias has no influence on those results. 

\subsection{Temperature}
The temperature of galaxy clusters is crucial for the hydrostatic mass determination. Not only the value of the temperature is important but also the slope of the profile at the radius of interest enters, which makes an appropriate parametrization essential. We start by describing how we extract and model the spectra of the Chandra observations.
\subsubsection{Regions}
We create a counts image (OBS) and a background counts image (BKG) from the exposure corrected blank-sky background file in the $\SIrange{0.7}{7}{keV}$ band for every observations of all the clusters. Both images have point source and substructure removed. Spherical rings around the emission weighted cluster center (taken from \citealp{zhang_hiflugcs:_2010}) are created based on a signal to noise threshold $\mathrm{S/N}$,
\begin{equation}
\mathrm{S/N} = \frac{\mathrm{OBS}_i - \mathrm{BKG}_i }{\sqrt{\mathrm{OBS}_i}} ~,
\end{equation}
where $i$ denotes a certain region. 
The minimum S/N threshold was set to 50 for all clusters, except for A3581, S1101 to 30 and for RXCJ1504, A400 and A1795 to 40, in order to get enough regions. For the bright cluster A2052 the threshold was set to 70. 
For a fixed aperture bright clusters would give very high S/N values, or in turn make the central regions very tiny. A high number of regions in the center would then give too much weight for the profile fit on the cluster center, which is not the primary target in this study. To avoid this we set a minimum size for a region of $\SI{25}{arcsec}$.
With this setup, on average 30 regions are extracted per cluster, 17 regions on average per observation. The minimum number of regions per cluster is 8 (A1736 and EXO0422). The average maximum extraction region is $\SI{12.1}{arcmin}$ or $\SI{670}{kpc}$, or compared to the $r_{500}$ determined later, it is 66\% on average. This means that most cluster profiles have to be extrapolated to calculate the $M_\mathrm{tot, 500}$. 

\subsubsection{Spectral modeling}
The $\SIrange{0.7}{7}{keV}$ energy band used in this study makes up 94\% of the total effective area. For galaxy cluster analyses this energy band is an appropriate choice since at energies below $\SI{0.5}{keV}$ the calibration is uncertain and between 0.5 and 0.7 $\si{keV}$ the effects of the uncertain Galactic absorption or solar wind charge exchange (SWCX) lines might bias temperature estimates. Also at energies above $\SI{7}{keV}$ the particle background becomes more and more dominant so the signal to noise ratio will decrease.

To create the spectra and response files the \verb|specextract| task from the CIAO 4.6 software package was used. Spectra were grouped to have at least 30 counts per bin. 

For the spectral fitting the astrophysical background components are determined from a simultaneous fit to data from the ROSAT All-Sky survey\footnote{\url{http://heasarc.gsfc.nasa.gov/cgi-bin/Tools/xraybg/xraybg.pl}} (\citealp{snowden_rass}). The extraction region for the ROSAT All-Sky Survey (RASS) data is an annulus from 0.7 to 1 degree around the cluster center (for NGC4636, NGC1399 and A3526 $r_{500}$ is larger than 0.7 degree, so the RASS data was extracted from 1.5 to 2 degree). The particle background was directly subtracted from the Chandra spectra using the stow events files from an epoch close to the observation date. The stow events files are created when the ACIS detector is in a position where it is not exposed to the sky and the HRC-I camera is in the field of view. This configuration is also called event histogram mode (EHM). As shown by comparisons to dark moon observations only particle events are recorded in the stow position (\citealp{2003ApJ...583...70M,2004ApJ...607..596W}). For each annulus the same detector region was used to extract the particle background spectra. These background spectra are normalized by the ratio of the $\SIrange{9.5}{12}{keV}$ band count rate of the observation and the stow events file to account for variations of the quiescent particle background component. The cluster emission is modeled by an absorbed thermal model (\verb|phabs*apec|), where all parameters apart from the redshift and the $N_\mathrm{H}$ are left free to vary.
Following \cite{2013MNRAS.tmp..859W} the Hydrogen columns density used as a tracer for the X-ray absorption,
\begin{equation}
N_\mathrm{H\, tot} = N_\mathrm{HI} + 2\cdot  N_{\mathrm{H}_2\mathrm{m}} \cdot \left[ 1-\exp \left(-N_\mathrm{HI}\cdot \frac{ E\left(B-V\right) }{ N_\mathrm{c}} \right) \right]^\alpha~,
\end{equation}
where the parameters $N_{\mathrm{H}_2\mathrm{m}} = \SI{7.2(3)e20}{cm^{-2}}$, $N_\mathrm{c} = \SI{3.0(3)e20}{cm^{-2}}$ and $\alpha = \num{1.1(1)}$ were calibrated using X-ray afterglows of Gamma ray bursts in the aforementioned reference. Both, the absorption $E(B-V)$ from the IRAS and COBE/DIRBE infrared dust maps (\citealp{1998ApJ...500..525S}) and the $N_\mathrm{HI}$ from \cite{2005A&A...440..775K} are computed at each cluster position. 
The combined effect of the uncertainties of these parameters, the scatter of this scaling relation (0.087) plus accounting for a 10\% uncertainty on $N_\mathrm{HI}$ and $E(B-V)$ has only an 11\% effect on $N_\mathrm{H\, tot}$, which typically affects best fit temperatures by 1\%. Since this is much smaller than the typical statistical uncertainties, any statistical uncertainty of $N_\mathrm{H\, tot}$ is neglected. 
For the relative abundance of heavy elements the \cite{2009ARA&A..47..481A} abundance table was used. 
For each observation all spectra from the different regions and chips\footnote{The ACIS-I chips (0-3) are grouped together into a single spectrum.} are fit simultaneously. The temperature and metallicity of spectra from the same region but different chips are linked together, while the normalizations are not because of spatial variations of the density distribution. For all observations the \verb|steppar| command was run on the temperatures. This task calculates the $\chi^2$ for the parameter within a given range of values in order not to get best fit parameters of a local minimum of the likelihood distribution. The reduced $\chi^2$ of all spectral fits was on average 1.03, while in 95\% of the cases it was below 1.17. This gives a hint that the spectral modeling is appropriate.

\subsubsection{Parametrization}
For most clusters the signal to noise threshold results in the largest region being smaller than $r_{500}$, so an extrapolation for the temperature and surface brightness, or mass profile has to be performed. We use several parametrizations to robustly model different types of temperature profiles, e.g., where a cool core as well as the decreasing behavior to the outskirts is present. One set of models which is used in this work is given in \cite{2007ApJ...669..158G}:

The first model is the connection of two powerlaws,
\begin{eqnarray}
\label{eq:model1}
T(r) &=& \left( \frac{1}{t_1(r)^{s}} + \frac{1}{t_2(r)^{s}} \right)^{-\frac{1}{s}}~,\\
t_i(r) &=& T_{i, 100} \left( \frac{r}{\SI{100}{kpc}} \right)^{p_i}~, ~ i =1, 2~.\nonumber
\end{eqnarray}
This model has 5 free parameters and can be used for low quality data or when the measurements are not reaching outer regions of the cluster (e.g, due to a very low redshift and the limited FOV). It was applied to three cluster profiles (A1795, NGC5044 and NGC4636). 

The second model is a composition of two powerlaws smoothed by an exponential function,
\begin{eqnarray}
\label{eq:model2}
T(r) &=& T_0 + t_1(r) e^{-(r/r_p)^\gamma} + t_2(r) \left( 1 - e^{-(r/r_p)^\gamma}  \right)~, \\
t_i(r) &=& T_i \cdot \left( \frac{r}{r_0} \right)^{p_i}~, i=1,2 ~.\nonumber
\end{eqnarray}
This model has 8 free parameters and can be applied to most of the clusters due to its many degrees of freedom. It was used for 46 clusters, which have a median of 19 independent temperature bins per cluster. 

\begin{figure*}
	\centering
	\includegraphics[width=220px]{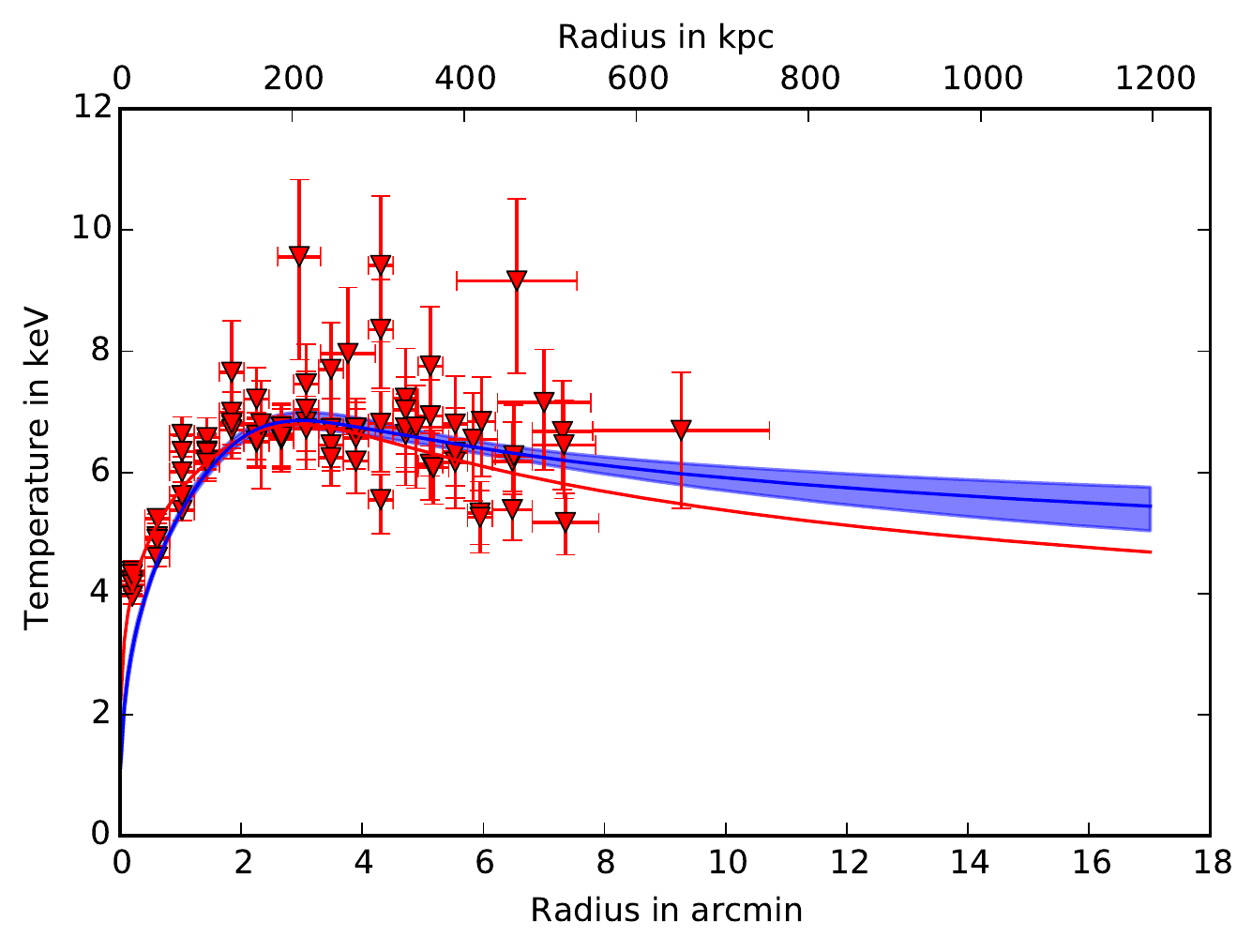}
	\includegraphics[width=220px]{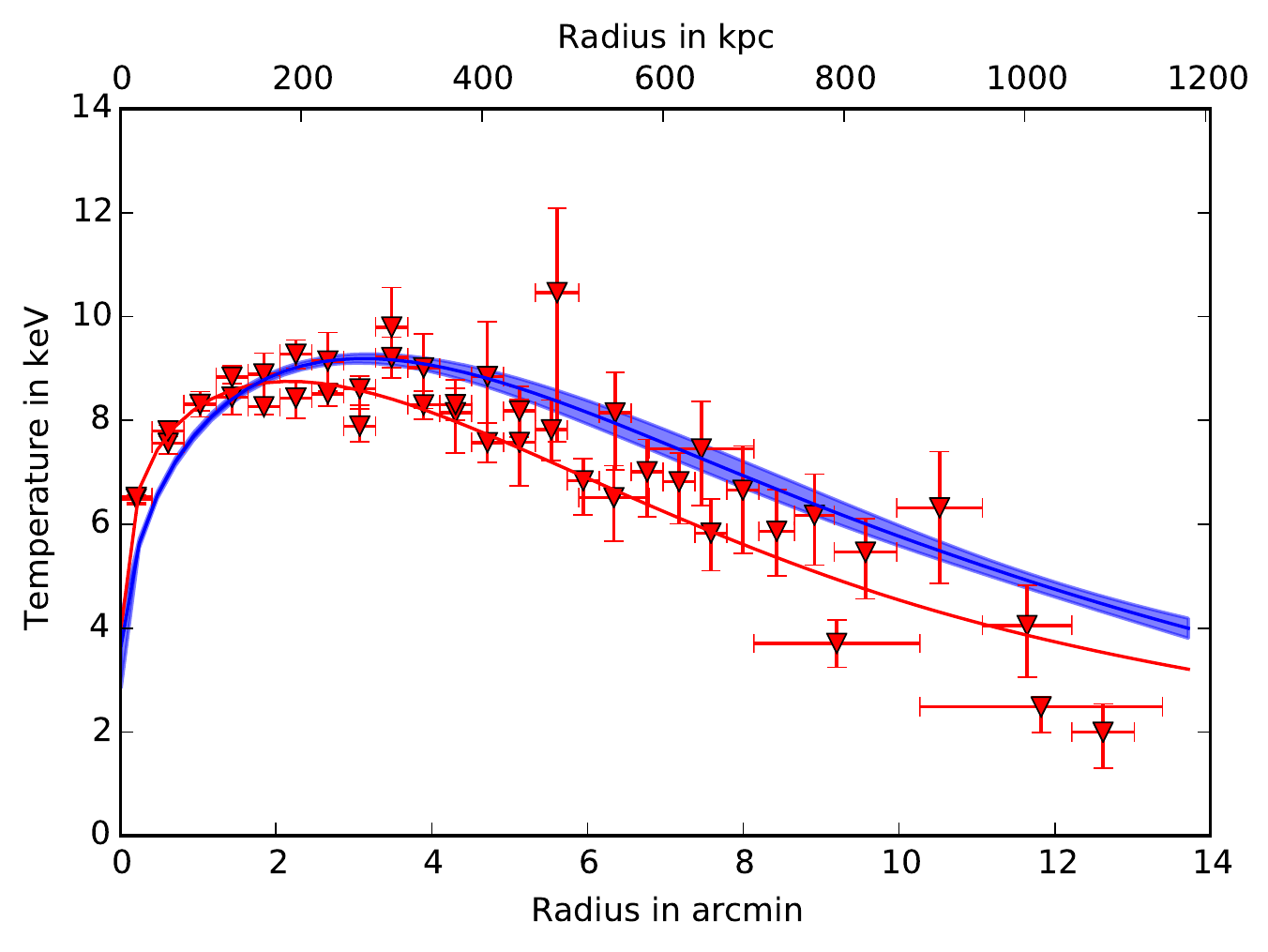}
	\includegraphics[width=220px]{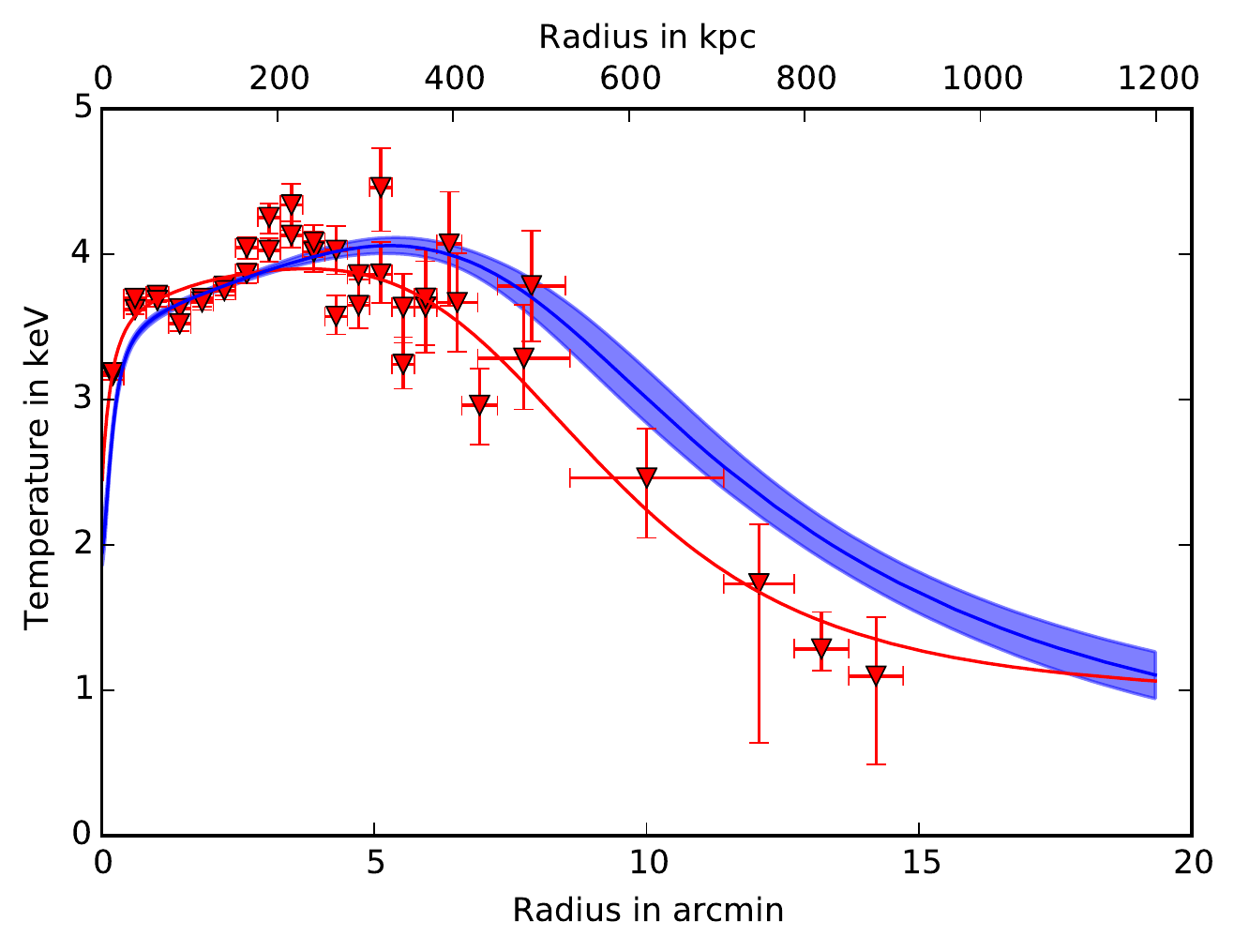}
	\caption[Temperature profile for A1795, A2029, HydraA]{Temperature profile for A1795 (top left, model \ref{eq:model1} with 5 free parameters), A2029 (top right, model \ref{eq:model2} with 8 free parameters) and HydraA (bottom, model \ref{eq:model3} with 10 free parameters). Red datapoints mark the observations, red lines the projected model fit and dark blue lines with the blue shaded area the deprojected profile (see Section \ref{ch:deprojection}) and its uncertainty.}
	\label{fig:tmodels}
\end{figure*}

The third model is a combination of the \cite{2001MNRAS.328L..37A} rising profile and an a falling profile.
\begin{eqnarray}
\label{eq:model3}
T(r) &=& t_1(r) e^{-(r/r_p)^\gamma} + t_2(r) \left( 1-e^{-(r/r_p)^\gamma } \right) ~, \\
t_1(r) &=& a + T_1 \left( \frac{(r/r_1)^{p_1}}{1+(r/r_1)^{p_1}} \right)~, \nonumber \\
t_2(r) &=& b + T_2 \left( \frac{(r/r_2)^{p_2}}{1+(r/r_2)^{p_2}} \right)~. \nonumber 
\end{eqnarray}
This model has 10 free parameters and is only applied to 15 clusters (with a median of 37 temperature bins per cluster). The three different models are shown on specific examples in Figure \ref{fig:tmodels}.

\subsubsection{Deprojection}
\label{ch:deprojection}
The measured spectra contain information from all emitting sources along the line of sight. While the foreground and background components (e.g., particle background, Galactic emission, unresolved AGNs) are either removed or modeled, the cluster emission within an annulus at an apparent radius from the cluster center is summed into the measured spectrum. 

Assuming a spherical, symmetric cluster and that the outermost projected annulus is not significantly contaminated by emission from outer cluster shells, one can start to remove or account for this emission when fitting the next inner annulus. This is usually referred as the onion-peeling-technique (\citealp{1981ApJ...248...47F,2000ApJ...532L.113B,2007ApJ...669..158G}).
For high quality data, this approach can lead to oscillations in the deprojected temperature profile (\citealp{2008MNRAS.390.1207R}).

\cite{2006ApJ...640..710V} developed an averaging code that takes several plasma components (with each temperature, metallicity and emission measure) and calculates the corresponding single temperature that would be measured within a certain energy band for either Chandra ACIS, XMM-Newton EPIC or ASCA SIS/GIS. The emission measure, EM, in a cluster shell can be easily  calculated from the density model, 
\begin{equation}
\mathrm{EM} = \int_V n_\mathrm{e} n_\mathrm{H}\, \mathrm{d}V~,
\end{equation}
where $n_\mathrm{e}$ and $n_\mathrm{H}$ are the electron and Hydrogen number densities, respectively, and $V$ is the emitting volume.
\begin{figure}
	\centering
	\includegraphics[width=220px]{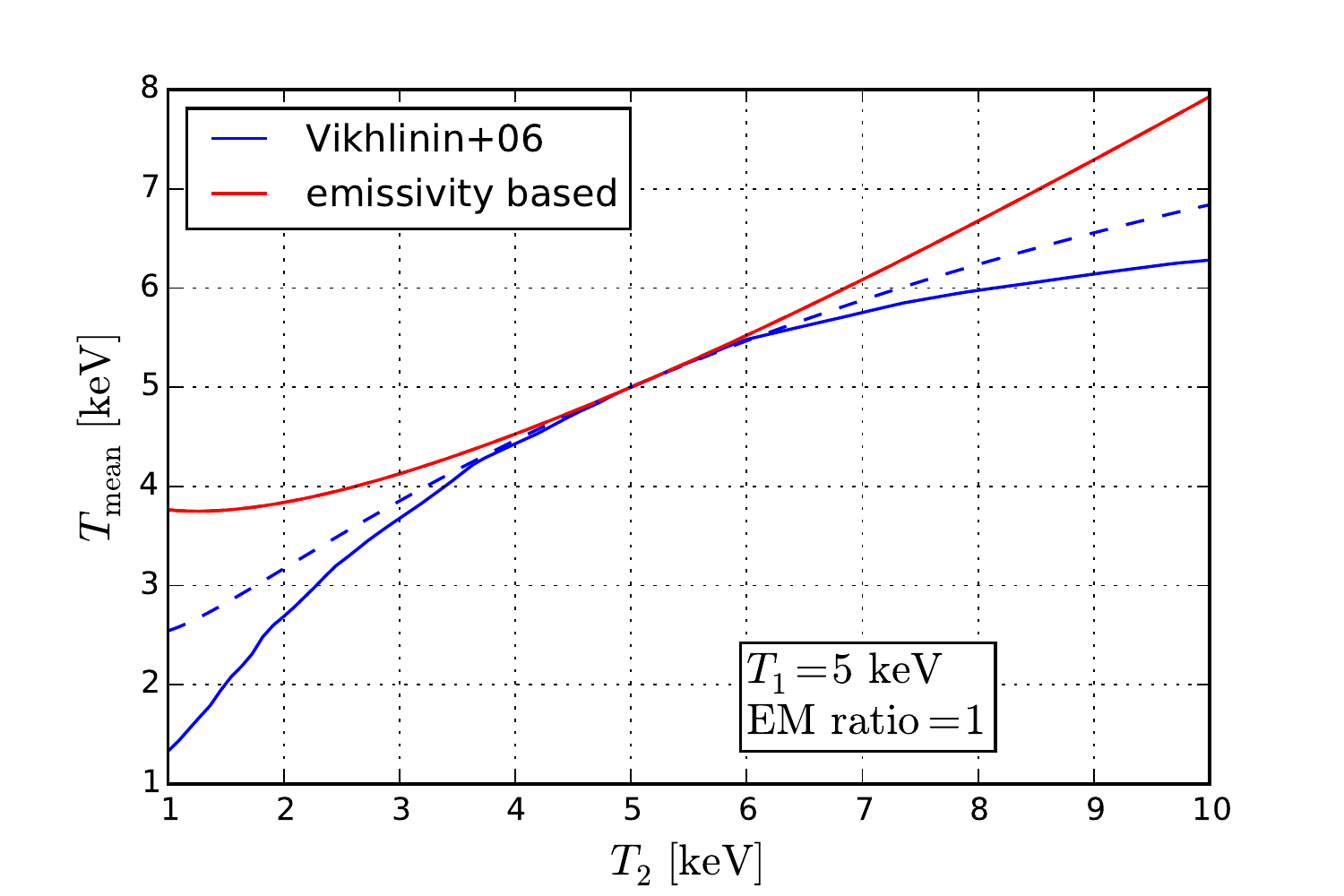}
	\caption{Comparison of the emissivity weighted temperature averaging, and the method used for the deprojection here assuming a two component plasma (with equal emission measure). The metallicity is 1 (solid) and 0.001 (dashed).}
	\label{fig:Tmean}
\end{figure}
The difference to a simple emissivity weighted averaging can be seen in Fig. \ref{fig:Tmean}: For a two component plasma with $T_1 = \SI{5}{keV}$ and an EM ratio of 1 the emissivity weighting method overpredicts (with respect to the \citealp{2006ApJ...640..710V} method, which we assume here to be the expected reference curve) the ``projected'' temperature not only due to the missing line emission (see dashed line for $Z\approx 0$) but also because of the instrumental sensitivity, which for ACIS is highest between 1 and 2 keV.
Other effects like absorption, redshift, the energy band and the redistribution (RMF), which are also taken into account by the \cite{2006ApJ...640..710V} method, play a minor role.

We use the \cite{2006ApJ...640..710V} method to deproject all cluster temperature profiles.
In practice the parameters for the deprojected temperature profile are determined using a Markov Chain Monte Carlo with an adaptive Metropolis-Hastings algorithm as implemented in PyMC (\citealp{2015ascl.soft06005F}). In each step the input temperature profile is projected along the line of sight at every radius that has a measurement. Each projected and measured temperature (of all observations used for the cluster) are then compared assuming a Gaussian probability distribution of the measured uncertainties. After the first 50\% of the samples are removed (``burn-in'', samples that are needed to achieve convergence) each cluster still has $\num{200000}$ samples, which are used to characterize the deprojected temperature profile and its uncertainty.

\subsection{Hydrostatic equilibrium}
\label{ch:he}
The total gravitating mass of a galaxy cluster is the most important parameter in a cosmological study. Assuming that the ICM is in hydrostatic equilibrium, the total gravitational potential determines temperature and density of the gas, which results in the hydrostatic mass equation,
\begin{equation}
\label{eq:hydro5}
M(<r) = -\frac{r k_\mathrm{B}T}{G m_\mathrm{p}\mu} \left( \frac{\mathrm{d}\ln \rho}{\mathrm{d}\ln r} + \frac{\mathrm{d}\ln T}{\mathrm{d}\ln r} \right)~,
\end{equation}
where $k_\mathrm{B}$ is the Boltzmann constant, $m_\mathrm{p}$ the proton mass, $\mu$ the mean molecular weight, and $G$ the gravitational constant.
The assumptions that enter here are that gravity is the only external force (e.g., no magnetic fields), the cluster is spherically symmetric and the pressure behaves as in an ideal gas. Also $\mu$ must not depend on the radius $r$. Outside the core of galaxy clusters and excluding strong merging events these assumptions are generally fulfilled. 

Using Eq. \ref{eq:hydro5} the mass can be calculated within any radius. In order to have a comparable quantity one usually defines $M_\Delta$, which is the total mass within a sphere of radius $R_\Delta$, inside which the mean density is $\Delta$ times the critical density of the Universe at cluster redshift $\rho_\mathrm{crit}(z)$ (see e.g., \citealp{Bocquet2015a}). $\Delta$ is also called the overdensity and typical values are $500$ or $200$, sometimes also $2500$. Note that with increasing values of $\Delta$ the radius decreases. In the following the cluster masses will be compared to the prediction of the \cite{2008ApJ...688..709T} halo mass function. Therein the halo mass function is given for 9 different overdensities in the range between 200 and 3200 with respect to the mean density of the Universe $\rho_\mathrm{mean}(z)$. The conversion between the critical and mean density is given by
\begin{equation}
\rho_\mathrm{mean} (z) = \Omega_\mathrm{m}(z) \cdot \rho_\mathrm{crit}(z) = \rho_\mathrm{mean}(0) \cdot (1+z)^3~. 
\end{equation}
This means that $\rho_\mathrm{mean}$ has to be recalculated each time a new $\Omega_\mathrm{m}$ is being tested for a fixed overdensity of $\Delta_{500c}$.
For all parameters of the \cite{2008ApJ...688..709T} halo mass function, which depend on the overdensity, second derivatives are provided that a spline interpolation (\citealp{1992nrca.book.....P}) can be performed. For typical values of $\Omega_\mathrm{m}$ and low redshifts the interpolation allows to calculate values up to 1000 times the critical density, which means that even with the highest overdensity given for the halo mass function many HIFLUGCS masses still need to be extrapolated.
Note that with increasing overdensity deviations from the universality of the mass function increase.

\subsection{HIFLUGCS luminosities}
Many studies show that there is a correlation between the X-ray luminosity $L_x$ and the total mass of a galaxy cluster that can be described by a powerlaw (e.g., \citealp{reiprich_hiflugcs,2009A&A...498..361P,2009ApJ...692.1033V,2010MNRAS.406.1773M,zhang_hiflugcs:_2010,2011A&A...535A...4R}). Also other quantities like the temperature of the gas or its mass show a correlation with the total mass and can be used as tracers for the total mass. Here we will focus on the luminosity, since it is the most simple quantity that can be derived for any galaxy cluster with known redshift. Especially for the all-sky survey by eROSITA (\citealp{2012arXiv1209.3114M}) most of the expected $\num{100000}$ galaxy clusters will only have a luminosity (with the redshift coming from optical follow-up observations). So this quantity is crucial and it needs to be understood in detail. 
The relation between the cluster mass and its luminosity can also be predicted from simple self-similar relations (\cite{1999MNRAS.305..631A,reiprich_hiflugcs,2009A&A...498..361P}):
Following \cite{1999MNRAS.305..631A} the luminosity can be written as
\begin{equation}
\label{eq:selfsimilar1}
L(T) =  f_\mathrm{gas}^2(T) \cdot M_\mathrm{tot}(T)  \cdot \frac{<\rho_\mathrm{gas}^2>}{<\rho_\mathrm{gas}>^2} \cdot \Lambda(T)~,
\end{equation}
where $<>$ denotes the volume average.
Using virial equilibrium, $2E_\mathrm{kin} = - E_\mathrm{pot}$, and the isothermal $\beta$ model, $\beta = \frac{\mu m_\mathrm{p} \sigma^2}{k_\mathrm{B} T}$, one gets $M \propto T\,R^{-1}$, where $R$ is the virial radius. Since $R\propto M^{\frac{1}{3}}$ one concludes
\begin{equation}
M \propto T^{\frac{3}{2}}~.
\end{equation}
From the free-free emission it follows that $\Lambda_\mathrm{bolo} \propto T^{\frac{1}{2}}$, while in the ROSAT band the cooling function does not depend strongly on $T$ for $T > \SI{2}{keV}$. For the $0.5-2.0\,\mathrm{keV}$ band a similar behavior as in the ROSAT band is expected. If the gas mass fraction in Eq. \ref{eq:selfsimilar1} is constant and the fraction $\frac{<\rho_\mathrm{gas}^2>}{<\rho_\mathrm{gas}>^2}$, which describes internal substructure or clumping, is assumed to be independent of temperature, one immediately can conclude
\begin{eqnarray}
L_\mathrm{bolo} &\propto& M^{\frac{4}{3}}~, \\
L_\mathrm{0.1-2.4 \mathrm{keV}} &\propto& M
\end{eqnarray}

As indicated in \cite{2009A&A...498..361P} observed slopes of the $L_\mathrm{bolo}-M$ as well as the $L_{0.1-2.4 \mathrm{keV}}-M$ relation (1.8 for the bolometric case, 1.5 for the $0.1-2.4\,$keV case) are much steeper than expected from the self-similar prediction. 

Furthermore, the luminosities are important for the cosmological analysis, since the cluster masses from the mass function need to have assigned a flux in order to follow the selection function (flux limit). A way to do this is to use a scaling relation. In \cite{2010MNRAS.406.1773M} it is demonstrated how the selection effects can bias the slope and normalization of the $L_x - M$ relation of a luminosity limited sample: Due to the intrinsic scatter fainter objects get detected with a lower probability, which makes the observed $L_x-M$ relation shallower. The prediction on how the $L_x-M$ relation of a flux limited sample would be affected is not trivial. 

Several remarks have to be made on the luminosities that were used in the end (\citealp{reiprich_hiflugcs}):

\begin{itemize}
	\item The cosmological analysis requires consistency between the selection of objects and the assigned luminosities in the $L_x -M$ relation. The HIFLUGCS clusters were selected based on redetermined fluxes of a bigger sample with a lower flux limit, which ensures a homogeneous selection instead of just using fluxes from the catalogs. So to stay consistent with the selection function we use the luminosities from \cite{reiprich_hiflugcs}. Those luminosities are tabulated for the $\SIrange{0.1}{2.4}{keV}$ band (source rest frame) in a flat Universe with $H_0 = \SI{50}{km\,s^{-1}\,Mpc^{-1}}$ and $\Omega_\mathrm{m} = 1$. So every time the cosmological parameters are varied the luminosities have to be recalculated. Additionally a K-correction has to be performed to account for the difference between the observed and the emitted energy band due to the redshift (see e.g., \citealp[App. B]{1998ApJ...495..100J}).
	\item The Chandra FOV is too small to directly measure the luminosity of the clusters from the observed count rates, especially for the low redshift objects. 
	\item After more detailed X-ray data was available, RXCJ1504 was added later to the sample, since it was first falsely classified to have a strong AGN boosting the X-ray luminosity just above the flux limit (see \citealp[Appendix C. 43]{hudson_what_2009} for details).
\end{itemize}

\section{Results}
\subsection{HIFLUGCS masses}
\label{ch:hiflugcs_masses}
Following the hydrostatic equation, all one needs for the total gravitating mass of a galaxy cluster is detailed (deprojected) temperature and density information. Both quantities are parametrized, as described before, and the parameters are determined using a Markov Chain Monte Carlo approach. So it is straightforward to calculate the total mass and its uncertainty by just using the saved Markov chains. The mass at a given overdensity (for HIFLUGCS at 2500, 500 and 200 times the critical density) is determined in an iterative way. To easily compare the masses to literature we use $500 \rho_\mathrm{crit}$ as the default overdensity (apart from other reasons described before), which will be denoted as $M_{500}$ in the following.

\begin{figure*}
	\centering
	\includegraphics[width=0.98\textwidth]{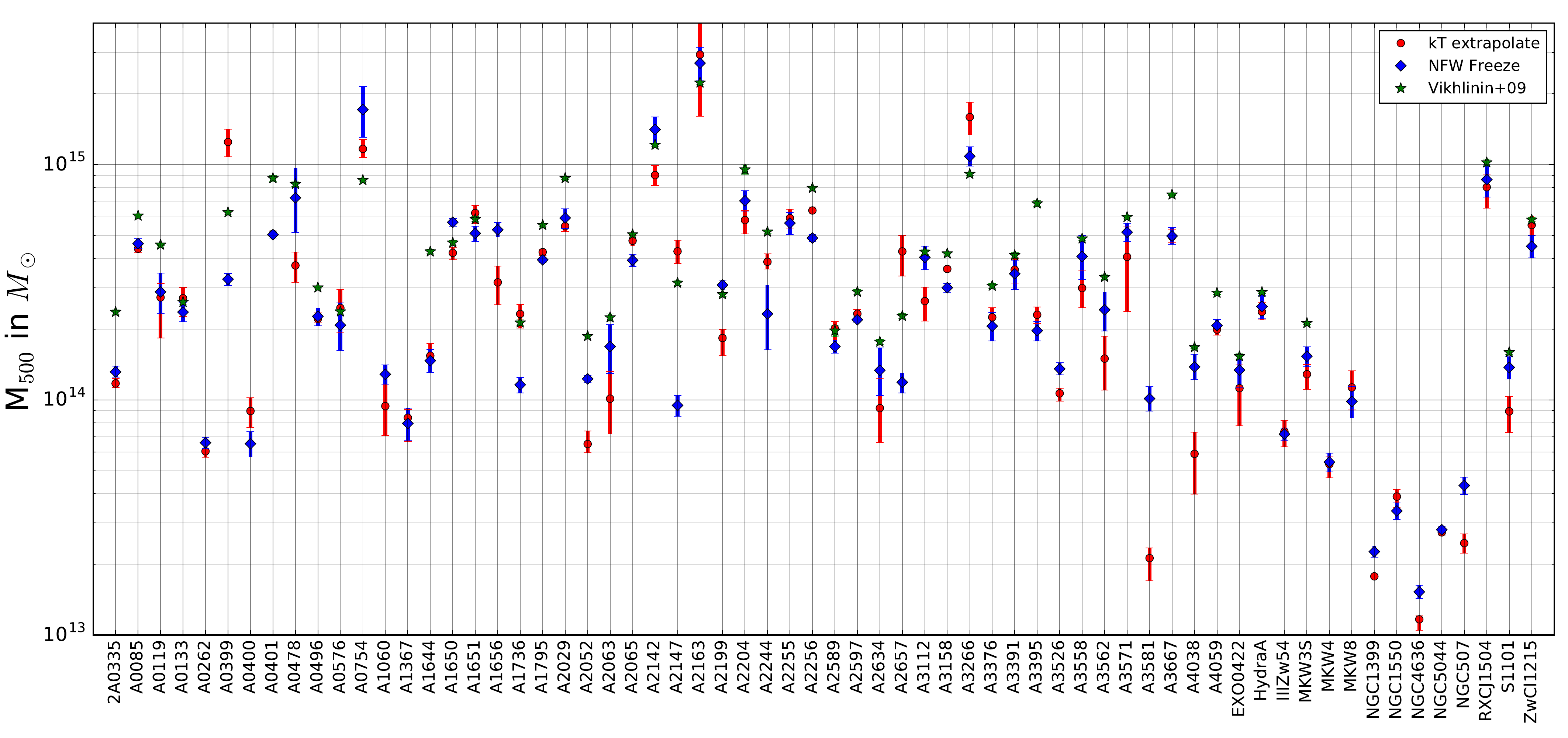}
	\caption[Total masses $M_{500}$ of all HIFLUGCS galaxy clusters]{Total masses $M_{500}$ of all HIFLUGCS galaxy clusters from the Chandra data using different extrapolation methods (see text for details). Green stars indicate the results from \cite{2009ApJ...692.1033V} (using $Y_X$ scaling relation). For a discussion see Section \ref{ch:discussion}.}
	\label{fig:allmasses}
\end{figure*}

In the following we describe the four different extrapolation methods used for all HIFLUGCS clusters:
\begin{itemize}
	\item ``kT extrapolate'': This is the most simple extrapolation by just using the temperature and surface brightness model. For small extrapolations this might provide still robust estimates but for cases where only $50\% r_{500}$ or less can be covered by the temperature profile, the uncertainties and systematics are probably underestimated, since the model is not physically motivated and at larger radii the galaxy cluster temperature does not necessarily need to follow the inferred behavior from inner regions. 
	\item  ``NFW All'': This extrapolation fits an NFW profile (\citealp{1996ApJ...462..563N,1997ApJ...490..493N}) to the cluster mass profile within a radius, where temperature measurements exist. 
	The NFW profile is a simulation motivated, well established parametrization of the Dark Matter density profile,
	\begin{equation}
	\label{eq:nfw}
	\frac{\rho (r)}{\rho_\mathrm{crit}} = \frac{\delta_c}{(r/r_s) (1 + r/r_s)^2}~,
	\end{equation}
	where $r_s$ is a scaling radius and $\delta_s$ a scale overdensity (see also \citealp{2009ApJ...707..354Z,2013MNRAS.432.1103L,2014MNRAS.441..378L}). The scale parameters can be rewritten using a concentration parameter,
	\begin{equation}
	c_\Delta = \frac{r_\Delta}{r_s}~,
	\end{equation}
	where $\Delta$ is an overdensity with respect to $\rho_\mathrm{crit}$. The use of the overdensity 200 is a convention, but it is possible to use the NFW profile with respect to any overdensity, only then $c$ should be labeled accordingly, e.g., $c_{500}$. It is then easy to conclude
	\begin{equation}
	\label{eq:nfw2}
	\frac{c_{500}}{c_{200}} = \frac{r_{500}}{r_{200}}~.
	\end{equation}
	\cite{2014MNRAS.441..378L} have shown that one can write the total mass in terms of the concentration parameter,
	\begin{equation}
	M(<r) = \frac{Y\left( c_{200} \frac{r}{r_{200}} \right)}{Y\left(c_{200} \right) }  \underbrace{200 \rho_\mathrm{crit} \frac{4}{3}\pi r^3 \left(\frac{r_{200}}{r}\right)^3}_{M_{200}}~,
	\end{equation}
	where $Y(u) = \ln(1+u) - u/(1+u)$.
	So the final fitting formula to extrapolate any radius to $r_{500}$ is given by
	\begin{equation}
	\label{eq:nfw6}
	M(<r) = \underbrace{M_{500}}_{\frac{4}{3}\pi r_{500}^3 \cdot 500\rho_\mathrm{crit}} \cdot \frac{Y\left( \frac{r}{r_{500}}c_{500} \right)}{Y(c_{500})}~,
	\end{equation}
	where only $r_{500}$ and $c_{500}$ are free parameters during the fit (again using an MCMC to account for the degeneracy between the parameters). 
	\item ``NFW Hudson'': This method is almost identical to the ``NFW All'' case, only that the central region of the cluster mass profile is not taken into account for the NFW fit. In detail, radii smaller than the cool core radius as defined for HIFLUGCS in \cite{hudson_what_2009} are excluded. This is motivated by the fact that in the cool core region hydrostatic equilibrium deviations (e.g., due to the central AGN) might exist. Excluding this region from the NFW fit is expected to improve the results.
	\item ``NFW Freeze'': In this case an NFW model is fit to the outermost measured mass profile (i.e., the last 3-5 bins in the temperature profile), but a relation from \cite{2013ApJ...766...32B} between $c_{200}$ and $r_{200}$, is used to decrease the degrees of freedom:
	\begin{equation}
	\label{eq:nfw3}
	c_{200} = \left( \frac{M_{200}}{\SI{2.519e22}{M_\odot}} \right)^{-0.08}~.
	\end{equation}
	Equation \ref{eq:nfw2} can be written as
	\begin{equation}
	\frac{c_{500}}{c_{200}} = \left(\frac{Y(c_{500})}{Y(c_{200})}\right)^{1/3} \left(\frac{200}{500} \right)^{1/3}~,
	\end{equation}	
	which can be approximated numerically to derive a conversion between $c_{500}$ and $c_{200}$,
	\begin{equation}
	\label{eq:nfw4}
	c_{500} = \num{0.7027}c_{200} - \num{0.0245}~.
	\end{equation}
	Inserting \ref{eq:nfw3} in \ref{eq:nfw4} one can derive a relation between $c_{500}$ and $M_{500}$, which we again approximate numerically,
	\begin{equation}
	\label{eq:nfw5}
	c_{500} = \num{0.056} \left(\log_{10} M_{500} \right)^2 - 2.18 \log_{10} M_{500} + 22.566~.
	\end{equation}
	By applying \ref{eq:nfw5} to \ref{eq:nfw6} the only free parameter is $r_{500}$. These mass estimates are used as the default to derive cosmological parameters.
\end{itemize}
Figure \ref{fig:allmasses} shows the determined total masses. 
\textit{kT extrapolate} and \textit{NFW Freeze} masses give realistic mass estimates, i.e. not too many masses above $\SI{e15}{M_\odot}$, none of them above $\SI{e16}{M_\odot}$, and also no mass estimate below $\SI{e13}{M_\odot}$. \textit{NFW Hudson} and \textit{NFW All} (not shown in Fig. \ref{fig:allmasses}) give, in very few cases, unrealistically high masses (in two cases $> \SI{e16}{M_\odot}$, one of them Coma), while one case (NGC4636) has a suspiciously low mass.
The average uncertainty estimates are 14\% (kT extrapolate), 13\% (NFW All), 11\% (NFW Hudson), and 11\% (NFW Freeze). Most likely this is due to the degrees of freedom that the different cases imply: The NFW model sets constraints and leaves less possibilities for scatter as the extrapolated temperatures. NFW Hudson excludes the cool core and reduces the scatter. The NFW Freeze model introduces more uncertainty for $M_{500}$ by just using the outermost mass measurements for the extrapolation, but also restricts the concentration parameter, so the final uncertainty is comparable to the other methods.

\begin{figure*}
	\centering
	\includegraphics[width=0.8\textwidth]{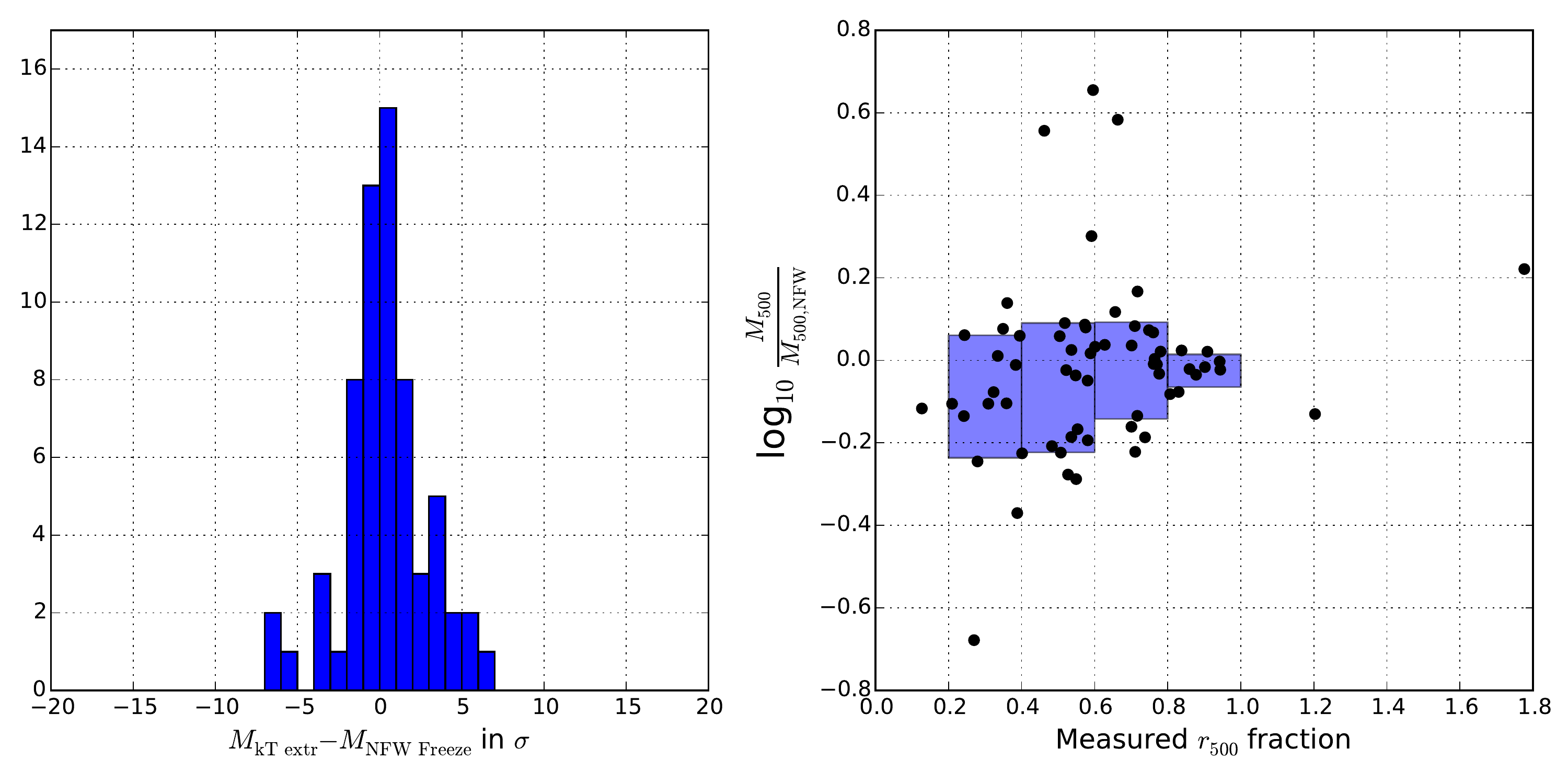}
	\caption{Comparison of the derived masses for HIFLUGCS. The left panel shows the difference in $\sigma = \sqrt{\sigma^2_\mathrm{NFW} + \sigma^2_\mathrm{kT}}$ between kT extrapolated and NFW Freeze masses. A negative significance means the kT extrapolated mass is larger. The right panel shows the ratio of the kT extrapolated and NFW Freeze masses as a function of the largest radius that has temperature measurements ($r_{500}$ from the NFW Freeze method). Blue shaded areas show the median and the scatter.}
	\label{fig:masses_comparison}
\end{figure*}
Figure \ref{fig:masses_comparison} shows the kT extrapolated masses compared to the NFW Freeze masses, which show often agreement and the smallest scatter, while the other two methods give on average higher masses. Moreover, NFW Freeze (the default in the following) approach the kT extrapolated masses when the extrapolation is small or non existing (Fig. \ref{fig:masses_comparison}, bottom right).  This is a consistency check that the measured hydrostatic masses (without extrapolation) more or less match the NFW Freeze masses.

The NFW Freeze masses will be chosen as the default cluster mass in the following because, on the one hand, it is consistent at smaller extrapolation scales with the temperature measurements, on the other hand it is based on a physically motivated model and does not give extreme mass estimates and, therefore, required extrapolations should be more robust.

\subsection{$L_x-M$ relation}
\label{ch:lxm}
The results of the $L_x-M$ relation are of particular interest since they do not only enter in the cosmological analysis, but can be compared directly with references. Furthermore, the $L_x-M$ relation is an important tracer for total masses, and cosmological results of, for example, eROSITA (\citealp{2012arXiv1209.3114M}), will rely on it. Indications for a steepening at the galaxy group scale have been raised in the past (e.g., \citealp{Lovisari2015}), but usually just a simple powerlaw is used to model the $L_x-M$ relation in a cosmological context. First we will start with this simple powerlaw description, also to be consistent with what has been done before (e.g., \citealp{reiprich_hiflugcs,2009ApJ...692.1060V,2010MNRAS.406.1773M}). 

Equation \ref{eq:lxm} describes the relation between mass and luminosity,
\begin{equation}
\label{eq:lxm}
\log_{10} \left( \frac{L_x}{h^{-2}\,\SI{e44}{erg\,s^{-1}}} \right) = A_\mathrm{LM} + B_\mathrm{LM} \cdot \log_{10} \left( \frac{M}{h^{-1}\,\SI{e15}{M_\odot}} \right)~.
\end{equation}
We do not include a redshift evolution term, $\propto E(z)$, since we are dealing with a low-redshift sample. 
The observed distribution of mass and luminosity can easily be fit using a linear regression code that accounts for intrinsic scatter like BCES (\citealp{1996ApJ...470..706A}) or the Bayesian code by \cite{2007ApJ...665.1489K}.
Furthermore, these algorithms take (symmetric) uncertainties of both parameters ($x$ and $y$) into account. To minimize the bias when turning the original probability distribution of each mass and luminosity into logspace and identify it with a normal distribution (for symmetric errorbars), we first calculate the values of the upper and lower boundary in logspace and assign then the logspace uncertainty by taking the arithmetic mean of the individual uncertainties. In case of the luminosity, $y = \log_{10} L_x$,
\begin{equation}
\Delta y = 0.5 \cdot (\log_{10} L_x^u  - \log_{10} L_x^l ) ~,
\end{equation}
where $L_x^u$ and $L_x^l$ are the upper and lower boundaries, respectively, and for the mass, $x$, the procedure is accordingly.
The intrinsic scatter is always measured according to $y$, $\sigma_\mathrm{intr}^{y} = \sqrt{ (\sigma_\mathrm{tot}^y )^2  - (\sigma_\mathrm{stat}^y)^2 - B^2 (\sigma_\mathrm{stat}^x )^2}$, where $\sigma_\mathrm{stat}^y = \langle \Delta y \rangle$, $\sigma_\mathrm{stat}^x = \langle \Delta x \rangle$ and $\sigma_\mathrm{tot}^y = \sqrt{ \langle \left(  y-A-B\cdot x \right)^2 \rangle }$, $A$ and $B$ are the intercept and slope, respectively. So the scatter is assumed to be normal distributed in logspace.

The BCES code includes four different minimization schemes: 
\begin{itemize}
	\item $Y|X$: Regression of y on x, i.e. minimization in y direction.
	\item $X|Y$: Minimization in x direction.
	\item Bisector: Taking the linear relation that bisects the functions from $Y|X$ and $X|Y$.
	\item Orthogonal: Minimization in orthogonal direction on the best fit relation.
\end{itemize}
\begin{figure*}
	\centering
	\includegraphics[trim=50px 0px 100px 50px,clip,width=0.99\textwidth]{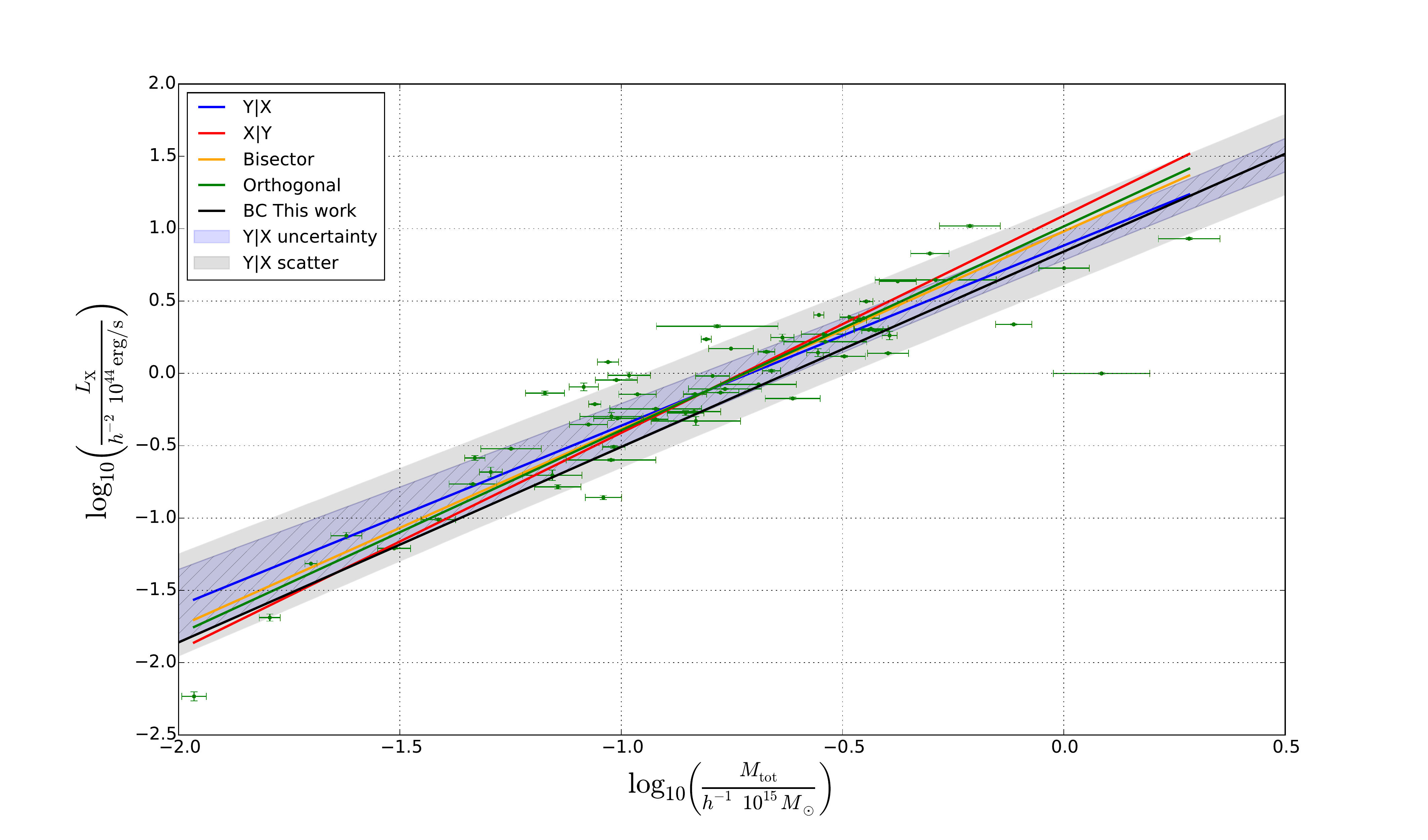}
	\caption{Comparison of the different BCES minimization methods for the observed $L_x-M$ relation. The blue shaded area (with hatching) shows the 68\% uncertainty region for the Y|X method using only the normalization and slope uncertainties, while the gray shaded uncertainty regions include the scatter. Also shown is the bias-corrected result based on the method described in this work (see paper II).}
	\label{fig:lxm_bces}
\end{figure*}
\begin{table}
	\renewcommand{\arraystretch}{1.5}
	\centering
	\begin{tabular}{ccccc}
		\hline
		BCES & Slope $B_\mathrm{LM}$  & Norm $A_\mathrm{LM}$ & $\sigma_\mathrm{tot}^y$ & $\sigma_\mathrm{intr}^y$\\
		\hline
		\hline
		$Y|X$ & $\num{1.24(11)}$ & $\num{0.88(10)}$ & 0.26 & 0.25 \\
		$X|Y$ & $\num{1.50(08)}$ & $\num{1.09(08)}$ & 0.29 & 0.28 \\
		Bisector & $\num{1.37(10)}$ & $\num{0.98(09)}$ & 0.27 & 0.26 \\
		Orthogonal & $\num{1.41(19)}$ & $\num{1.02(09)}$ & 0.27 & 0.26 \\
		Bayesian & $\num{1.27(8)}$ & $\num{0.91(7)}$ & 0.26 & 0.25 \\
		$Y|X$ Groups & $\num{1.81(19)}$ & $\num{1.58(26)}$ & 0.24 & 0.23 \\
		$Y|X$ Clusters &  $\num{0.91(20)}$ & $\num{0.69(13)}$ & 0.22 & 0.22 \\
		\hline
		\textbf{bias corrected} & $\num{1.35(7)}$ & $\num{0.84(6)}$ & 0.26 & 0.25 \\
		\textbf{bias corr. groups}$^{*}$ & $\num{1.88(16)}$ & $\num{1.56(18)}$ & 0.24 & 0.23 \\
		\textbf{bias corr. clusters}$^{*}$ & $\num{1.06(9)}$ & $\num{0.74(18)}$ & 0.22 & 0.22 \\
		\hline
		\hline
	\end{tabular}
	\caption{Resulting parameters of the $L_x-M$ relation. For the setups marked with $^{*}$ a broken powerlaw was used to model the scaling relation (see text for details). Note that the main results here are the bias corrected values.}
	\label{tab:bces}
\end{table}
All these methods take into account the errors of both parameters and intrinsic scatter. For any linear regression of the $L_x-M$ relation we assume the uncertainties of the luminosity and mass to be uncorrelated.  Uncertainties on the best fit slope and intercept are taken from $\num{10000}$ bootstrap realizations\footnote{The bootstrap method picks $N$ elements randomly out of a dataset of length $N$, so the distribution function of the dataset is not necessary.}. Figure \ref{fig:lxm_bces} and Tab. \ref{tab:bces} show that the different minimization methods can give different results. The regression of y on x gives slightly shallower slopes than the minimization in x. These two methods mark the two extremes in terms of slopes and normalizations, the bisector (by definition) and the orthogonal method are in between and very similar. 
The uncertainty range based on the slope and intercept errors (not accounting for the correlation) of the $Y|X$ minimization is shown in Fig. \ref{fig:lxm_bces} by the blue hatched area, which just includes the bisector and orthogonal methods but not the $X|Y$ case. The gray shaded area represents the uncertainty of the best fit $L_x-M$ relation including the intrinsic scatter, which is so large that all other best fit relations are within this region.

If we use the ``kT ext'' masses instead of our default choice, we find that the $L_x-M$ relation is driven by several clusters with large mass uncertainties (especially for the $Y|X$ method). Excluding clusters with a relative mass uncertainty above 20\% results in very similar values for slope and normalization ($\sim 0.5\%$ change).

We also tested the Bayesian linear regression code by \cite{2007ApJ...665.1489K} which was implemented in python by Josh Meyers\footnote{\url{https://github.com/jmeyers314/linmix}}. \cite{2012arXiv1210.6232A} have shown that a Bayesian regression seems to be more unbiased than least square methods like BCES. Table \ref{tab:bces} also contains the results of this method using $\num{50000}$ MCMC steps, which show very good agreement with the BCES $Y|X$ method. 
The BCES code is clearly superior to the Bayesian method in terms of required computational time, because the Bayesian code runs an MCMC chain.

\section{Discussion}
\subsection{Selection effects in the $L_x-M$ relation}
Intrinsically brighter objects have a higher chance to be included in a flux or luminosity limited sample. The observed relation between mass and luminosity is expected to be biased. For the simple case of a luminosity limited sample, \cite{2010MNRAS.406.1773M} demonstrated that the observed scaling relation will be shallower. For a flux limited sample, the situation is more complex.

We take selection effects into account when calculating the likelihood function for our cosmological analysis (see paper II), where the parameters of the scaling relation are free parameters. An alternative approach to correct for selection effects has been shown in \cite{Lovisari2015}, where fake samples were constructed from the halo mass function using the sample selection criteria and an assumed scaling relation, where the parameters are varied until the observed scaling relation is matched by the fake samples after applying the selection.

\begin{figure*}
	\centering
	\includegraphics[trim=50px 0px 100px 50px,clip,width=0.99\textwidth]{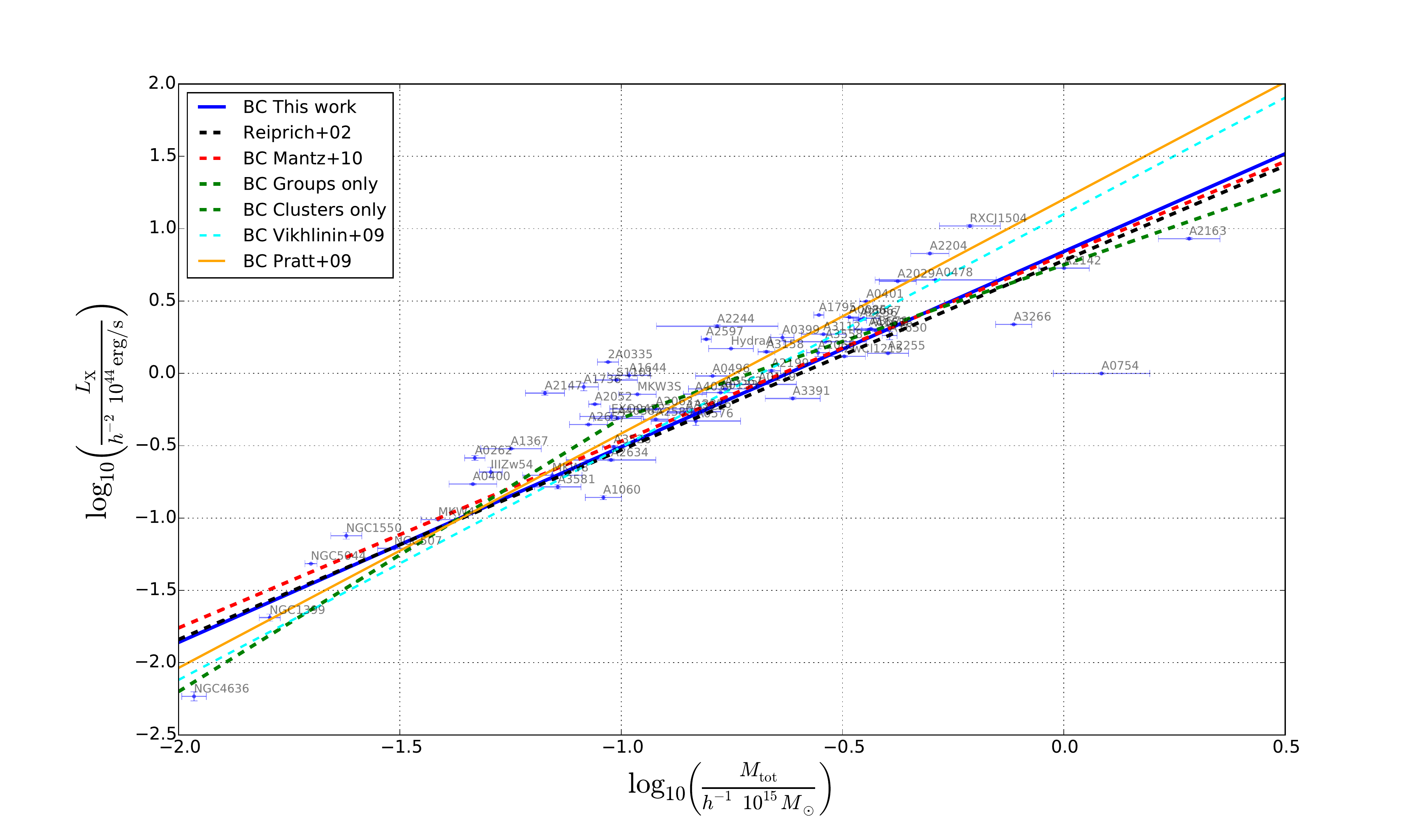}
	\caption{$L_x-M_{500}$ scaling relations. ``BC'' indicates bias corrected relations.}
	\label{fig:lxm_scaling}
\end{figure*}

Compared to the observed $L_x-M$ relation the bias corrected one (both shown in Fig. \ref{fig:lxm_scaling}) is slightly steeper, while the normalizations are formally in agreement (although for most cluster masses, the normalization of the corrected relation lies actually below the uncorrected one). This behavior fulfills the naive expectations of a Malmquist and Eddington bias for luminosity cut. 
The bias corrected $L_x-M$ relation (see Tab. \ref{tab:bces}) is in agreement with \cite{reiprich_hiflugcs,2010MNRAS.406.1773M} (Fig. \ref{fig:lxm_scaling}), while \cite{2009ApJ...692.1033V,2009A&A...498..361P} find significantly steeper relations. 
Note that \cite{2009ApJ...692.1033V} used a different energy band to derive luminosities, which are converted to the ROSAT band in Fig. \ref{fig:lxm_scaling}.

We tested the effects of the sample selection on the scatter of the observed sample by leaving this parameter free in the likelihood analysis, but we did not find any significant change with respect to the observed value of the scatter. 

\begin{figure*}
	\centering
	\includegraphics[width=0.99\textwidth]{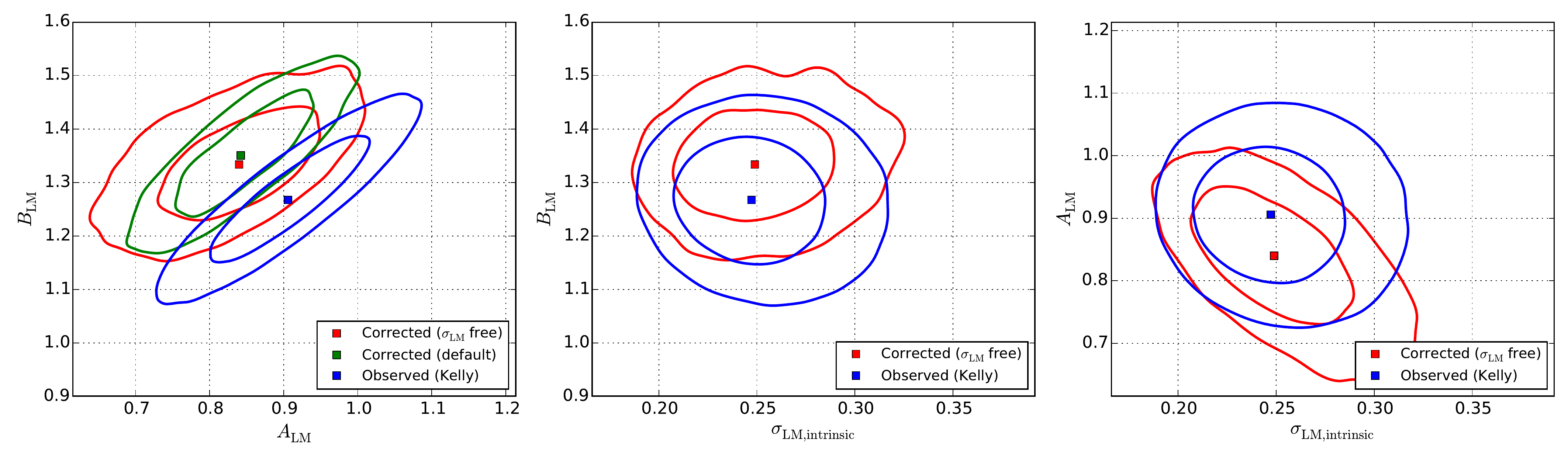}
	\caption[Confidence regions for the $L_x-M$ parameters]{68.3\% and 95.4\% confidence regions for the $L_x-M$ parameters of the observed (using \citealp{2007ApJ...665.1489K} algorithm) and corrected $L_x-M$ relation. \textit{Left:} Slope vs. normalization for the observed and corrected (also with scatter free to vary) scaling relation. \textit{Center/Right:} Slope/Normalization vs. scatter only for the cases with variable scatter.}
	\label{fig:kelly}
\end{figure*}
Since the Bayesian code from \cite{2007ApJ...665.1489K} runs an MCMC, parameter confidence levels can be plotted, as shown in Fig. \ref{fig:kelly}. The center and right panel of this figure demonstrate that the scatter is consistent between the observed and bias corrected sample, which is also found, e.g., by \cite{Lovisari2015} for galaxy groups. The slope and normalization are not in agreement within the 68.3\% range (left panel).

\subsection{Differences for galaxy groups}
Clusters of galaxies with a mass below $\SI{e14}{M_\odot}$ are called galaxy groups. 
These objects are known to have different scaling properties compared to their more massive counterparts (e.g., \citealp{2009ApJ...693.1142S,2011A&A...535A.105E,Lovisari2015,Bharadwaj2015b}).
Especially there are strong indications that the $L_x-M$ relation will steepen on the galaxy group scale. For a complete sample like HIFLUGCS, which consists also of several galaxy groups, a single powerlaw model might not be sufficient and cause further biases in the cosmological analysis. Splitting the observed mass and luminosity distribution into groups and clusters, with the threshold at a total mass of $\SI{e14}{M_\odot}$ results in a confirmation of this trend: The 23 low mass objects have a slope of $\num{1.81(19)}$ in the $L_x-M$ relation, while the rest of the clusters show a much flatter distribution, $\num{0.91(20)}$ (see Tab. \ref{tab:bces}). For both subsamples, the scatter is reduced by about 15\%. 

In this case, the selection effects were corrected again within the likelihood analysis, but due to the relatively large uncertainties in the mass, the sample was not split but the $L_x-M$ relation was modeled by a broken powerlaw (see Fig. \ref{fig:lxm_scaling}). We smoothed the model locally around the transition of the two powerlaws. For the (bias corrected) two slopes we found values, which are in good agreement with what was found by \cite{Lovisari2015}. The high mass slope is in agreement with the self-similar prediction, while there is strong tension with the low mass slope. When excluding the group with the lowest mass and luminosity (NGC4636), the observed slope is slightly shallower, but still in agreement with the overall slope of the groups.

\subsection{Independent mass estimates}
\label{ch:discussion}
Here we explore the use of independent masses of the HIFLUGCS sample, in particular the mass from the velocity dispersion $\sigma_\mathrm{v}$ and the mass from the Planck SZ measurements. The SZ masses were obtained using a scaling relation calibrated with XMM-Newton data by \cite{2010A&A...517A..92A}. From three different pipelines (Matched Multi-Filter MMF1, MMF3 and PowellSnakes, PwS) clusters are detected in the Planck maps. All detections with the corresponding integrated Comptonization $Y$ parameter and total masses, $M_{500}$, are summarized in the ``Union catalogue'' (\citealp{2015arXiv150201598P,PlanckCollaboration2015ae,2015arXiv150201596P}, which is available on the Planck website\footnote{\url{http://wiki.cosmos.esa.int/planckpla2015/index.php/Catalogues}}).

We compared this catalog with HIFLUGCS and detected 50 out of the 64 objects. The missing ones are mainly low mass objects: NGC1399, A2052, A3581, EXO0422, NGC1550, MKW4, NGC4636, A0400, ZwCl1215, NGC5044, NGC507, A1060, HydraA, and IIIZw54. 90\% of the detected objects have an SZ signal-to-noise ratio (SNR) above 6, which was used for the Planck SZ study. Here we use all detected objects (minimum SNR is 4.5) and set the mass uncertainties of the undetected objects to a very broad range, which gives them no weight. Out of these 14 undetected clusters, 12 are in the low redshift subsample of HIFLUGCS. 
\begin{figure*}
	\centering
	\includegraphics[width=0.49\textwidth]{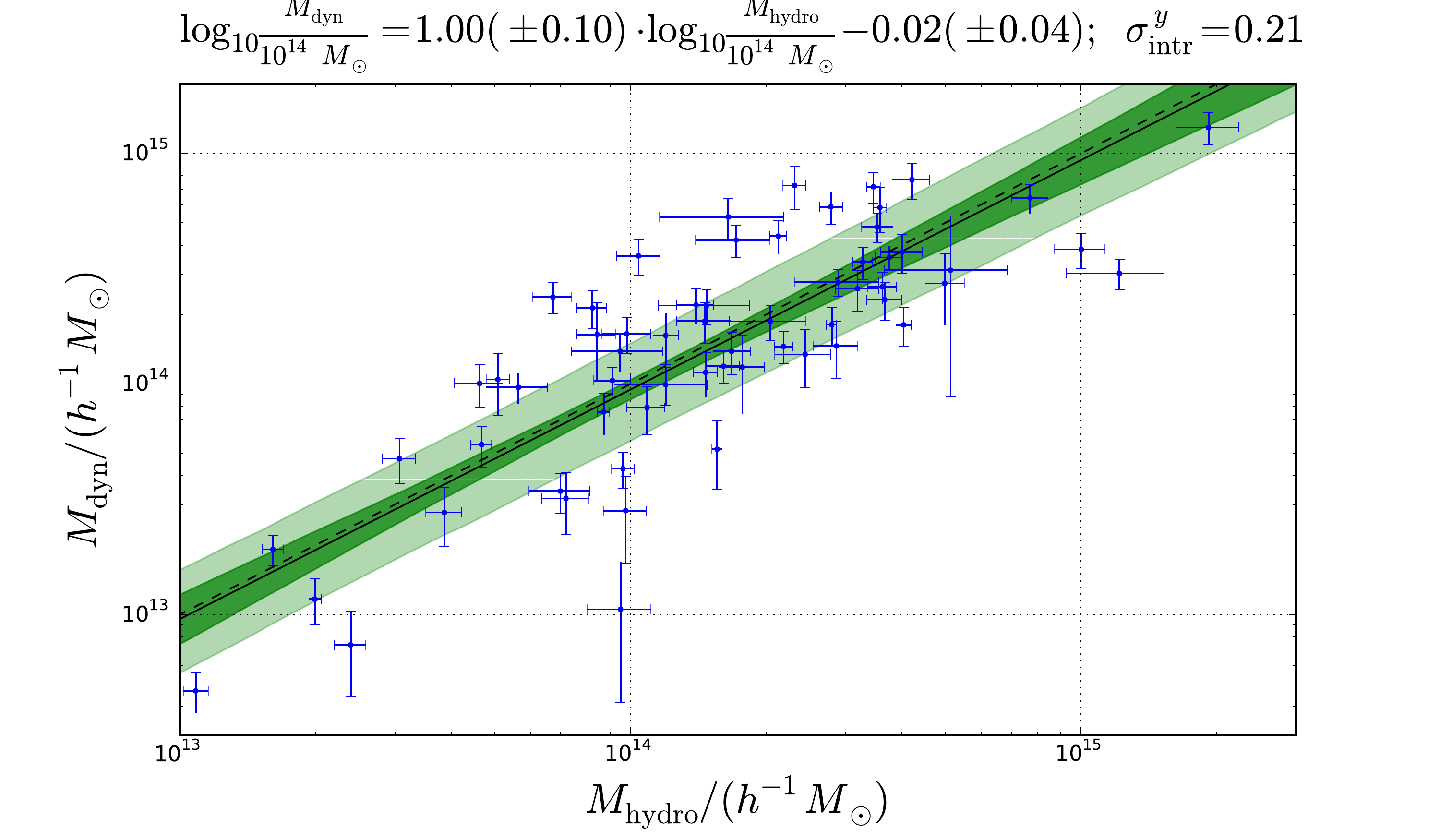}
	\includegraphics[width=0.49\textwidth]{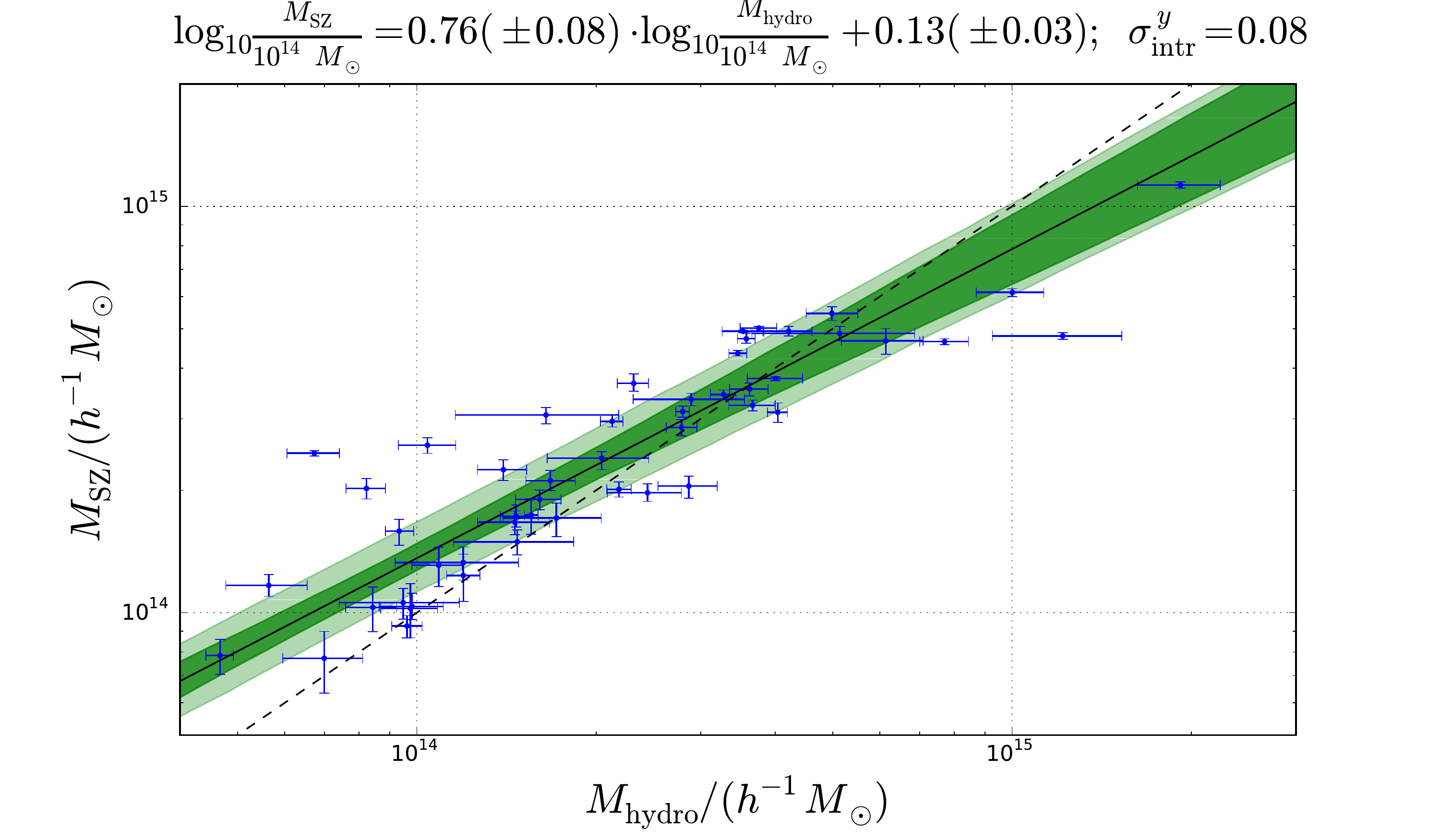}
	\caption{Direct comparison of the hydrostatic X-ray masses (default) and 62 dynamical mass estimates (left; \citealp{2016arXiv160806585Z}), and the hydrostatic masses and 50 Planck SZ masses (right). The dark green region is the 68.3\% uncertainty region of the scaling relation (the light green region including the intrinsic scatter). The black dashed line is equality of masses. The best fit relations above each graph are calculated using the BCES $Y|X$ estimator.} 
	\label{fig:diff_masses}
\end{figure*}
The average bias between the Planck SZ and our Chandra masses is determined by an MCMC analysis, resulting in 
\begin{equation}
\frac{M_\mathrm{CXO}}{M_\mathrm{SZ}} = \num{0.86(1)}~.
\end{equation}
Therefore, we construct the likelihood function that the log of the X-ray masses are the \textit{observations} and the log of the SZ masses plus the log of the bias are the \textit{model}. The SZ mass priors are set to their observed values and uncertainties, the bias has a uniform distribution with 0.5 and 2 as lower and upper boundaries.

Also parameterizing the mass difference by a linear regression fit (BCES $Y|X$) shows a slope of $\num{0.76(8)}$, so Planck SZ masses are higher at low masses ($\sim \SI{e14}{M_\odot}$) and slightly lower at high masses ($\sim \SI{e15}{M_\odot}$), compared to our Chandra masses (details see Fig. \ref{fig:diff_masses}, right).
The $L_x-M$ relation constrained from the Planck masses (using the same luminosities) deviates strongly from the default analysis: The slope and normalization are about 20\% larger. 
A higher normalization of the $L_x-M$ relation could be due to a systematic bias that lowers all Planck masses. Following the degeneracy between slope and normalization, the slope has to increase as well. The Chandra$-$XMM-Newton cross calibration contributes to this effect.
The trend of the Planck vs hydro masses in Fig. \ref{fig:diff_masses} will steepen the $L_x-M$ relation when using Planck SZ masses. 
But it is not immediately clear why the Planck masses are larger at the low mass end. One possibility is, that the $Y-M_{500}$ scaling relation from \cite{2010A&A...517A..92A}, which is not corrected for selection effects, does not reflect the real behavior.
Furthermore, the uncertainties of the Planck masses are much smaller than the X-ray constraints, because no uncertainties on the pressure profile and no uncertainties or scatter of the $Y-M_{500}$ relation enter in the calculation and the errorbars represent only statistical uncertainties (\citealp{2015arXiv150201598P}).
Applying the Chandra -- XMM-Newton total mass scaling relation from \cite{Schellenberger2015} on the derived Chandra HIFLUGCS masses results in XMM-Newton mass estimates for the HIFLUGCS sample. Comparing those with the Planck SZ masses, which were also derived with a scaling relation calibrated using XMM-Newton data, shows still the same slope as in Fig. \ref{fig:diff_masses} (right), but a $\sim 20\%$ higher normalization.
Clearly, the Planck selection function misses low mass systems (with low $Y_\mathrm{SZ}$) because they fall below the 4.5 S/N cut, which will also bias the Planck vs hydro mass relation at the low mass end.
This is because for low mass clusters only those will end up in the Planck sample whose $Y_\mathrm{SZ}$ scatters up for a given true mass. Since masses are then derived from this upscattered $Y_\mathrm{SZ}$, the resulting masses will on average be overestimated compared to the true underlying masses. Since this effect will be less strong for the more massive high-S/N clusters, this selection effect results in a slope $< 1$ in our mass-mass comparison.
On the other hand it is clearly visible in Fig. \ref{fig:diff_masses} (right) that the six highest mass clusters are all biased toward lower SZ masses with respect to the hydrostatic masses. 
In particular, taking only all clusters with $M_\mathrm{hydro} > \SI{5e14}{M_\odot\, \textit{h}^{-1} }$ results in a mass ratio
\begin{equation}
\frac{M_\mathrm{CXO}}{M_\mathrm{SZ}} = \num{1.46(8)}~.
\end{equation}
This is also the mass range, where most weak lensing comparisons have been done (WtG \citealp{2015arXiv150902162A}, CCCP \citealp{2015MNRAS.449..685H}, LoCuSS \citealp{2015arXiv151101919S}, but not 400d \citealp{2014A&A...564A.129I}). 
Therefore, here we confirm a trend that at these high masses, the Planck masses are low compared to other estimates.
Together this creates a biased slope as observed and does not indicate that $M_\mathrm{hydro}$ are biased low at low masses.

We verified this explanation by creating $\num{10000}$ fake samples for the Planck SZ masses from our Chandra X-ray masses by adding scatter (0.17 in logscale). 

After selecting according to the Planck SZ cluster selection function, we find a median slope of $\num{0.86(6)}$. 
Our fake samples consist on average of about 50 objects, as it is the case for the real sample. 
So the bias arising only from the Planck SZ selection function is consistent with the mass-dependent bias found in our mass-mass comparison.
But other effects might also contribute to the observed bias, such as
the fact that Planck SZ masses have been calibrated with a $Y_\mathrm{X}-M$ scaling relation. As indicated in \cite{2015MNRAS.450.3649S}, already a $M_{Y_\mathrm{X}}$ vs. hydrostatic mass comparison exhibits a mass dependent bias similar to our finding. Moreover, the instrumental cross-calibration uncertainties between Chandra and XMM-Newton (e.g., \citealp{Schellenberger2015}) enter in our Planck SZ vs. Chandra hydrostatic mass comparison as well, which may explain the higher Chandra masses at the high-mass end.

A connection between the galaxy velocity dispersion and the total mass follows from simple considerations of the virial theorem,
\begin{equation}
\sigma_\mathrm{v} = \sqrt{\frac{M G}{r}}~,
\end{equation}
where $r$ is the virial radius and $M$ is called the dynamic mass. Since this radius is usually unknown, one can use an NFW model to describe the density distribution, as done by \cite{2006A&A...456...23B}. 
\cite{2013ApJ...772...47S} use simulations to calibrate the dynamic mass. We use the mass estimates by \cite{2016arXiv160806585Z} based on the velocity dispersion of the HIFLUGCS sample from  \cite{zhang_hiflugcs:_2010} with some updates. 
The direct comparison to the (default) hydrostatic masses in Fig. \ref{fig:diff_masses} (left) shows that despite the scatter the dynamic and hydrostatic masses are in good agreement (overall ratio using Bayesian analysis $\num{0.97(3)}$). This independent mass estimator shows that the hydrostatic masses derived here are robust. The BCES fit (Fig. \ref{fig:diff_masses}, left) shows only a very small difference between the masses and no mass dependency of this insignificant bias.

Although the SZ signal might be a very unbiased mass proxy, see \cite{2015MNRAS.450.3649S}, the selection of clusters has to be carefully taken into account, since the intrinsic scatter, especially for hydrostatic masses is of the order of 25\% (\citealp{2015MNRAS.450.3633S}). Even weak lensing masses have up to 10\% bias and notable scatter, which makes them not ideal calibrators for scaling relations.

In Fig. \ref{fig:allmasses} we also showed the \cite{2009ApJ...692.1033V} masses for comparison. These masses have been obtained using a scaling relation calibrated with relaxed galaxy clusters. Statistical uncertainties in this sample are very small ($\sim 2\%$ on the masses). Compared with the NFW extrapolated masses, the ones by \cite{2009ApJ...692.1033V} are $\sim 15\%$ larger with a scatter of about 10\%, but 16 clusters from HIFLUGCS, especially the low mass groups (due to the redshift cut in \citealp{2009ApJ...692.1033V}) are missing in this comparison. The effects on the cosmological parameter estimation will be discussed in Paper II.
The ''kT extrapolated`` masses show no significant bias with respect to the \cite{2009ApJ...692.1033V} masses.

\section{Conclusion}
The aim of this project is to perform a cosmological analysis using a complete, X-ray selected, purely flux limited sample of 64 local galaxy clusters and calibrate the $L_x-M$ scaling relation simultaneously. In this work we have demonstrated how we obtain the total cluster masses, compare them with independent estimates and quantify the luminosity-mass scaling-relation for a complete sample including a bias correction.

Masses of the galaxy clusters have been calculated individually for each cluster from the temperature and surface brightness profiles. A crucial step was to perform the extrapolation of the mass, since the Chandra FOV and, for some clusters, the limited exposure time, does not allow us to measure the temperature and its gradient at $r_{500}$. Several different methods are proposed to perform the extrapolation, either simply by using models for the temperature profiles, which can produce (unphysical) decreasing total mass profiles in the outer regions, or by using an NFW model. The NFW fit with a concentration parameter linked to the total mass, was found to be most reliable and used as default in the end. 

Our main results can be summarized as follows:
\begin{itemize}
	\item The observed $L_x-M$ relation exhibits a slope of $\num{1.24(11)}$. The (logarithmic) intrinsic scatter of 0.25 is expected for a typical $L_x-M$ relation using a sample of relaxed and disturbed clusters of galaxies.
	\item The bias corrected $L_x-M$ relation shows an insignificantly steeper slope ($\num{1.35(7)}$) and a lower normalization at a median cluster mass range, which is in agreement with expectations for selection effects on a luminosity limited sample.
	\item Using a broken powerlaw to parameterize the $L_x-M$ relation, we find a steepening to the galaxy group scale and a flattening toward the high-mass end.
	\item Independently obtained masses from galaxy velocity dispersion and from the Planck Sunyaev Zeldovich effect measurements show overall agreement with our hydrostatic masses. While we see very good agreement between our hydrostatic masses and the dynamic masses ($\frac{M_\mathrm{hydro}}{M_\mathrm{dyn}} = 0.97$) with large scatter (0.21), we notice a mass dependent bias between the SZ and the hydrostatic masses (slope 0.75) but smaller scatter (0.09). The overall bias between hydrostatic and SZ masses is $\frac{M_\mathrm{hydro}}{M_\mathrm{SZ}} = 0.86$, while for clusters with a hydrostatic mass $> \SI{5e14}{M_\odot\, \textit{h}^{-1}}$ it is $\num{1.46}$.
	This can be explained with the Planck selection function and the intrinsic scatter of the hydrostatic mass estimates.
\end{itemize}

In paper II we will describe and discuss the results of the cosmological analysis. The $L_x-M$ relation is of crucial importance for these constraints, so we will also show all degeneracies of cosmological parameters with those of the scaling relation.

\section*{Acknowledgements}
\renewcommand*{\thefootnote}{\fnsymbol{footnote}}
\setcounter{footnote}{0}
GS, THR acknowledge support by the German Research Association (DFG) through grant RE 1462/6. GS acknowledges support by the Bonn-Cologne Graduate School of Physics and Astronomy (BCGS) and the International Max Planck Research School (IMPRS) for Astronomy and Astrophysics at the Universities of Bonn and Cologne. THR acknowledges support by the DFG through Heisenberg grant RE 1462/5 and the Transregio 33 ''The Dark Universe`` sub-project B18.
The authors would like to thank Yu-Ying Zhang\footnote{We regret very much Yu-Ying Zhang's untimely death in December 2016.} for the helpful discussion and providing the dynamical masses of HIFLUGCS. 
This research made use of Astropy, a community-developed core Python package for Astronomy.

%%%%%%%%%%%%%%%%%%%%%%%%%%%%%%%%%%%%%%%%%%%%%%%%%%



%\bibliographystyle{mnras}
\bibliographystyle{mnras}
\bibliography{Astro}

%\begin{thebibliography}{99}
%\end{thebibliography}

%%%%%%%%%%%%%%%%%%%%%%%%%%%%%%%%%%%%%%%%%%%%%%%%%%

%%%%%%%%%%%%%%%%% APPENDICES %%%%%%%%%%%%%%%%%%%%%
\appendix
\onecolumn
\section{Planck ESZ mass comparison}
\begin{figure}
	\centering
	\includegraphics[width=0.75\textwidth]{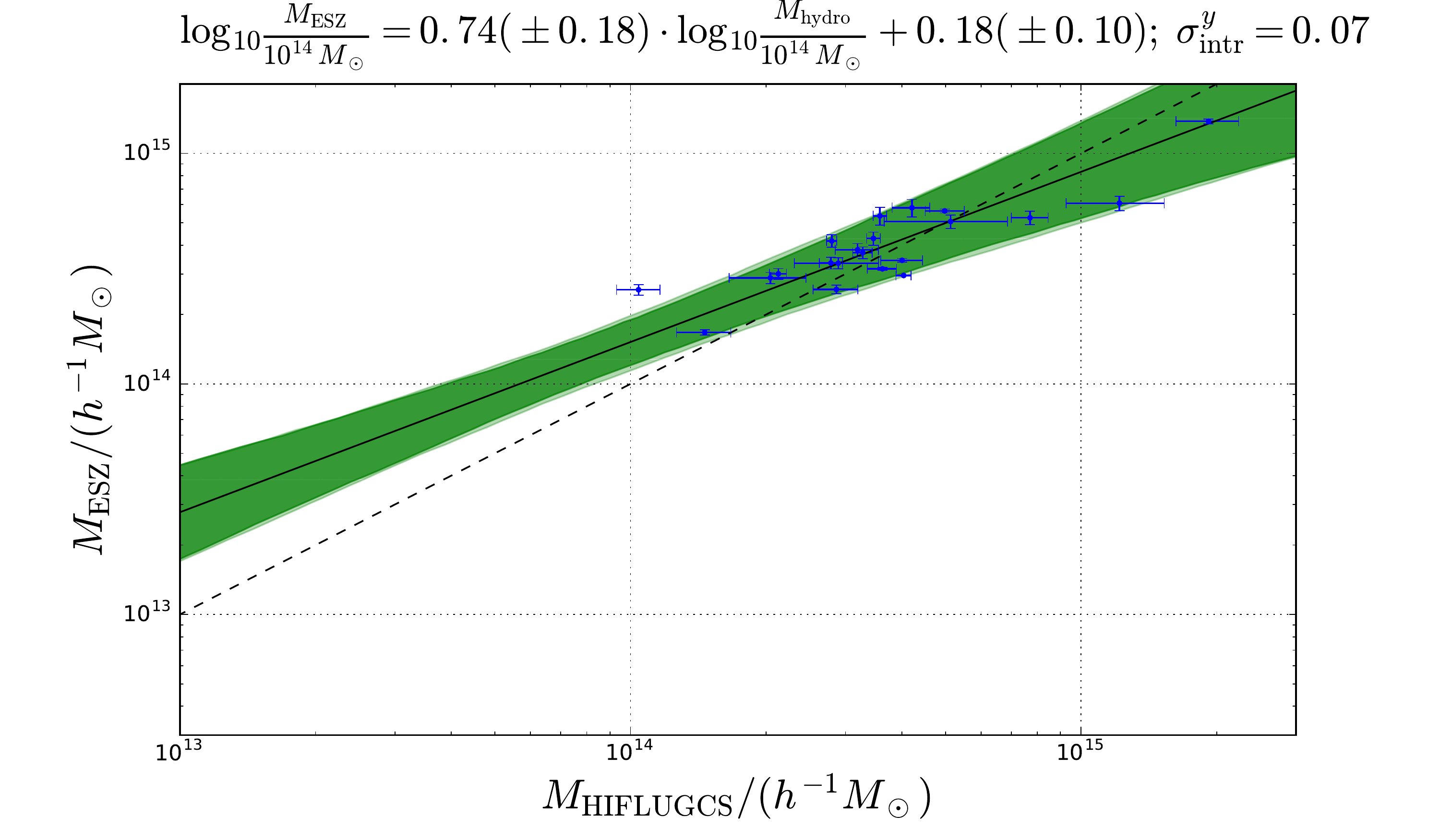}
	\caption{Direct comparison of the hydrostatic X-ray masses (default) and 21 XMM-Newton mass estimates from the Planck ESZ sample (\citealp{2011A&A...536A..11P}). The dark green region is the 68.3\% uncertainty region of the scaling relation (the light green region including the intrinsic scatter). The black dashed line is equality of masses. The best fit relations above each graph are calculated using the BCES $Y|X$ estimator.}
	\label{fig:esz}
\end{figure}

In order to complement the discussion of the Planck SZ mass comparison with our hydrostatic Chandra masses (Section \ref{ch:discussion}) we also show here a comparison with the Planck ESZ masses derived from XMM-Newton data using the $Y_\mathrm{X}-M$ scaling-relation (\citealp{2011A&A...536A..11P}, see Fig. \ref{fig:esz}).
Slope and normalization of the direct mass comparison have almost the same values as the comparison of the Planck SZ masses with our default (slope 0.74 vs
0.76, intercept 0.18 vs 0.13, scatter 0.07 vs 0.08). This seems plausible since the ESZ masses helped to
calibrate scaling relations used for the full Planck SZ sample.

The overlap of the Planck ESZ sample with HIFLUGCS is limited to only 21 clusters because 
a) there existed no useful XMM-Newton observation,
b) the Planck SZ signal-to-noise was below 6.
The latter point raises the suspicion, that Fig. \ref{fig:esz} does not only show a purely X-ray mass-mass comparison, but the Planck SZ selection function enters as well. 
Assuming the intrinsic scatter in the $Y_\mathrm{X}-M$ and $Y_\mathrm{SZ}-M$ relations are correlated, then clusters entering the sample have a $Y_\mathrm{SZ}$ that is biased high for a given mass (due to intrinsic scatter). These clusters will also have a high $Y_\mathrm{X}$, and so the mass given in the ESZ paper is biased high. 
We conclude that many factors enter in the ESZ mass comparison, such as a more complicated selection function than the Planck SZ clusters, XMM-Newton derived masses or the use of the  $Y_\mathrm{X}$ scaling relation instead of hydrostatic masses (e.g., \citealp{2015MNRAS.450.3649S}) introducing biases. 
In the case of the Planck SZ masses  (Section \ref{ch:discussion}), the overlap with HIFLUGCS is much larger and the effects of the SZ selection function are found to be consistent with the observed bias.

\section{Tables}
\begin{table}
\caption{\label{tab:observations}Chandra observation IDs for the 64 HIFLUGCS clusters, which were used here. The bold number gives the main ID (see text for details).}
\footnotesize
\centering
\begin{tabular}{cccc}
\hline 
Cluster & Chandra Observation ID & Total exposure time [ks] & \\ 
\hline 
2A0335 & \textbf{7939} 9792 919  & 104 & \\ 
A0085 & \textbf{15173} 15174 904 16263 16264 4887  & 208 & \\ 
A0119 & \textbf{7918} 4180  & 58 & \\ 
A0133 & \textbf{9897} 13442 13456 14333 13454 3710 3183  & 706 & \\ 
A0262 & \textbf{7921} 2215  & 141 & \\ 
A0399 & \textbf{3230}  & 49 & \\ 
A0400 & \textbf{4181}  & 22 & \\ 
A0401 & \textbf{14024}  & 137 & \\ 
A0478 & \textbf{1669} 7217 7231 7232  & 95 & \\ 
A0496 & \textbf{4976}  & 76 & \\ 
A0576 & \textbf{3289}  & 39 & \\ 
A0754 & \textbf{10743} 577  & 140 & \\ 
A1060 & \textbf{2220}  & 32 & \\ 
A1367 & \textbf{514} 4189  & 89 & \\ 
A1644 & \textbf{7922} 2206  & 71 & \\ 
A1650 & \textbf{5823} 7242 6356 6357 6358 4178 5822  & 228 & \\ 
A1651 & \textbf{4185}  & 10 & \\ 
A1656 & \textbf{13996} 13994 14410 13995 2941  & 415 & \\ 
A1736 & \textbf{4186}  & 15 & \\ 
A1795 & \textbf{10899} 493 494 10900 10898 10901 12028 5289 5290 12026 14274 14275 15487 15490 15491  & 874 & \\ 
 &  16434 16468 16469 16471 16472 16466 6159 6163 13108 13412 13413 13417 6160 12027 12029  & & \\ 
 &  16438 13111 13113 14272 14273 16437 13415 13109 13110 13112 13416 15488 15489 15492 16435  & & \\ 
 &  16436 16439 16467 16470 13414 5288 3666 5286 5287 14270 14271 6162 6161  & & \\ 
A2029 & \textbf{4977} 891  & 99 & \\ 
A2052 & \textbf{10478} 890  & 158 & \\ 
A2063 & \textbf{6263} 6262  & 31 & \\ 
A2065 & \textbf{3182}  & 50 & \\ 
A2142 & \textbf{5005} 17168 15186 17492  & 313 & \\ 
A2147 & \textbf{3211}  & 18 & \\ 
A2163 & \textbf{1653}  & 72 & \\ 
A2199 & \textbf{10748} 10805 10803  & 102 & \\ 
A2204 & \textbf{7940}  & 78 & \\ 
A2244 & \textbf{4179}  & 58 & \\ 
A2255 & \textbf{894}  & 40 & \\ 
A2256 & \textbf{2419} 16514 16516 16129 16515  & 191 & \\ 
A2589 & \textbf{7190} 7340  & 69 & \\ 
A2597 & \textbf{7329} 6934  & 114 & \\ 
A2634 & \textbf{4816}  & 50 & \\ 
A2657 & \textbf{4941}  & 16 & \\ 
A3112 & \textbf{13135} 6972 7323 7324  & 128 & \\ 
A3158 & \textbf{3712} 3201  & 56 & \\ 
A3266 & \textbf{899}  & 30 & \\ 
A3376 & \textbf{3202} 3450  & 65 & \\ 
A3391 & \textbf{4943} 13525  & 68 & \\ 
A3395 & \textbf{4944} 13522  & 72 & \\ 
A3526 & \textbf{4954} 5310 4955 4190 4191 504 8179  & 317 & \\ 
A3558 & \textbf{1646}  & 15 & \\ 
A3562 & \textbf{4167}  & 20 & \\ 
A3571 & \textbf{4203}  & 34 & \\ 
A3581 & \textbf{12884}  & 86 & \\ 
A3667 & \textbf{5751} 5753 5752 6295 6296 6292  & 444 & \\ 
A4038 & \textbf{4992}  & 34 & \\ 
A4059 & \textbf{5785}  & 93 & \\ 
EXO0422 & \textbf{4183}  & 10 & \\ 
HydraA & \textbf{4969} 4970  & 198 & \\ 
IIIZw54 & \textbf{4182}  & 24 & \\ 
MKW3S & \textbf{900}  & 58 & \\ 
MKW4 & \textbf{3234}  & 30 & \\ 
MKW8 & \textbf{4942}  & 23 & \\ 
NGC1399 & \textbf{9530} 4176 4174 4168 4173 4171 4172 14527  & 364 & \\ 
NGC1550 & \textbf{5800} 5801  & 90 & \\ 
NGC4636 & \textbf{3926} 4415  & 151 & \\ 
NGC5044 & \textbf{9399} 798  & 105 & \\ 
NGC507 & \textbf{2882} 10536  & 63 & \\ 
RXCJ1504 & \textbf{5793}  & 40 & \\ 
S1101 & \textbf{11758}  & 99 & \\ 
ZwCl1215 & \textbf{4184}  & 12 & \\ 
\hline
\end{tabular}
\end{table}

\clearpage
{\small
	\begin{longtable}{ccccccc}
		\caption{\label{tab:hiflugcs_details} Properties of the 64 HIFLUGCS clusters. The last column is the average radius of the largest annulus for the temperature measurements.}\\
		\hline
		\parbox[0pt][1.6em][c]{0cm}{}Name & \multicolumn{2}{c}{Emission weighted center} & Redshift & $N_\mathrm{H}$ & $L_x$ &$R_\mathrm{kT}^\mathrm{max}$\\
		\parbox[0pt][1.6em][c]{0cm}{} & RA [h:m:s] & DEC [d:m:s] &   & $\SI{e22}{cm^{-2}}$ & $\SI{e43}{erg\,s^{-1}\,cm^{-2}}$ & arcmin\\
		\hline
		\endfirsthead
		\caption{continued.}\\
		\hline
		\parbox[0pt][1.6em][c]{0cm}{}Name & \multicolumn{2}{c}{Emission weighted center} & Redshift & $N_\mathrm{H}$ & $L_x$ &$R_\mathrm{kT}^\mathrm{max}$\\
		\parbox[0pt][1.6em][c]{0cm}{} & RA [h:m:s] & DEC [d:m:s] &   & $\SI{e22}{cm^{-2}}$ & $\SI{e43}{erg\,s^{-1}\,cm^{-2}}$ &arcmin\\
		\hline
		\endhead
		\hline
		\endfoot
\parbox[0pt][1.6em][c]{0cm}{} 2A0335 & 3:38:40.2941 & 9:58:4.7892 & 0.035 & 0.307 & $11.972^{+0.090}_{-0.090}$ & 10.7 \\
\parbox[0pt][1.6em][c]{0cm}{} A0085 & 0:41:50.3064 & -9:18:11.1240 & 0.056 & 0.031 & $24.472^{+0.150}_{-0.150}$ & 15.4 \\
\parbox[0pt][1.6em][c]{0cm}{} A0119 & 0:56:16.0400 & -1:15:20.6000 & 0.044 & 0.033 & $8.384^{+0.076}_{-0.076}$ & 10.0 \\
\parbox[0pt][1.6em][c]{0cm}{} A0133 & 1:2:43.1400 & -21:52:47.0280 & 0.057 & 0.017 & $7.360^{+0.062}_{-0.062}$ & 7.1 \\
\parbox[0pt][1.6em][c]{0cm}{} A0262 & 1:52:45.6096 & 36:9:3.9240 & 0.016 & 0.071 & $2.601^{+0.100}_{-0.100}$ & 17.1 \\
\parbox[0pt][1.6em][c]{0cm}{} A0399 & 2:57:51.6360 & 13:2:49.5240 & 0.071 & 0.171 & $17.675^{+0.956}_{-0.956}$ & 8.3 \\
\parbox[0pt][1.6em][c]{0cm}{} A0400 & 2:57:41.3496 & 6:1:36.9480 & 0.024 & 0.131 & $1.716^{+0.019}_{-0.019}$ & 7.6 \\
\parbox[0pt][1.6em][c]{0cm}{} A0401 & 2:58:57.2160 & 13:34:46.5600 & 0.075 & 0.152 & $31.383^{+0.338}_{-0.338}$ & 10.7 \\
\parbox[0pt][1.6em][c]{0cm}{} A0478 & 4:13:25.2960 & 10:27:57.9600 & 0.085 & 0.257 & $44.225^{+0.252}_{-0.252}$ & 9.9 \\
\parbox[0pt][1.6em][c]{0cm}{} A0496 & 4:33:37.8192 & -13:15:38.5560 & 0.033 & 0.060 & $9.591^{+0.071}_{-0.071}$ & 18.1 \\
\parbox[0pt][1.6em][c]{0cm}{} A0576 & 7:21:26.1144 & 55:45:34.2360 & 0.038 & 0.071 & $4.679^{+0.319}_{-0.319}$ & 14.8 \\
\parbox[0pt][1.6em][c]{0cm}{} A0754 & 9:9:18.1872 & -9:41:15.9360 & 0.053 & 0.058 & $9.975^{+0.162}_{-0.162}$ & 16.1 \\
\parbox[0pt][1.6em][c]{0cm}{} A1060 & 10:36:42.8592 & -27:31:42.0960 & 0.011 & 0.062 & $1.386^{+0.045}_{-0.045}$ & 13.3 \\
\parbox[0pt][1.6em][c]{0cm}{} A1367 & 11:44:44.5008 & 19:43:55.8120 & 0.022 & 0.020 & $3.015^{+0.023}_{-0.023}$ & 12.3 \\
\parbox[0pt][1.6em][c]{0cm}{} A1644 & 12:57:10.7352 & -17:24:10.2960 & 0.047 & 0.051 & $9.690^{+0.496}_{-0.496}$ & 12.9 \\
\parbox[0pt][1.6em][c]{0cm}{} A1650 & 12:58:41.8848 & -1:45:32.9040 & 0.085 & 0.014 & $18.270^{+1.205}_{-1.205}$ & 15.6 \\
\parbox[0pt][1.6em][c]{0cm}{} A1651 & 12:59:22.3512 & -4:11:46.6080 & 0.086 & 0.016 & $20.000^{+0.239}_{-0.239}$ & 7.0 \\
\parbox[0pt][1.6em][c]{0cm}{} A1656 & 12:59:45.3408 & 27:57:5.6160 & 0.023 & 0.009 & $19.791^{+0.272}_{-0.272}$ & 9.6 \\
\parbox[0pt][1.6em][c]{0cm}{} A1736 & 13:26:53.7120 & -27:10:35.4000 & 0.046 & 0.055 & $8.056^{+0.504}_{-0.504}$ & 8.0 \\
\parbox[0pt][1.6em][c]{0cm}{} A1795 & 13:48:52.7904 & 26:35:34.3680 & 0.062 & 0.012 & $25.310^{+0.086}_{-0.086}$ & 9.3 \\
\parbox[0pt][1.6em][c]{0cm}{} A2029 & 15:10:55.9896 & 5:44:33.6480 & 0.077 & 0.037 & $43.282^{+0.262}_{-0.262}$ & 12.6 \\
\parbox[0pt][1.6em][c]{0cm}{} A2052 & 15:16:43.5100 & 7:1:19.8000 & 0.035 & 0.027 & $6.123^{+0.063}_{-0.063}$ & 9.5 \\
\parbox[0pt][1.6em][c]{0cm}{} A2063 & 15:23:5.7720 & 8:36:25.3800 & 0.035 & 0.030 & $5.681^{+0.073}_{-0.073}$ & 14.0 \\
\parbox[0pt][1.6em][c]{0cm}{} A2065 & 15:22:29.0832 & 27:43:14.3760 & 0.072 & 0.030 & $13.901^{+0.848}_{-0.848}$ & 9.4 \\
\parbox[0pt][1.6em][c]{0cm}{} A2142 & 15:58:19.7760 & 27:14:0.9600 & 0.090 & 0.044 & $53.363^{+0.463}_{-0.463}$ & 9.3 \\
\parbox[0pt][1.6em][c]{0cm}{} A2147 & 16:2:16.3056 & 15:58:18.4440 & 0.035 & 0.034 & $7.297^{+0.237}_{-0.237}$ & 9.8 \\
\parbox[0pt][1.6em][c]{0cm}{} A2163 & 16:15:46.3920 & -6:8:36.9600 & 0.201 & 0.205 & $85.320^{+1.316}_{-1.316}$ & 7.0 \\
\parbox[0pt][1.6em][c]{0cm}{} A2199 & 16:28:37.1256 & 39:32:53.3040 & 0.030 & 0.009 & $10.412^{+0.190}_{-0.190}$ & 11.3 \\
\parbox[0pt][1.6em][c]{0cm}{} A2204 & 16:32:47.0592 & 5:34:32.0160 & 0.152 & 0.073 & $67.345^{+1.092}_{-1.092}$ & 6.6 \\
\parbox[0pt][1.6em][c]{0cm}{} A2244 & 17:2:41.9760 & 34:3:28.0800 & 0.097 & 0.020 & $21.171^{+0.454}_{-0.454}$ & 15.0 \\
\parbox[0pt][1.6em][c]{0cm}{} A2255 & 17:12:54.5376 & 64:3:51.4440 & 0.080 & 0.027 & $13.765^{+0.166}_{-0.166}$ & 10.6 \\
\parbox[0pt][1.6em][c]{0cm}{} A2256 & 17:3:52.4688 & 78:40:19.1280 & 0.060 & 0.050 & $23.304^{+0.320}_{-0.320}$ & 11.0 \\
\parbox[0pt][1.6em][c]{0cm}{} A2589 & 23:23:56.7720 & 16:46:33.2040 & 0.042 & 0.035 & $4.809^{+0.063}_{-0.063}$ & 5.9 \\
\parbox[0pt][1.6em][c]{0cm}{} A2597 & 23:25:20.0088 & -12:7:27.1920 & 0.085 & 0.028 & $17.206^{+0.210}_{-0.210}$ & 5.0 \\
\parbox[0pt][1.6em][c]{0cm}{} A2634 & 23:38:29.0448 & 27:1:51.6720 & 0.031 & 0.062 & $2.519^{+0.039}_{-0.039}$ & 14.5 \\
\parbox[0pt][1.6em][c]{0cm}{} A2657 & 23:44:56.7432 & 9:11:52.9440 & 0.040 & 0.084 & $4.426^{+0.039}_{-0.039}$ & 7.2 \\
\parbox[0pt][1.6em][c]{0cm}{} A3112 & 3:17:58.7136 & -44:14:8.3760 & 0.075 & 0.014 & $18.640^{+0.205}_{-0.205}$ & 6.9 \\
\parbox[0pt][1.6em][c]{0cm}{} A3158 & 3:42:53.5824 & -53:37:51.7080 & 0.059 & 0.014 & $14.095^{+0.214}_{-0.214}$ & 8.4 \\
\parbox[0pt][1.6em][c]{0cm}{} A3266 & 4:31:14.9088 & -61:26:54.1320 & 0.059 & 0.017 & $21.794^{+0.144}_{-0.144}$ & 16.0 \\
\parbox[0pt][1.6em][c]{0cm}{} A3376 & 6:2:10.1088 & -39:57:35.7480 & 0.045 & 0.058 & $5.435^{+0.079}_{-0.079}$ & 10.4 \\
\parbox[0pt][1.6em][c]{0cm}{} A3391 & 6:26:24.2232 & -53:41:24.0360 & 0.053 & 0.076 & $6.702^{+0.129}_{-0.129}$ & 10.0 \\
\parbox[0pt][1.6em][c]{0cm}{} A3395 & 6:26:46.0800 & -54:32:43.0800 & 0.050 & 0.098 & $5.327^{+0.204}_{-0.204}$ & 11.4 \\
\parbox[0pt][1.6em][c]{0cm}{} A3526 & 12:48:50.6424 & -41:18:15.2640 & 0.010 & 0.122 & $3.102^{+0.068}_{-0.068}$ & 22.1 \\
\parbox[0pt][1.6em][c]{0cm}{} A3558 & 13:28:0.4104 & -31:30:0.7920 & 0.048 & 0.049 & $16.539^{+0.076}_{-0.076}$ & 14.2 \\
\parbox[0pt][1.6em][c]{0cm}{} A3562 & 13:33:36.4872 & -31:40:25.5360 & 0.050 & 0.045 & $7.792^{+0.073}_{-0.073}$ & 7.7 \\
\parbox[0pt][1.6em][c]{0cm}{} A3571 & 13:47:27.8688 & -32:51:37.6560 & 0.037 & 0.051 & $20.330^{+0.151}_{-0.151}$ & 8.4 \\
\parbox[0pt][1.6em][c]{0cm}{} A3581 & 14:7:30.6264 & -27:0:47.3400 & 0.021 & 0.053 & $1.642^{+0.053}_{-0.053}$ & 7.3 \\
\parbox[0pt][1.6em][c]{0cm}{} A3667 & 20:12:40.7088 & -56:50:27.0600 & 0.056 & 0.052 & $24.059^{+0.164}_{-0.164}$ & 17.2 \\
\parbox[0pt][1.6em][c]{0cm}{} A4038 & 23:47:44.6520 & -28:8:42.4680 & 0.028 & 0.016 & $4.889^{+0.062}_{-0.062}$ & 8.9 \\
\parbox[0pt][1.6em][c]{0cm}{} A4059 & 23:57:1.6992 & -34:45:29.1240 & 0.046 & 0.013 & $7.180^{+0.095}_{-0.095}$ & 14.8 \\
\parbox[0pt][1.6em][c]{0cm}{} EXO0422 & 4:25:51.2232 & -8:33:40.3560 & 0.039 & 0.124 & $5.037^{+0.311}_{-0.311}$ & 5.4 \\
\parbox[0pt][1.6em][c]{0cm}{} HydraA & 9:18:5.9880 & -12:5:36.1680 & 0.054 & 0.055 & $14.824^{+0.083}_{-0.083}$ & 14.2 \\
\parbox[0pt][1.6em][c]{0cm}{} IIIZw54 & 3:41:18.7296 & 15:24:13.8960 & 0.031 & 0.267 & $2.077^{+0.160}_{-0.160}$ & 5.6 \\
\parbox[0pt][1.6em][c]{0cm}{} MKW3S & 15:21:50.2776 & 7:42:11.7720 & 0.045 & 0.030 & $7.162^{+0.074}_{-0.074}$ & 12.6 \\
\parbox[0pt][1.6em][c]{0cm}{} MKW4 & 12:4:27.6600 & 1:53:41.4960 & 0.020 & 0.019 & $0.975^{+0.016}_{-0.016}$ & 18.0 \\
\parbox[0pt][1.6em][c]{0cm}{} MKW8 & 14:40:42.1512 & 3:28:17.8680 & 0.027 & 0.027 & $1.974^{+0.165}_{-0.165}$ & 8.5 \\
\parbox[0pt][1.6em][c]{0cm}{} NGC1399 & 3:38:28.7904 & -35:27:4.5000 & 0.005 & 0.016 & $0.205^{+0.011}_{-0.011}$ & 15.9 \\
\parbox[0pt][1.6em][c]{0cm}{} NGC1550 & 4:19:38.0208 & 2:24:33.3720 & 0.012 & 0.162 & $0.754^{+0.041}_{-0.041}$ & 7.9 \\
\parbox[0pt][1.6em][c]{0cm}{} NGC4636 & 12:42:50.2656 & 2:41:30.6240 & 0.004 & 0.021 & $0.058^{+0.004}_{-0.004}$ & 10.5 \\
\parbox[0pt][1.6em][c]{0cm}{} NGC5044 & 13:15:23.7816 & -16:23:11.6880 & 0.009 & 0.062 & $0.483^{+0.002}_{-0.002}$ & 16.0 \\
\parbox[0pt][1.6em][c]{0cm}{} NGC507 & 1:23:38.5680 & 33:15:2.0880 & 0.017 & 0.064 & $0.617^{+0.008}_{-0.008}$ & 7.4 \\
\parbox[0pt][1.6em][c]{0cm}{} RXCJ1504 & 15:4:7.8024 & -2:48:10.2892 & 0.215 & 0.084 & $104.370^{+1.960}_{-1.960}$ & 5.0 \\
\parbox[0pt][1.6em][c]{0cm}{} S1101 & 23:13:58.3128 & -42:43:36.1200 & 0.058 & 0.012 & $8.992^{+0.082}_{-0.082}$ & 8.5 \\
\parbox[0pt][1.6em][c]{0cm}{} ZwCl1215 & 12:17:40.6368 & 3:39:29.6640 & 0.075 & 0.019 & $13.100^{+0.166}_{-0.166}$ & 6.9 \\
\end{longtable}
}

\clearpage
{\small
\begin{longtable}{cccccccc}
\caption{Masses for HIFLUGCS clusters with different extrapolation methods and for two overdensities using $h = 0.7$ and $\Omega_\mathrm{m}=0.3$.}\\
\hline
\parbox[0pt][1.6em][c]{0cm}{}Name & $M_{500}^\mathrm{NFW\,Freeze}$ & $M_{500}^\mathrm{kT\,extrp}$  & $M_{500}^\mathrm{NFW\,Hudson}$  & $M_{500}^\mathrm{NFW\,All}$ & $M_{200}^\mathrm{NFW\,Freeze}$ & $M_{200}^\mathrm{kT\,extrp}$ & $M_{500}^\mathrm{Planck\,SZ}$\\
\hline
\endfirsthead
\caption{continued.}\\
\hline
\parbox[0pt][1.6em][c]{0cm}{}Name & $M_{500}^\mathrm{NFW\,Freeze}$ & $M_{500}^\mathrm{kT\,extrp}$  & $M_{500}^\mathrm{NFW\,Hudson}$  & $M_{500}^\mathrm{NFW\,All}$ & $M_{200}^\mathrm{NFW\,Freeze}$ & $M_{200}^\mathrm{kT\,extrp}$ & $M_{500}^\mathrm{Planck\,SZ}$\\
\hline
\endhead
\hline
\endfoot
 %\parbox[0pt][1.6em][c]{0cm}{} & $\SI{e14}{M_\odot}$ & $\SI{e14}{M_\odot}$  & $\SI{e14}{M_\odot}$  & $\SI{e14}{M_\odot}$ & $\SI{e14}{M_\odot}$ & $\SI{e14}{M_\odot}$  & $\SI{e14}{M_\odot}$\\
 %\parbox[0pt][1.6em][c]{0cm}{}Name & $M_{500}^\mathrm{NFW\,Freeze}$ & $M_{500}^\mathrm{kT\,extrapolate}$  & $M_{500}^\mathrm{NFW\,Hudson}$  & $M_{500}^\mathrm{NFW\,All}$ & $M_{200}^\mathrm{NFW\,Freeze}$ & $M_{200}^\mathrm{kT\,extrapolate}$ & $M_{500}^\mathrm{Planck\,SZ}$\\
\parbox[0pt][1.6em][c]{0cm}{} & $\SI{e14}{M_\odot}$ & $\SI{e14}{M_\odot}$  & $\SI{e14}{M_\odot}$  & $\SI{e14}{M_\odot}$ & $\SI{e14}{M_\odot}$ & $\SI{e14}{M_\odot}$  & $\SI{e14}{M_\odot}$\\
\parbox[0pt][1.6em][c]{0cm}{} 2A0335 & $0.934^{+0.054}_{-0.048}$ & $0.852^{+0.029}_{-0.036}$ & $1.177^{+0.018}_{-0.018}$ & $2.246^{+0.032}_{-0.033}$ & $1.664^{+0.121}_{-0.111}$ & $1.062^{+0.053}_{-0.062}$ & $1.588^{+0.108}_{-0.122}$ \\
\parbox[0pt][1.6em][c]{0cm}{} A0085 & $3.276^{+0.163}_{-0.162}$ & $3.946^{+0.179}_{-0.218}$ & $4.956^{+0.041}_{-0.040}$ & $6.537^{+0.058}_{-0.054}$ & $5.093^{+0.327}_{-0.296}$ & $3.494^{+0.208}_{-0.244}$ & $3.443^{+0.089}_{-0.102}$ \\
\parbox[0pt][1.6em][c]{0cm}{} A0119 & $2.044^{+0.409}_{-0.388}$ & $8.141^{+7.152}_{-4.696}$ & $20.735^{+3.455}_{-5.327}$ & $20.980^{+3.678}_{-5.411}$ & $3.956^{+1.018}_{-0.898}$ & $2.378^{+0.713}_{-1.304}$ & $2.401^{+0.090}_{-0.152}$ \\
\parbox[0pt][1.6em][c]{0cm}{} A0133 & $1.676^{+0.170}_{-0.152}$ & $2.063^{+0.378}_{-0.169}$ & $1.997^{+0.138}_{-0.123}$ & $2.074^{+0.149}_{-0.135}$ & $3.348^{+0.437}_{-0.406}$ & $2.105^{+0.541}_{-0.659}$ & $2.112^{+0.125}_{-0.115}$ \\
\parbox[0pt][1.6em][c]{0cm}{} A0262 & $0.467^{+0.025}_{-0.025}$ & $0.485^{+0.021}_{-0.021}$ & $0.789^{+0.020}_{-0.020}$ & $0.838^{+0.023}_{-0.021}$ & $0.824^{+0.053}_{-0.054}$ & $0.520^{+0.041}_{-0.034}$ & $0.785^{+0.075}_{-0.080}$ \\
\parbox[0pt][1.6em][c]{0cm}{} A0399 & $2.314^{+0.135}_{-0.142}$ & $4.139^{+1.476}_{-0.946}$ & $2.318^{+0.123}_{-0.117}$ & $2.165^{+0.124}_{-0.111}$ & $4.026^{+0.319}_{-0.296}$ & $16.629^{+4.071}_{-4.236}$ & $3.668^{+0.204}_{-0.162}$ \\
\parbox[0pt][1.6em][c]{0cm}{} A0400 & $0.462^{+0.058}_{-0.056}$ & $7.403^{+4.658}_{-4.272}$ & $6.498^{+4.941}_{-2.834}$ & $7.204^{+4.770}_{-3.687}$ & $1.043^{+0.184}_{-0.158}$ & $0.843^{+0.204}_{-0.245}$ & $-$ \\
\parbox[0pt][1.6em][c]{0cm}{} A0401 & $3.578^{+0.125}_{-0.121}$ & $3.483^{+0.111}_{-0.142}$ & $4.720^{+0.074}_{-0.069}$ & $4.911^{+0.083}_{-0.083}$ & $5.922^{+0.278}_{-0.264}$ & $4.792^{+0.146}_{-0.400}$ & $4.722^{+0.155}_{-0.121}$ \\
\parbox[0pt][1.6em][c]{0cm}{} A0478 & $5.133^{+1.729}_{-1.478}$ & $5.521^{+0.622}_{-0.360}$ & $18.025^{+1.349}_{-1.194}$ & $28.216^{+2.555}_{-2.158}$ & $6.082^{+3.029}_{-2.398}$ & $1.589^{+0.464}_{-0.243}$ & $4.867^{+0.198}_{-0.198}$ \\
\parbox[0pt][1.6em][c]{0cm}{} A0496 & $1.609^{+0.138}_{-0.144}$ & $1.710^{+0.117}_{-0.185}$ & $2.088^{+0.046}_{-0.042}$ & $2.217^{+0.040}_{-0.035}$ & $2.561^{+0.293}_{-0.260}$ & $1.971^{+0.168}_{-0.126}$ & $1.900^{+0.102}_{-0.108}$ \\
\parbox[0pt][1.6em][c]{0cm}{} A0576 & $1.475^{+0.360}_{-0.323}$ & $1.739^{+0.567}_{-0.436}$ & $5.962^{+2.867}_{-1.531}$ & $19.255^{+5.308}_{-5.860}$ & $2.372^{+0.680}_{-0.594}$ & $3.175^{+1.028}_{-1.034}$ & $1.494^{+0.103}_{-0.107}$ \\
\parbox[0pt][1.6em][c]{0cm}{} A0754 & $12.166^{+3.128}_{-2.898}$ & $11.269^{+3.486}_{-1.702}$ & $188.530^{+3.126}_{-3.303}$ & $189.006^{+3.133}_{-3.418}$ & $24.998^{+8.809}_{-7.010}$ & $13.164^{+1.515}_{-1.273}$ & $4.798^{+0.087}_{-0.089}$ \\
\parbox[0pt][1.6em][c]{0cm}{} A1060 & $0.912^{+0.089}_{-0.083}$ & $1.265^{+0.109}_{-0.117}$ & $1.141^{+0.094}_{-0.078}$ & $1.144^{+0.097}_{-0.080}$ & $2.966^{+0.486}_{-0.417}$ & $0.894^{+0.446}_{-0.423}$ & $-$ \\
\parbox[0pt][1.6em][c]{0cm}{} A1367 & $0.564^{+0.090}_{-0.086}$ & $0.624^{+0.034}_{-0.034}$ & $0.692^{+0.031}_{-0.030}$ & $0.691^{+0.034}_{-0.030}$ & $0.878^{+0.193}_{-0.168}$ & $0.832^{+0.092}_{-0.386}$ & $1.167^{+0.075}_{-0.071}$ \\
\parbox[0pt][1.6em][c]{0cm}{} A1644 & $1.042^{+0.120}_{-0.111}$ & $1.077^{+0.052}_{-0.052}$ & $1.094^{+0.047}_{-0.046}$ & $1.179^{+0.048}_{-0.047}$ & $1.549^{+0.204}_{-0.197}$ & $1.717^{+0.077}_{-0.072}$ & $2.582^{+0.110}_{-0.116}$ \\
\parbox[0pt][1.6em][c]{0cm}{} A1650 & $4.039^{+0.156}_{-0.157}$ & $5.159^{+0.072}_{-0.062}$ & $5.208^{+0.059}_{-0.054}$ & $5.548^{+0.069}_{-0.066}$ & $6.595^{+0.346}_{-0.311}$ & $1.701^{+0.234}_{-0.230}$ & $3.113^{+0.169}_{-0.173}$ \\
\parbox[0pt][1.6em][c]{0cm}{} A1651 & $3.623^{+0.268}_{-0.271}$ & $4.648^{+0.584}_{-0.546}$ & $5.061^{+0.471}_{-0.410}$ & $5.872^{+0.864}_{-0.744}$ & $7.026^{+0.702}_{-0.657}$ & $6.602^{+0.788}_{-1.681}$ & $3.552^{+0.124}_{-0.140}$ \\
\parbox[0pt][1.6em][c]{0cm}{} A1656 & $3.754^{+0.271}_{-0.261}$ & $3.409^{+0.279}_{-0.294}$ & $12.742^{+0.590}_{-0.566}$ & $13.776^{+0.796}_{-0.607}$ & $13.158^{+2.096}_{-1.711}$ & $2.163^{+0.636}_{-0.800}$ & $5.016^{+0.047}_{-0.075}$ \\
\parbox[0pt][1.6em][c]{0cm}{} A1736 & $0.823^{+0.062}_{-0.063}$ & $2.659^{+1.383}_{-0.910}$ & $2.933^{+1.283}_{-1.011}$ & $2.299^{+1.686}_{-0.931}$ & $1.454^{+0.146}_{-0.141}$ & $3.077^{+0.512}_{-0.624}$ & $2.022^{+0.117}_{-0.117}$ \\
\parbox[0pt][1.6em][c]{0cm}{} A1795 & $2.797^{+0.073}_{-0.073}$ & $2.738^{+0.071}_{-0.065}$ & $2.961^{+0.049}_{-0.047}$ & $4.226^{+0.067}_{-0.064}$ & $5.168^{+0.224}_{-0.214}$ & $4.420^{+0.221}_{-0.283}$ & $3.126^{+0.099}_{-0.099}$ \\
\parbox[0pt][1.6em][c]{0cm}{} A2029 & $4.214^{+0.400}_{-0.406}$ & $4.194^{+0.252}_{-0.169}$ & $7.400^{+0.093}_{-0.093}$ & $7.869^{+0.096}_{-0.099}$ & $6.592^{+0.762}_{-0.757}$ & $3.664^{+0.152}_{-0.177}$ & $4.928^{+0.139}_{-0.135}$ \\
\parbox[0pt][1.6em][c]{0cm}{} A2052 & $0.872^{+0.027}_{-0.028}$ & $0.697^{+0.033}_{-0.017}$ & $1.670^{+0.026}_{-0.025}$ & $1.787^{+0.026}_{-0.026}$ & $1.629^{+0.082}_{-0.075}$ & $0.412^{+0.149}_{-0.142}$ & $-$ \\
\parbox[0pt][1.6em][c]{0cm}{} A2063 & $1.197^{+0.285}_{-0.277}$ & $2.142^{+0.249}_{-0.258}$ & $2.253^{+0.160}_{-0.137}$ & $2.373^{+0.174}_{-0.177}$ & $1.974^{+0.582}_{-0.499}$ & $0.418^{+0.192}_{-0.140}$ & $1.329^{+0.122}_{-0.126}$ \\
\parbox[0pt][1.6em][c]{0cm}{} A2065 & $2.784^{+0.171}_{-0.159}$ & $3.359^{+0.258}_{-0.281}$ & $3.830^{+0.214}_{-0.203}$ & $4.217^{+0.320}_{-0.277}$ & $4.729^{+0.372}_{-0.353}$ & $5.451^{+0.412}_{-0.546}$ & $2.857^{+0.130}_{-0.133}$ \\
\parbox[0pt][1.6em][c]{0cm}{} A2142 & $10.012^{+1.297}_{-1.307}$ & $23.180^{+22.428}_{-9.800}$ & $58.743^{+20.821}_{-18.587}$ & $7.844^{+0.244}_{-0.241}$ & $28.030^{+10.835}_{-8.686}$ & $7.928^{+1.755}_{-2.120}$ & $6.140^{+0.130}_{-0.147}$ \\
\parbox[0pt][1.6em][c]{0cm}{} A2147 & $0.673^{+0.068}_{-0.068}$ & $2.115^{+0.871}_{-0.755}$ & $1.447^{+1.010}_{-0.461}$ & $0.693^{+0.423}_{-0.191}$ & $1.172^{+0.153}_{-0.141}$ & $5.164^{+0.798}_{-0.791}$ & $2.469^{+0.036}_{-0.041}$ \\
\parbox[0pt][1.6em][c]{0cm}{} A2163 & $19.186^{+3.191}_{-2.942}$ & $27.380^{+13.377}_{-9.066}$ & $23.019^{+2.046}_{-1.824}$ & $22.208^{+2.214}_{-1.772}$ & $37.413^{+8.051}_{-6.685}$ & $10.768^{+6.350}_{-3.668}$ & $11.282^{+0.208}_{-0.205}$ \\
\parbox[0pt][1.6em][c]{0cm}{} A2199 & $2.185^{+0.105}_{-0.100}$ & $2.308^{+0.332}_{-0.357}$ & $10.294^{+0.541}_{-0.501}$ & $15.778^{+1.292}_{-1.416}$ & $5.070^{+0.417}_{-0.394}$ & $1.397^{+0.210}_{-0.268}$ & $2.011^{+0.087}_{-0.085}$ \\
\parbox[0pt][1.6em][c]{0cm}{} A2204 & $4.982^{+0.521}_{-0.472}$ & $4.878^{+0.406}_{-0.268}$ & $6.723^{+0.288}_{-0.271}$ & $7.671^{+0.295}_{-0.276}$ & $8.508^{+1.119}_{-1.053}$ & $3.583^{+0.571}_{-0.607}$ & $5.452^{+0.210}_{-0.211}$ \\
\parbox[0pt][1.6em][c]{0cm}{} A2244 & $1.648^{+0.536}_{-0.488}$ & $2.600^{+0.143}_{-0.130}$ & $3.464^{+0.101}_{-0.104}$ & $3.527^{+0.100}_{-0.102}$ & $2.084^{+0.712}_{-0.645}$ & $2.595^{+0.355}_{-0.701}$ & $3.066^{+0.133}_{-0.151}$ \\
\parbox[0pt][1.6em][c]{0cm}{} A2255 & $4.007^{+0.440}_{-0.415}$ & $5.360^{+1.445}_{-1.063}$ & $11.473^{+2.651}_{-2.022}$ & $16.336^{+1.647}_{-2.603}$ & $6.603^{+0.898}_{-0.796}$ & $5.264^{+0.975}_{-1.107}$ & $3.768^{+0.041}_{-0.043}$ \\
\parbox[0pt][1.6em][c]{0cm}{} A2256 & $3.461^{+0.122}_{-0.117}$ & $4.959^{+0.163}_{-0.181}$ & $5.022^{+0.124}_{-0.133}$ & $7.846^{+0.197}_{-0.184}$ & $10.172^{+2.962}_{-2.383}$ & $3.424^{+1.386}_{-1.528}$ & $4.348^{+0.071}_{-0.064}$ \\
\parbox[0pt][1.6em][c]{0cm}{} A2589 & $1.197^{+0.079}_{-0.075}$ & $1.742^{+0.323}_{-0.408}$ & $2.858^{+0.227}_{-0.216}$ & $3.168^{+0.290}_{-0.295}$ & $2.994^{+0.303}_{-0.290}$ & $2.058^{+0.291}_{-0.543}$ & $1.235^{+0.158}_{-0.169}$ \\
\parbox[0pt][1.6em][c]{0cm}{} A2597 & $1.557^{+0.041}_{-0.040}$ & $1.670^{+0.048}_{-0.055}$ & $1.692^{+0.037}_{-0.035}$ & $1.619^{+0.034}_{-0.034}$ & $4.259^{+0.198}_{-0.169}$ & $4.247^{+0.027}_{-0.030}$ & $1.738^{+0.169}_{-0.180}$ \\
\parbox[0pt][1.6em][c]{0cm}{} A2634 & $0.949^{+0.231}_{-0.208}$ & $1.526^{+0.190}_{-0.183}$ & $1.860^{+0.213}_{-0.166}$ & $1.911^{+0.203}_{-0.186}$ & $1.554^{+0.452}_{-0.386}$ & $0.432^{+0.227}_{-0.103}$ & $1.058^{+0.088}_{-0.095}$ \\
\parbox[0pt][1.6em][c]{0cm}{} A2657 & $0.843^{+0.082}_{-0.084}$ & $2.893^{+3.306}_{-1.164}$ & $0.691^{+0.084}_{-0.071}$ & $0.511^{+0.059}_{-0.052}$ & $1.691^{+0.225}_{-0.192}$ & $4.062^{+1.161}_{-1.061}$ & $1.030^{+0.126}_{-0.132}$ \\
\parbox[0pt][1.6em][c]{0cm}{} A3112 & $2.864^{+0.333}_{-0.323}$ & $4.027^{+0.654}_{-1.086}$ & $2.972^{+0.181}_{-0.166}$ & $3.418^{+0.199}_{-0.181}$ & $5.733^{+0.901}_{-0.857}$ & $1.679^{+0.379}_{-0.435}$ & $2.049^{+0.120}_{-0.138}$ \\
\parbox[0pt][1.6em][c]{0cm}{} A3158 & $2.129^{+0.091}_{-0.094}$ & $2.417^{+0.090}_{-0.090}$ & $2.715^{+0.072}_{-0.069}$ & $2.754^{+0.088}_{-0.081}$ & $3.971^{+0.246}_{-0.237}$ & $3.970^{+0.236}_{-0.277}$ & $2.957^{+0.103}_{-0.093}$ \\
\parbox[0pt][1.6em][c]{0cm}{} A3266 & $7.704^{+0.744}_{-0.704}$ & $19.488^{+4.142}_{-3.850}$ & $10.909^{+2.352}_{-1.651}$ & $8.334^{+1.882}_{-1.362}$ & $13.440^{+1.554}_{-1.516}$ & $13.080^{+4.147}_{-3.547}$ & $4.646^{+0.080}_{-0.081}$ \\
\parbox[0pt][1.6em][c]{0cm}{} A3376 & $1.461^{+0.208}_{-0.196}$ & $1.945^{+0.345}_{-0.419}$ & $2.248^{+0.248}_{-0.213}$ & $2.320^{+0.292}_{-0.239}$ & $2.582^{+0.453}_{-0.430}$ & $2.055^{+0.195}_{-0.208}$ & $1.669^{+0.111}_{-0.114}$ \\
\parbox[0pt][1.6em][c]{0cm}{} A3391 & $2.441^{+0.341}_{-0.350}$ & $4.088^{+1.370}_{-0.946}$ & $4.306^{+1.155}_{-0.853}$ & $6.246^{+4.352}_{-2.053}$ & $4.514^{+0.860}_{-0.789}$ & $3.167^{+0.692}_{-0.705}$ & $1.973^{+0.100}_{-0.094}$ \\
\parbox[0pt][1.6em][c]{0cm}{} A3395 & $1.398^{+0.131}_{-0.134}$ & $1.429^{+0.163}_{-0.131}$ & $1.492^{+0.122}_{-0.110}$ & $1.499^{+0.147}_{-0.123}$ & $2.236^{+0.256}_{-0.243}$ & $2.755^{+0.276}_{-0.250}$ & $2.248^{+0.129}_{-0.135}$ \\
\parbox[0pt][1.6em][c]{0cm}{} A3526 & $0.962^{+0.058}_{-0.055}$ & $0.996^{+0.043}_{-0.041}$ & $1.765^{+0.031}_{-0.028}$ & $1.965^{+0.020}_{-0.020}$ & $2.266^{+0.220}_{-0.194}$ & $0.865^{+0.085}_{-0.098}$ & $0.927^{+0.057}_{-0.061}$ \\
\parbox[0pt][1.6em][c]{0cm}{} A3558 & $2.891^{+0.661}_{-0.582}$ & $2.939^{+0.630}_{-0.312}$ & $13.372^{+2.764}_{-2.255}$ & $50.782^{+26.221}_{-17.731}$ & $4.886^{+1.296}_{-1.146}$ & $2.063^{+0.463}_{-0.570}$ & $3.350^{+0.116}_{-0.122}$ \\
\parbox[0pt][1.6em][c]{0cm}{} A3562 & $1.716^{+0.324}_{-0.321}$ & $2.741^{+0.588}_{-0.607}$ & $2.909^{+0.567}_{-0.426}$ & $3.145^{+0.597}_{-0.538}$ & $3.534^{+0.849}_{-0.747}$ & $1.047^{+0.446}_{-0.597}$ & $1.710^{+0.149}_{-0.171}$ \\
\parbox[0pt][1.6em][c]{0cm}{} A3571 & $3.666^{+0.333}_{-0.322}$ & $11.626^{+8.100}_{-4.137}$ & $13.024^{+2.625}_{-2.042}$ & $13.671^{+3.298}_{-2.119}$ & $10.904^{+1.825}_{-1.476}$ & $3.345^{+1.733}_{-1.672}$ & $3.239^{+0.097}_{-0.104}$ \\
\parbox[0pt][1.6em][c]{0cm}{} A3581 & $0.719^{+0.090}_{-0.084}$ & $1.876^{+0.814}_{-0.523}$ & $0.607^{+0.021}_{-0.018}$ & $0.728^{+0.023}_{-0.022}$ & $2.114^{+0.399}_{-0.364}$ & $0.235^{+0.026}_{-0.046}$ & $-$ \\
\parbox[0pt][1.6em][c]{0cm}{} A3667 & $3.532^{+0.295}_{-0.273}$ & $3.545^{+0.234}_{-0.183}$ & $5.614^{+0.132}_{-0.125}$ & $5.752^{+0.129}_{-0.130}$ & $5.301^{+0.516}_{-0.479}$ & $4.503^{+0.359}_{-0.205}$ & $4.925^{+0.036}_{-0.037}$ \\
\parbox[0pt][1.6em][c]{0cm}{} A4038 & $0.981^{+0.126}_{-0.117}$ & $0.934^{+0.253}_{-0.268}$ & $1.351^{+0.055}_{-0.050}$ & $1.362^{+0.055}_{-0.050}$ & $2.231^{+0.405}_{-0.351}$ & $0.433^{+0.183}_{-0.302}$ & $1.036^{+0.080}_{-0.075}$ \\
\parbox[0pt][1.6em][c]{0cm}{} A4059 & $1.468^{+0.091}_{-0.088}$ & $1.452^{+0.078}_{-0.065}$ & $2.088^{+0.040}_{-0.040}$ & $2.168^{+0.043}_{-0.042}$ & $2.194^{+0.153}_{-0.152}$ & $1.511^{+0.113}_{-0.103}$ & $1.728^{+0.114}_{-0.104}$ \\
\parbox[0pt][1.6em][c]{0cm}{} EXO0422 & $0.950^{+0.160}_{-0.150}$ & $0.917^{+0.268}_{-0.187}$ & $1.492^{+0.267}_{-0.225}$ & $2.072^{+0.611}_{-0.439}$ & $2.492^{+0.615}_{-0.540}$ & $1.227^{+0.389}_{-0.637}$ & $-$ \\
\parbox[0pt][1.6em][c]{0cm}{} HydraA & $1.771^{+0.210}_{-0.201}$ & $1.939^{+0.112}_{-0.105}$ & $4.264^{+0.184}_{-0.198}$ & $4.076^{+0.129}_{-0.128}$ & $2.604^{+0.366}_{-0.323}$ & $1.430^{+0.154}_{-0.155}$ & $-$ \\
\parbox[0pt][1.6em][c]{0cm}{} IIIZw54 & $0.508^{+0.031}_{-0.030}$ & $0.615^{+0.099}_{-0.082}$ & $0.756^{+0.081}_{-0.069}$ & $0.918^{+0.143}_{-0.127}$ & $1.229^{+0.115}_{-0.107}$ & $0.690^{+0.150}_{-0.186}$ & $-$ \\
\parbox[0pt][1.6em][c]{0cm}{} MKW3S & $1.088^{+0.103}_{-0.108}$ & $1.181^{+0.077}_{-0.076}$ & $2.027^{+0.048}_{-0.043}$ & $2.592^{+0.060}_{-0.066}$ & $1.659^{+0.195}_{-0.182}$ & $0.722^{+0.110}_{-0.217}$ & $1.309^{+0.141}_{-0.148}$ \\
\parbox[0pt][1.6em][c]{0cm}{} MKW4 & $0.386^{+0.035}_{-0.035}$ & $0.402^{+0.033}_{-0.030}$ & $0.522^{+0.026}_{-0.023}$ & $0.419^{+0.022}_{-0.019}$ & $0.589^{+0.066}_{-0.061}$ & $0.483^{+0.065}_{-0.092}$ & $-$ \\
\parbox[0pt][1.6em][c]{0cm}{} MKW8 & $0.699^{+0.112}_{-0.104}$ & $1.180^{+0.511}_{-0.299}$ & $1.014^{+0.318}_{-0.204}$ & $1.173^{+0.908}_{-0.347}$ & $1.549^{+0.321}_{-0.292}$ & $1.097^{+0.413}_{-0.376}$ & $0.772^{+0.127}_{-0.139}$ \\
\parbox[0pt][1.6em][c]{0cm}{} NGC1399 & $0.161^{+0.009}_{-0.009}$ & $0.110^{+0.008}_{-0.010}$ & $0.114^{+0.004}_{-0.003}$ & $0.115^{+0.003}_{-0.003}$ & $0.506^{+0.058}_{-0.050}$ & $0.184^{+0.025}_{-0.004}$ & $-$ \\
\parbox[0pt][1.6em][c]{0cm}{} NGC1550 & $0.239^{+0.019}_{-0.019}$ & $0.186^{+0.035}_{-0.035}$ & $1.445^{+0.239}_{-0.205}$ & $1.141^{+0.131}_{-0.112}$ & $0.693^{+0.093}_{-0.081}$ & $0.402^{+0.043}_{-0.079}$ & $-$ \\
\parbox[0pt][1.6em][c]{0cm}{} NGC4636 & $0.108^{+0.007}_{-0.007}$ & $0.048^{+0.002}_{-0.002}$ & $0.048^{+0.002}_{-0.002}$ & $0.033^{+0.001}_{-0.001}$ & $0.534^{+0.077}_{-0.064}$ & $0.119^{+0.006}_{-0.020}$ & $-$ \\
\parbox[0pt][1.6em][c]{0cm}{} NGC5044 & $0.199^{+0.007}_{-0.006}$ & $0.161^{+0.011}_{-0.010}$ & $0.246^{+0.007}_{-0.008}$ & $0.155^{+0.002}_{-0.002}$ & $0.411^{+0.022}_{-0.021}$ & $0.265^{+0.010}_{-0.011}$ & $-$ \\
\parbox[0pt][1.6em][c]{0cm}{} NGC507 & $0.307^{+0.026}_{-0.026}$ & $0.696^{+0.085}_{-0.068}$ & $0.713^{+0.085}_{-0.076}$ & $0.712^{+0.072}_{-0.070}$ & $0.815^{+0.100}_{-0.096}$ & $0.185^{+0.040}_{-0.028}$ & $-$ \\
\parbox[0pt][1.6em][c]{0cm}{} RXCJ1504 & $6.135^{+0.953}_{-0.968}$ & $7.044^{+0.739}_{-0.819}$ & $6.828^{+0.963}_{-0.826}$ & $7.147^{+0.498}_{-0.463}$ & $10.844^{+2.272}_{-1.899}$ & $6.632^{+1.741}_{-2.202}$ & $4.666^{+0.336}_{-0.338}$ \\
\parbox[0pt][1.6em][c]{0cm}{} S1101 & $0.975^{+0.108}_{-0.106}$ & $0.945^{+0.123}_{-0.063}$ & $1.110^{+0.023}_{-0.022}$ & $1.064^{+0.015}_{-0.016}$ & $1.568^{+0.209}_{-0.194}$ & $0.369^{+0.154}_{-0.126}$ & $1.025^{+0.154}_{-0.157}$ \\
\parbox[0pt][1.6em][c]{0cm}{} ZwCl1215 & $3.191^{+0.358}_{-0.343}$ & $5.259^{+1.488}_{-1.183}$ & $6.757^{+1.518}_{-1.121}$ & $10.779^{+10.573}_{-3.542}$ & $6.538^{+0.956}_{-0.925}$ & $5.579^{+0.936}_{-1.958}$ & $-$ \\
\end{longtable}
}

\clearpage
\section{Temperature and mass profiles (online only)}
\clearpage
\begin{figure}
	\centering
	\includegraphics[width=0.45\textwidth]{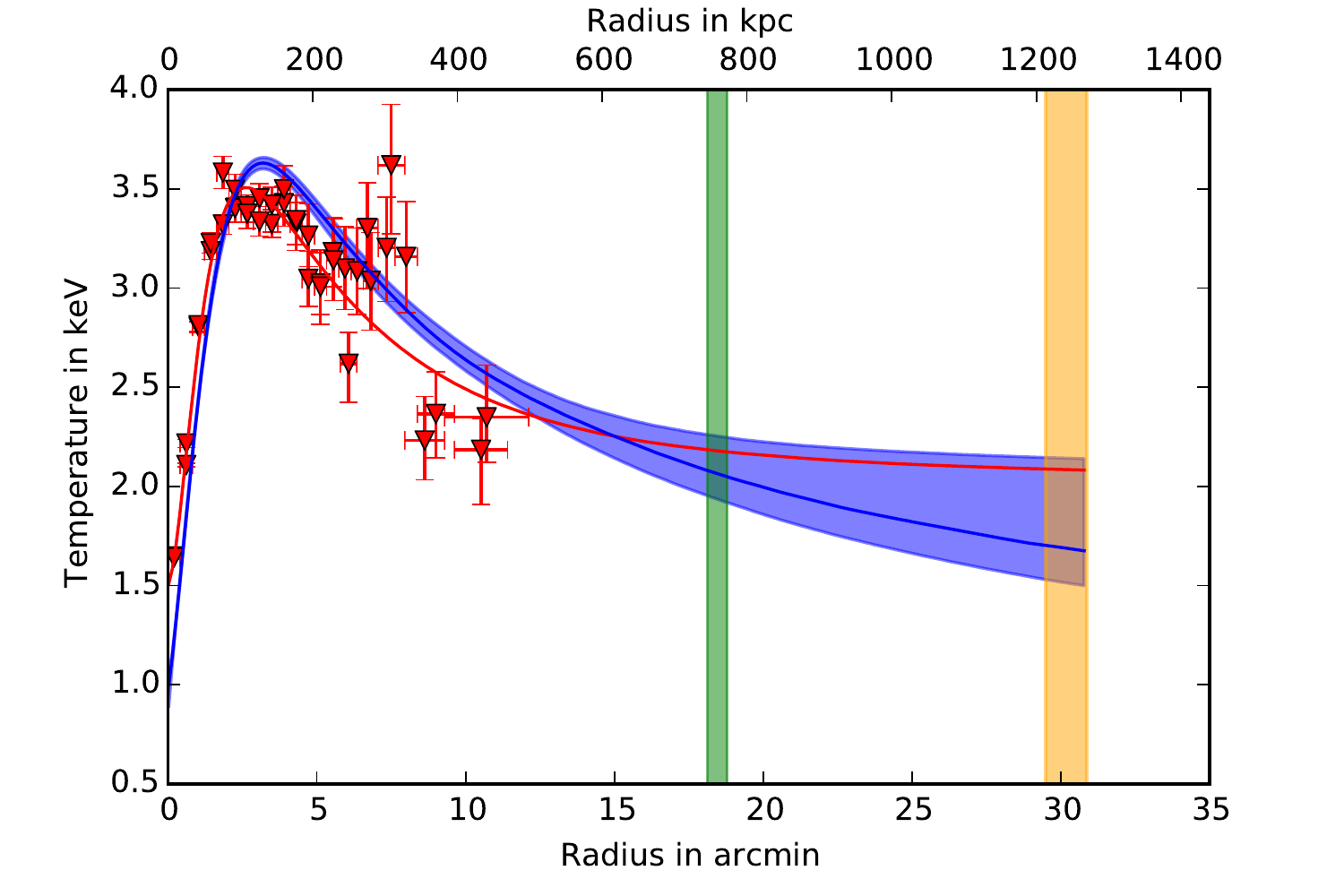}
	\includegraphics[width=0.45\textwidth]{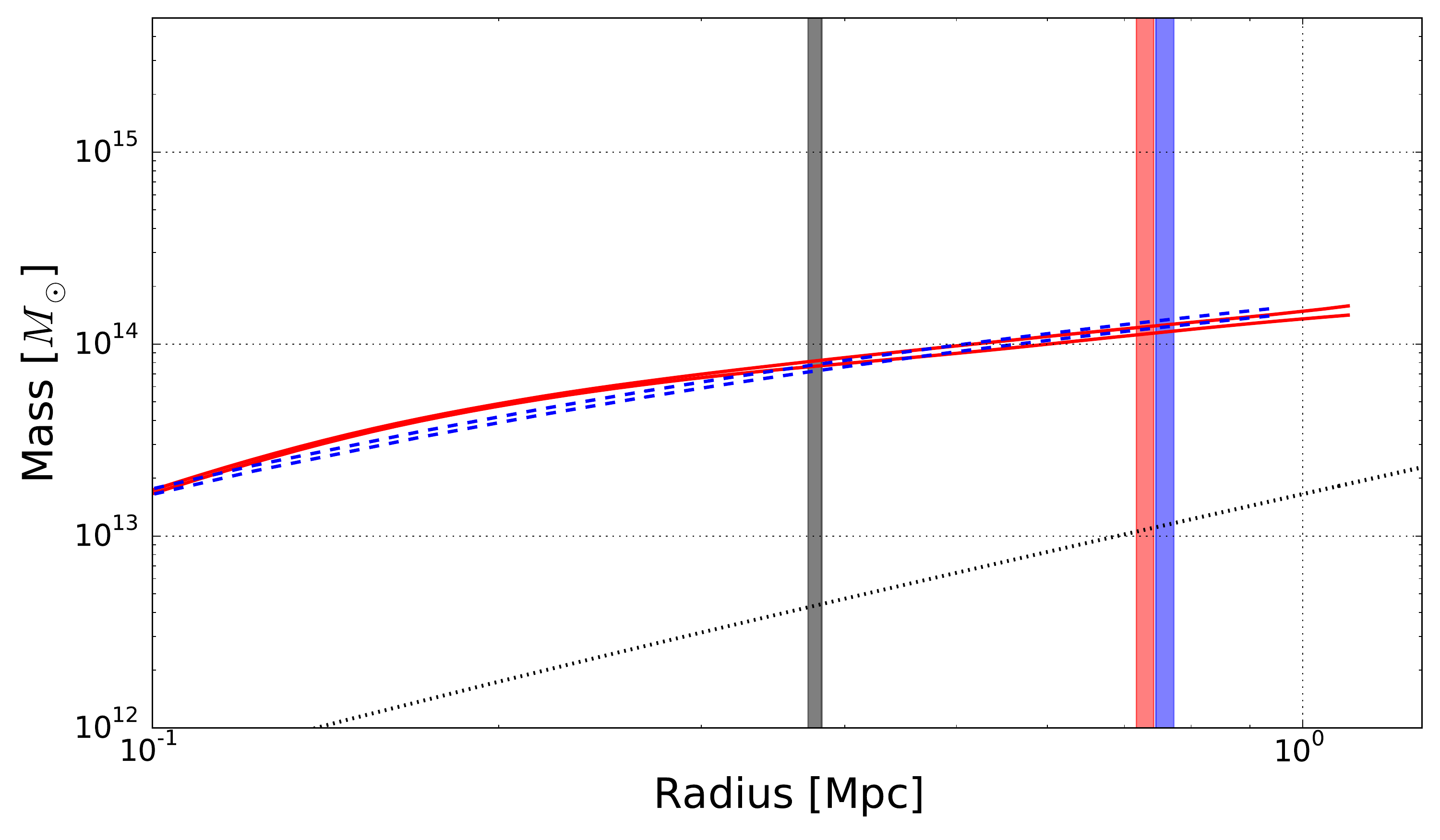}
	\caption{Temperature and mass profiles for 2A0335. The left panel shows the measured temperatures  (red datapoints) and the best fit model (red line), while the deprojected temperature is represented by the blue line. The green and orange regions show $r_{500}$ and $r_{200}$ (from the NFW-Freeze model) estimates, respectively. The right panel shows the total mass of the temperature profile extrapolation (red), and the NFW-Freeze model (blue). The black lines show the gas mass estimates. The red and blue vertical regions represent the corresponding $r_{500}$, the black region $r_{2500}$ from the temperature extrapolation method. All regions are 68.3\% confidence levels.}
	\label{fig:app_2A0335}
\end{figure}
\begin{figure}
	\centering
	\includegraphics[width=0.45\textwidth]{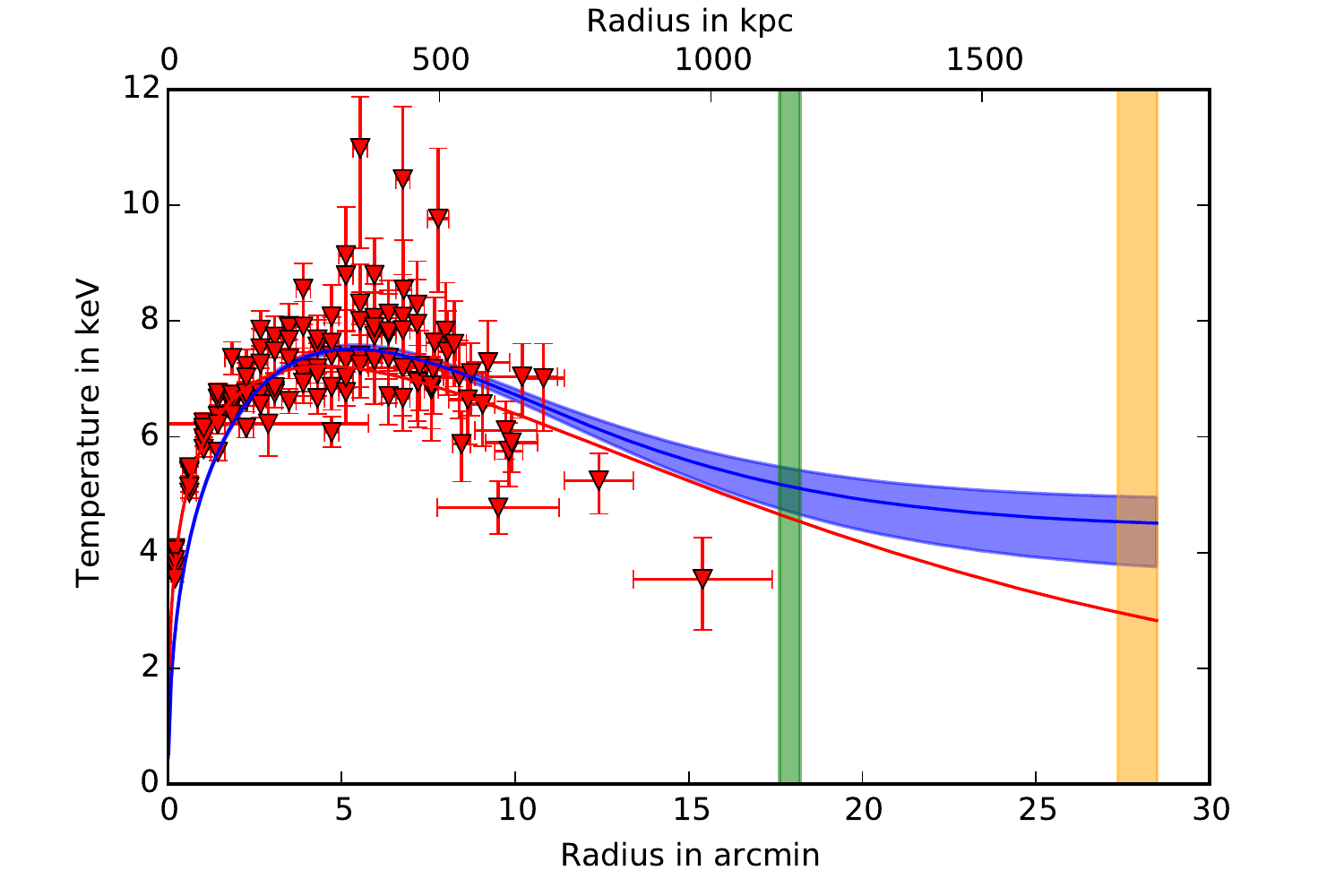}
	\includegraphics[width=0.45\textwidth]{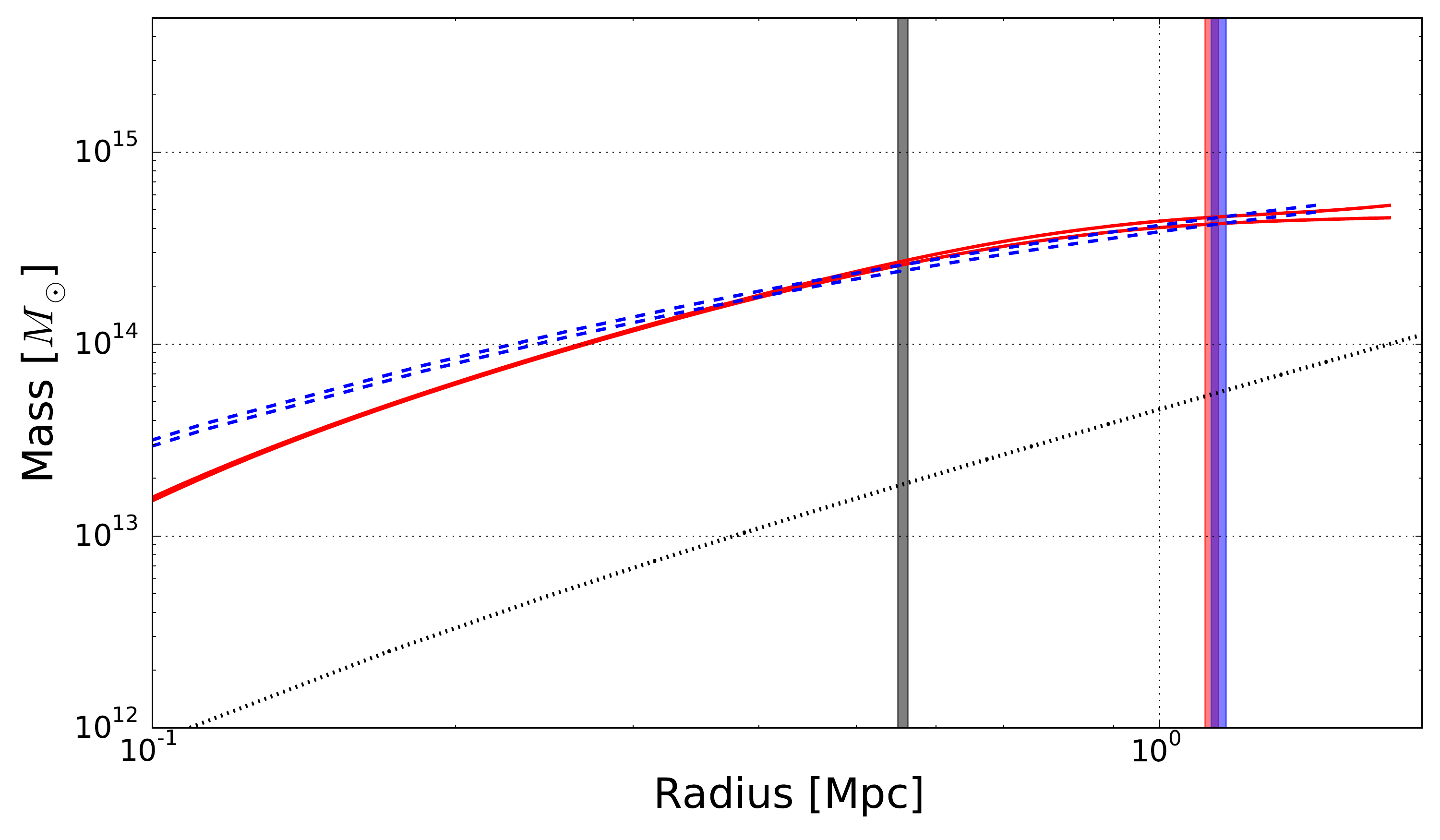}
	\caption{As Fig. \ref{fig:app_2A0335} but for A0085.}
	\label{fig:app_A0085}
\end{figure}
\begin{figure}
	\centering
	\includegraphics[width=0.45\textwidth]{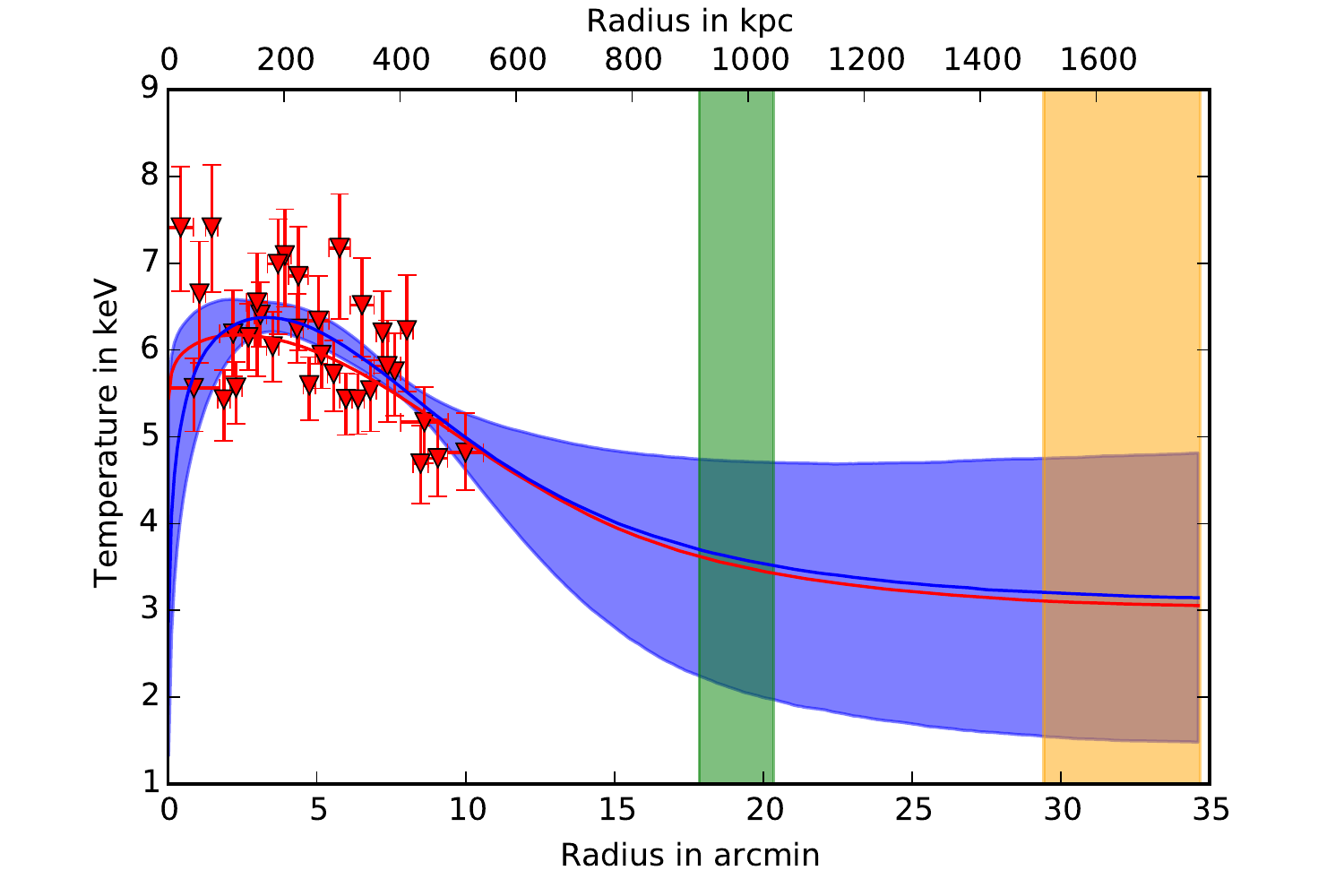}
	\includegraphics[width=0.45\textwidth]{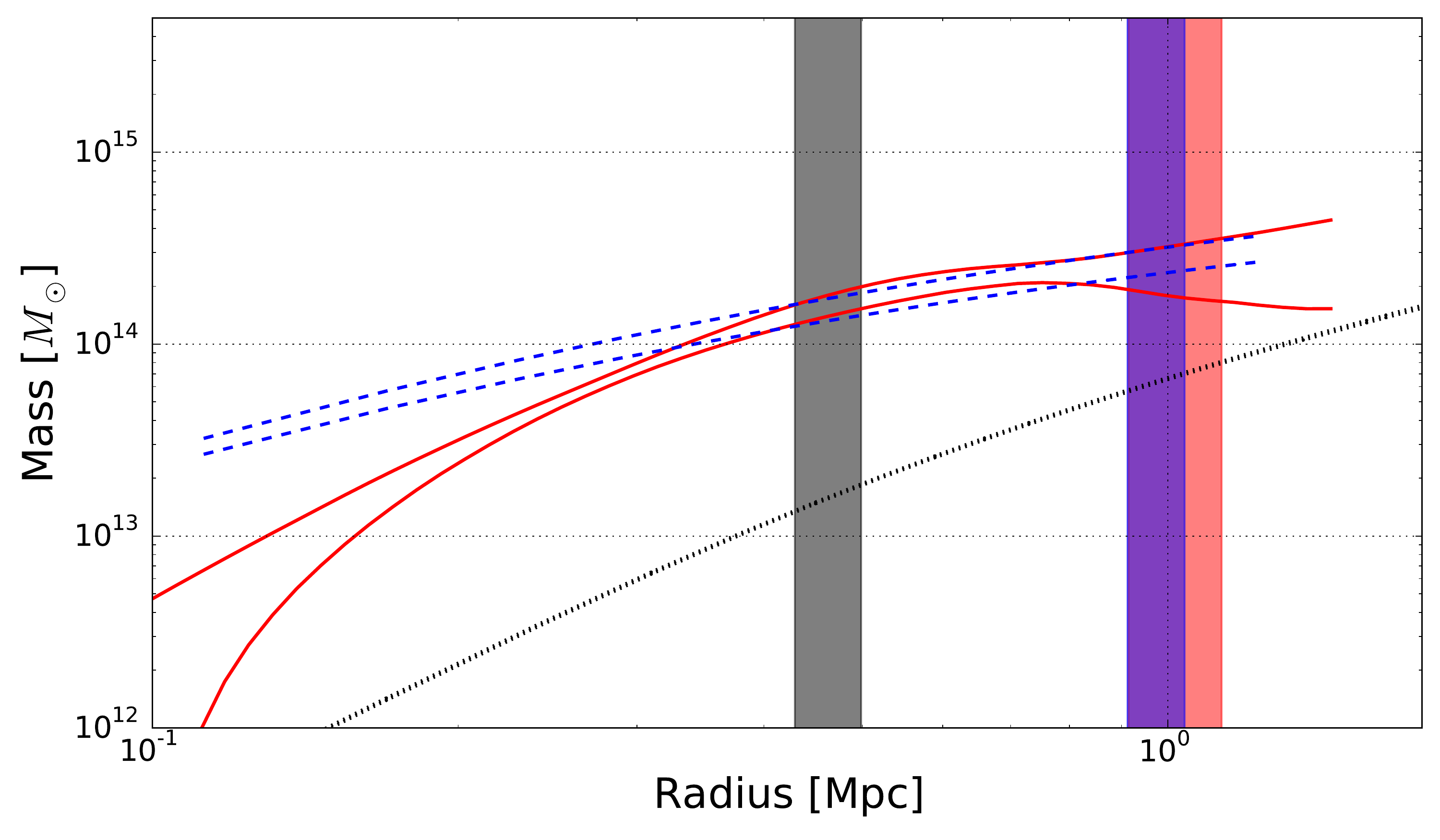}
	\caption{As Fig. \ref{fig:app_2A0335} but for A0119.}
	\label{fig:app_A0119}
\end{figure}
\clearpage
\begin{figure}
	\centering
	\includegraphics[width=0.45\textwidth]{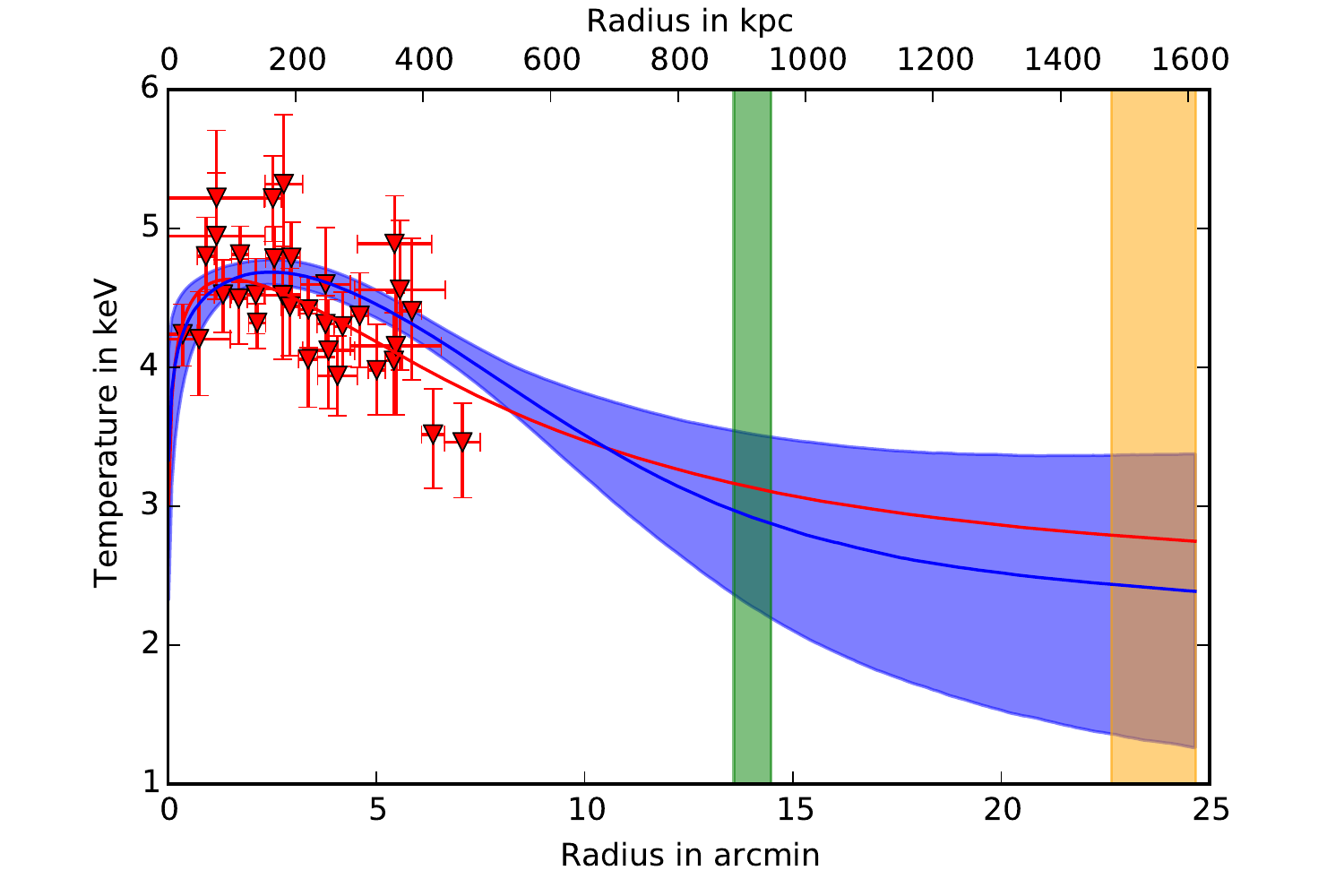}
	\includegraphics[width=0.45\textwidth]{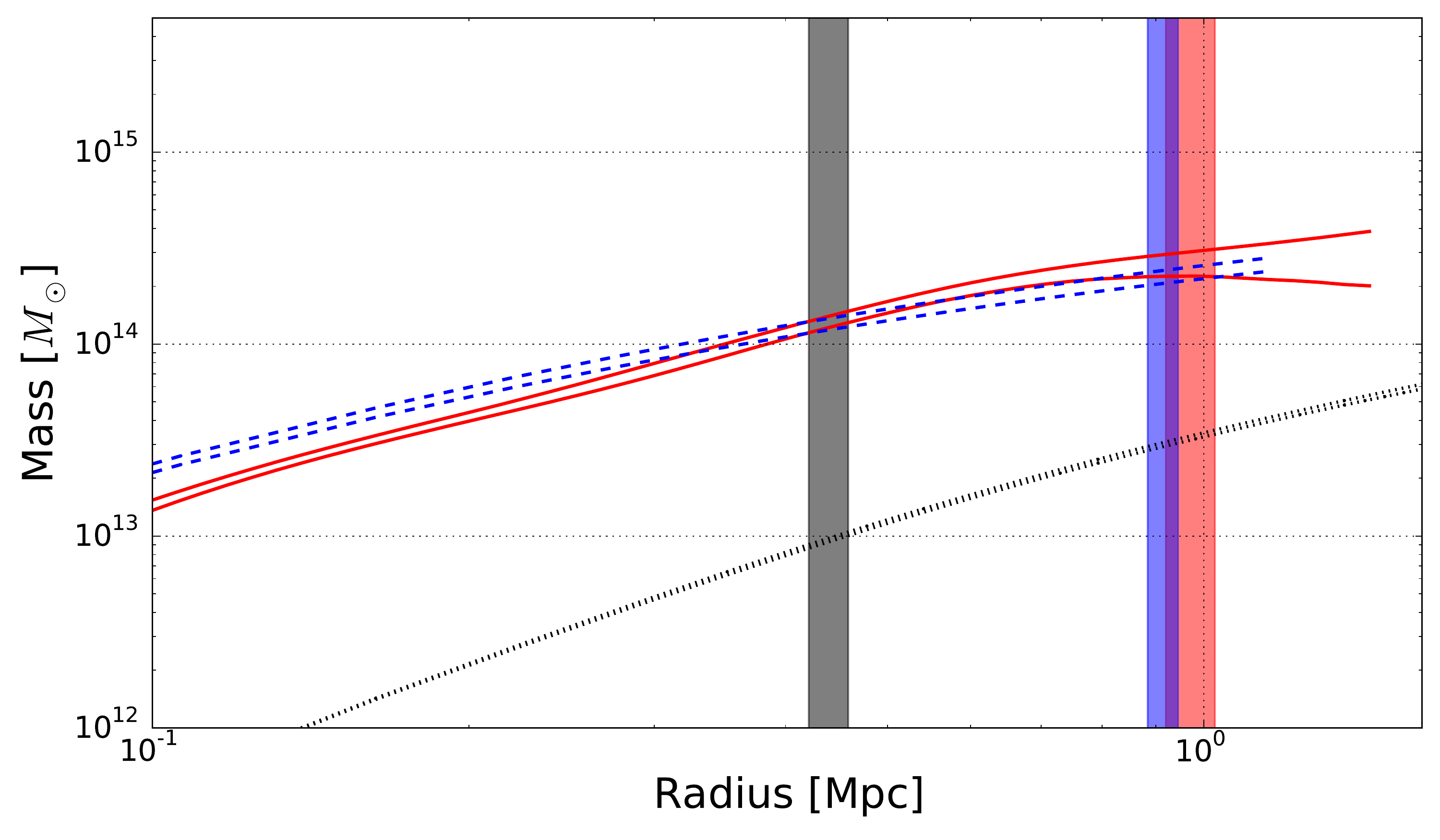}
	\caption{As Fig. \ref{fig:app_2A0335} but for A0133.}
	\label{fig:app_A0133}
\end{figure}
\begin{figure}
	\centering
	\includegraphics[width=0.45\textwidth]{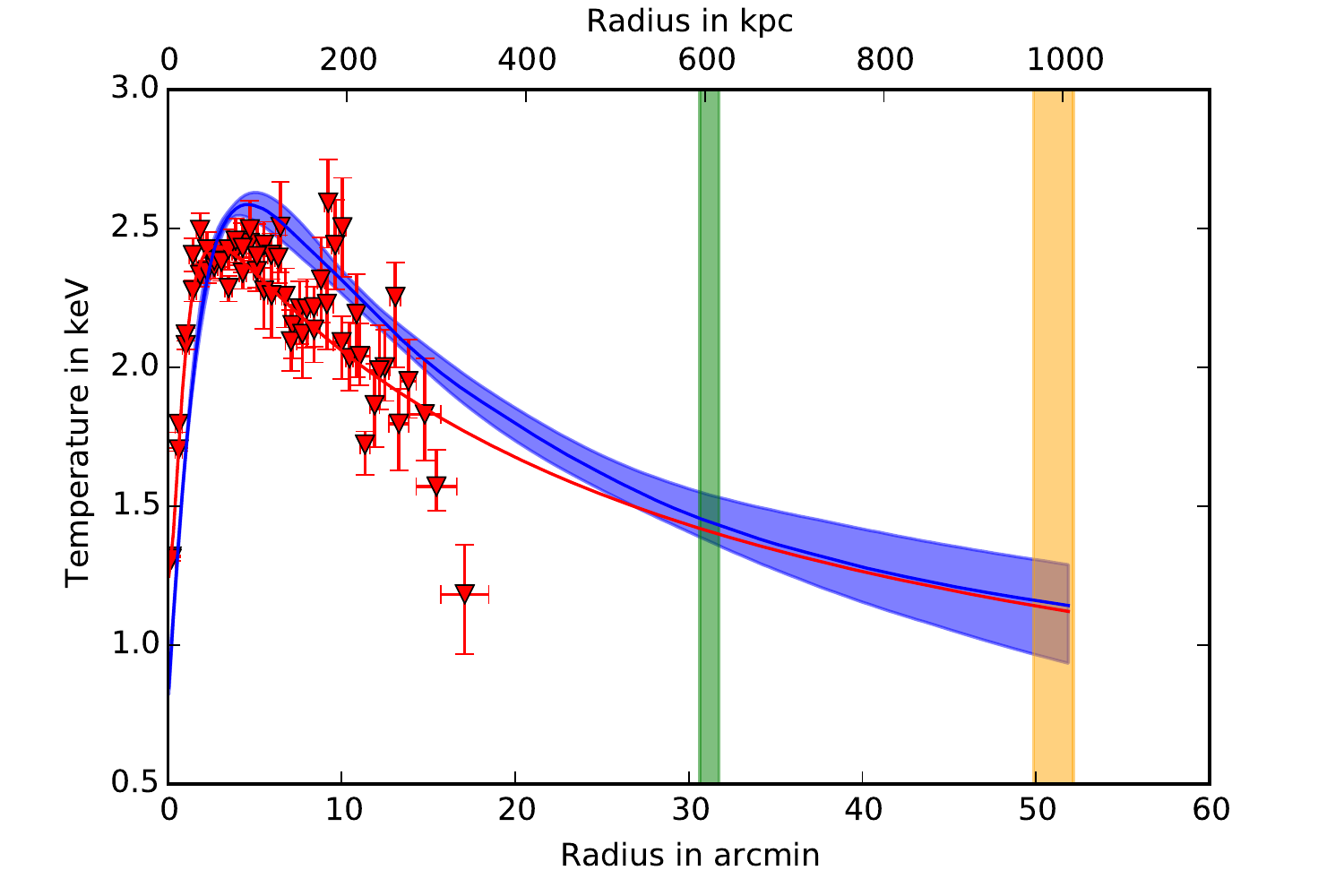}
	\includegraphics[width=0.45\textwidth]{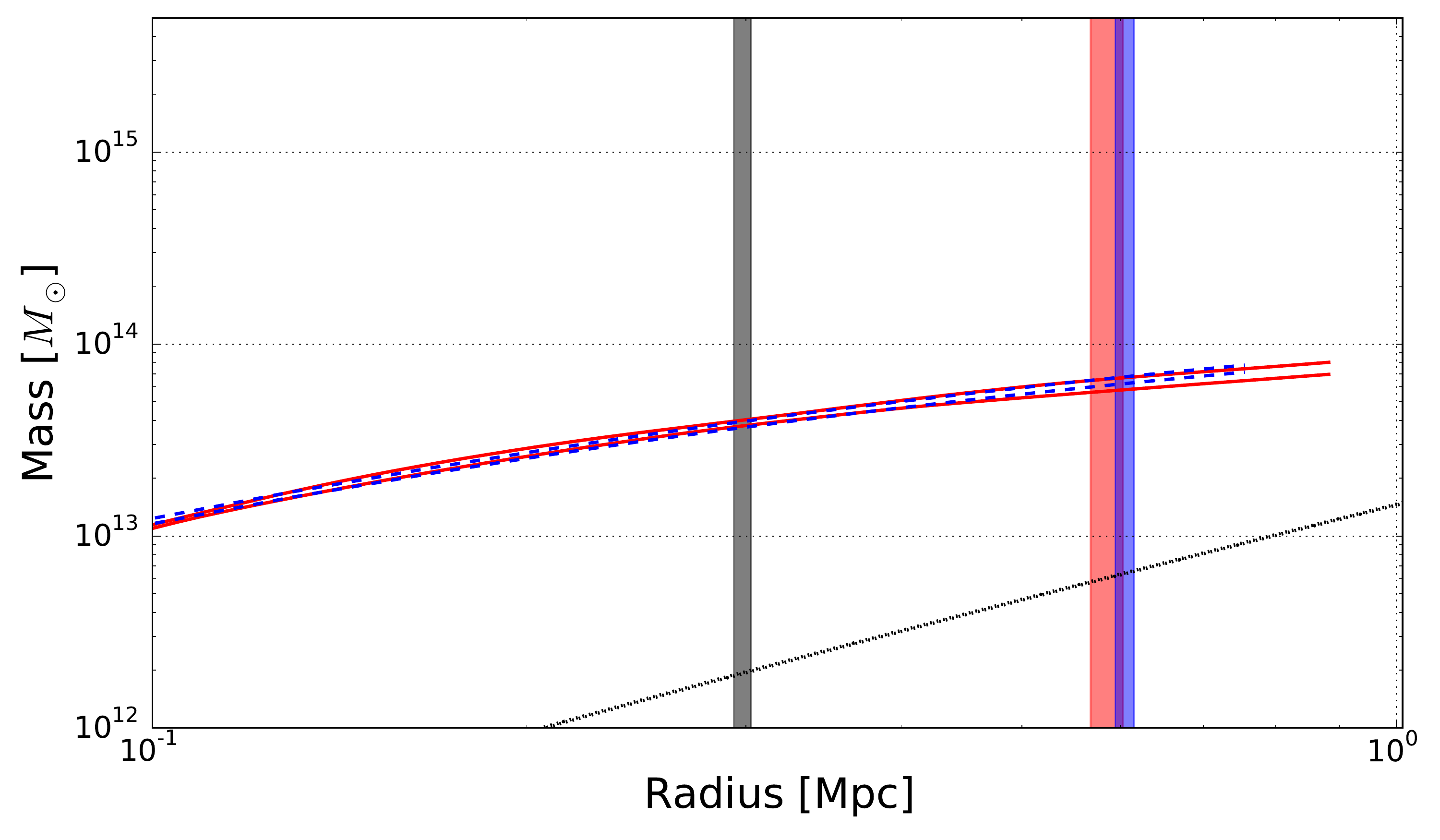}
	\caption{As Fig. \ref{fig:app_2A0335} but for A0262.}
	\label{fig:app_A0262}
\end{figure}
\begin{figure}
	\centering
	\includegraphics[width=0.45\textwidth]{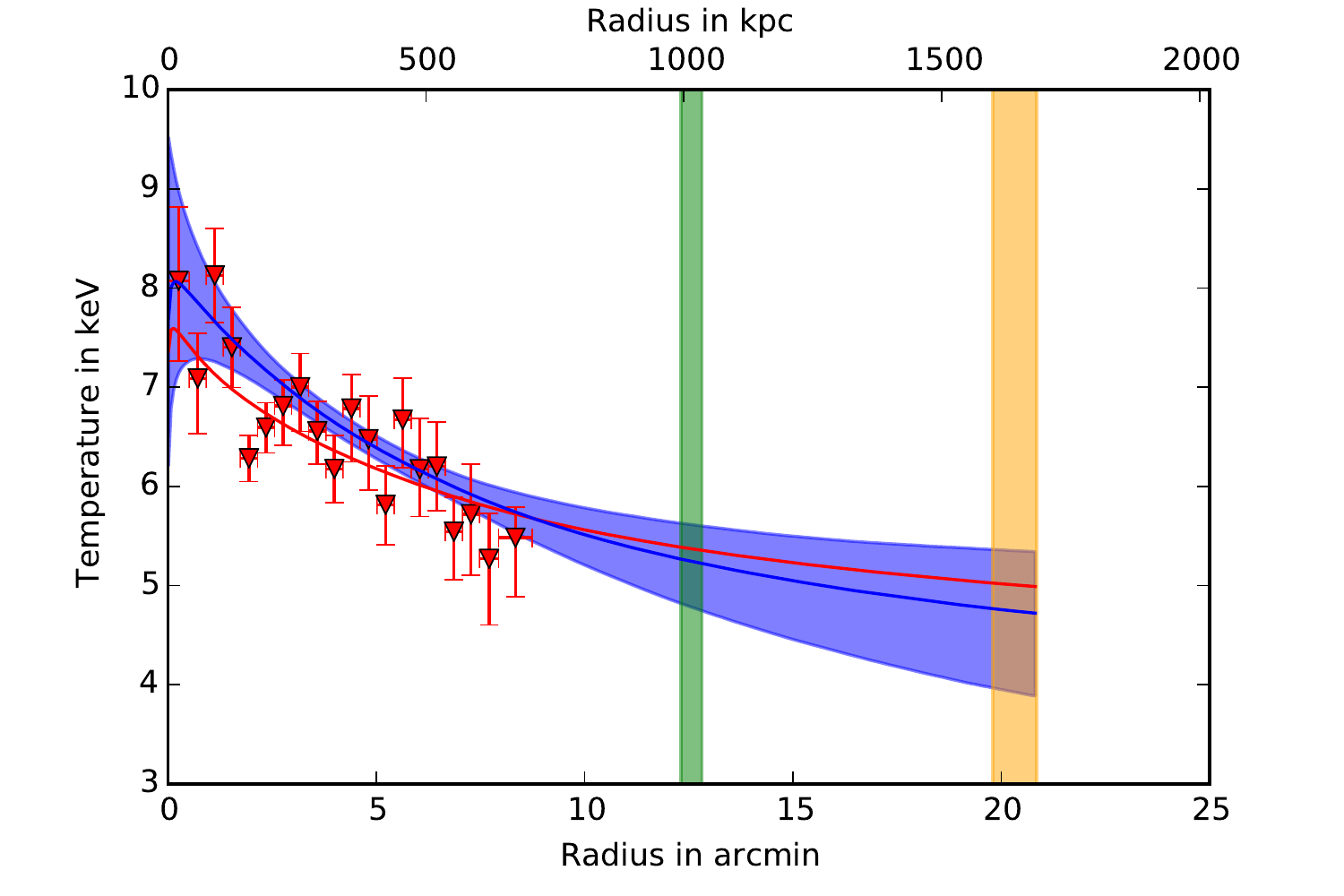}
	\includegraphics[width=0.45\textwidth]{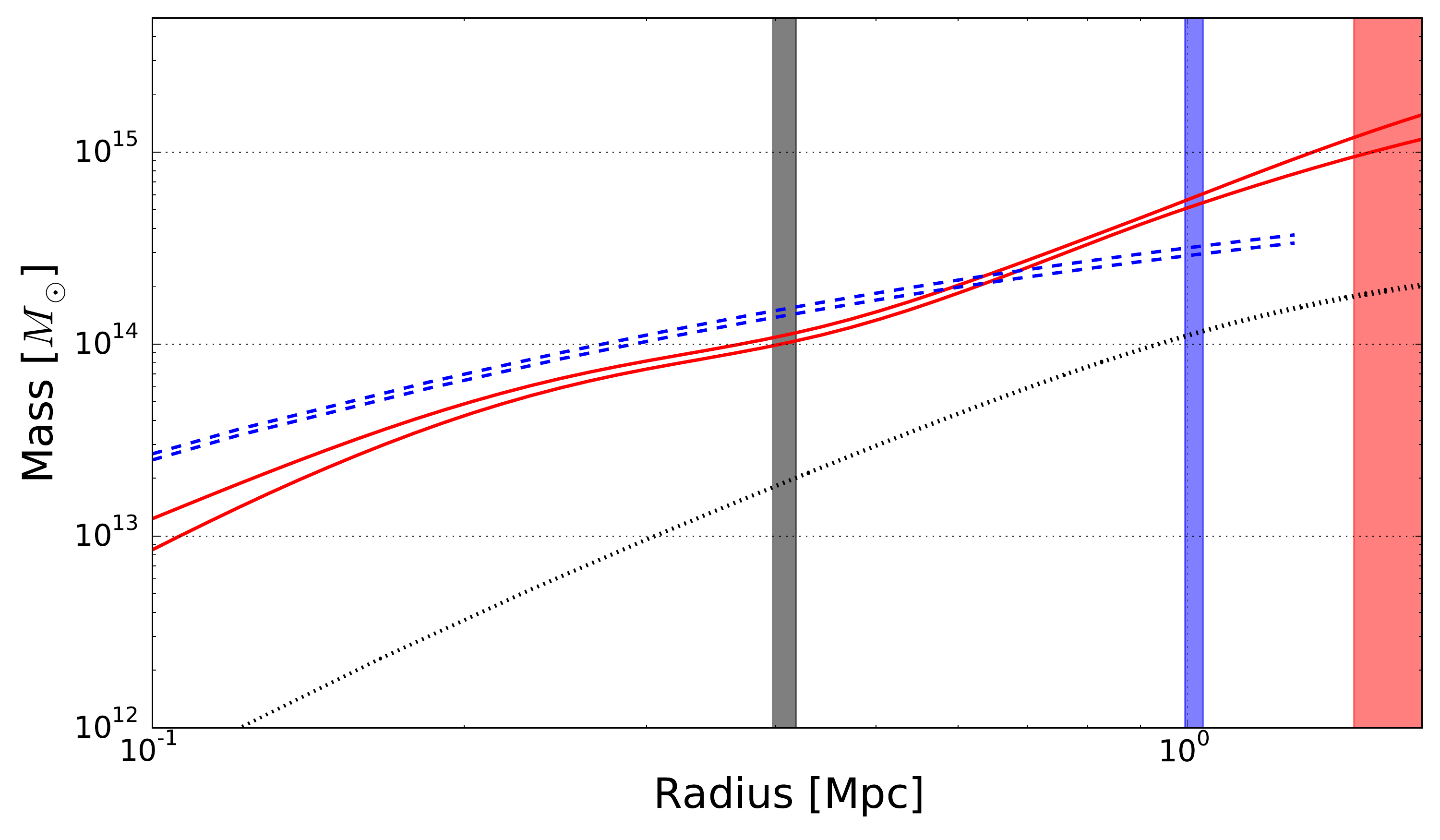}
	\caption{As Fig. \ref{fig:app_2A0335} but for A0399.}
	\label{fig:app_A0399}
\end{figure}
\clearpage
\begin{figure}
	\centering
	\includegraphics[width=0.45\textwidth]{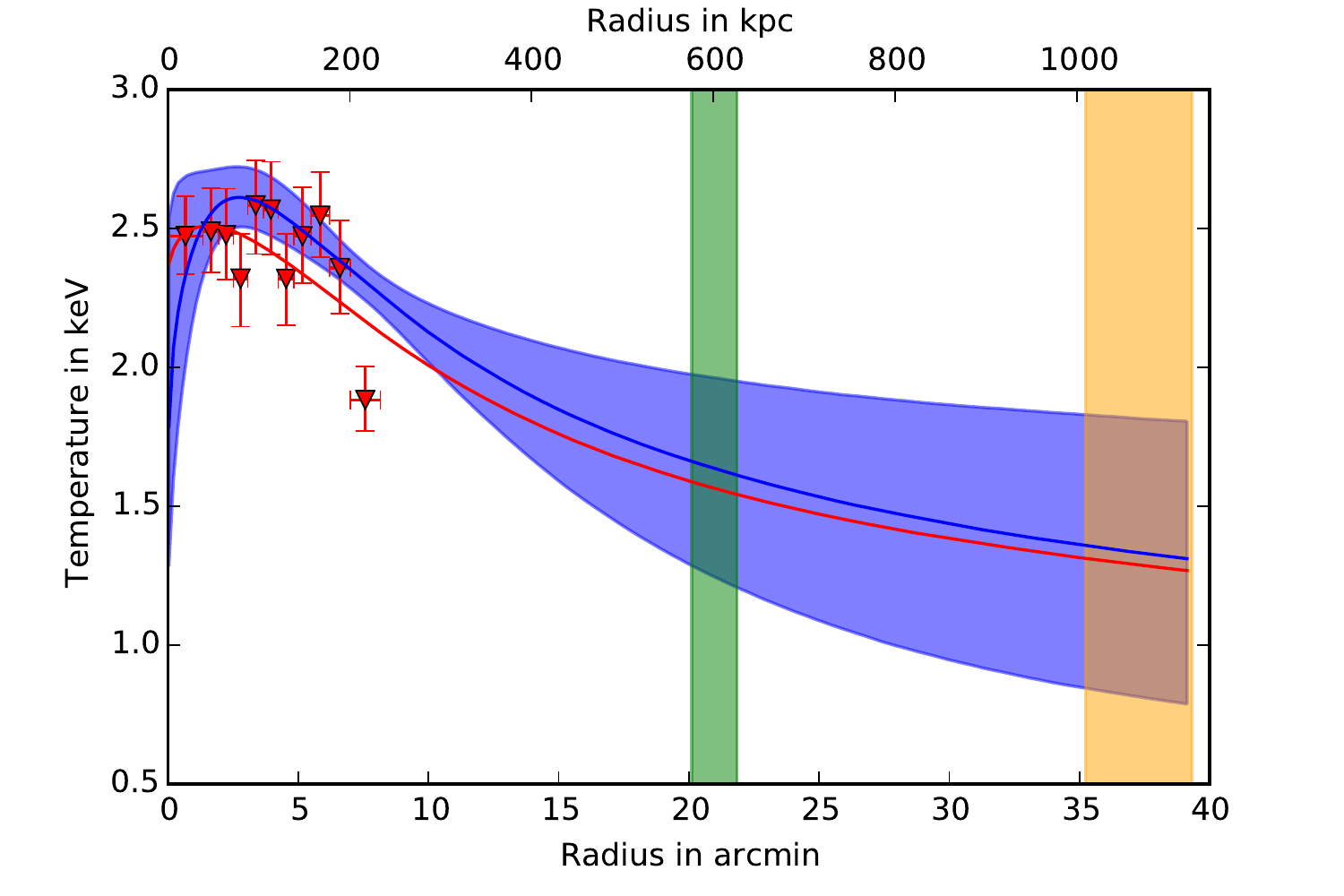}
	\includegraphics[width=0.45\textwidth]{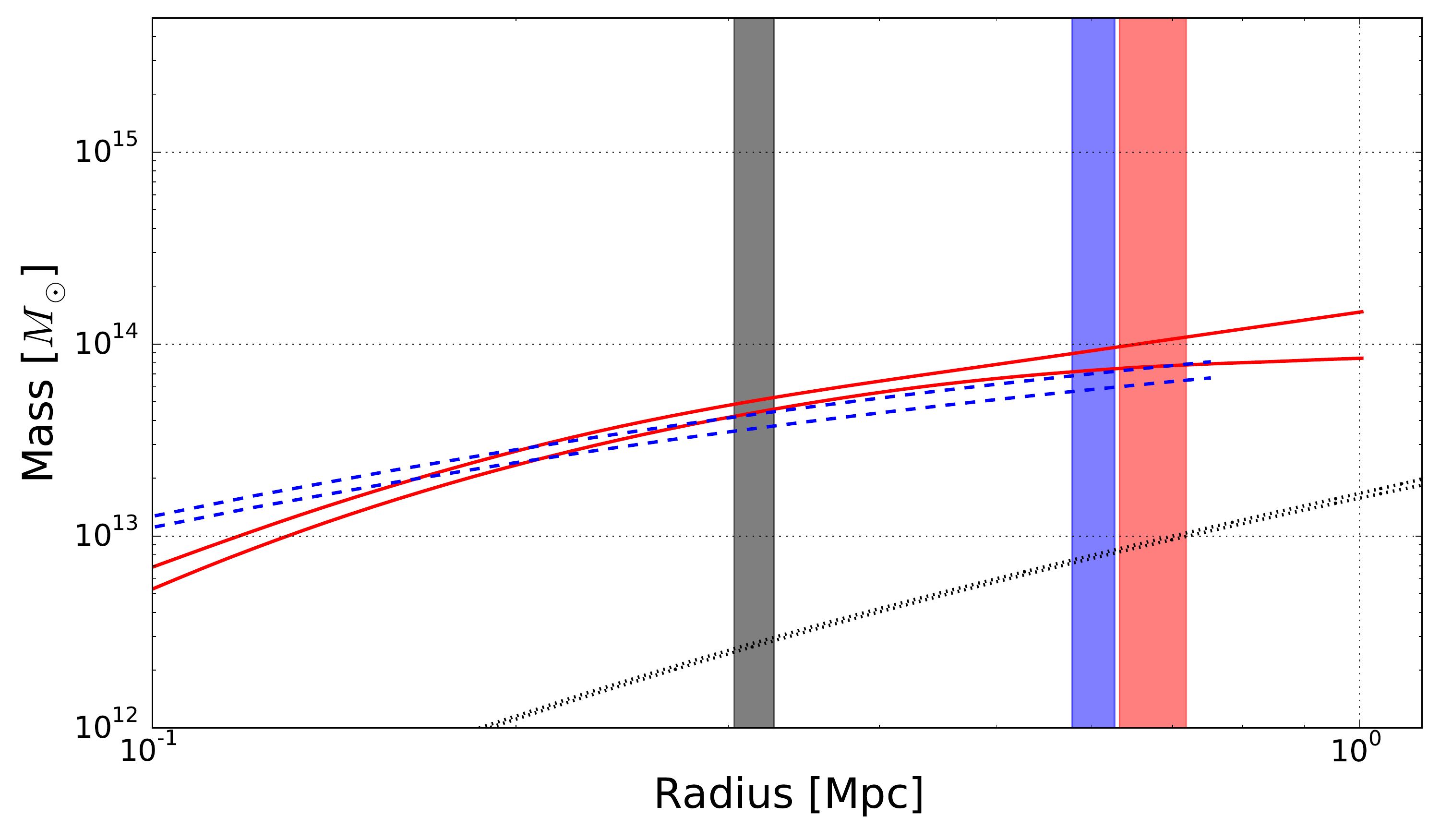}
	\caption{As Fig. \ref{fig:app_2A0335} but for A0400.}
	\label{fig:app_A0400}
\end{figure}
\begin{figure}
	\centering
	\includegraphics[width=0.45\textwidth]{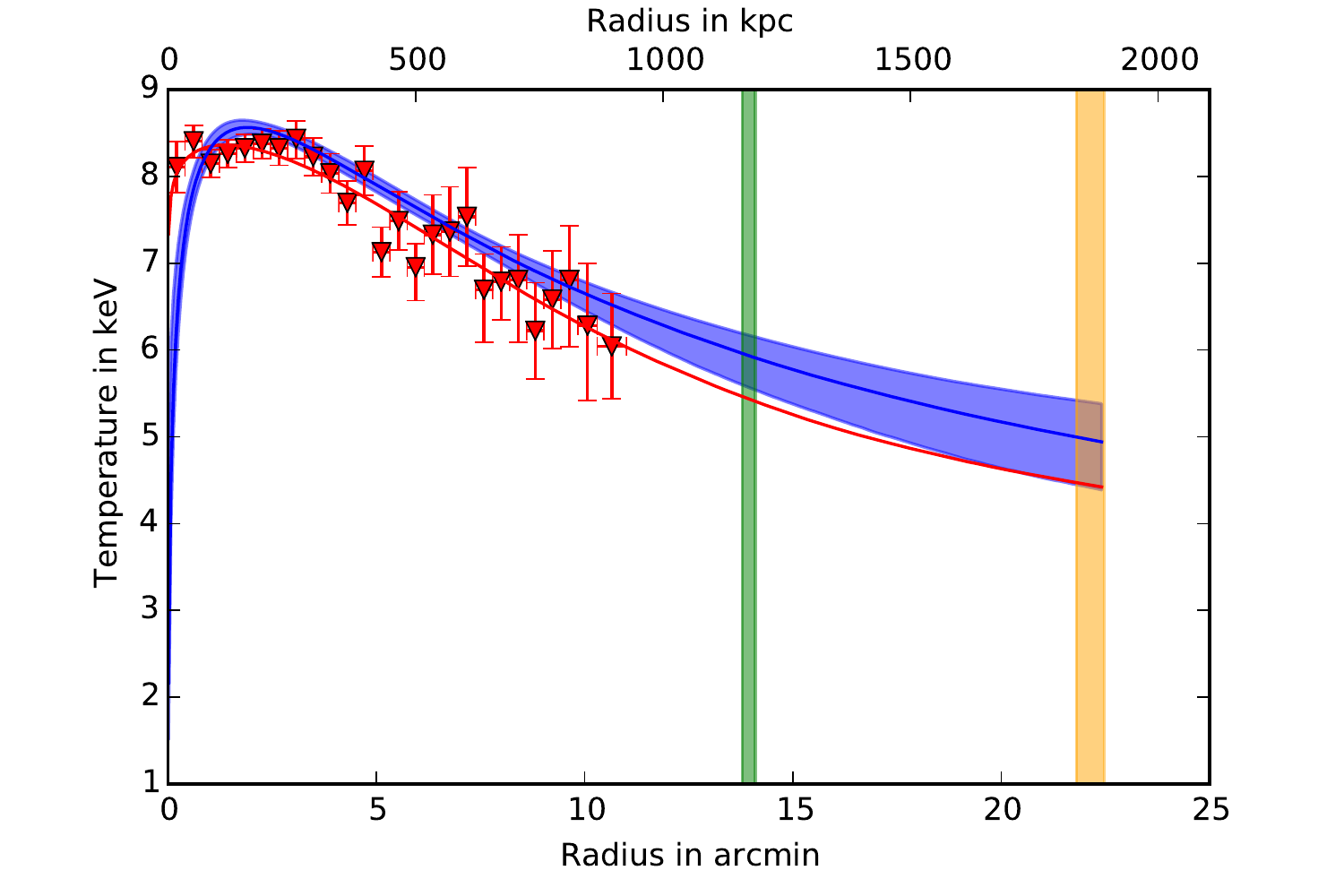}
	\includegraphics[width=0.45\textwidth]{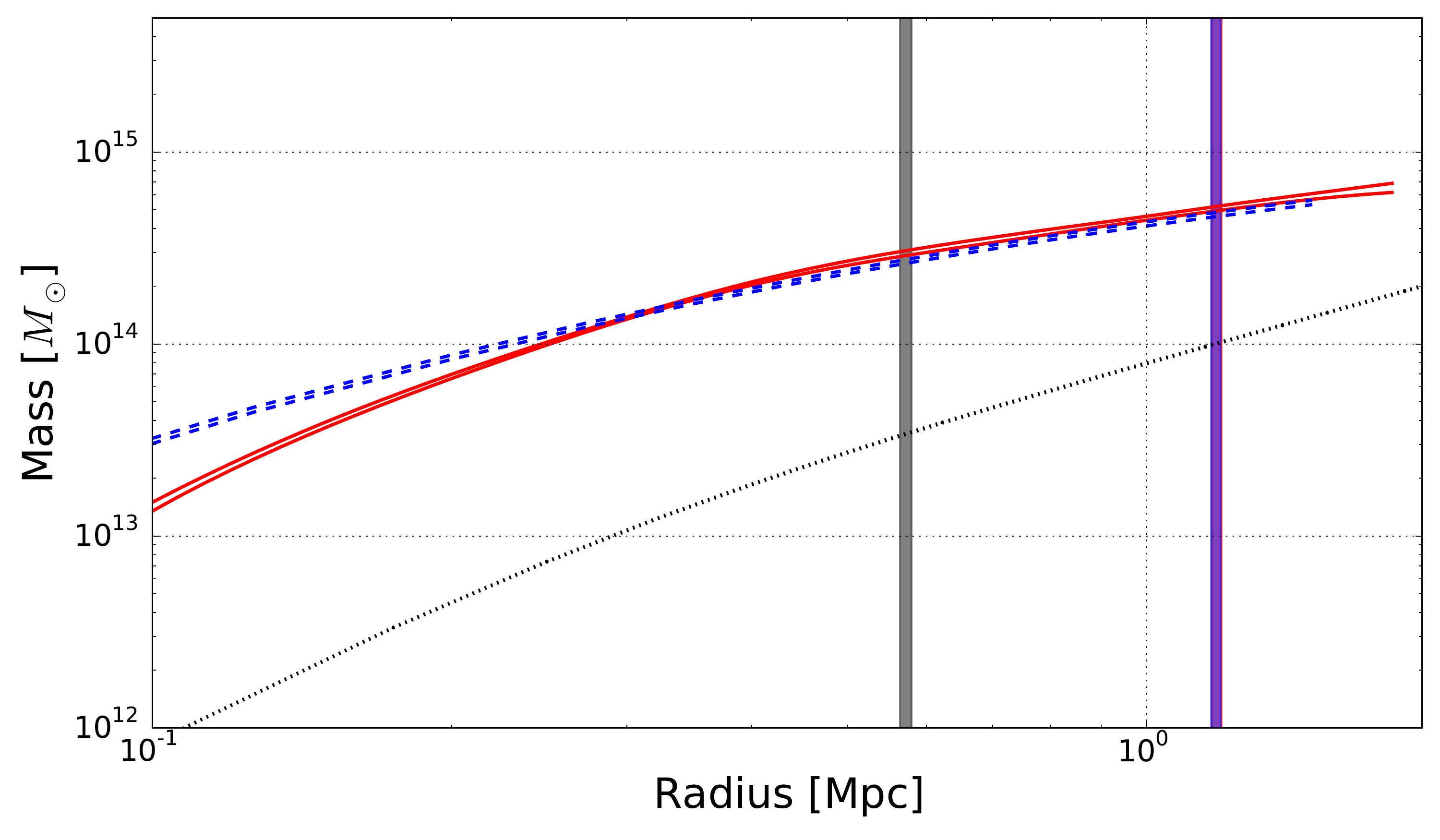}
	\caption{As Fig. \ref{fig:app_2A0335} but for A0401.}
	\label{fig:app_A0401}
\end{figure}
\begin{figure}
	\centering
	\includegraphics[width=0.45\textwidth]{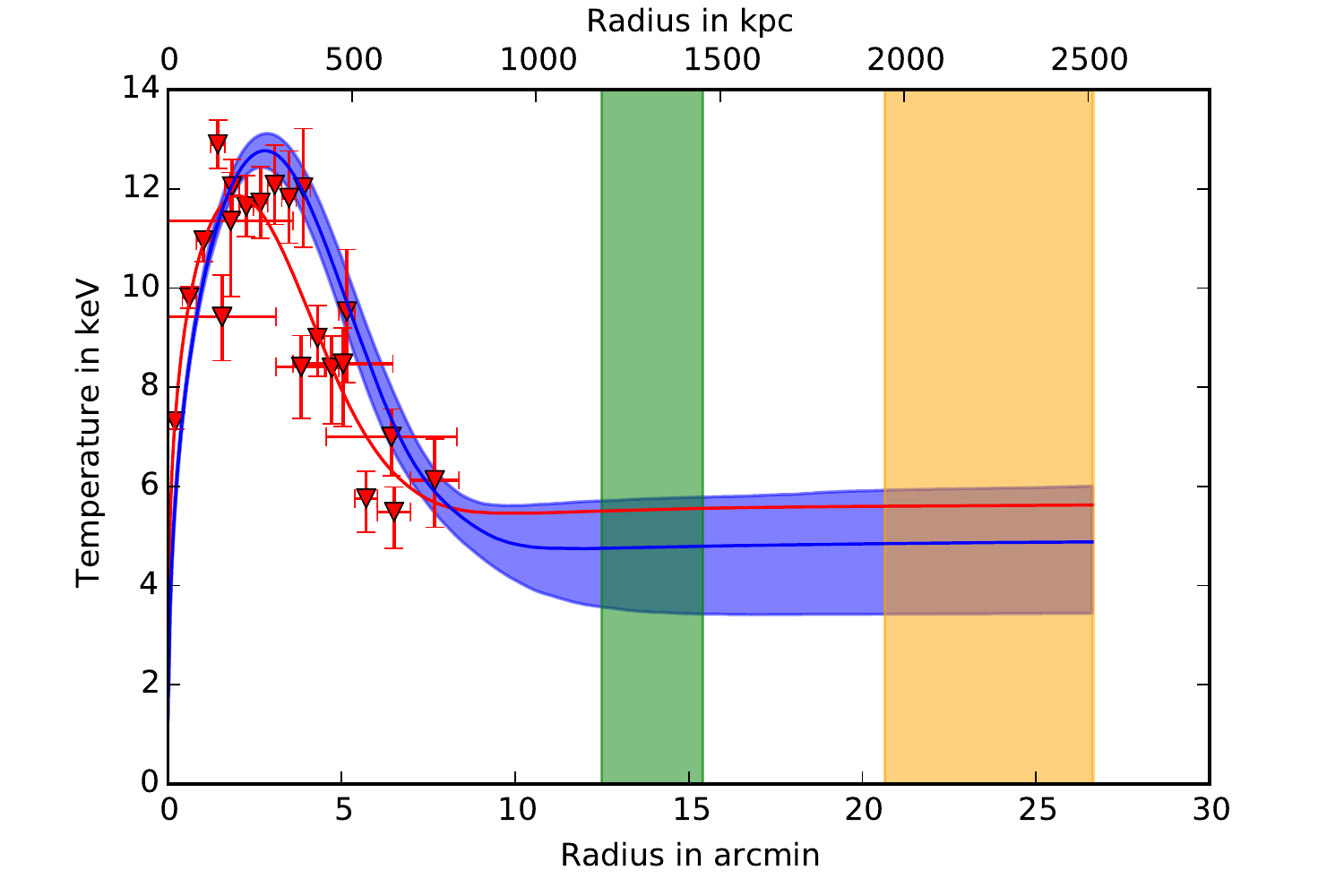}
	\includegraphics[width=0.45\textwidth]{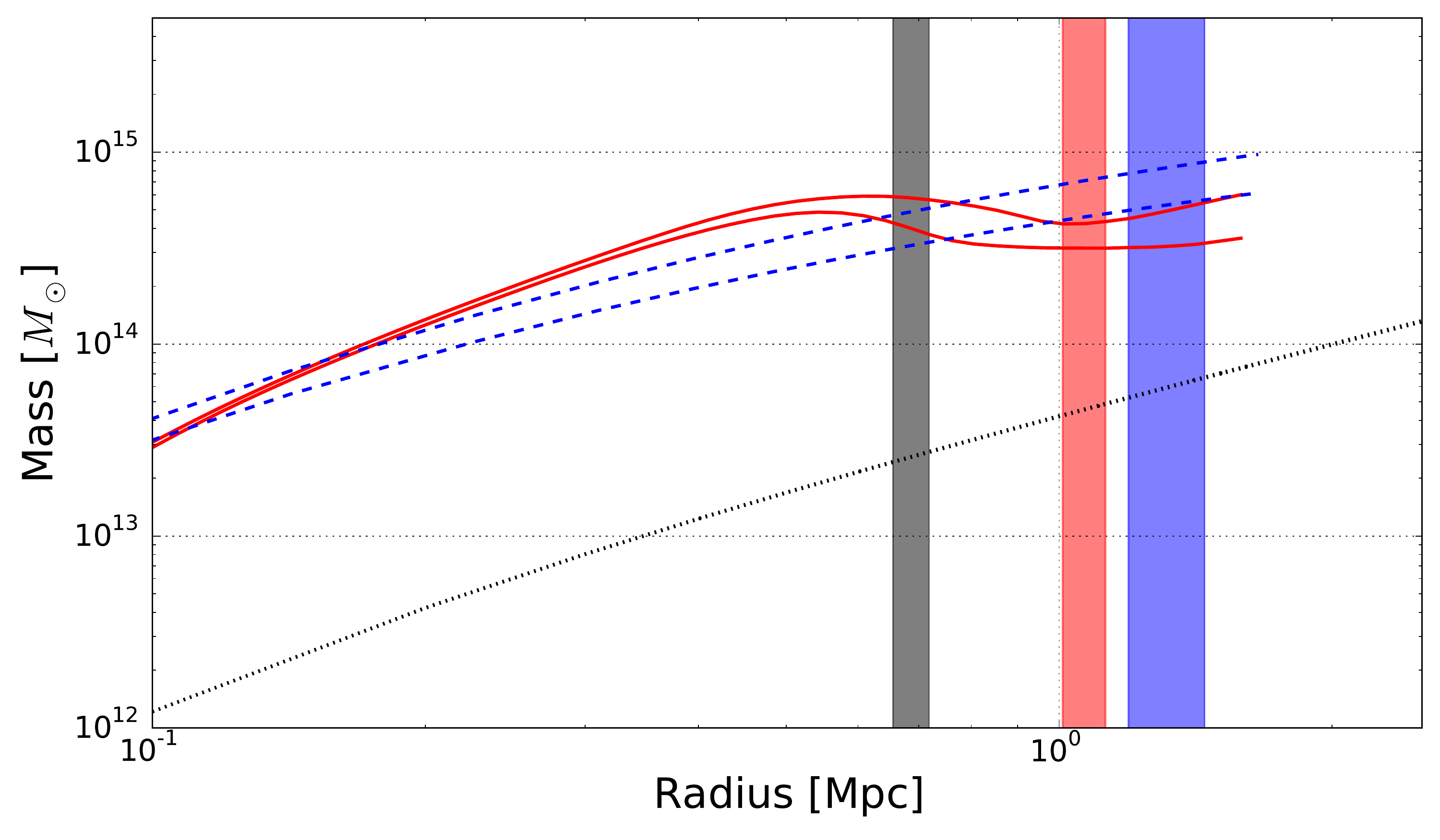}
	\caption{As Fig. \ref{fig:app_2A0335} but for A0478.}
	\label{fig:app_A0478}
\end{figure}
\clearpage
\begin{figure}
	\centering
	\includegraphics[width=0.45\textwidth]{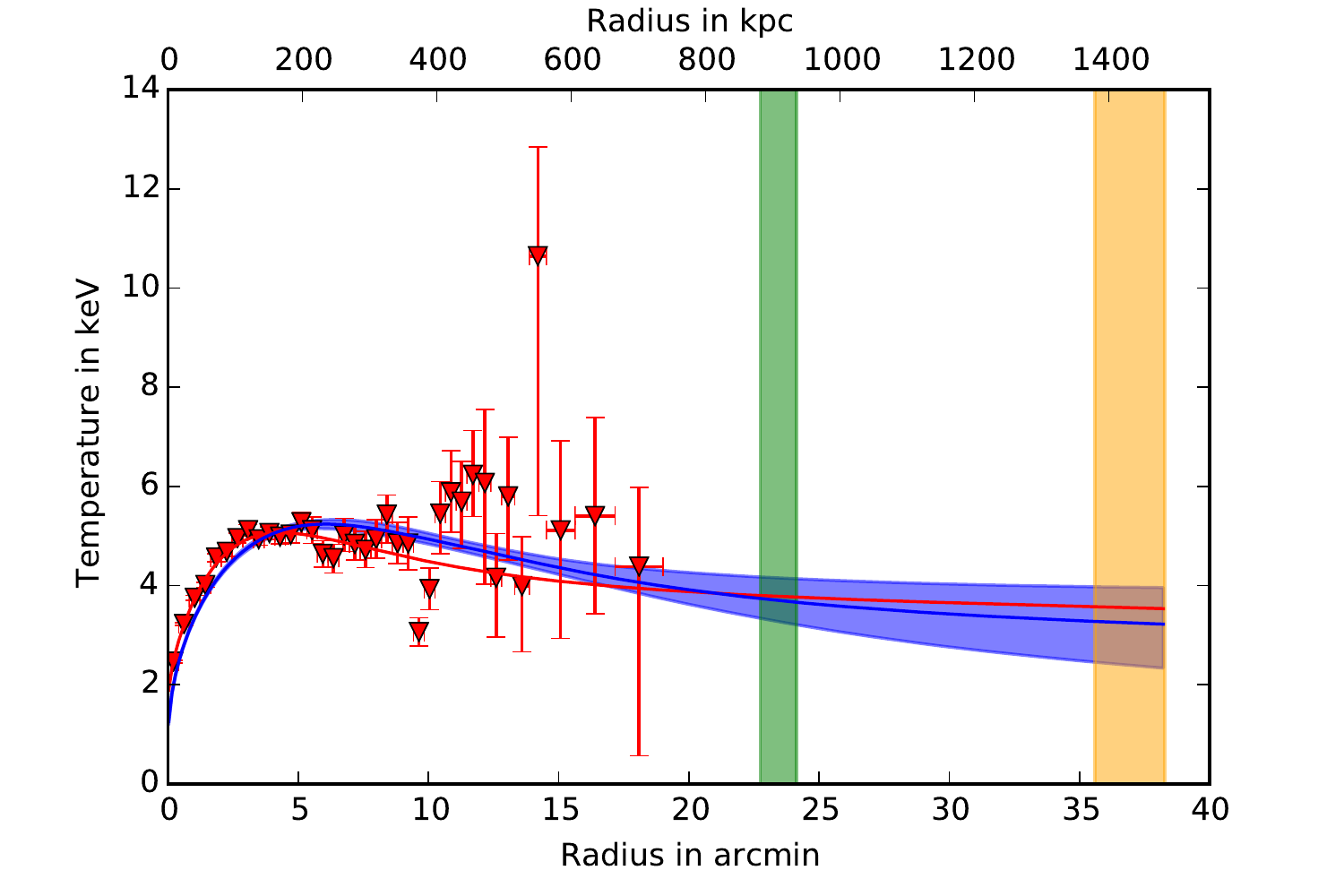}
	\includegraphics[width=0.45\textwidth]{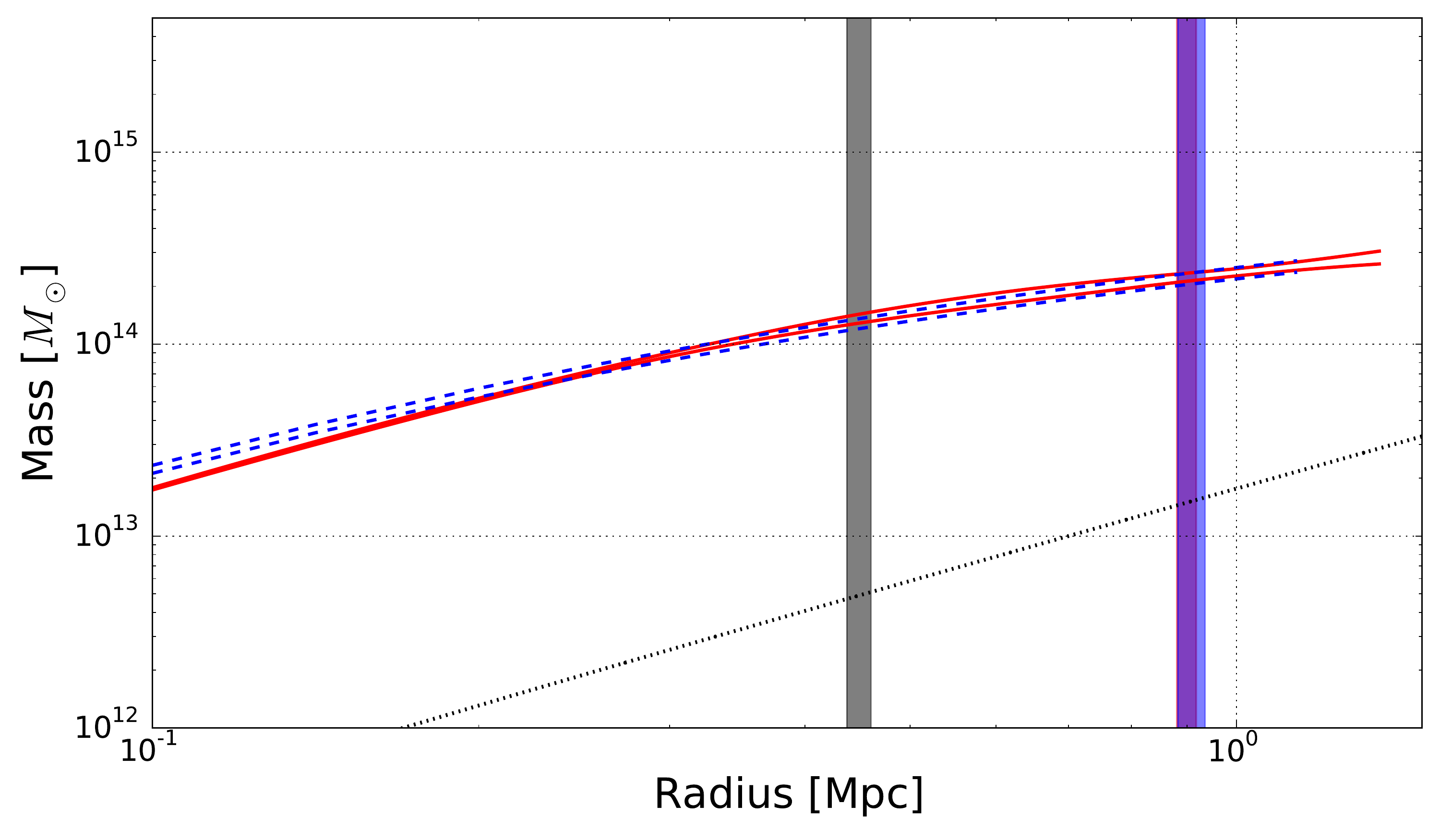}
	\caption{As Fig. \ref{fig:app_2A0335} but for A0496.}
	\label{fig:app_A0496}
\end{figure}
\begin{figure}
	\centering
	\includegraphics[width=0.45\textwidth]{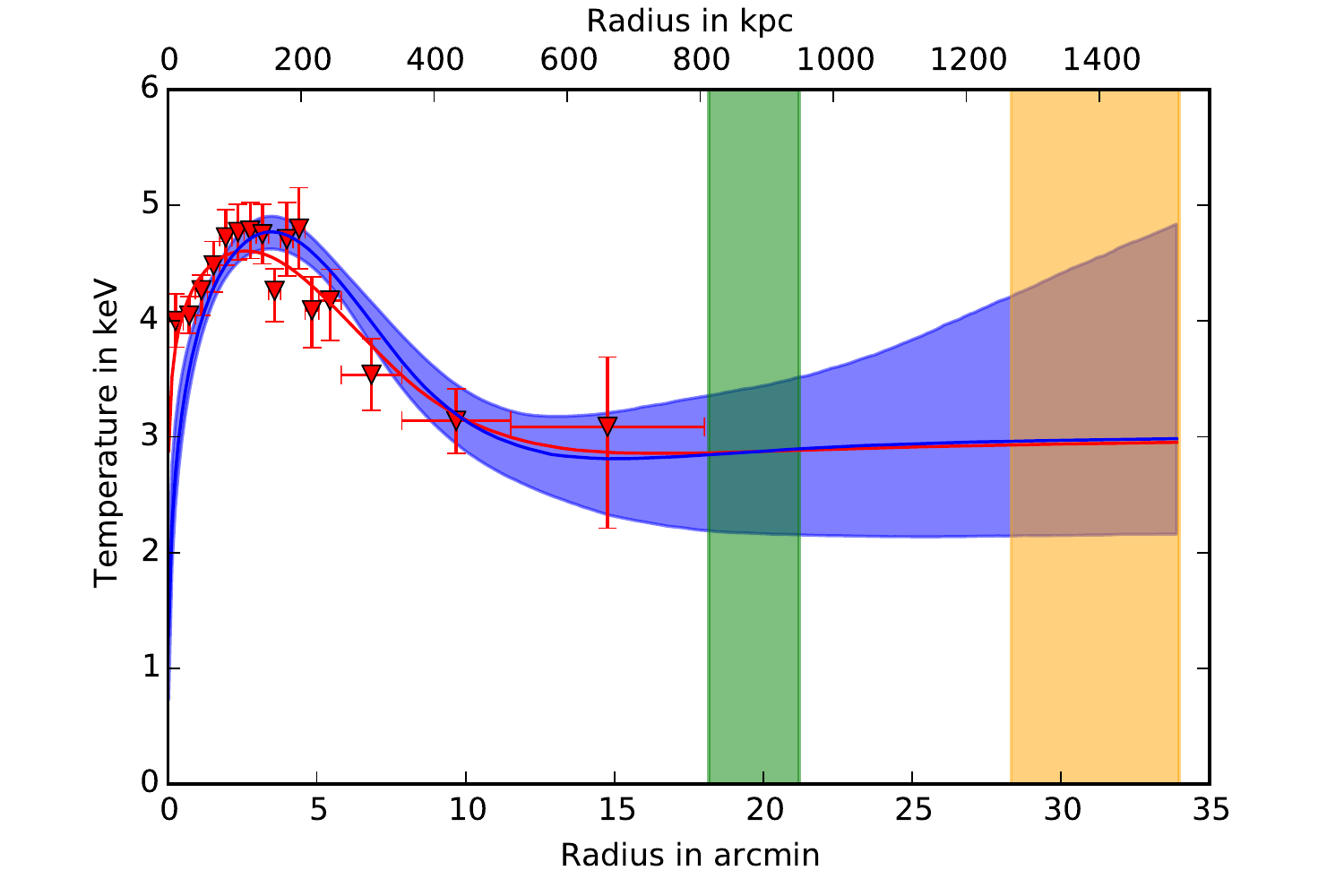}
	\includegraphics[width=0.45\textwidth]{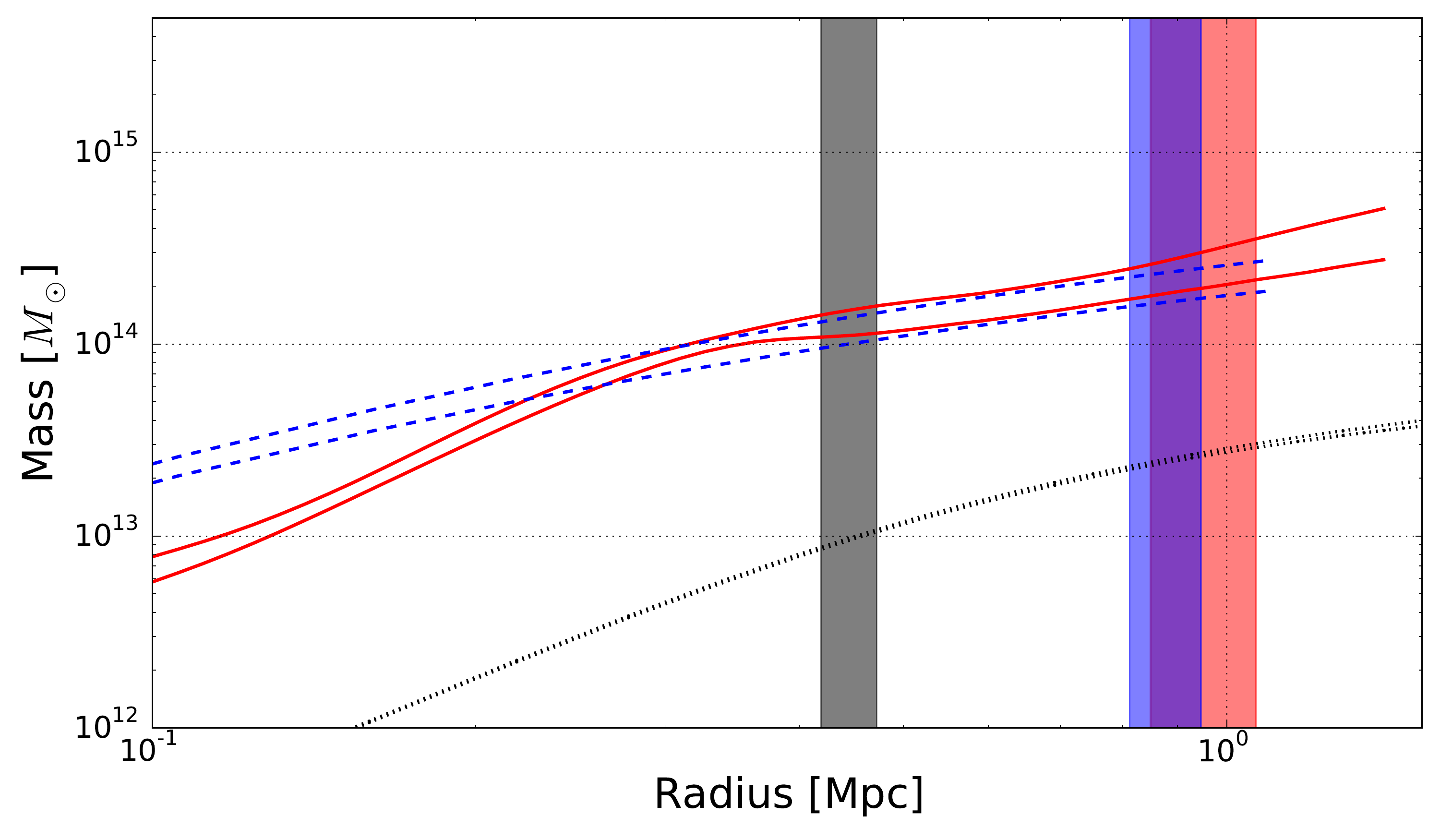}
	\caption{As Fig. \ref{fig:app_2A0335} but for A0576.}
	\label{fig:app_A0576}
\end{figure}
\begin{figure}
	\centering
	\includegraphics[width=0.45\textwidth]{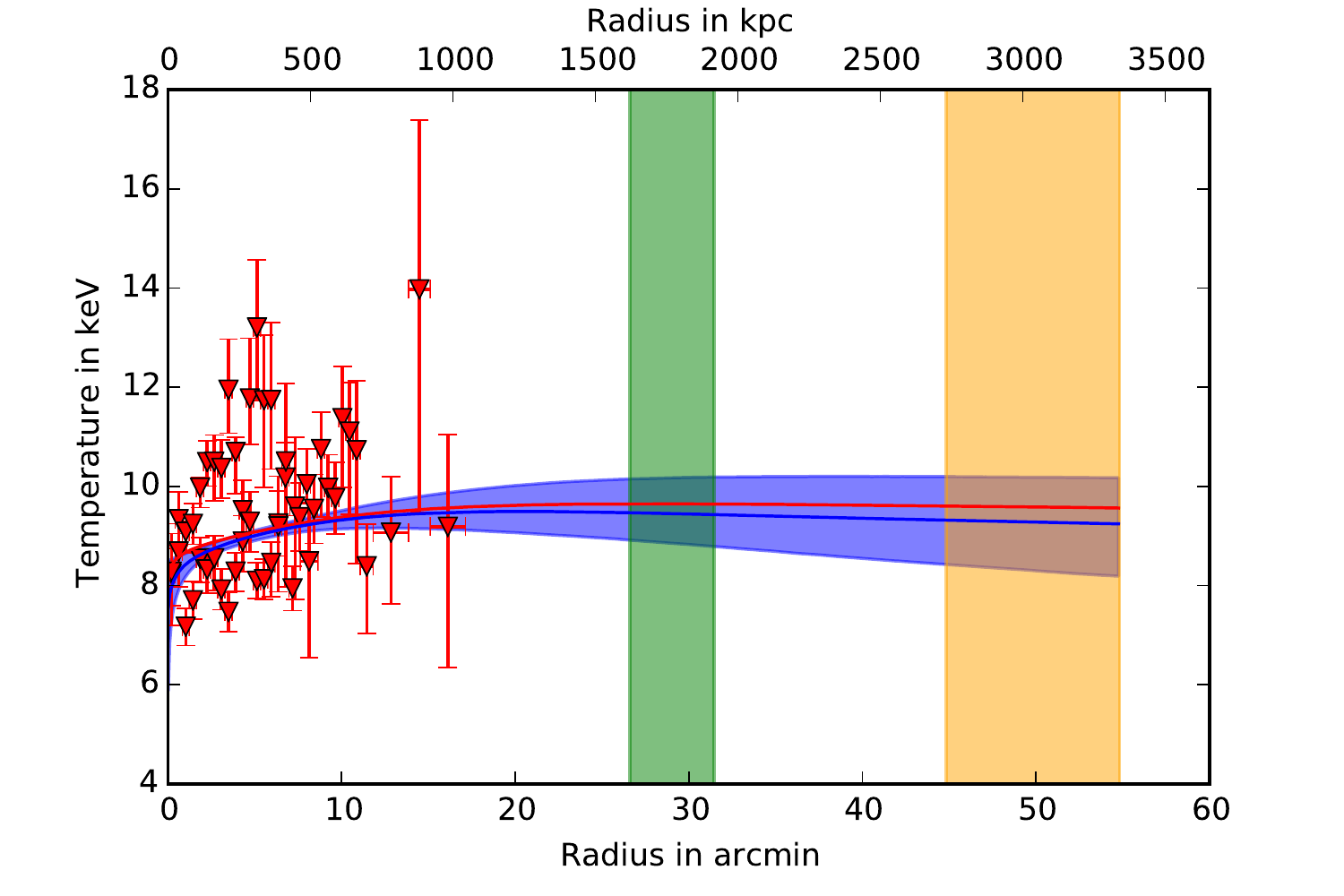}
	\includegraphics[width=0.45\textwidth]{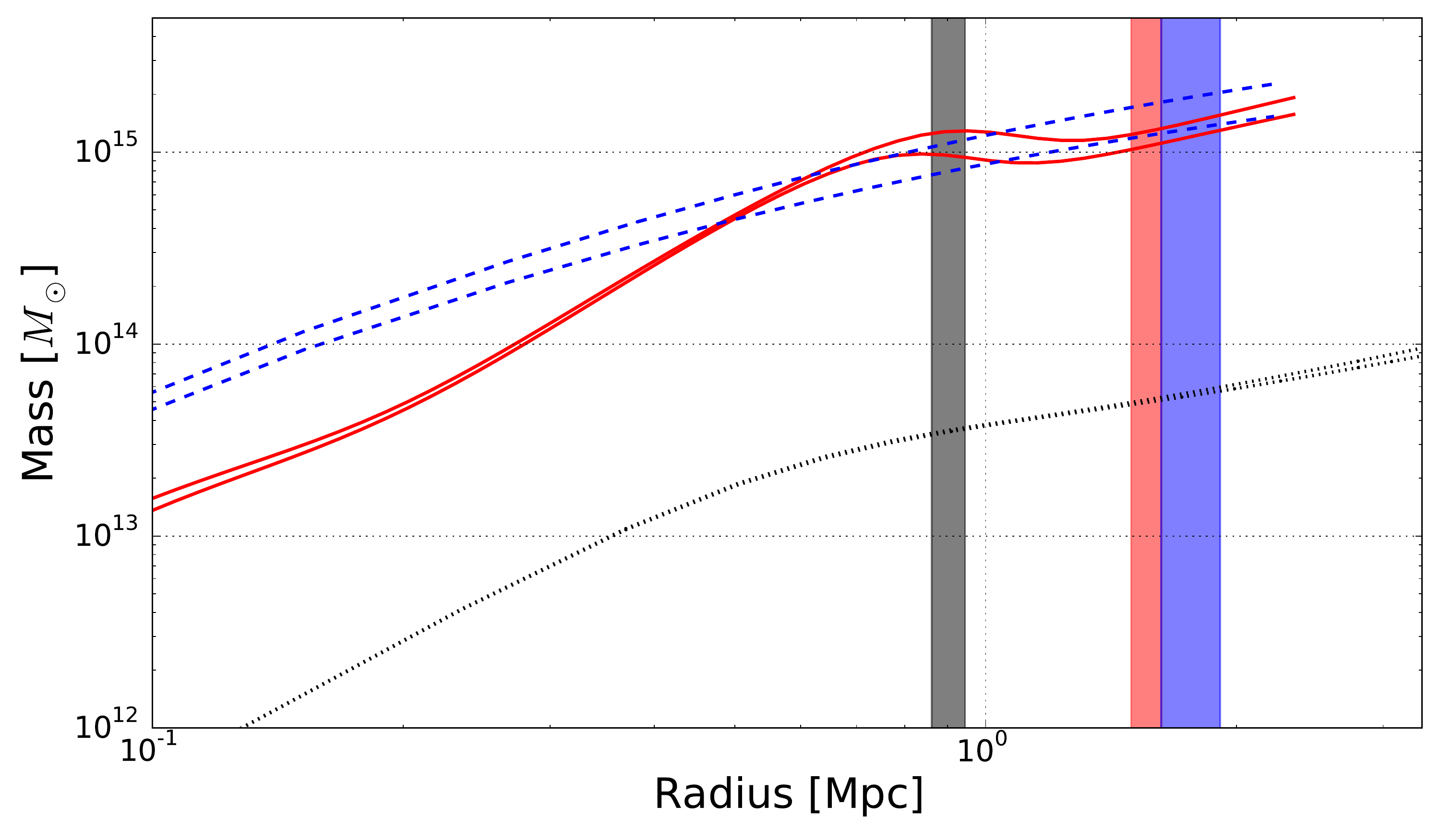}
	\caption{As Fig. \ref{fig:app_2A0335} but for A0754.}
	\label{fig:app_A0754}
\end{figure}
\clearpage
\begin{figure}
	\centering
	\includegraphics[width=0.45\textwidth]{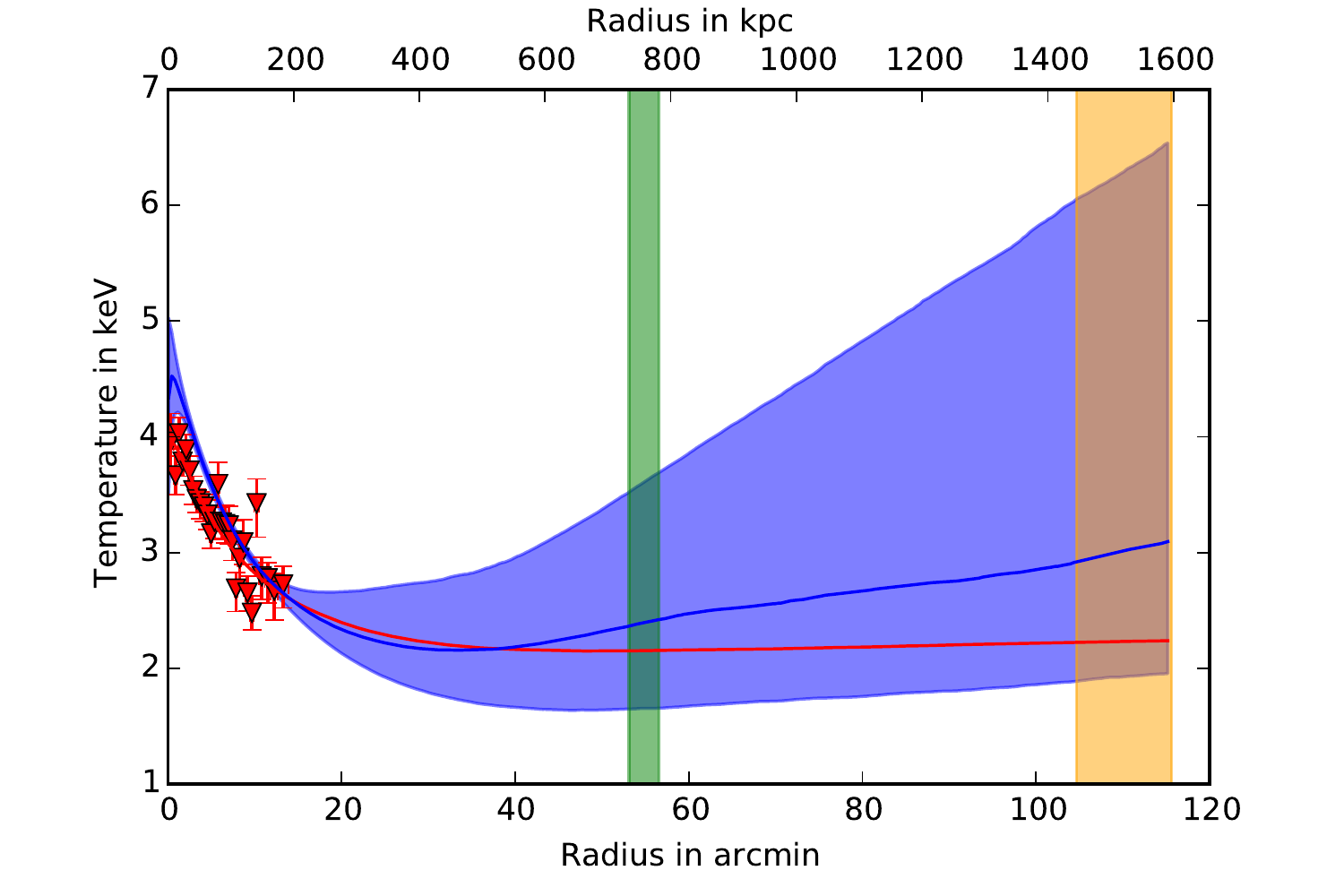}
	\includegraphics[width=0.45\textwidth]{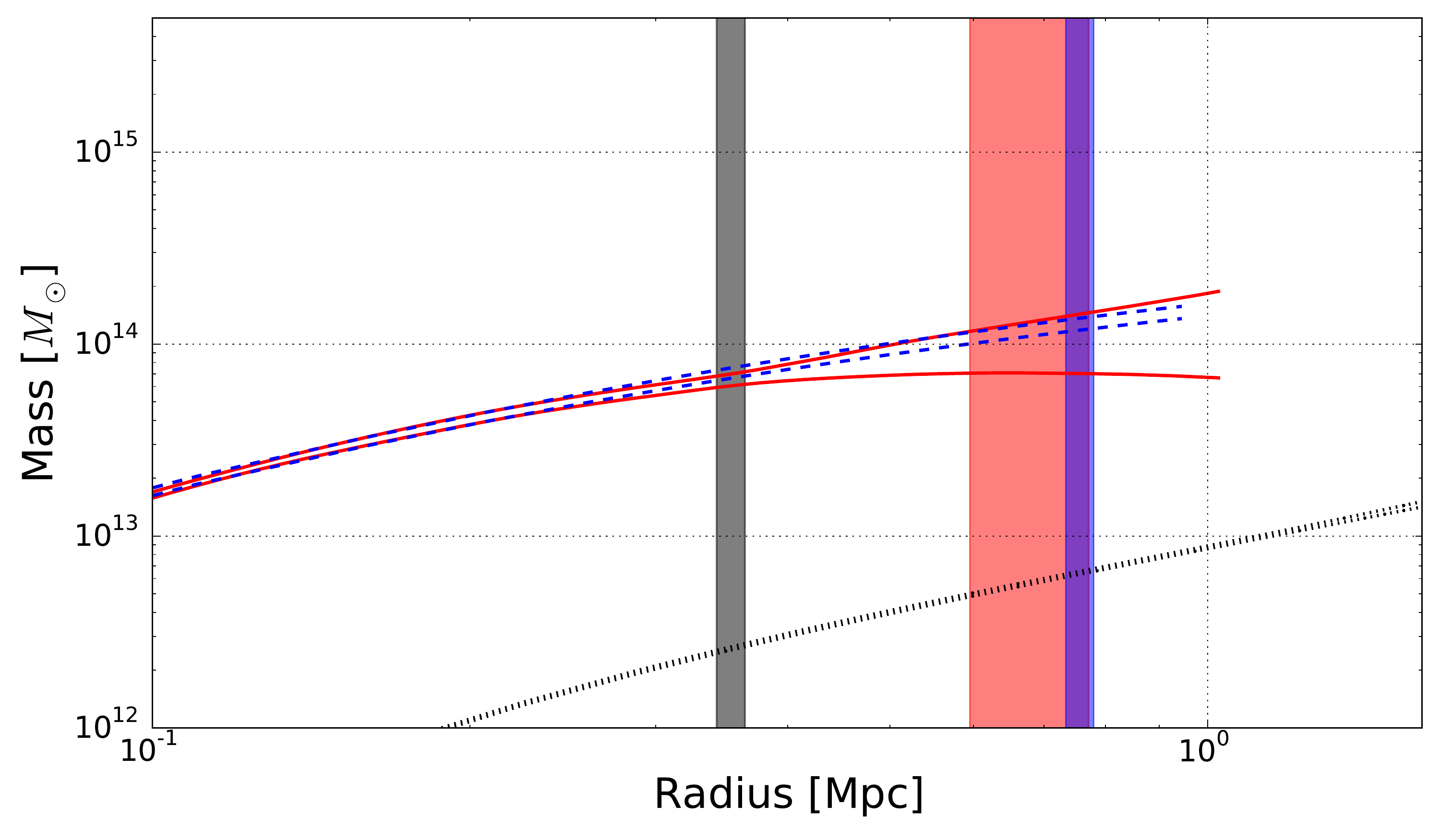}
	\caption{As Fig. \ref{fig:app_2A0335} but for A1060.}
	\label{fig:app_A1060}
\end{figure}
\begin{figure}
	\centering
	\includegraphics[width=0.45\textwidth]{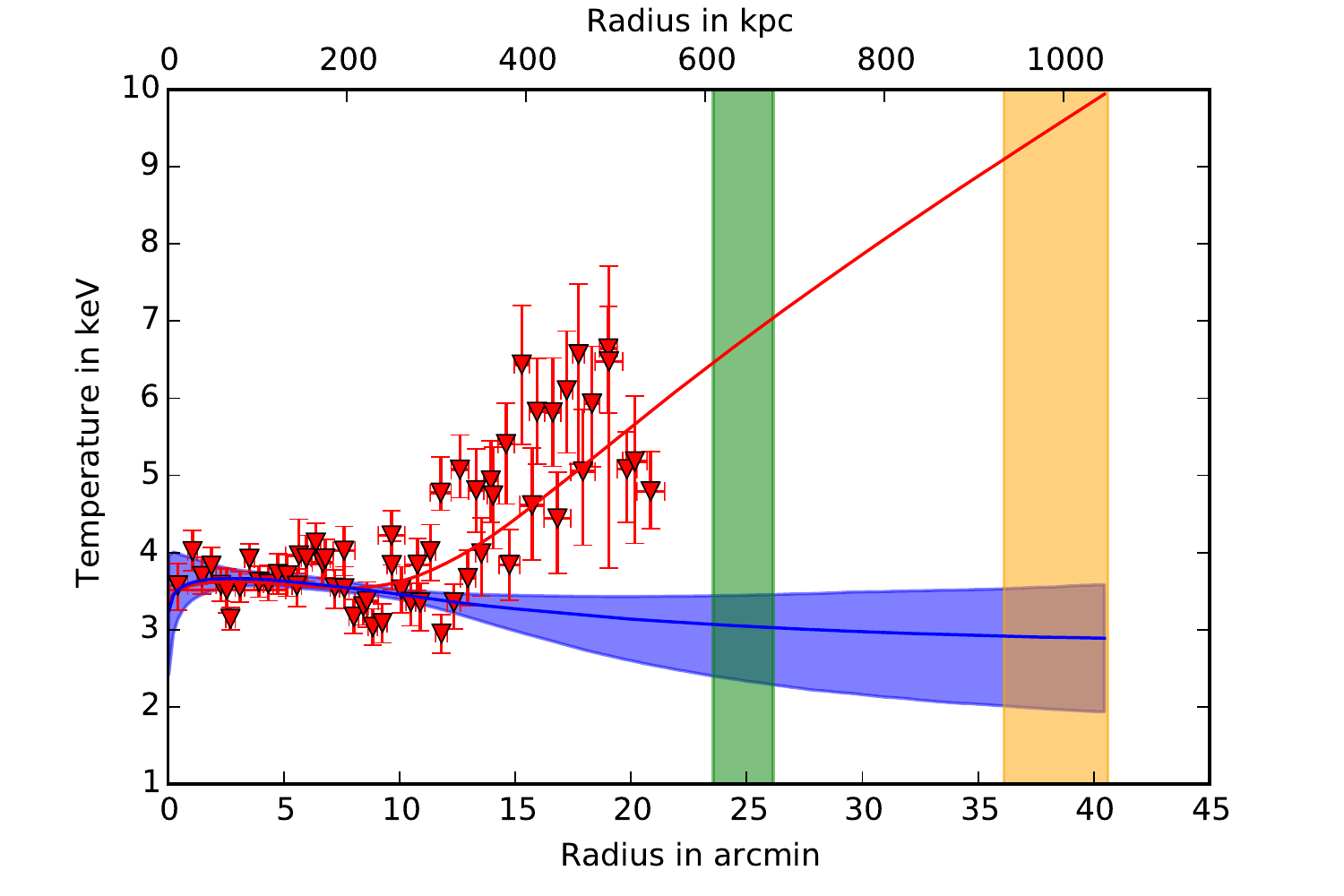}
	\includegraphics[width=0.45\textwidth]{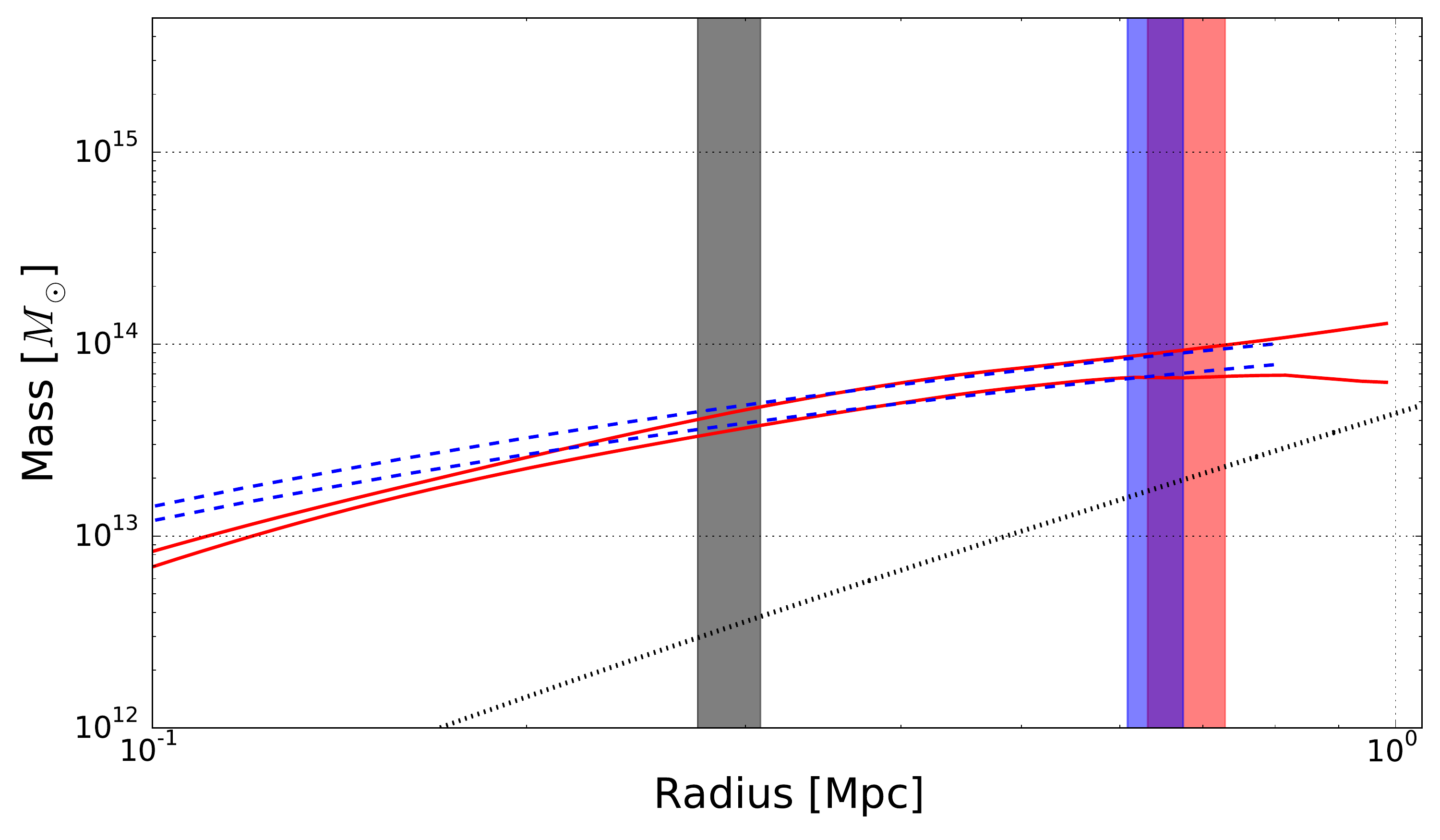}
	\caption{As Fig. \ref{fig:app_2A0335} but for A1367.}
	\label{fig:app_A1367}
\end{figure}
\begin{figure}
	\centering
	\includegraphics[width=0.45\textwidth]{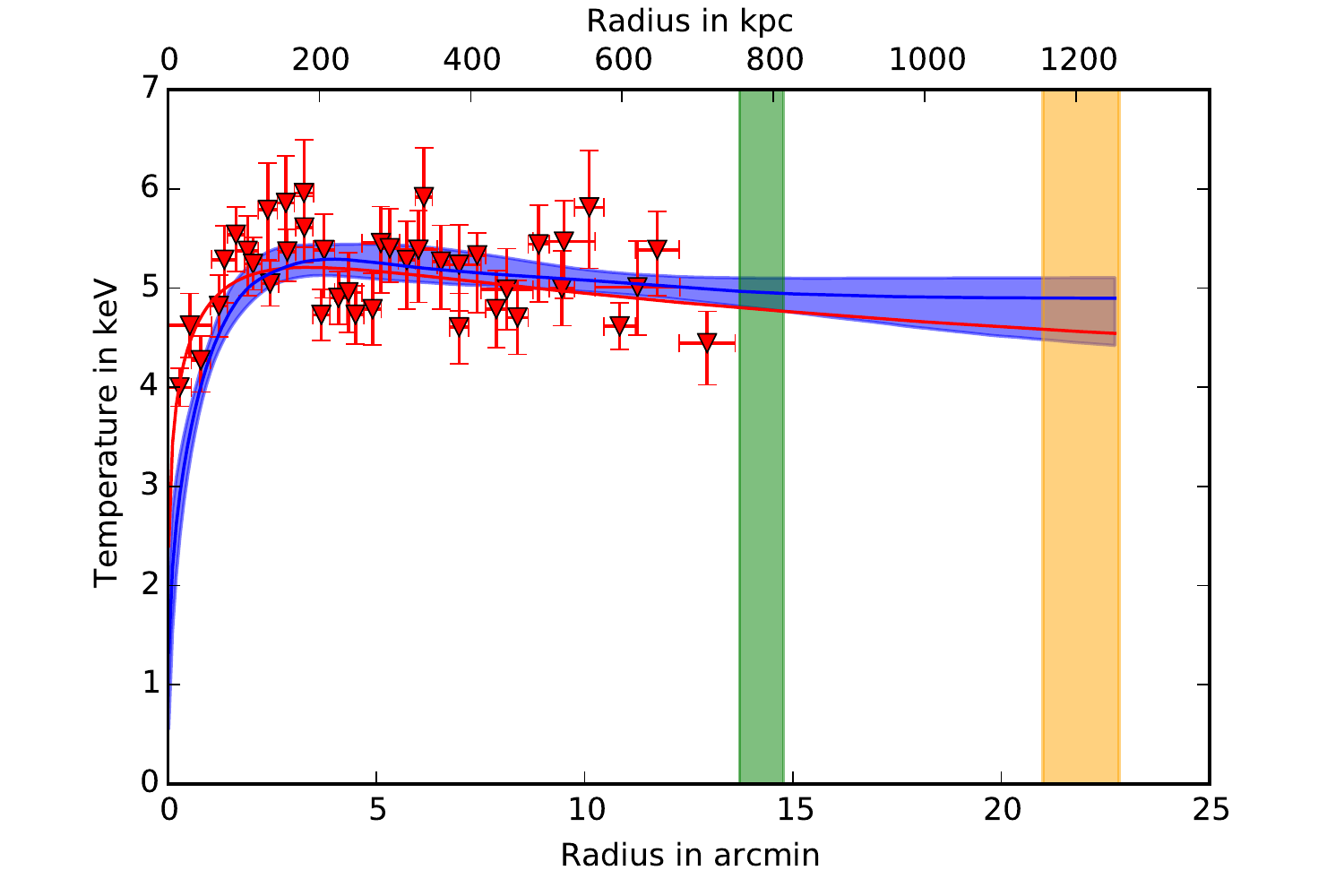}
	\includegraphics[width=0.45\textwidth]{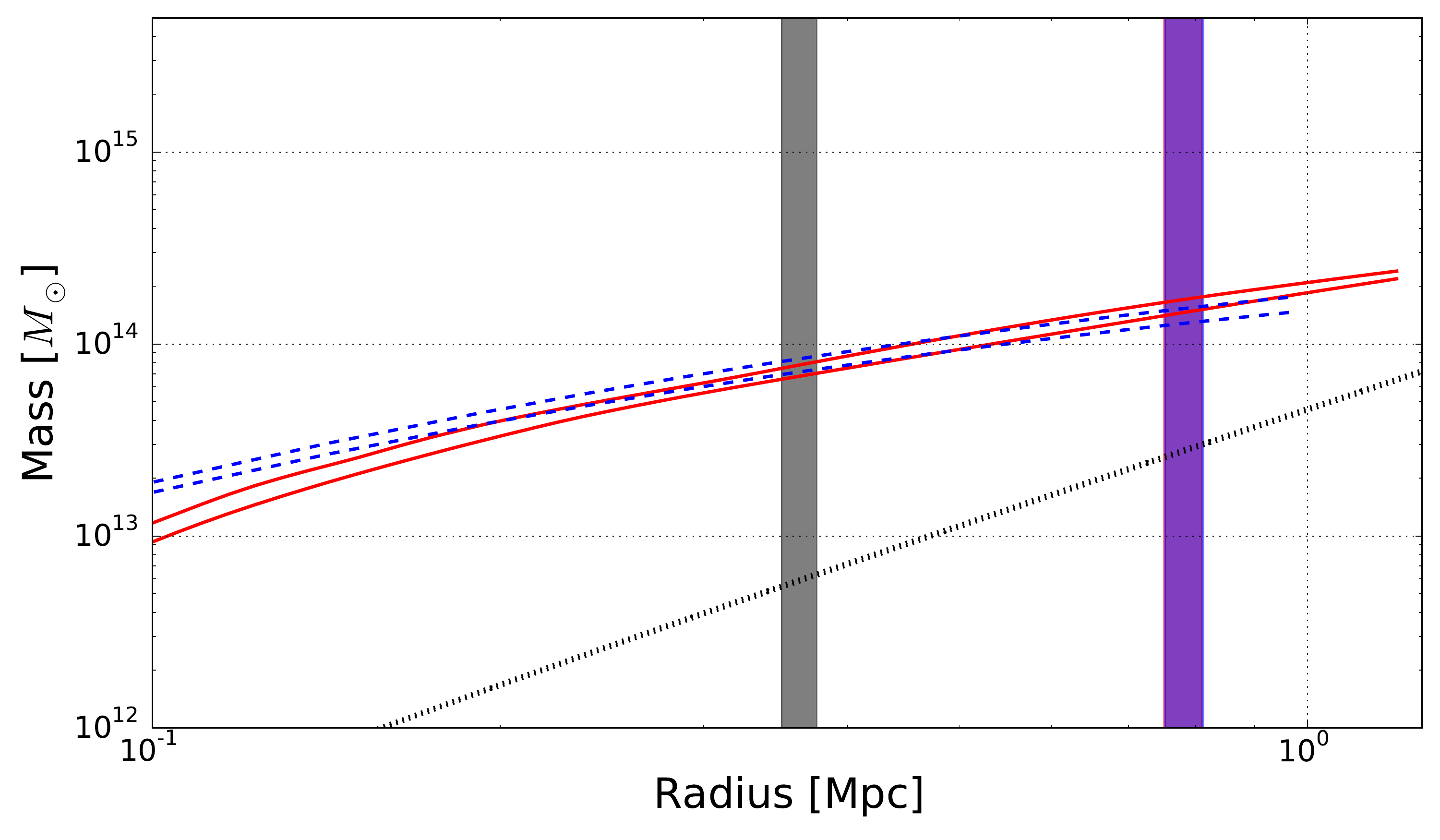}
	\caption{As Fig. \ref{fig:app_2A0335} but for A1644.}
	\label{fig:app_A1644}
\end{figure}
\clearpage
\begin{figure}
	\centering
	\includegraphics[width=0.45\textwidth]{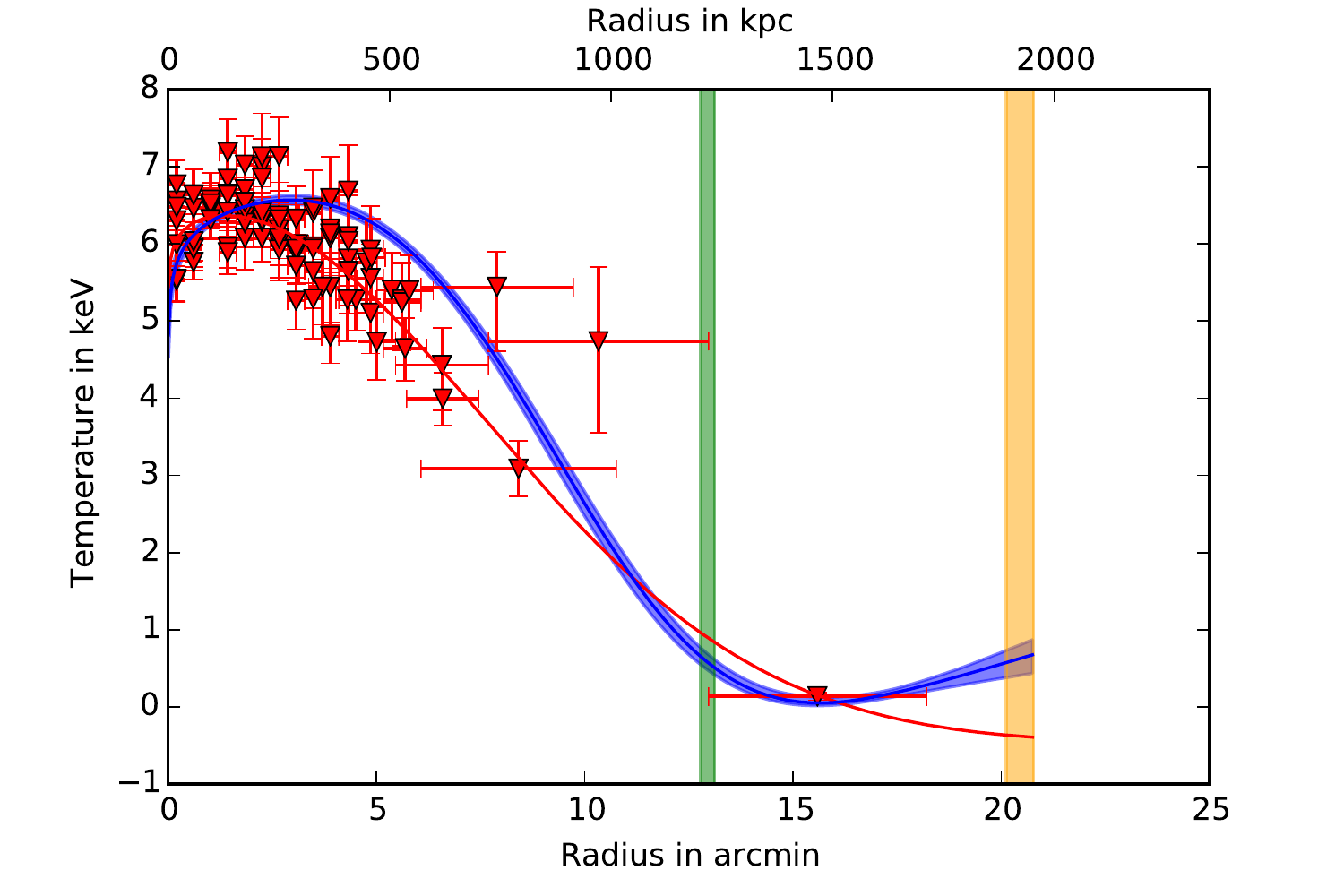}
	\includegraphics[width=0.45\textwidth]{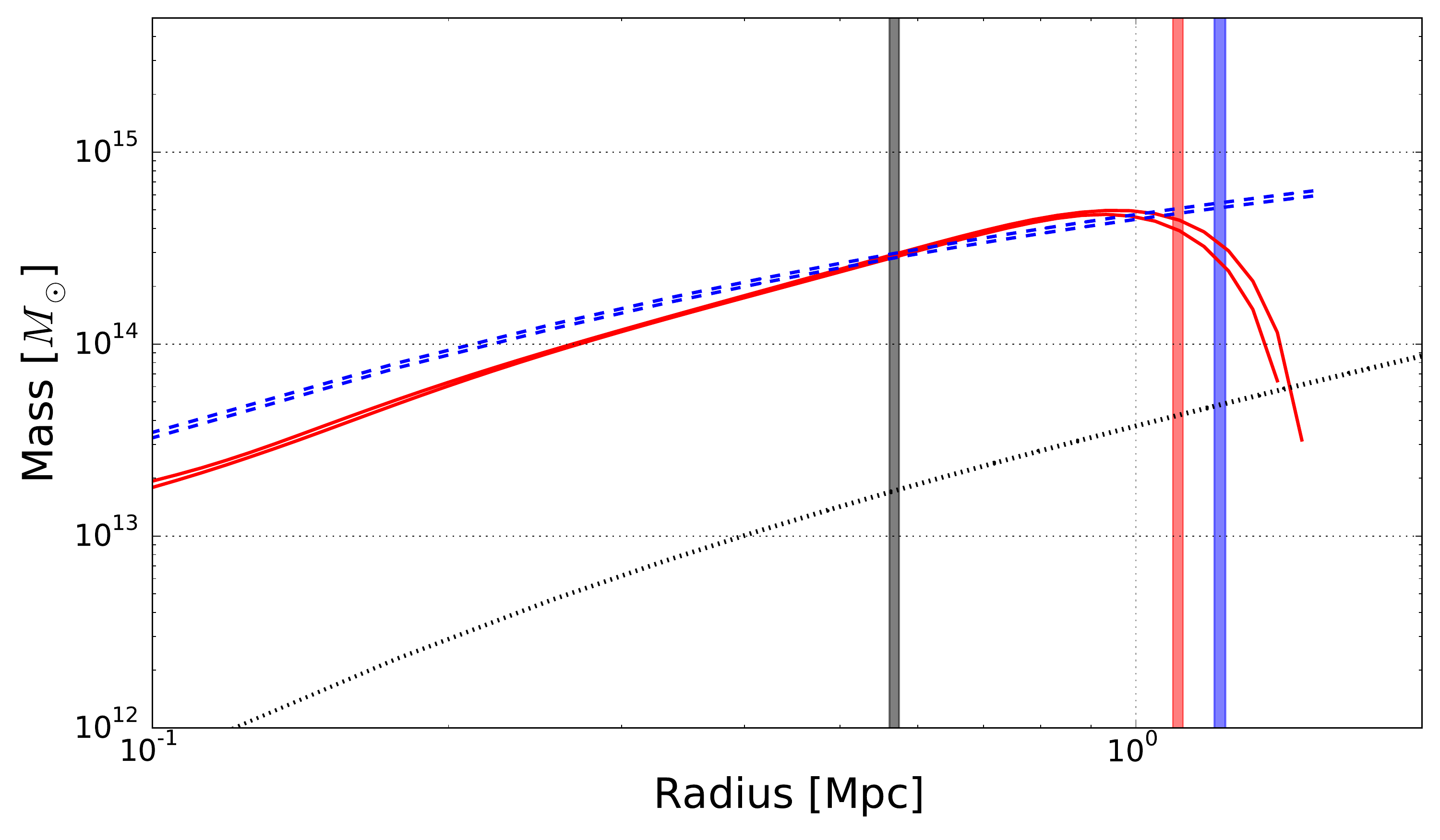}
	\caption{As Fig. \ref{fig:app_2A0335} but for A1650.}
	\label{fig:app_A1650}
\end{figure}
\begin{figure}
	\centering
	\includegraphics[width=0.45\textwidth]{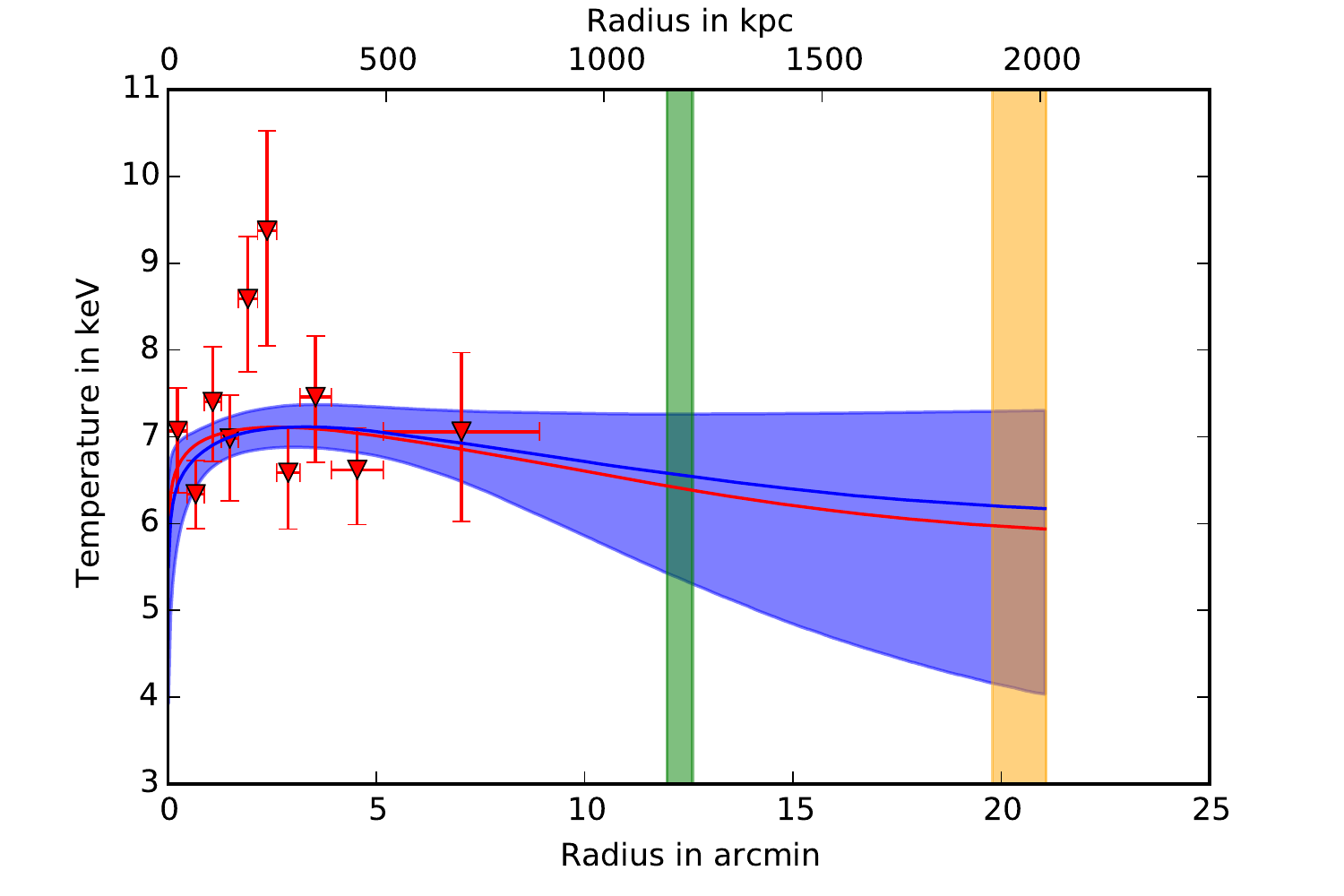}
	\includegraphics[width=0.45\textwidth]{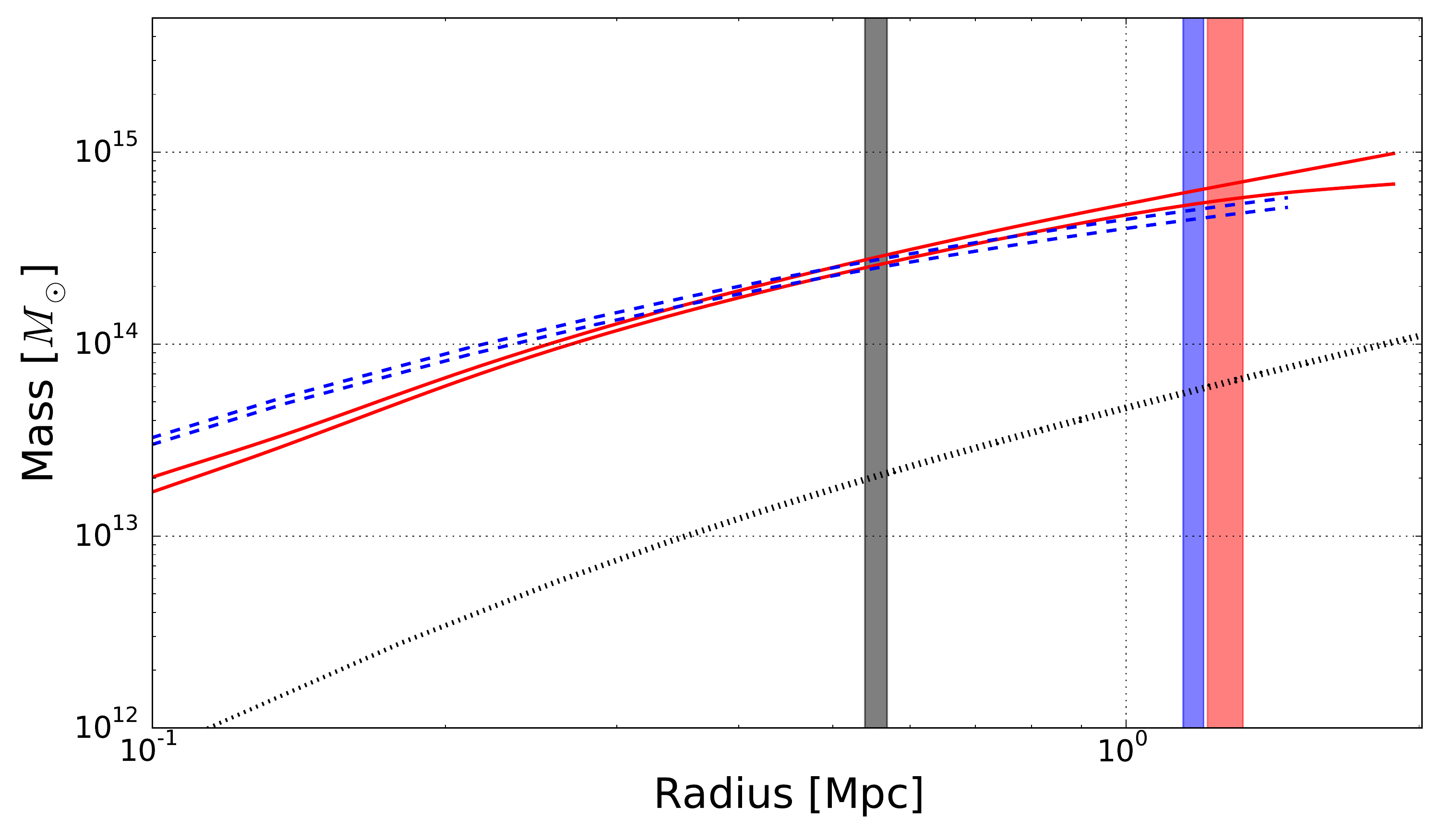}
	\caption{As Fig. \ref{fig:app_2A0335} but for A1651.}
	\label{fig:app_A1651}
\end{figure}
\begin{figure}
	\centering
	\includegraphics[width=0.45\textwidth]{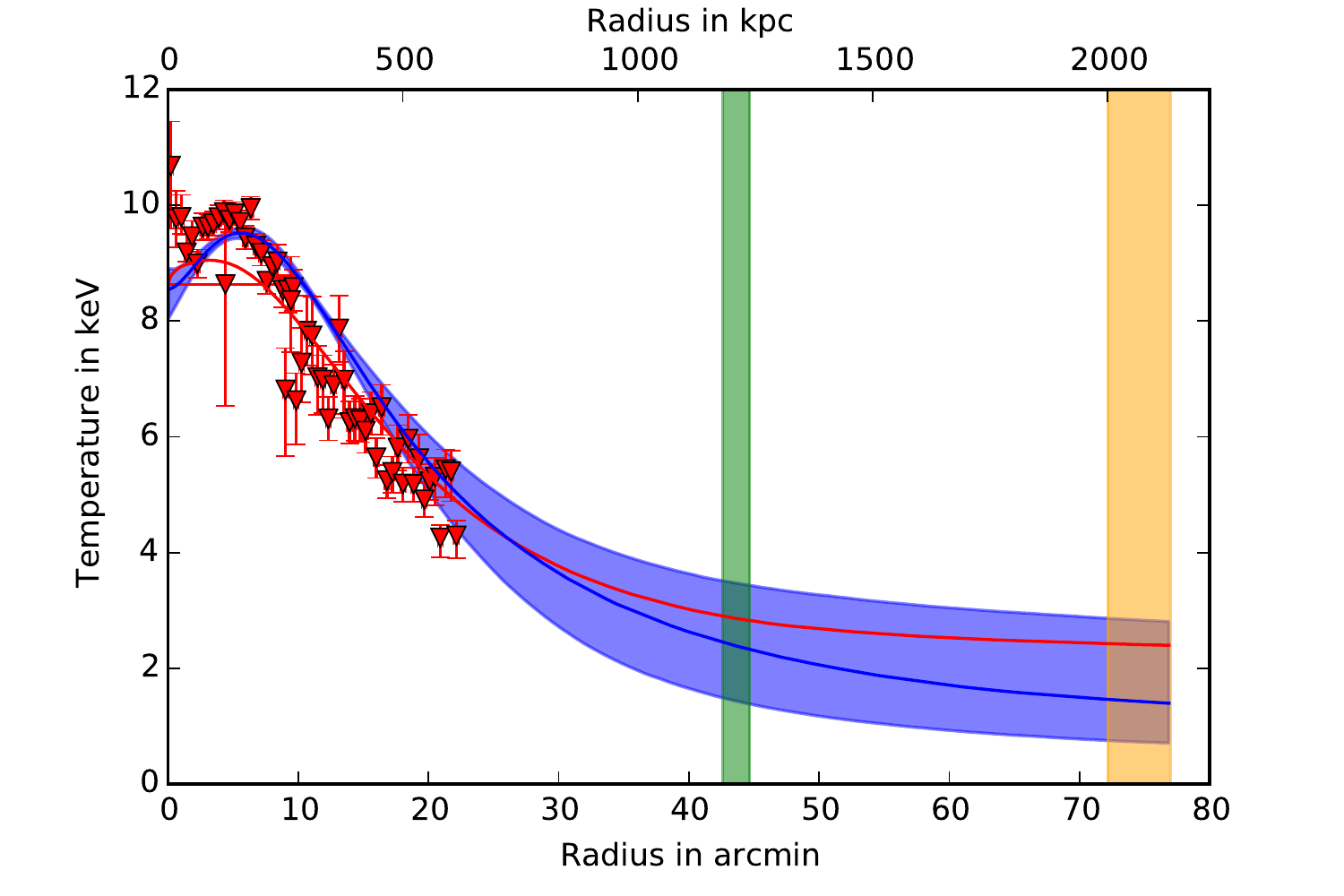}
	\includegraphics[width=0.45\textwidth]{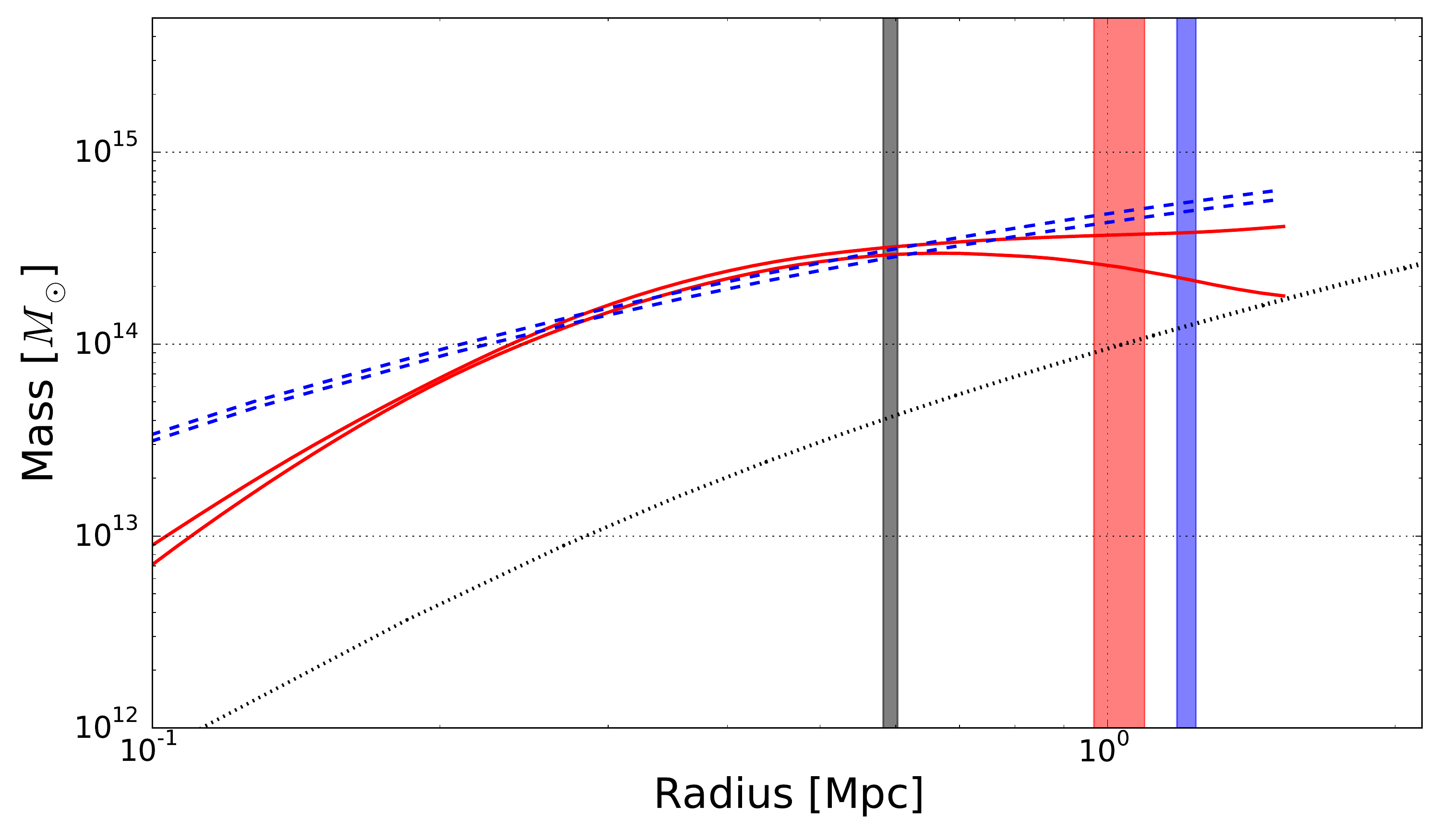}
	\caption{As Fig. \ref{fig:app_2A0335} but for A1656.}
	\label{fig:app_A1656}
\end{figure}
\clearpage
\begin{figure}
	\centering
	\includegraphics[width=0.45\textwidth]{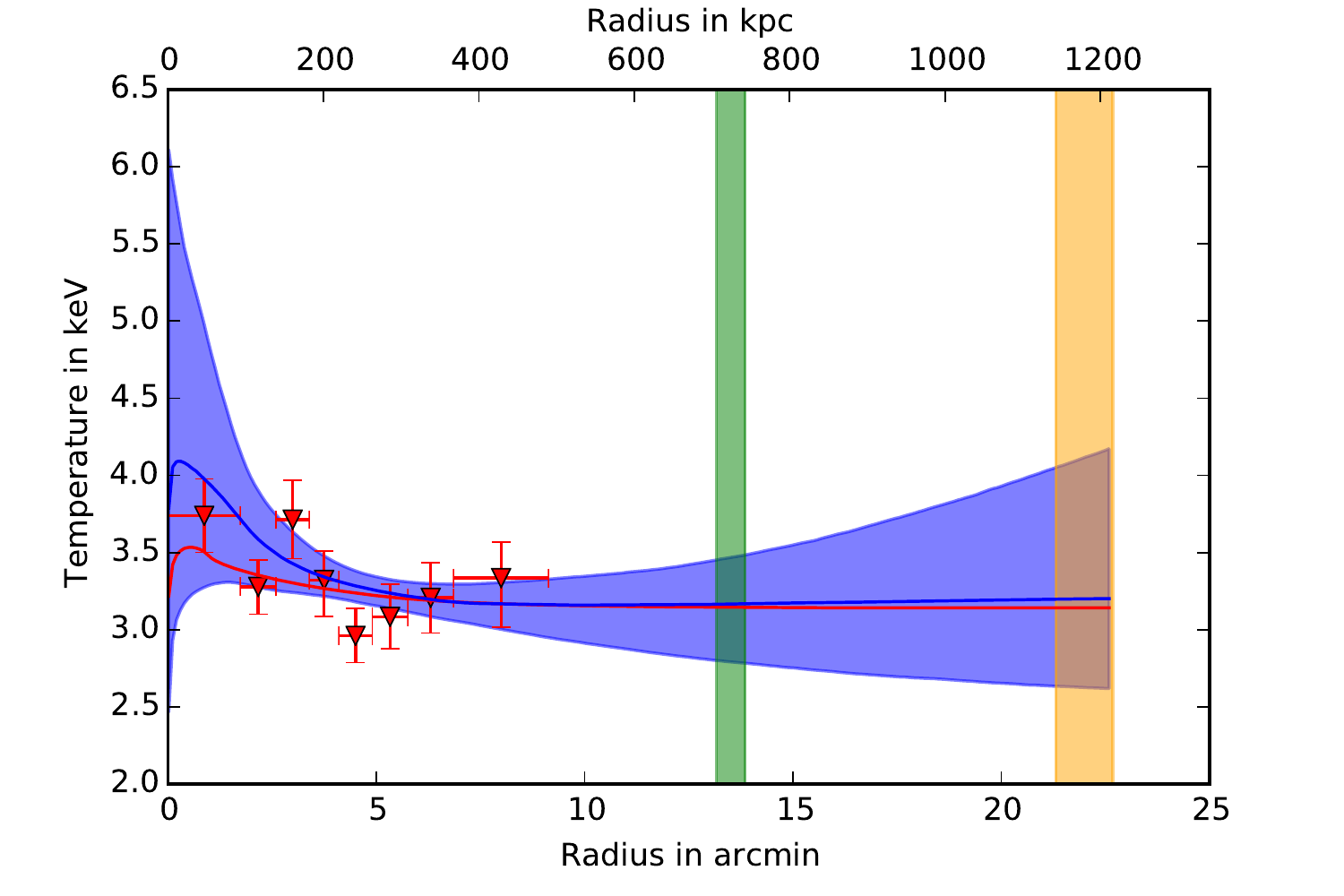}
	\includegraphics[width=0.45\textwidth]{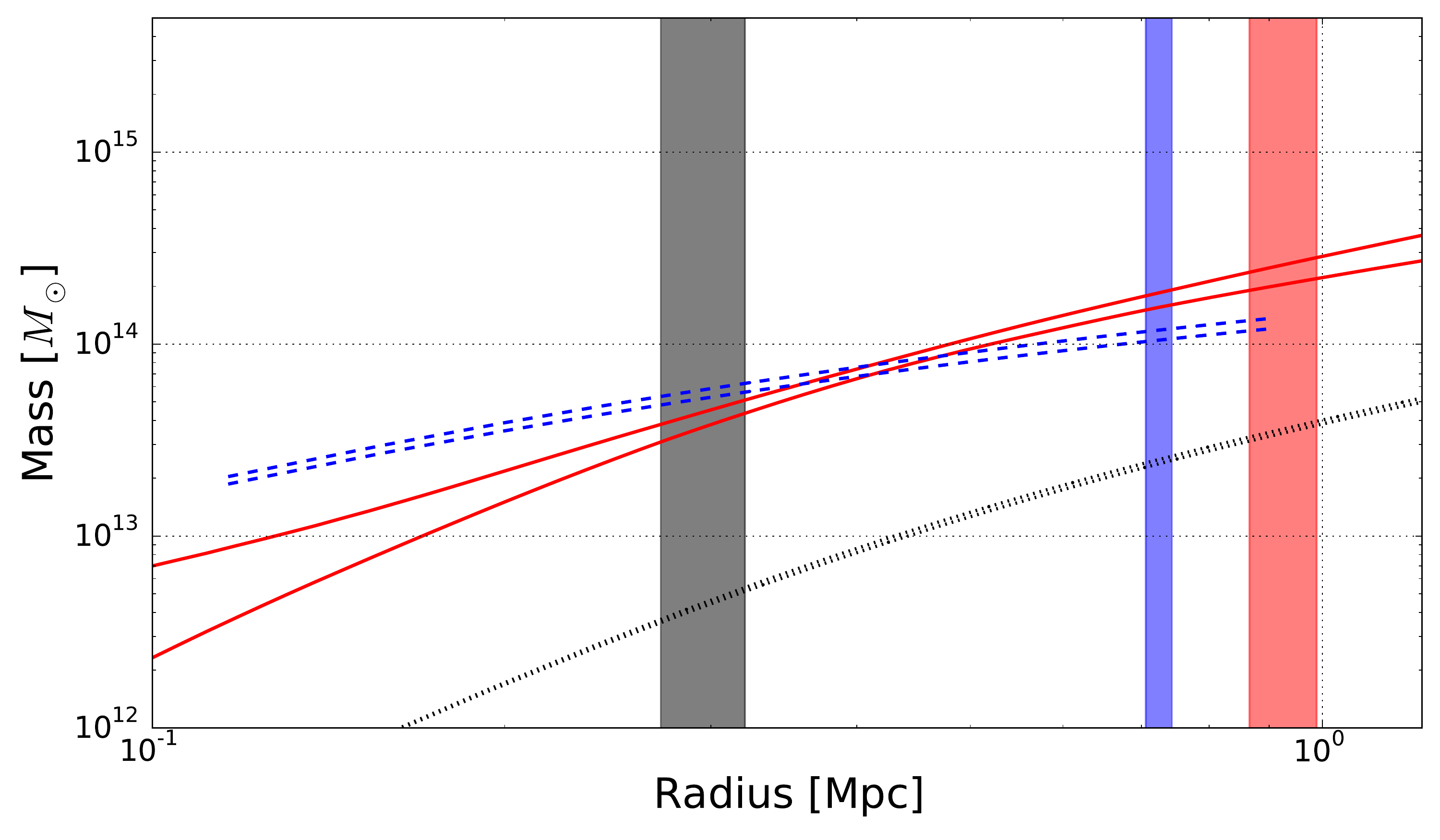}
	\caption{As Fig. \ref{fig:app_2A0335} but for A1736.}
	\label{fig:app_A1736}
\end{figure}
\begin{figure}
	\centering
	\includegraphics[width=0.45\textwidth]{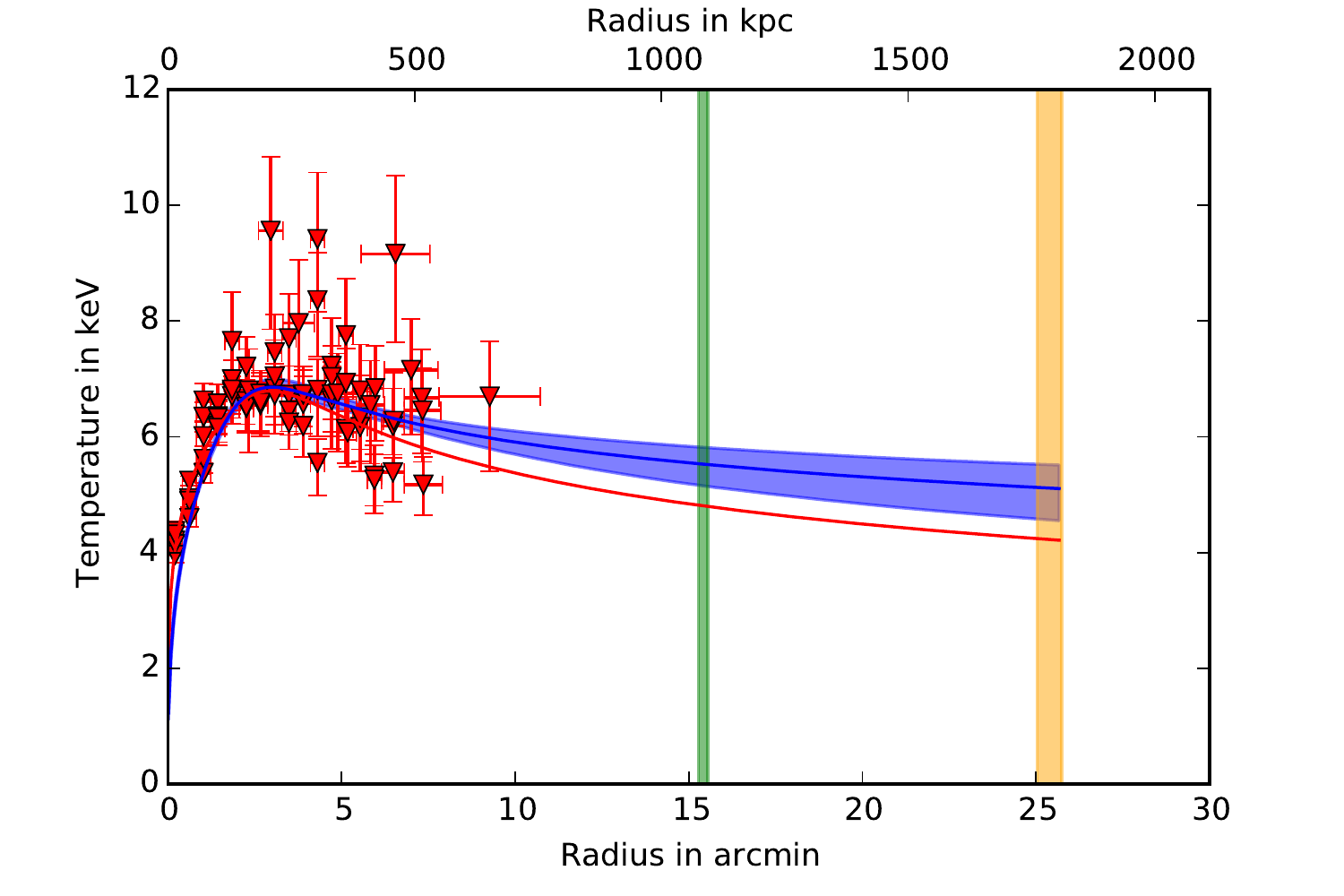}
	\includegraphics[width=0.45\textwidth]{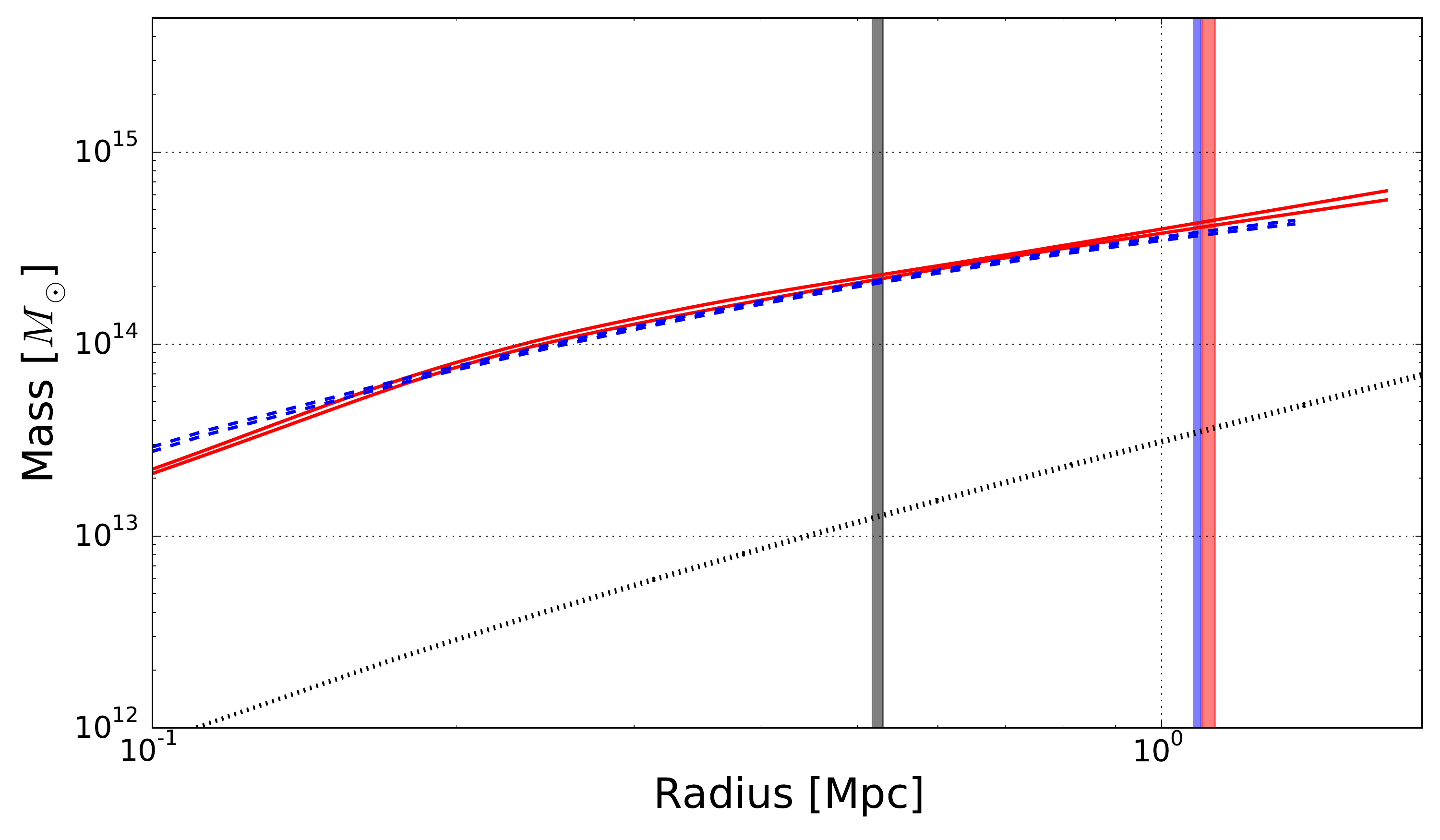}
	\caption{As Fig. \ref{fig:app_2A0335} but for A1795.}
	\label{fig:app_A1795}
\end{figure}
\begin{figure}
	\centering
	\includegraphics[width=0.45\textwidth]{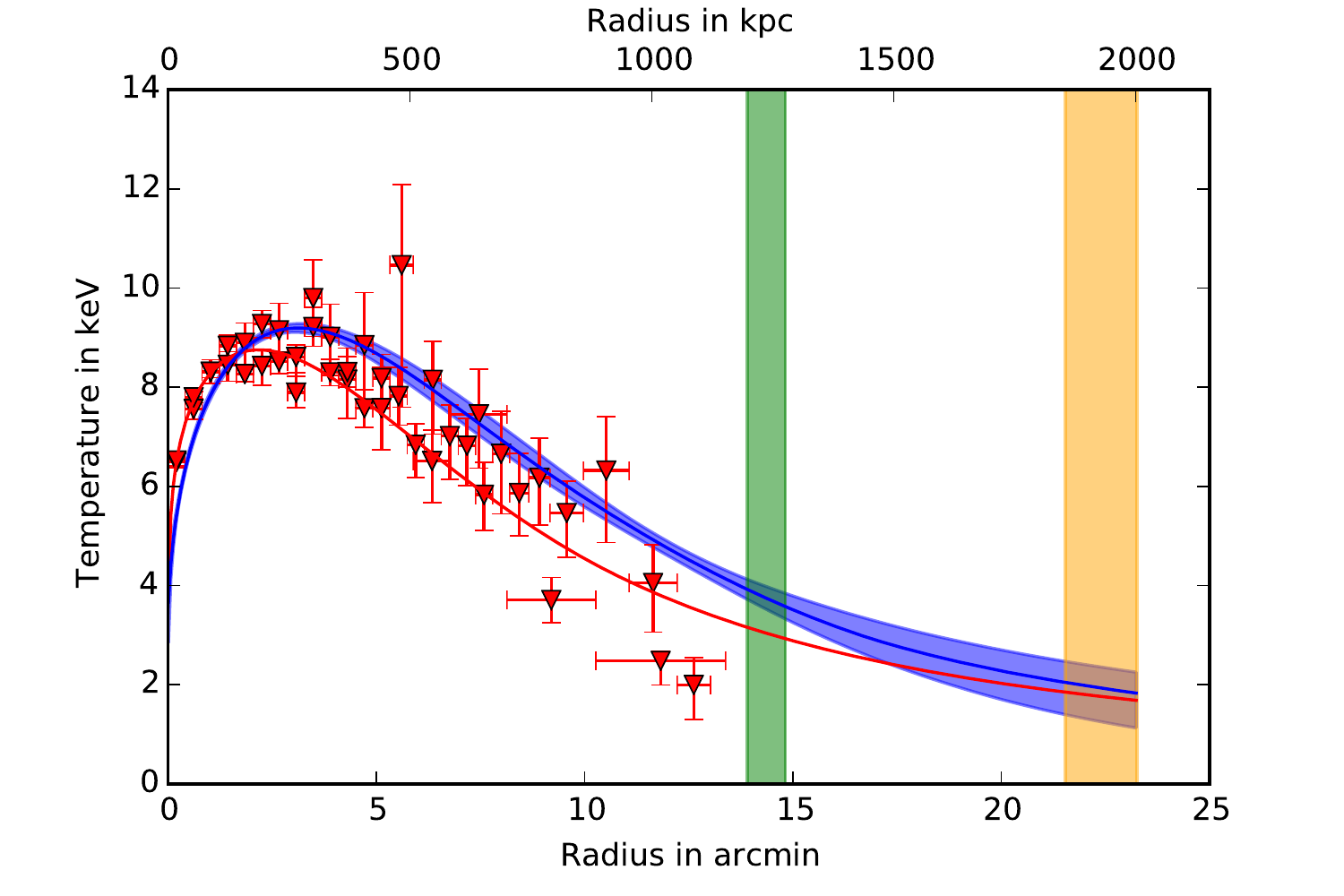}
	\includegraphics[width=0.45\textwidth]{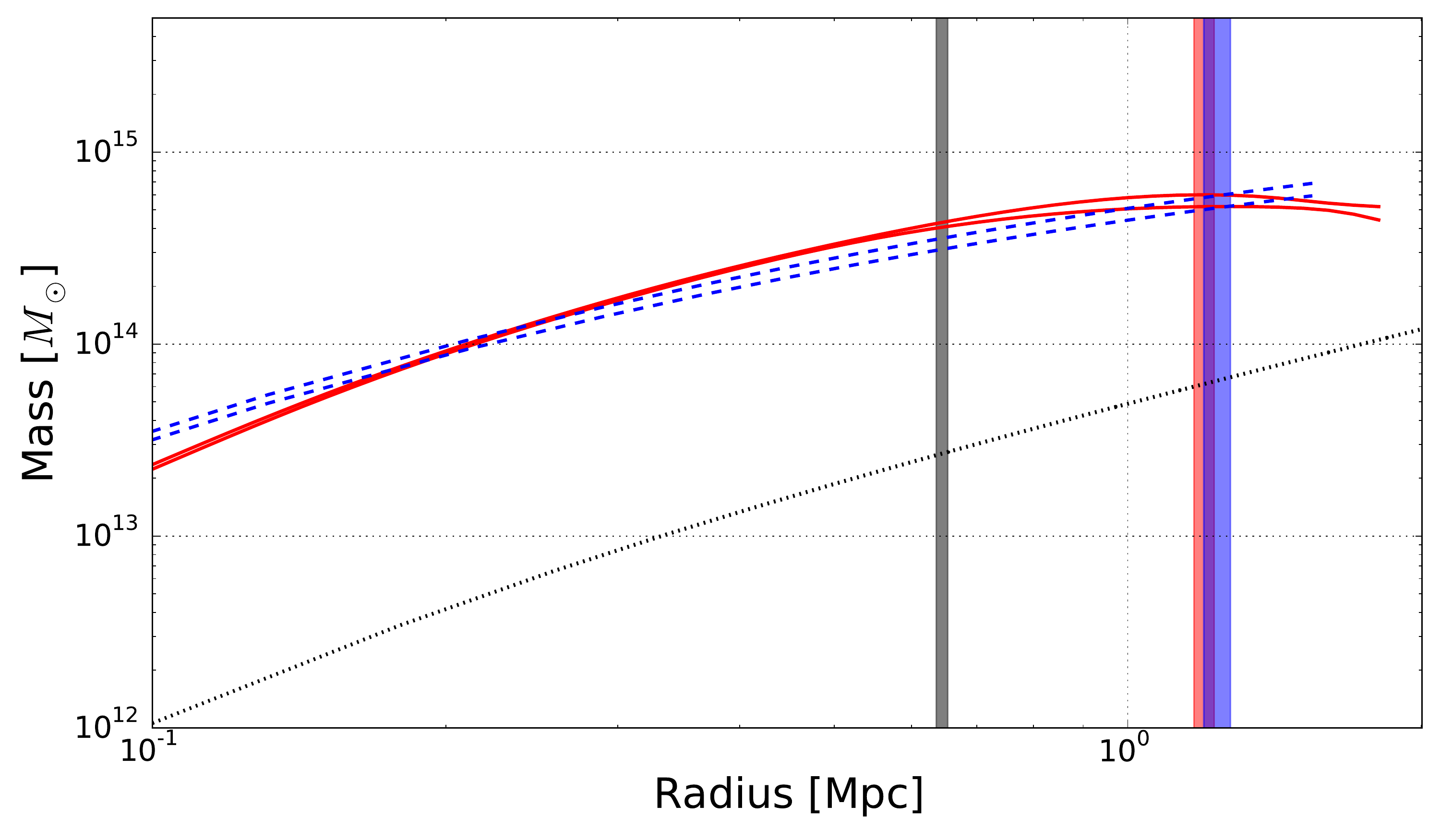}
	\caption{As Fig. \ref{fig:app_2A0335} but for A2029.}
	\label{fig:app_A2029}
\end{figure}
\clearpage
\begin{figure}
	\centering
	\includegraphics[width=0.45\textwidth]{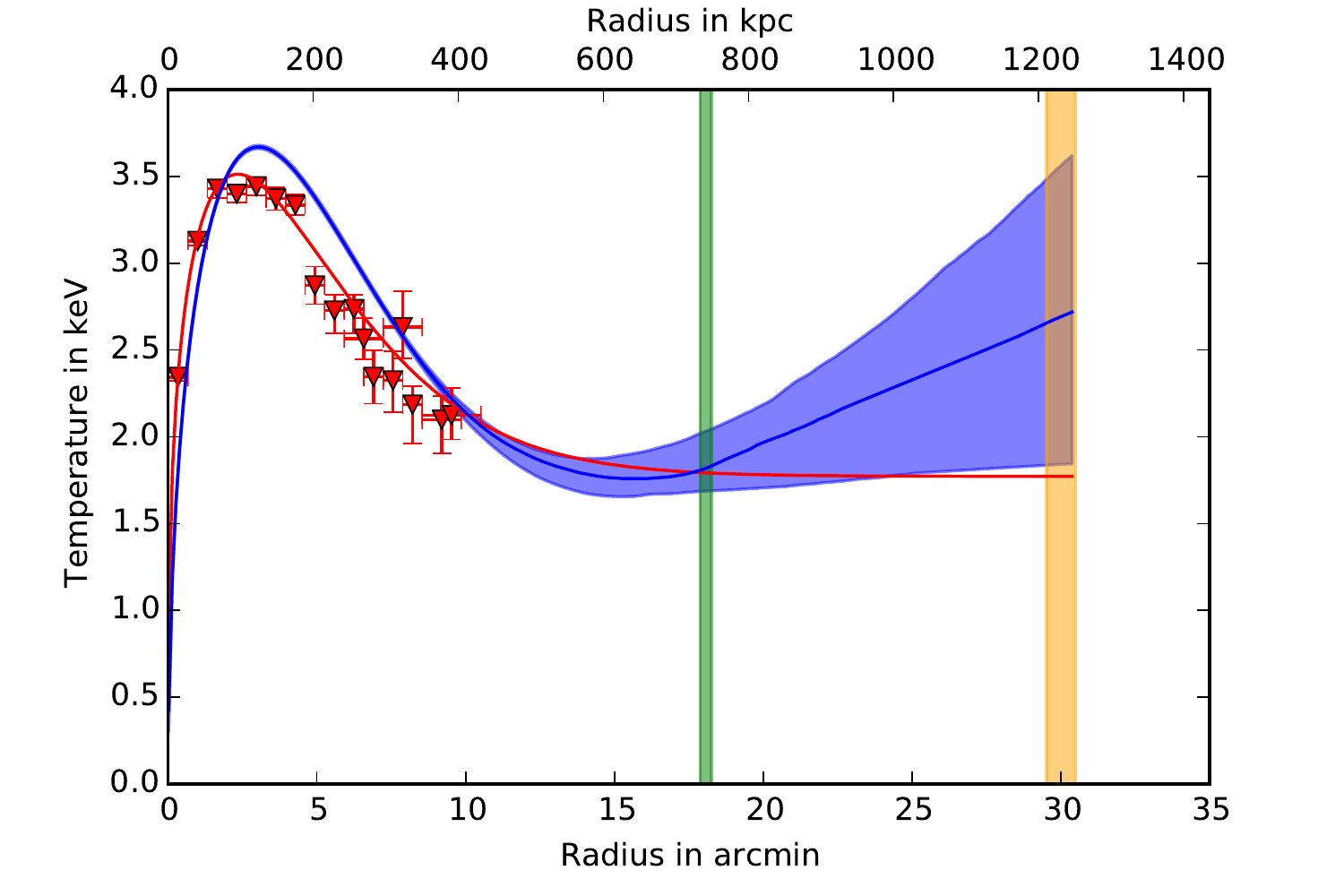}
	\includegraphics[width=0.45\textwidth]{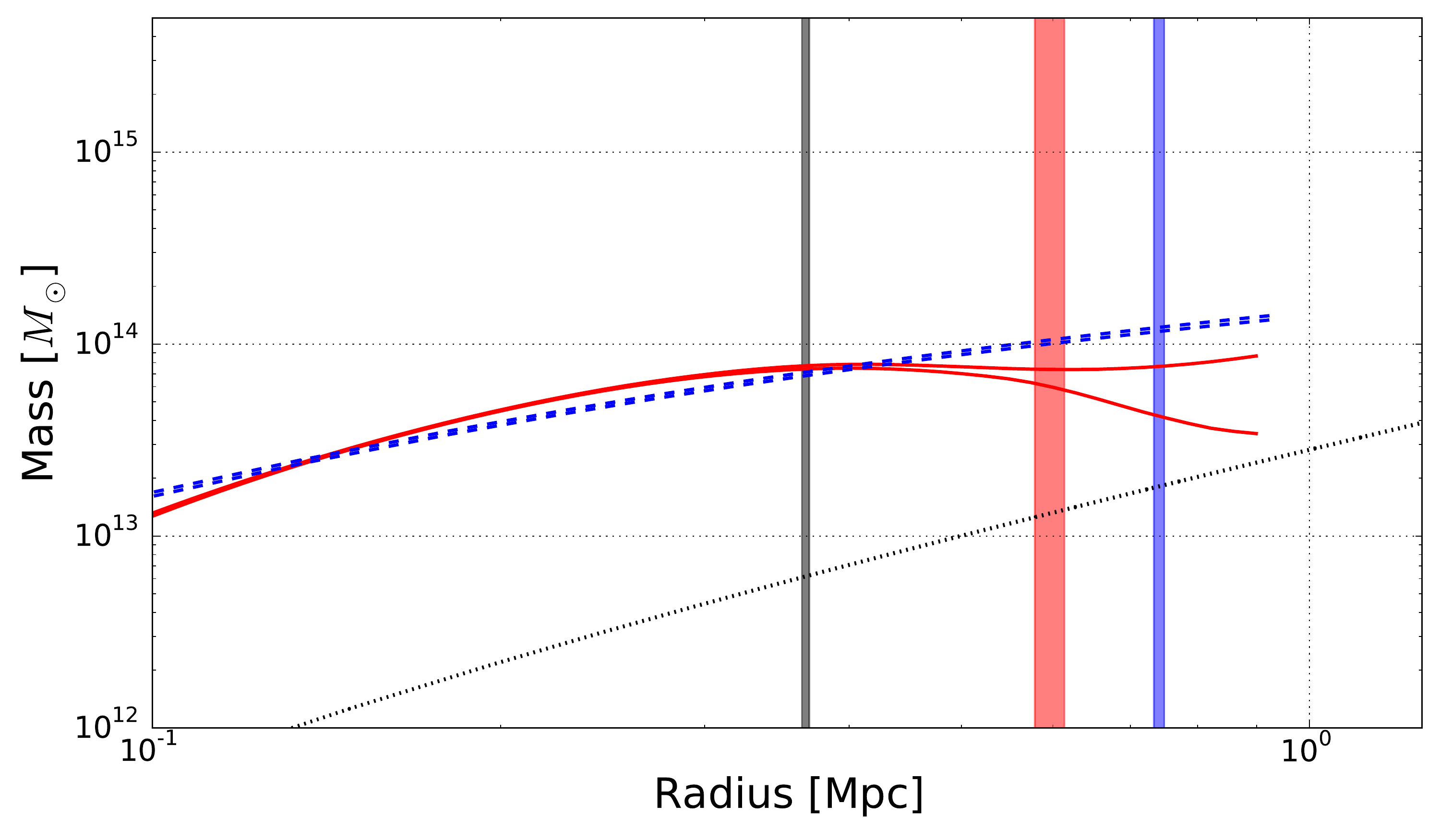}
	\caption{As Fig. \ref{fig:app_2A0335} but for A2052.}
	\label{fig:app_A2052}
\end{figure}
\begin{figure}
	\centering
	\includegraphics[width=0.45\textwidth]{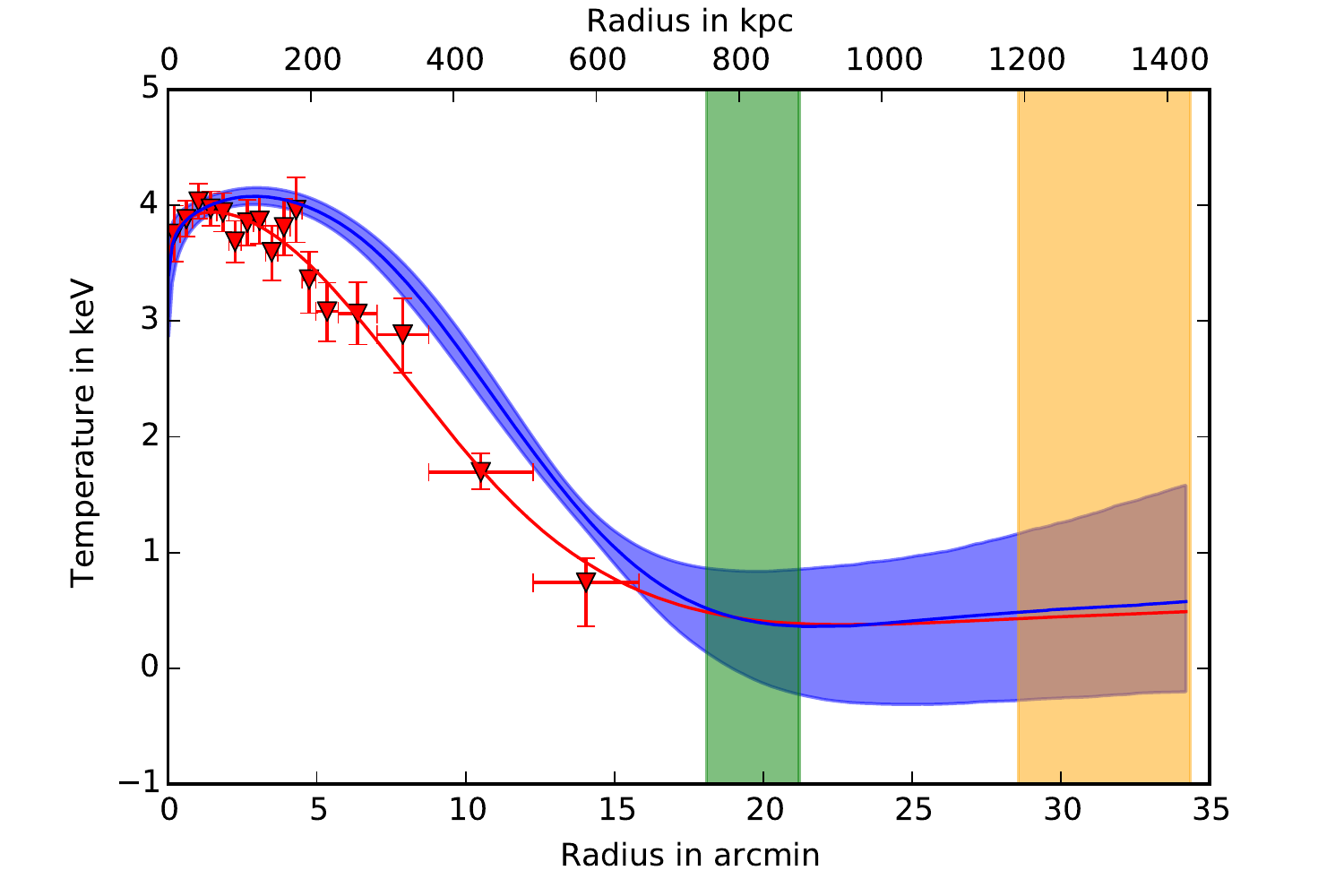}
	\includegraphics[width=0.45\textwidth]{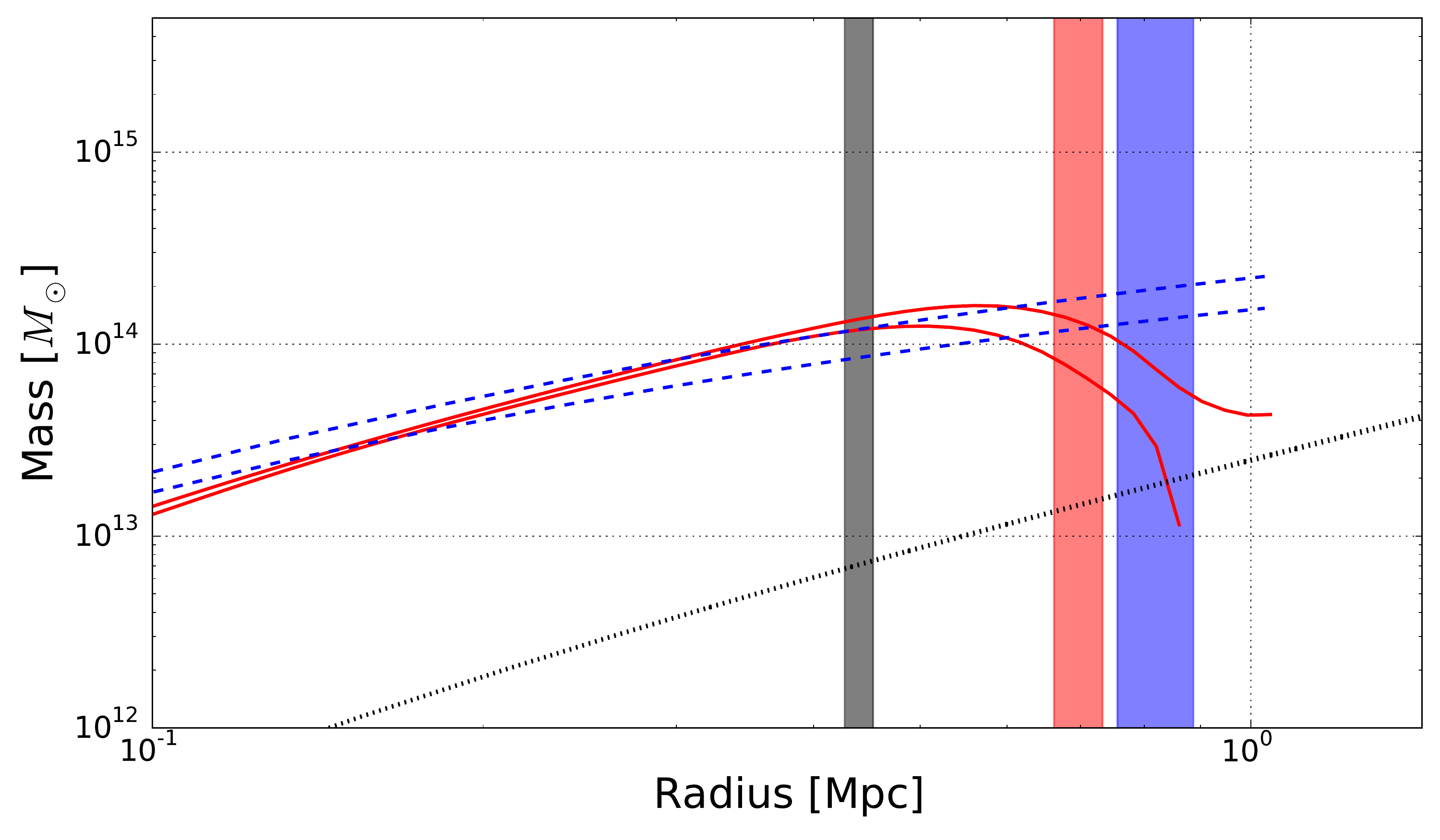}
	\caption{As Fig. \ref{fig:app_2A0335} but for A2063.}
	\label{fig:app_A2063}
\end{figure}
\begin{figure}
	\centering
	\includegraphics[width=0.45\textwidth]{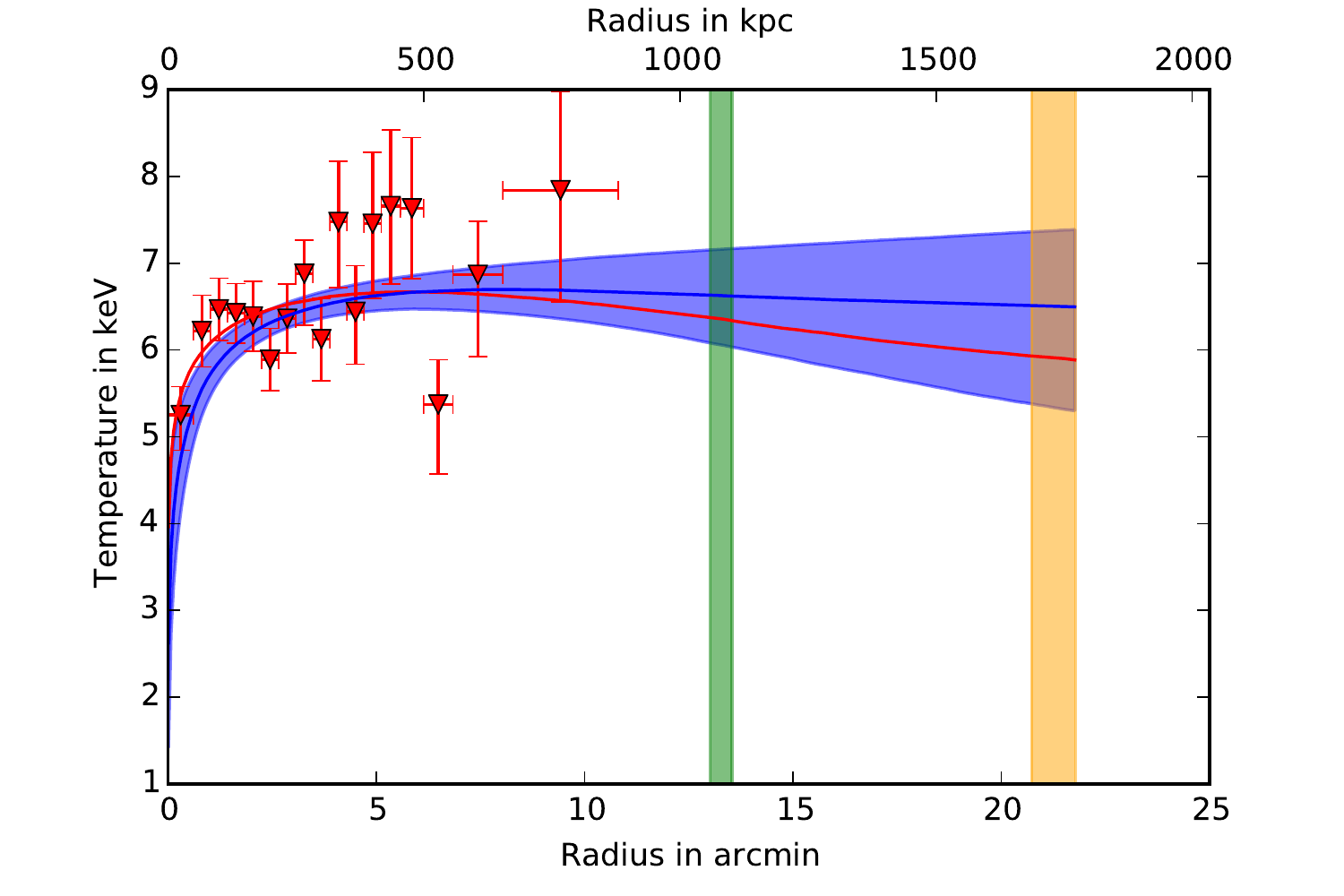}
	\includegraphics[width=0.45\textwidth]{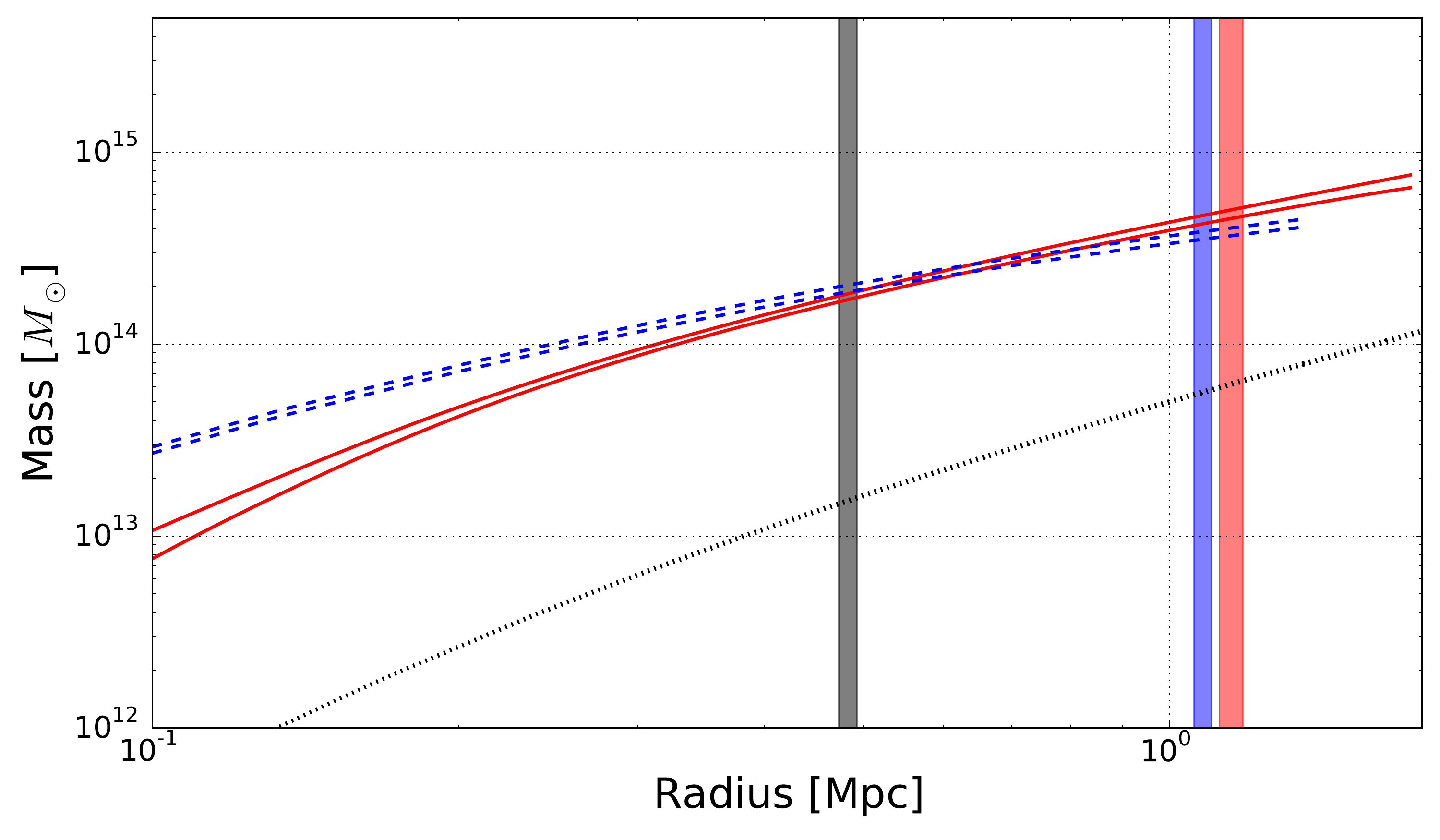}
	\caption{As Fig. \ref{fig:app_2A0335} but for A2065.}
	\label{fig:app_A2065}
\end{figure}
\clearpage
\begin{figure}
	\centering
	\includegraphics[width=0.45\textwidth]{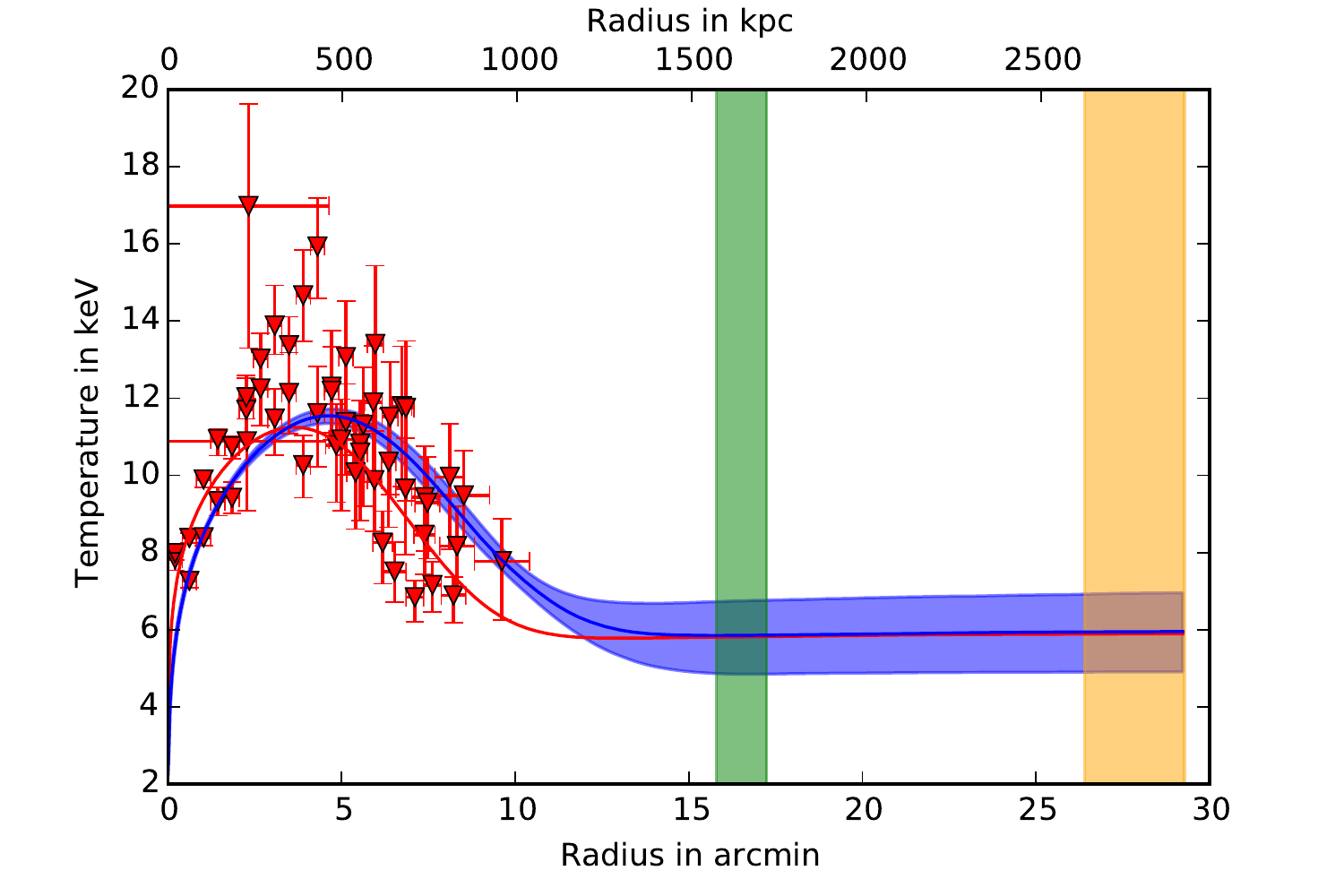}
	\includegraphics[width=0.45\textwidth]{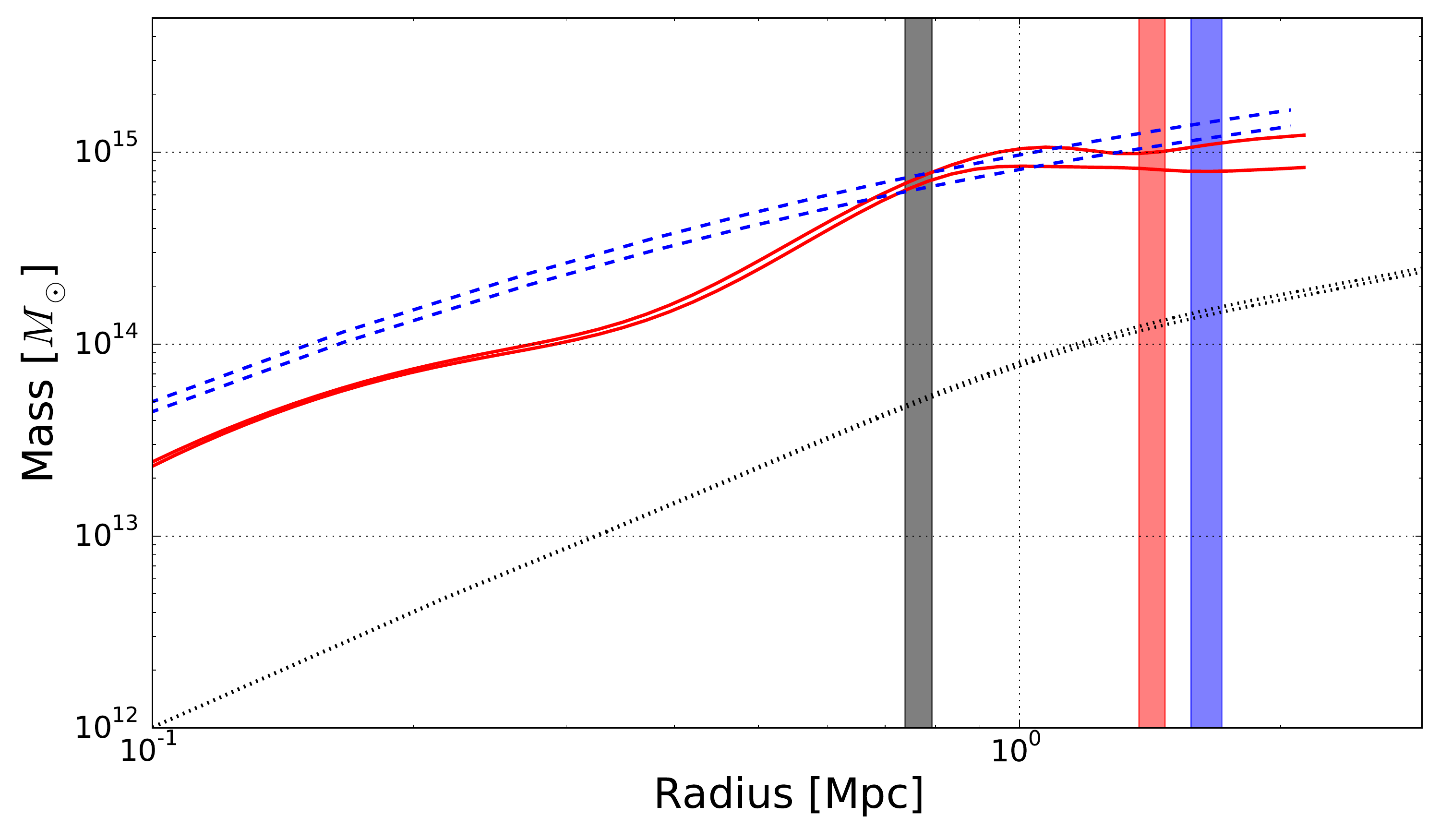}
	\caption{As Fig. \ref{fig:app_2A0335} but for A2142.}
	\label{fig:app_A2142}
\end{figure}
\begin{figure}
	\centering
	\includegraphics[width=0.45\textwidth]{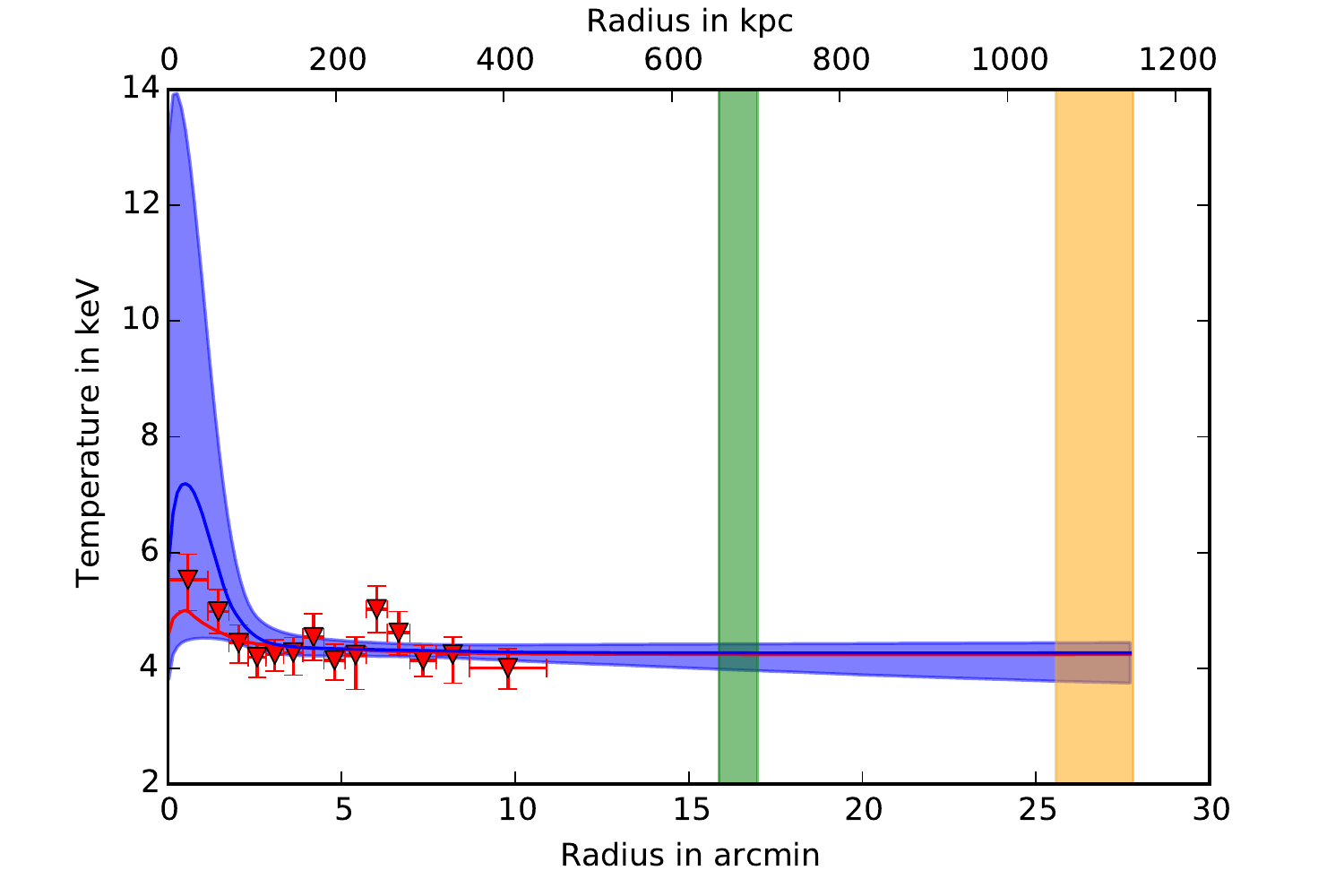}
	\includegraphics[width=0.45\textwidth]{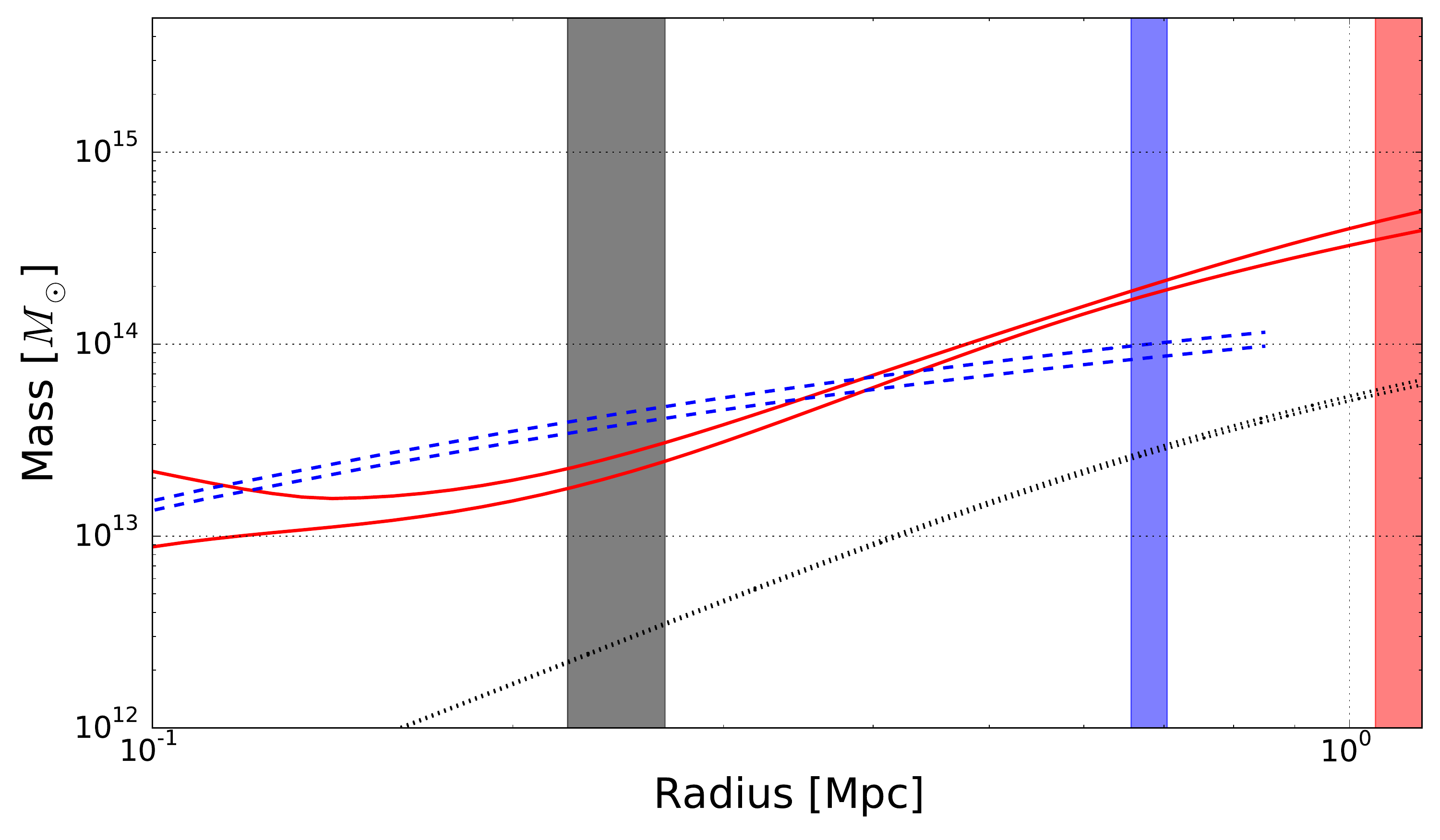}
	\caption{As Fig. \ref{fig:app_2A0335} but for A2147.}
	\label{fig:app_A2147}
\end{figure}
\begin{figure}
	\centering
	\includegraphics[width=0.45\textwidth]{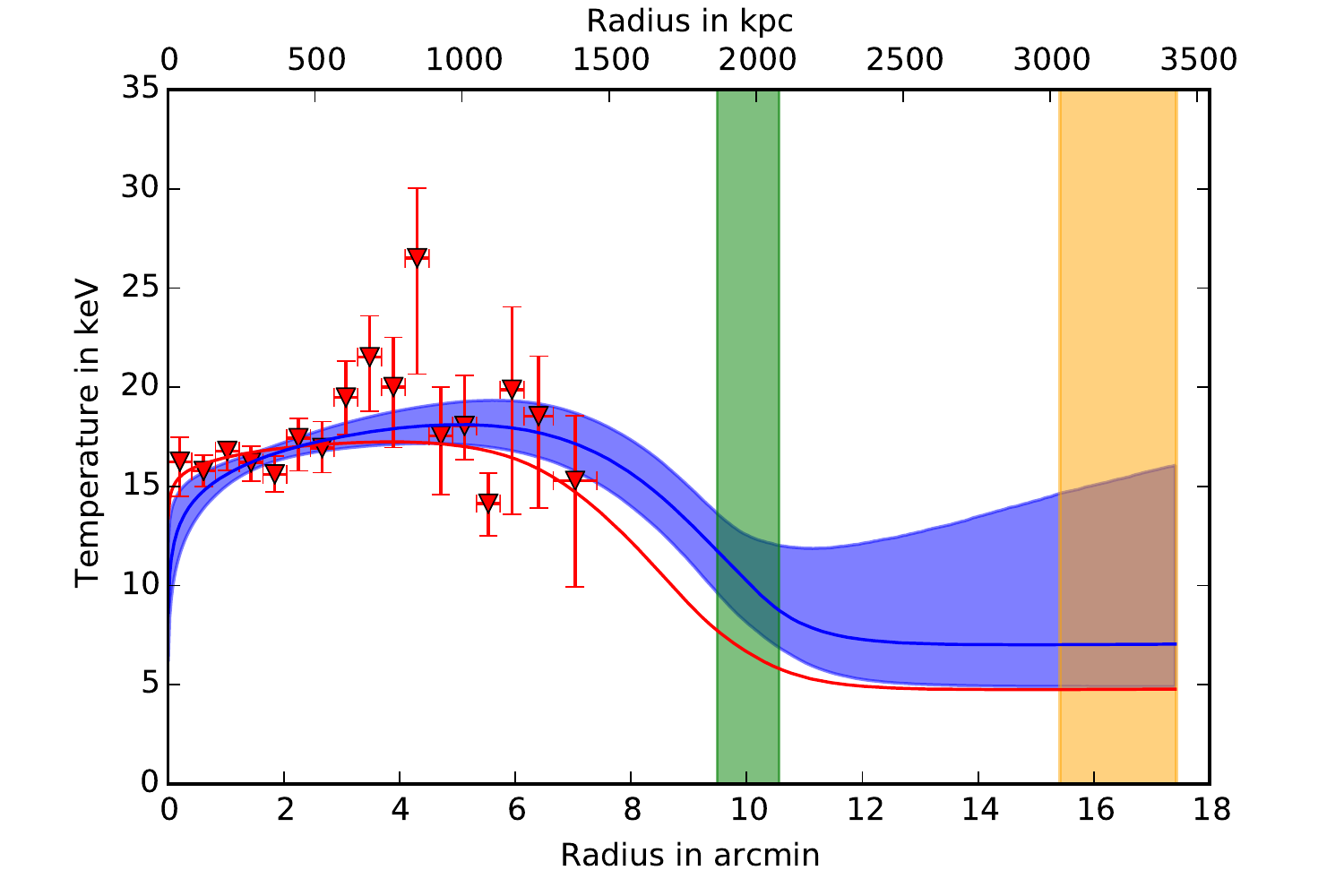}
	\includegraphics[width=0.45\textwidth]{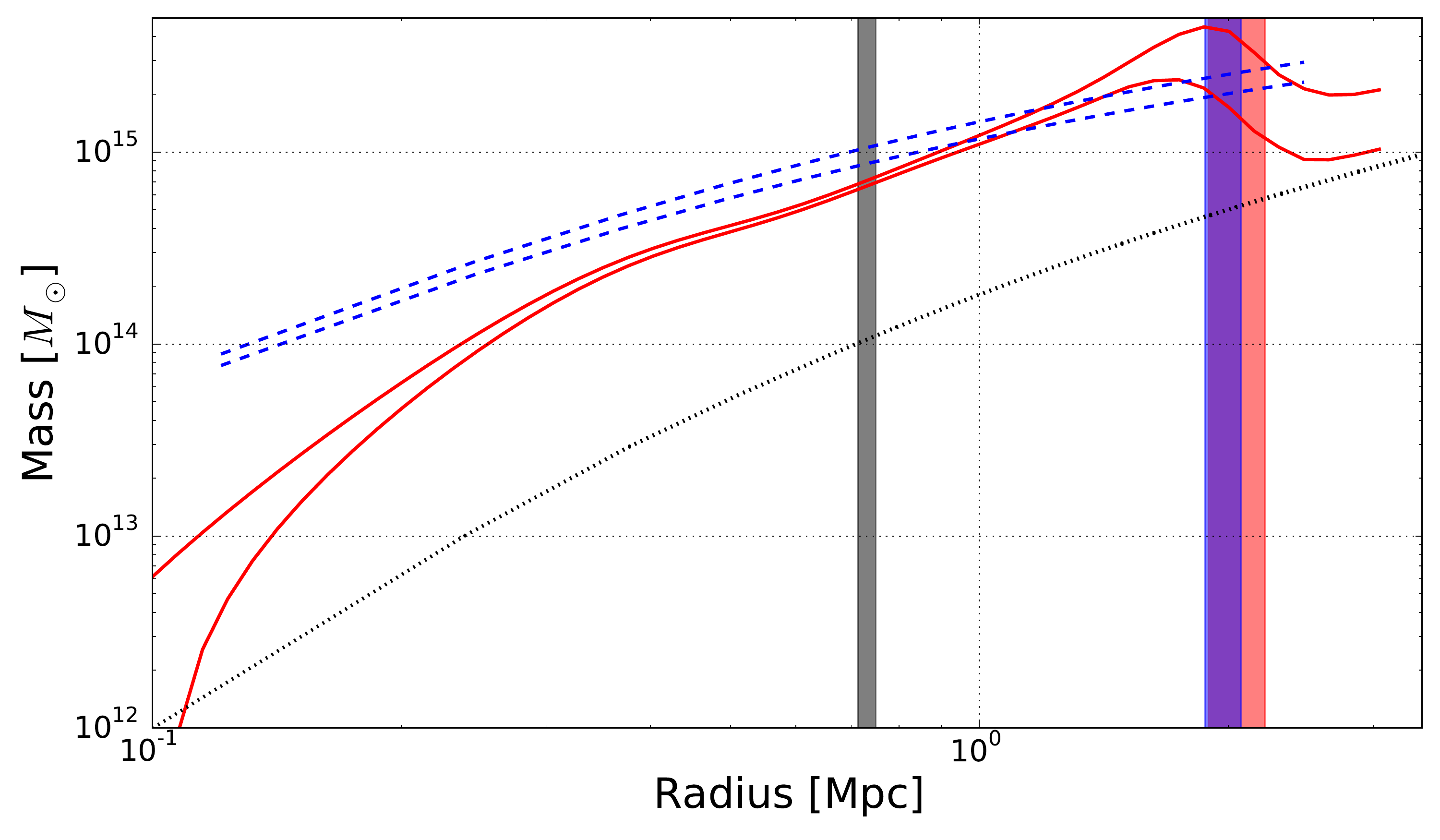}
	\caption{As Fig. \ref{fig:app_2A0335} but for A2163.}
	\label{fig:app_A2163}
\end{figure}
\clearpage
\begin{figure}
	\centering
	\includegraphics[width=0.45\textwidth]{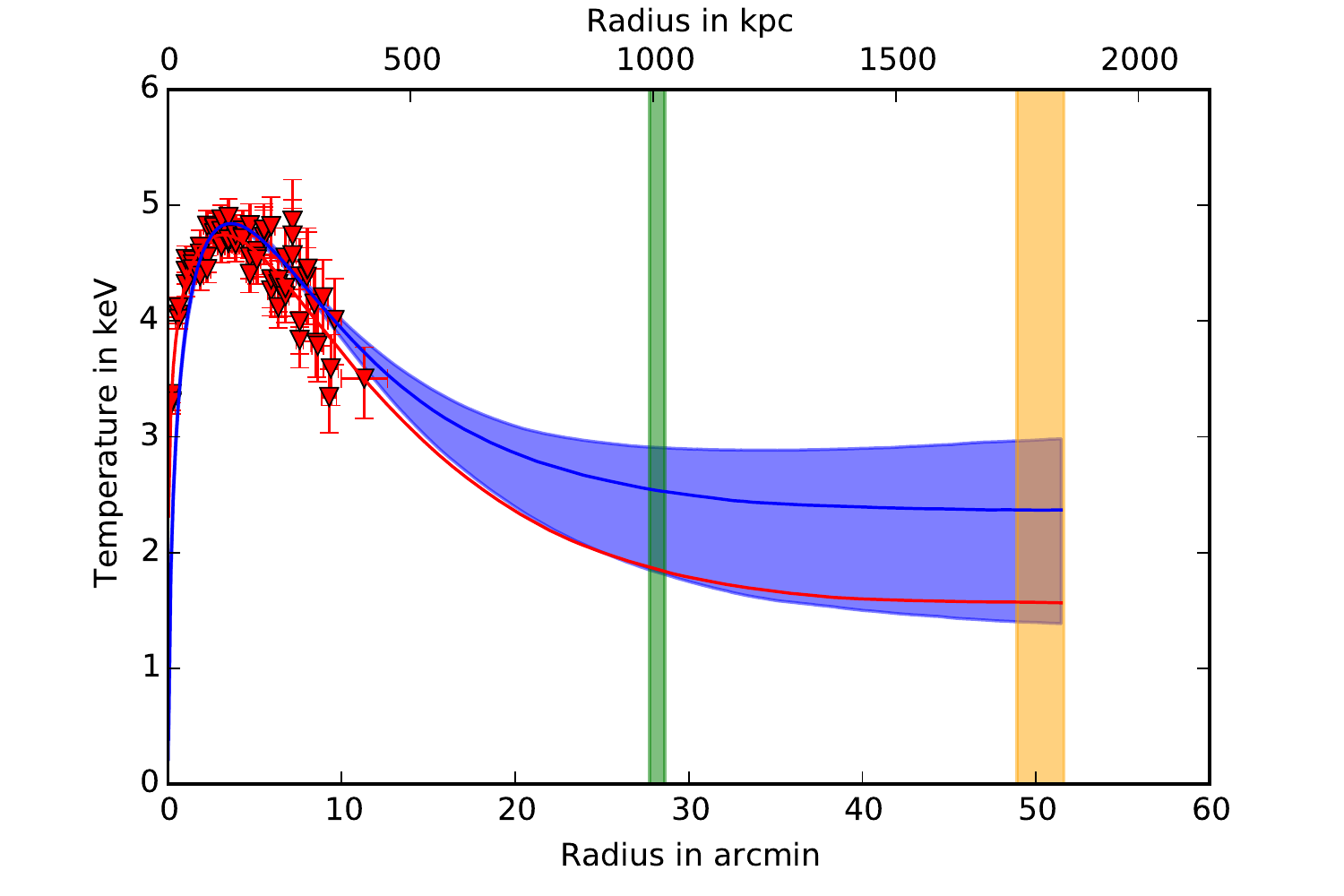}
	\includegraphics[width=0.45\textwidth]{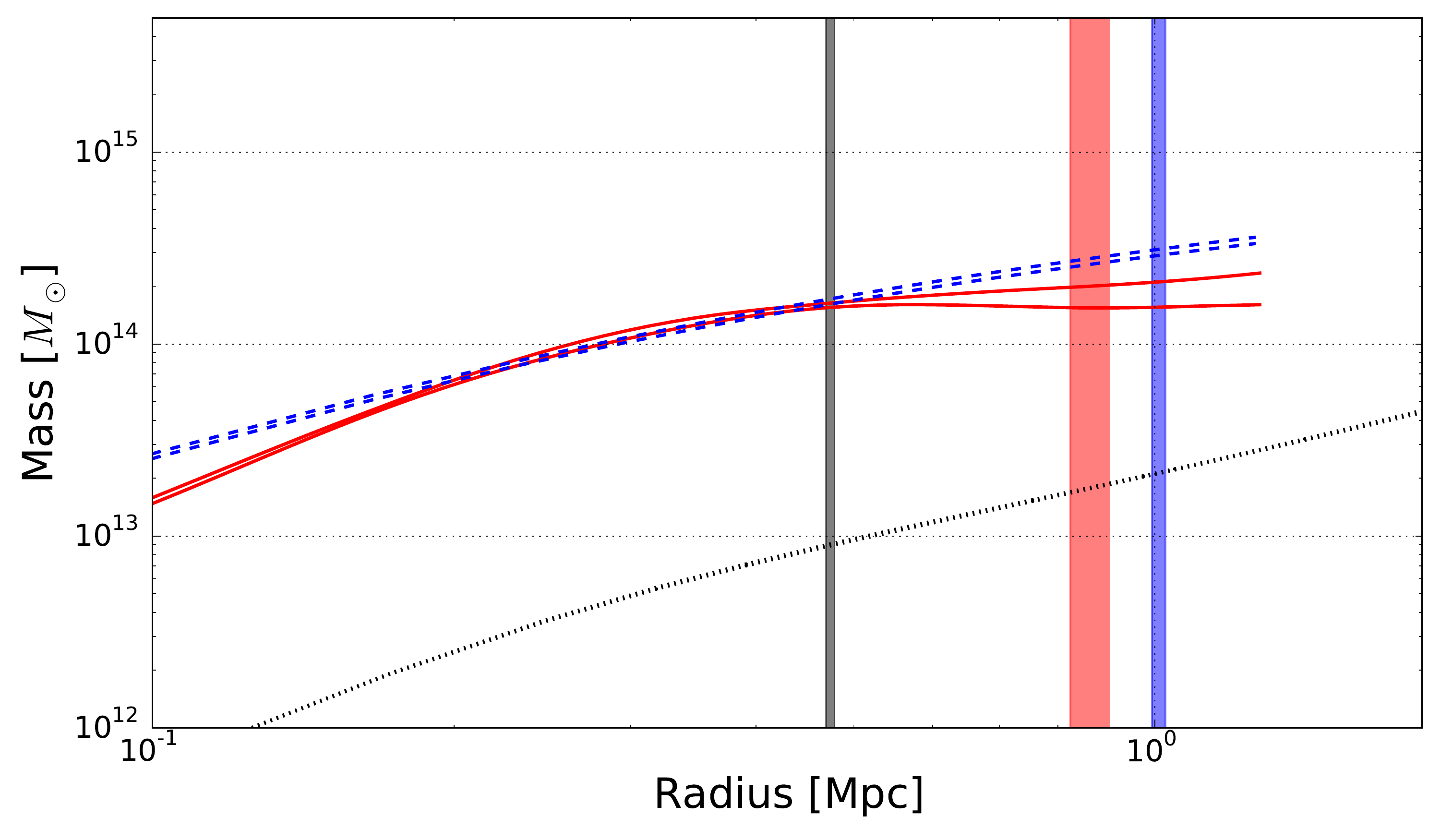}
	\caption{As Fig. \ref{fig:app_2A0335} but for A2199.}
	\label{fig:app_A2199}
\end{figure}
\begin{figure}
	\centering
	\includegraphics[width=0.45\textwidth]{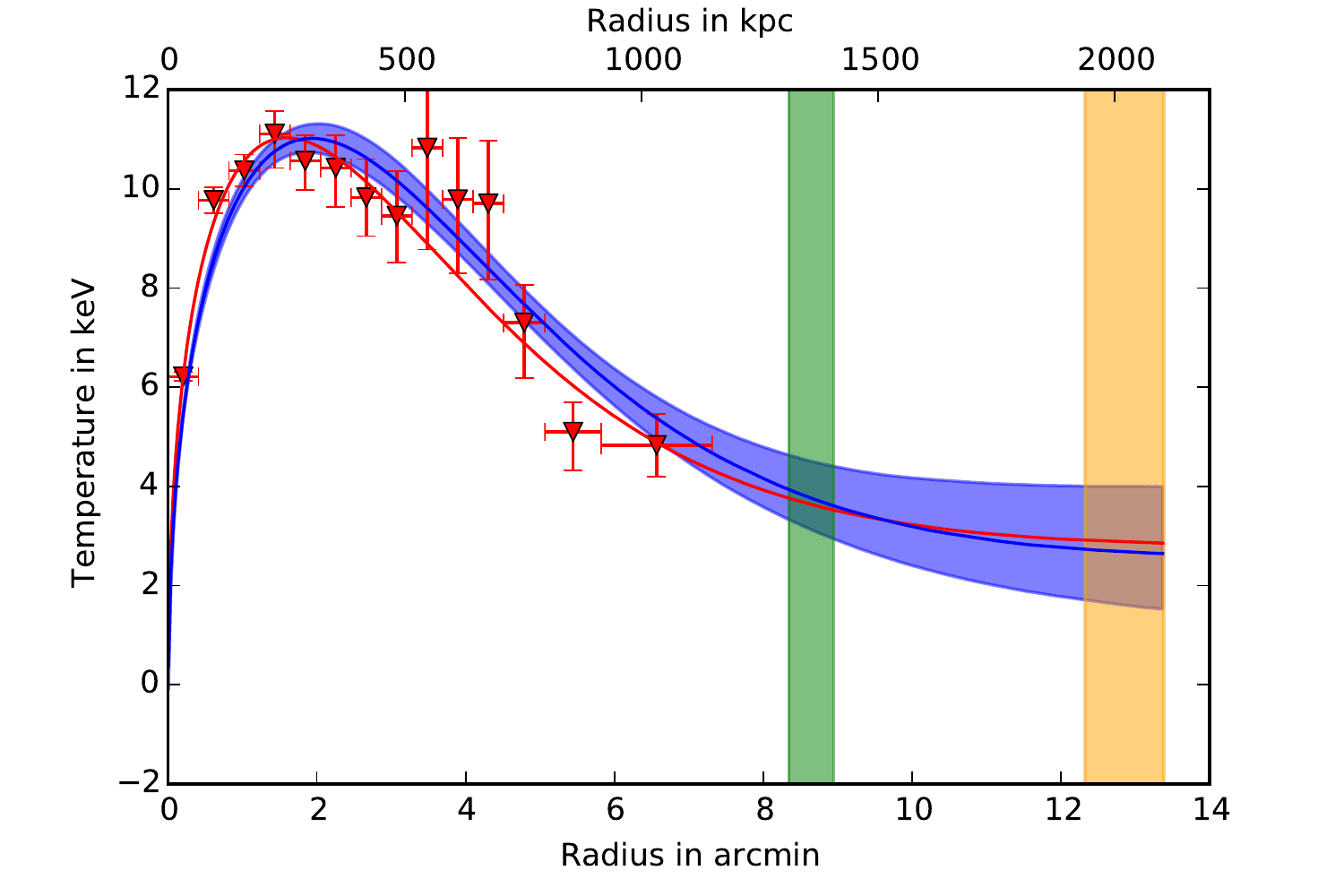}
	\includegraphics[width=0.45\textwidth]{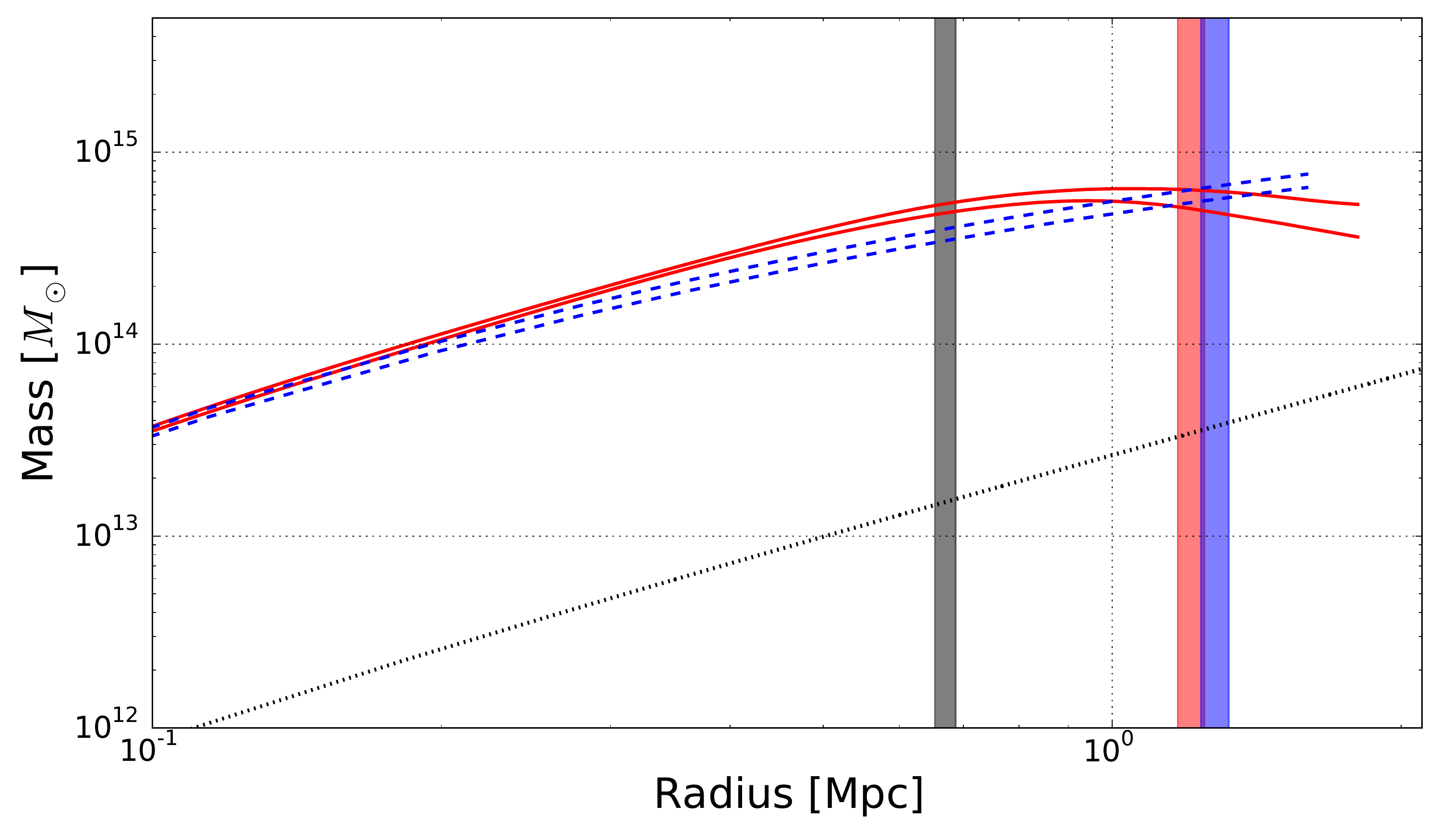}
	\caption{As Fig. \ref{fig:app_2A0335} but for A2204.}
	\label{fig:app_A2204}
\end{figure}
\begin{figure}
	\centering
	\includegraphics[width=0.45\textwidth]{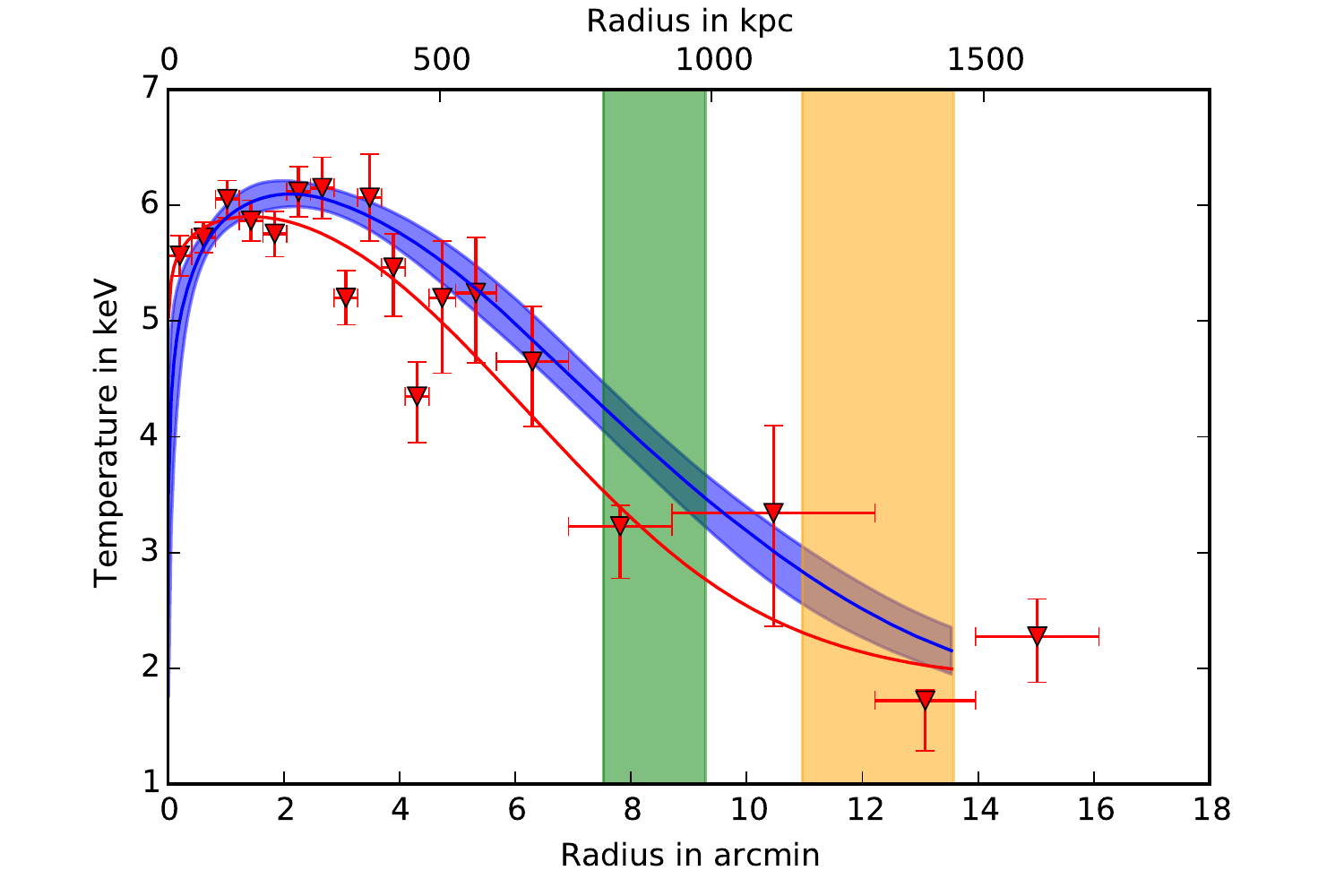}
	\includegraphics[width=0.45\textwidth]{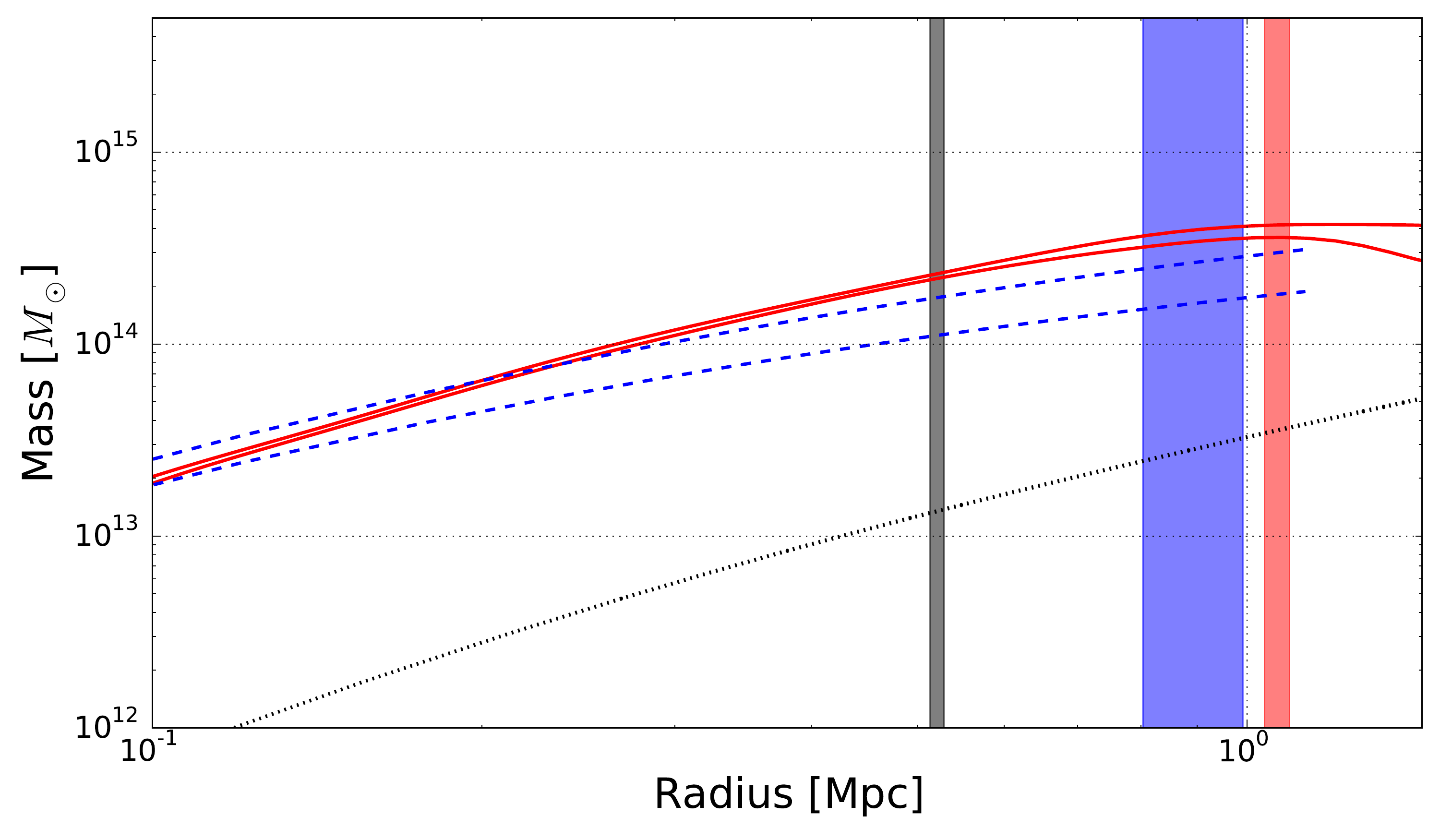}
	\caption{As Fig. \ref{fig:app_2A0335} but for A2244.}
	\label{fig:app_A2244}
\end{figure}
\clearpage
\begin{figure}
	\centering
	\includegraphics[width=0.45\textwidth]{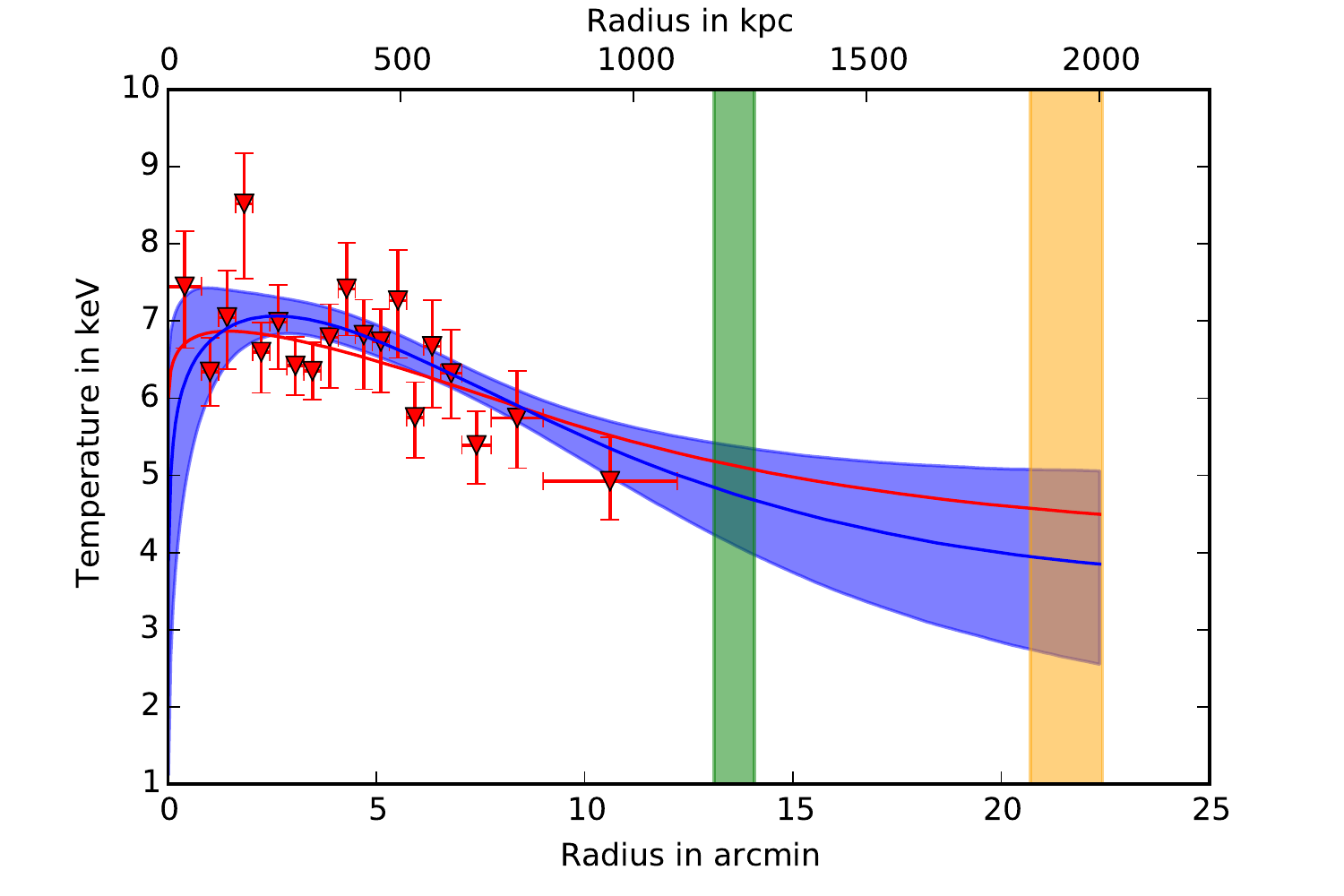}
	\includegraphics[width=0.45\textwidth]{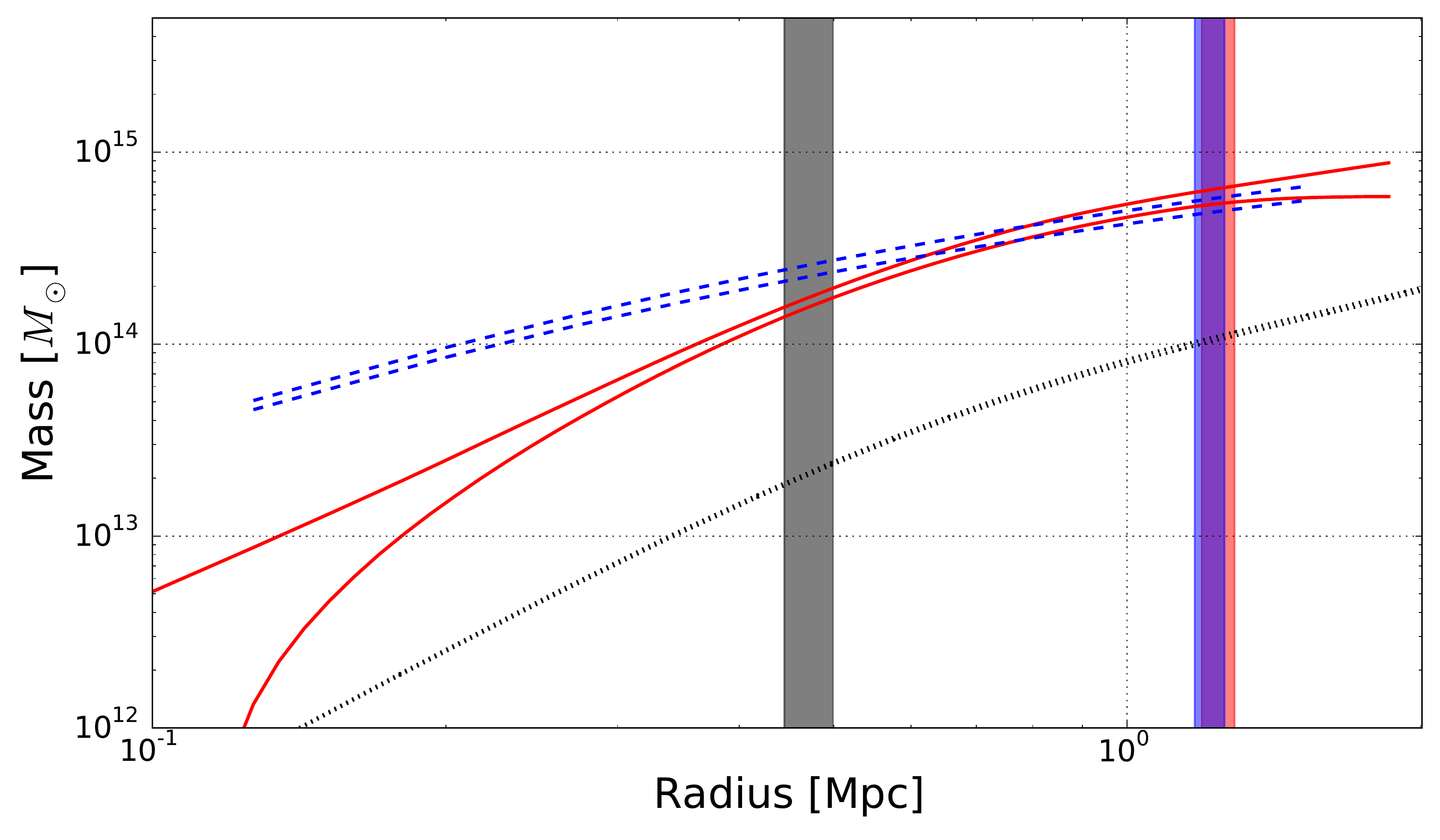}
	\caption{As Fig. \ref{fig:app_2A0335} but for A2255.}
	\label{fig:app_A2255}
\end{figure}
\begin{figure}
	\centering
	\includegraphics[width=0.45\textwidth]{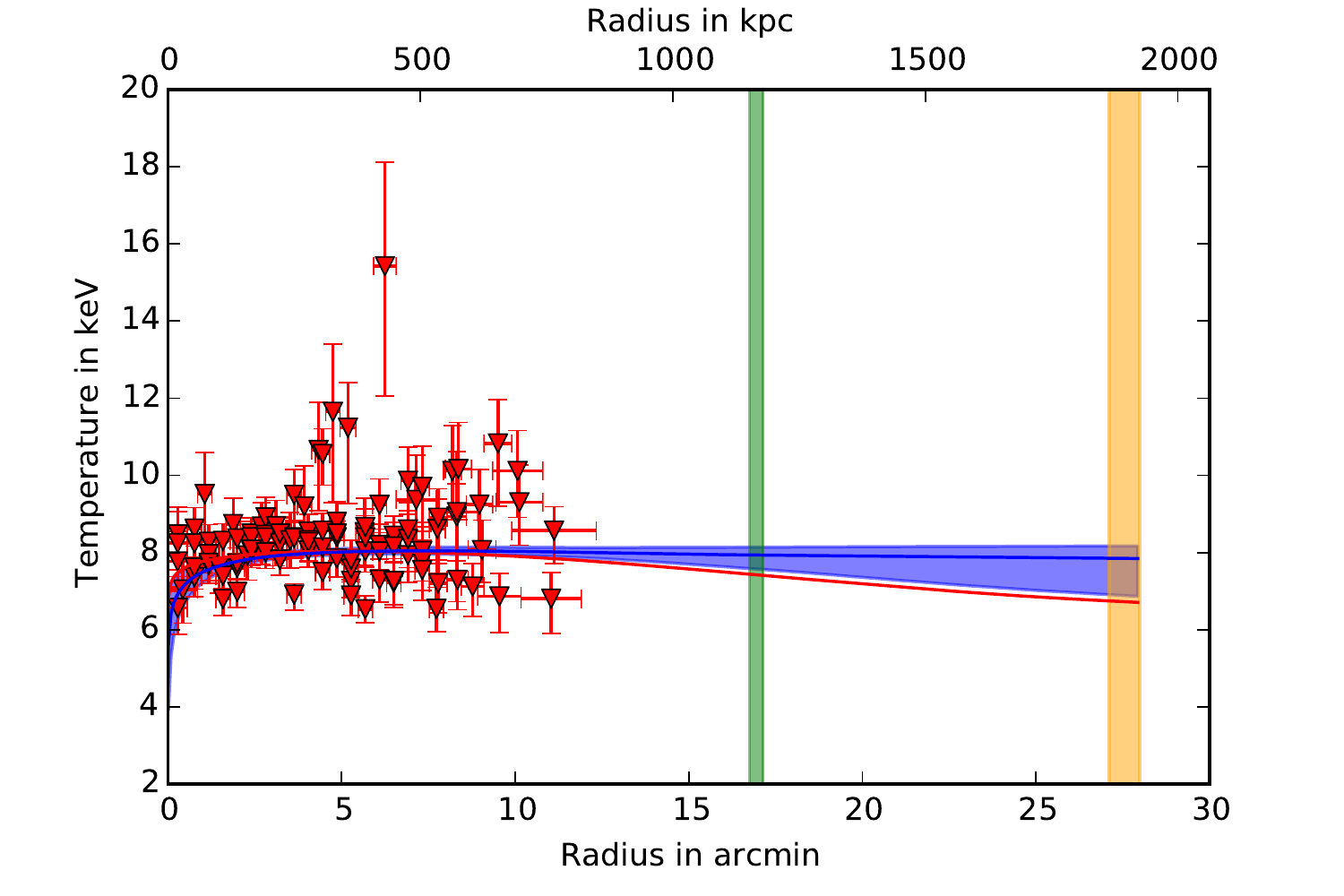}
	\includegraphics[width=0.45\textwidth]{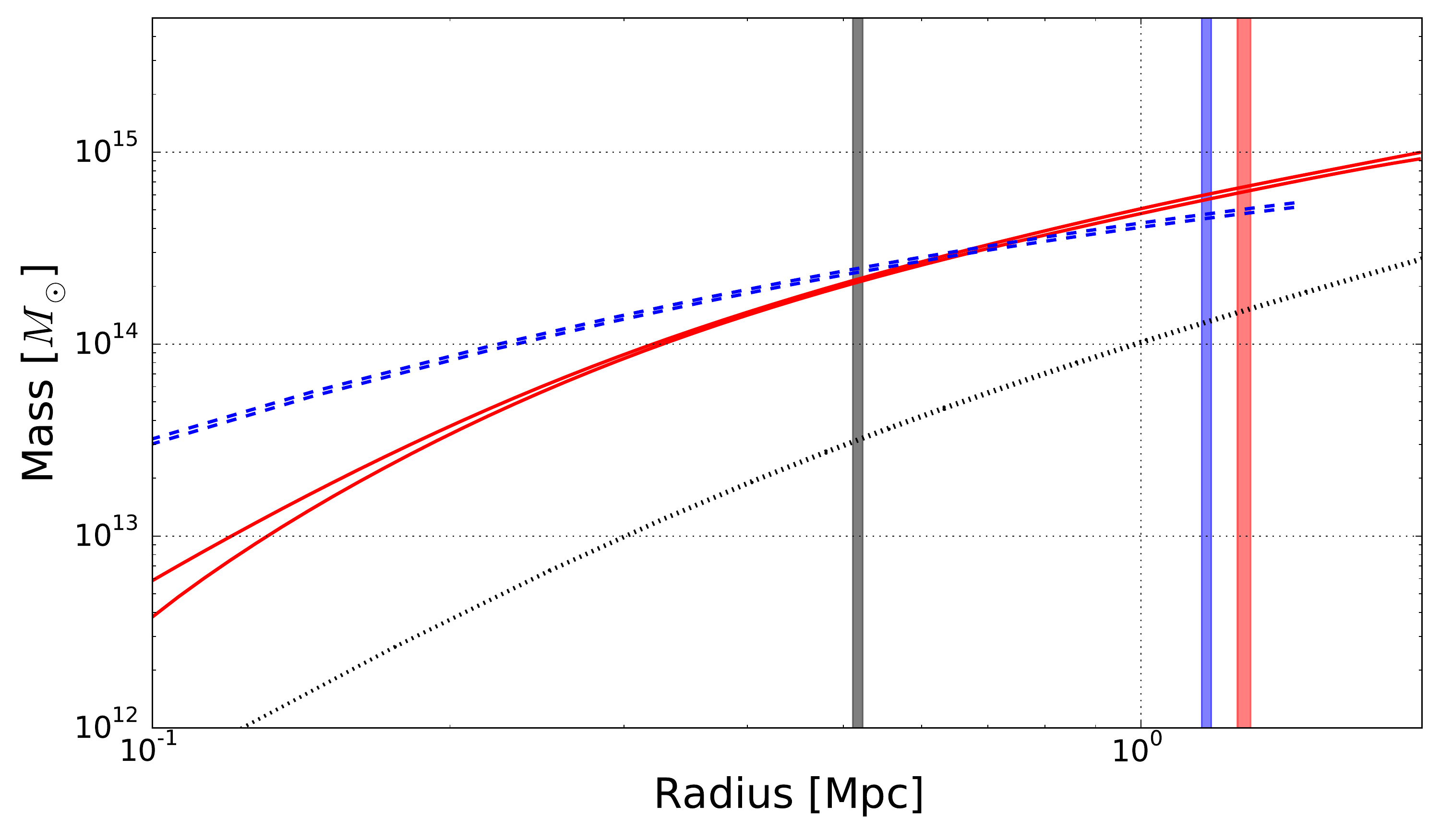}
	\caption{As Fig. \ref{fig:app_2A0335} but for A2256.}
	\label{fig:app_A2256}
\end{figure}
\begin{figure}
	\centering
	\includegraphics[width=0.45\textwidth]{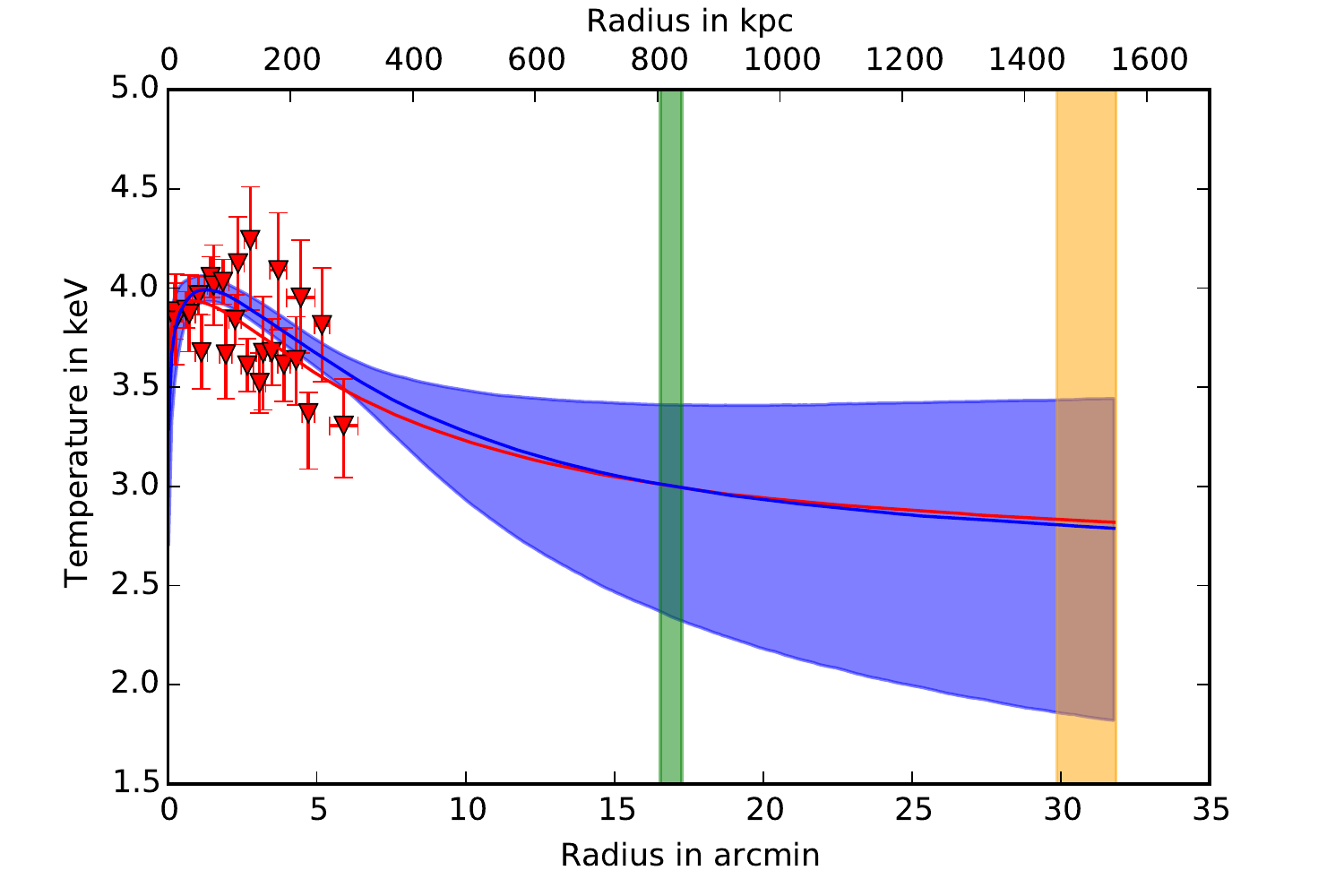}
	\includegraphics[width=0.45\textwidth]{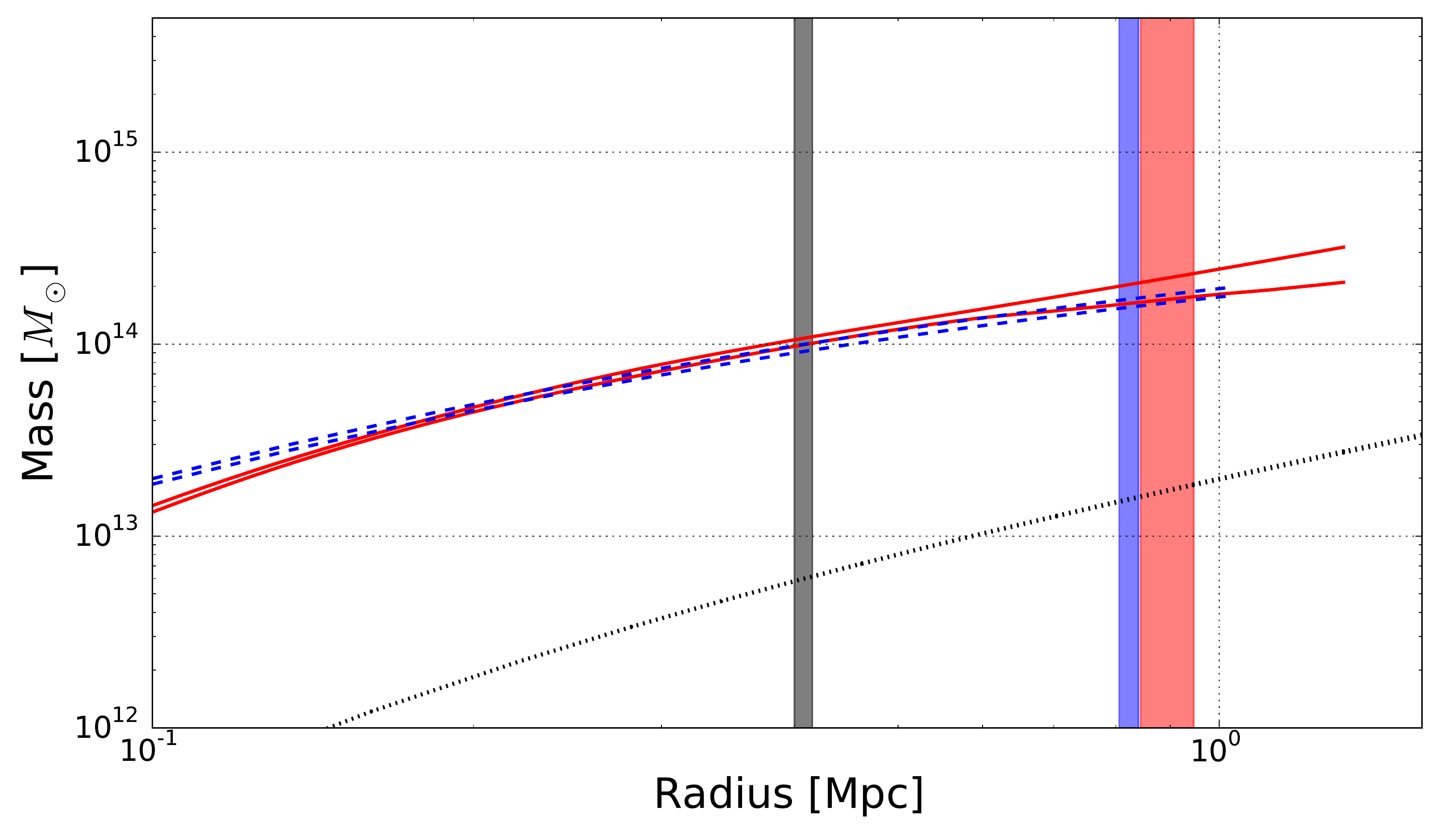}
	\caption{As Fig. \ref{fig:app_2A0335} but for A2589.}
	\label{fig:app_A2589}
\end{figure}
\clearpage
\begin{figure}
	\centering
	\includegraphics[width=0.45\textwidth]{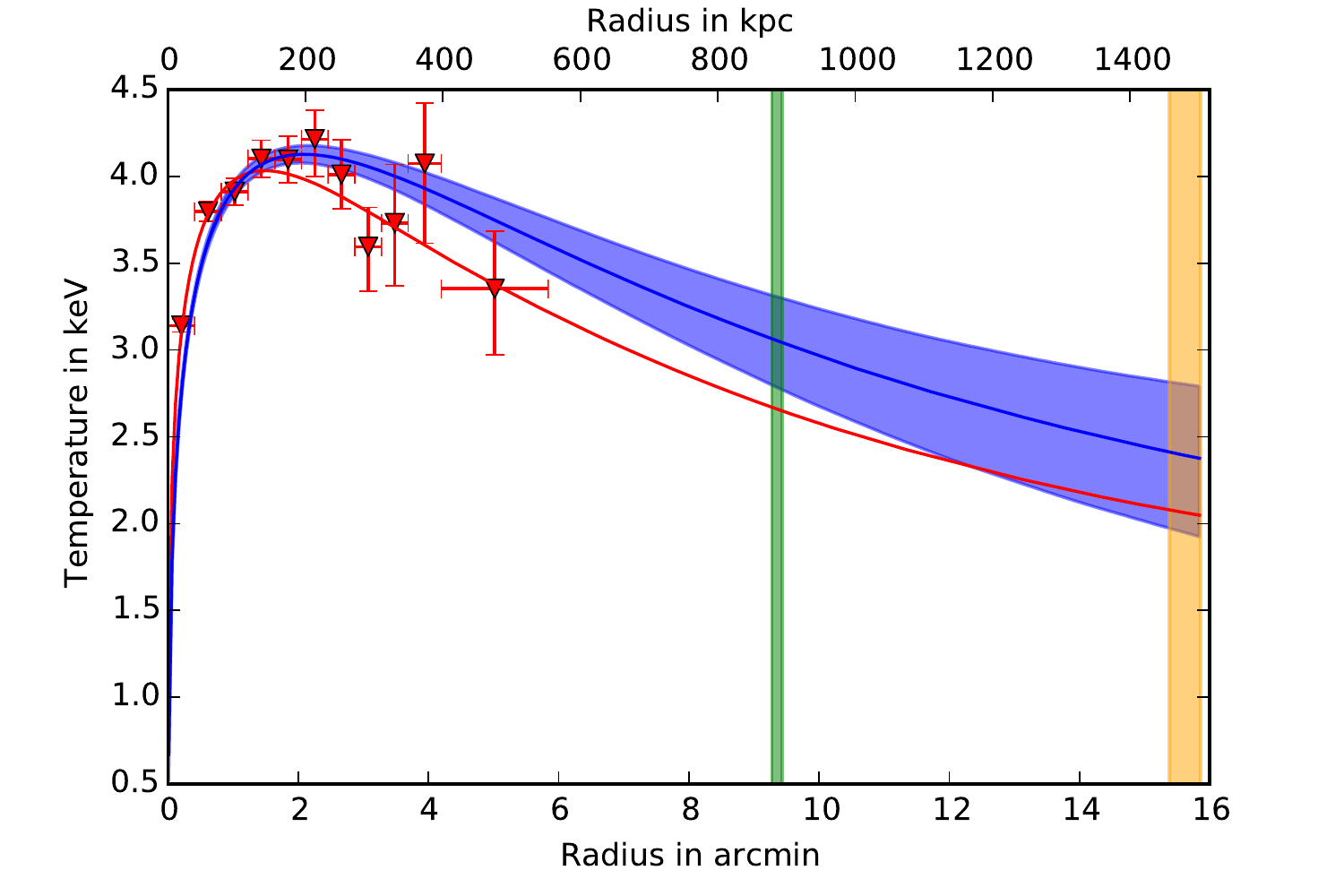}
	\includegraphics[width=0.45\textwidth]{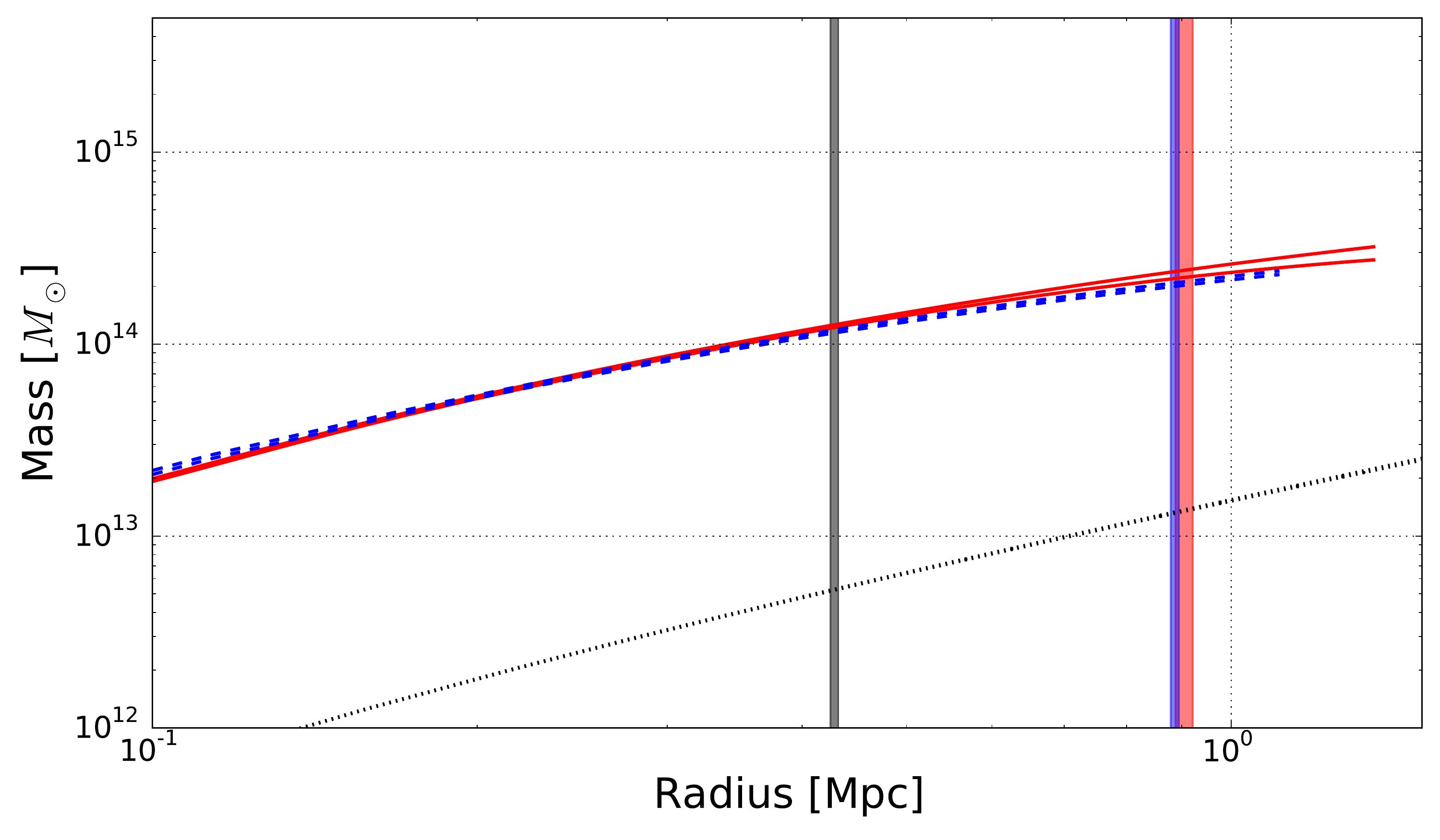}
	\caption{As Fig. \ref{fig:app_2A0335} but for A2597.}
	\label{fig:app_A2597}
\end{figure}
\begin{figure}
	\centering
	\includegraphics[width=0.45\textwidth]{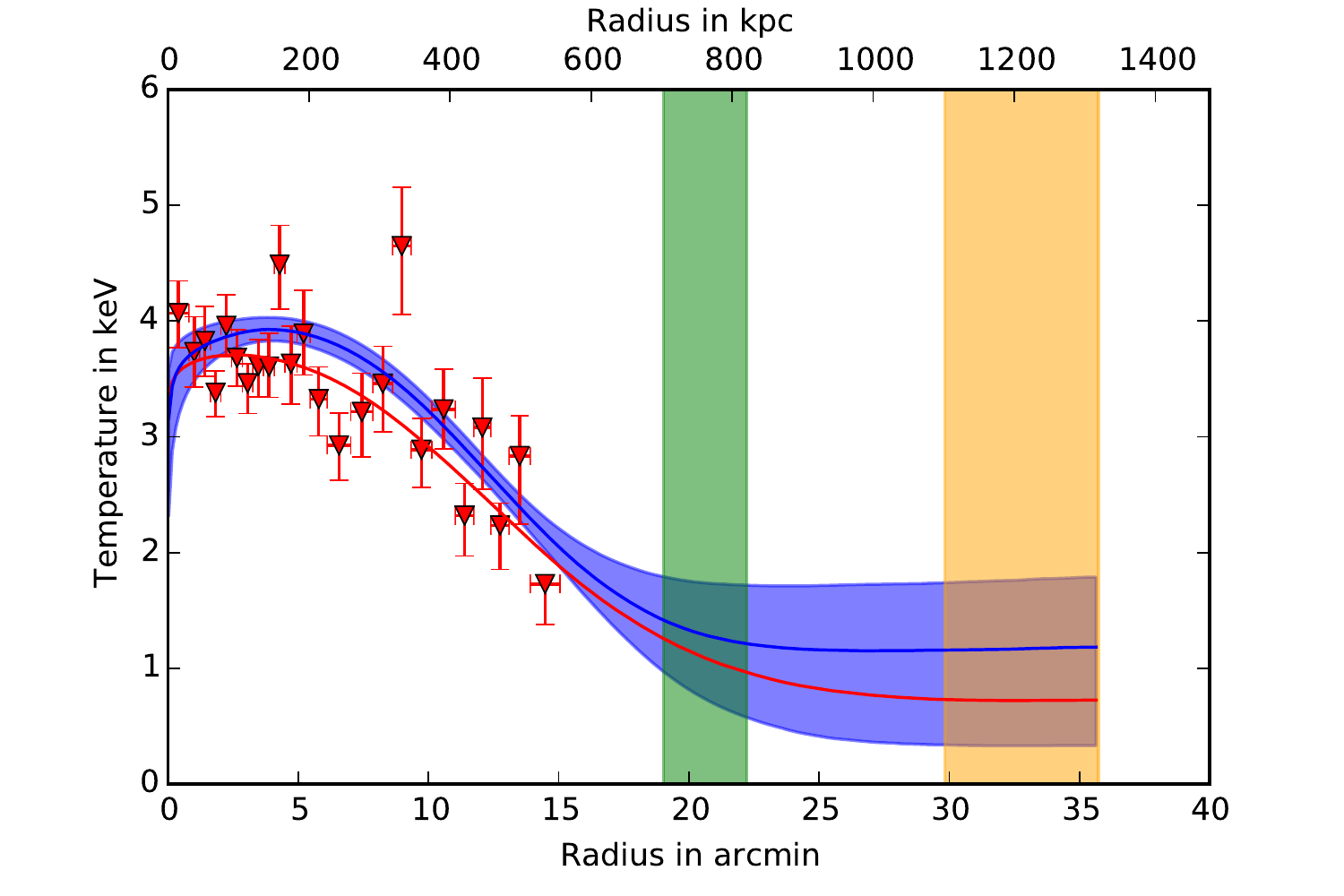}
	\includegraphics[width=0.45\textwidth]{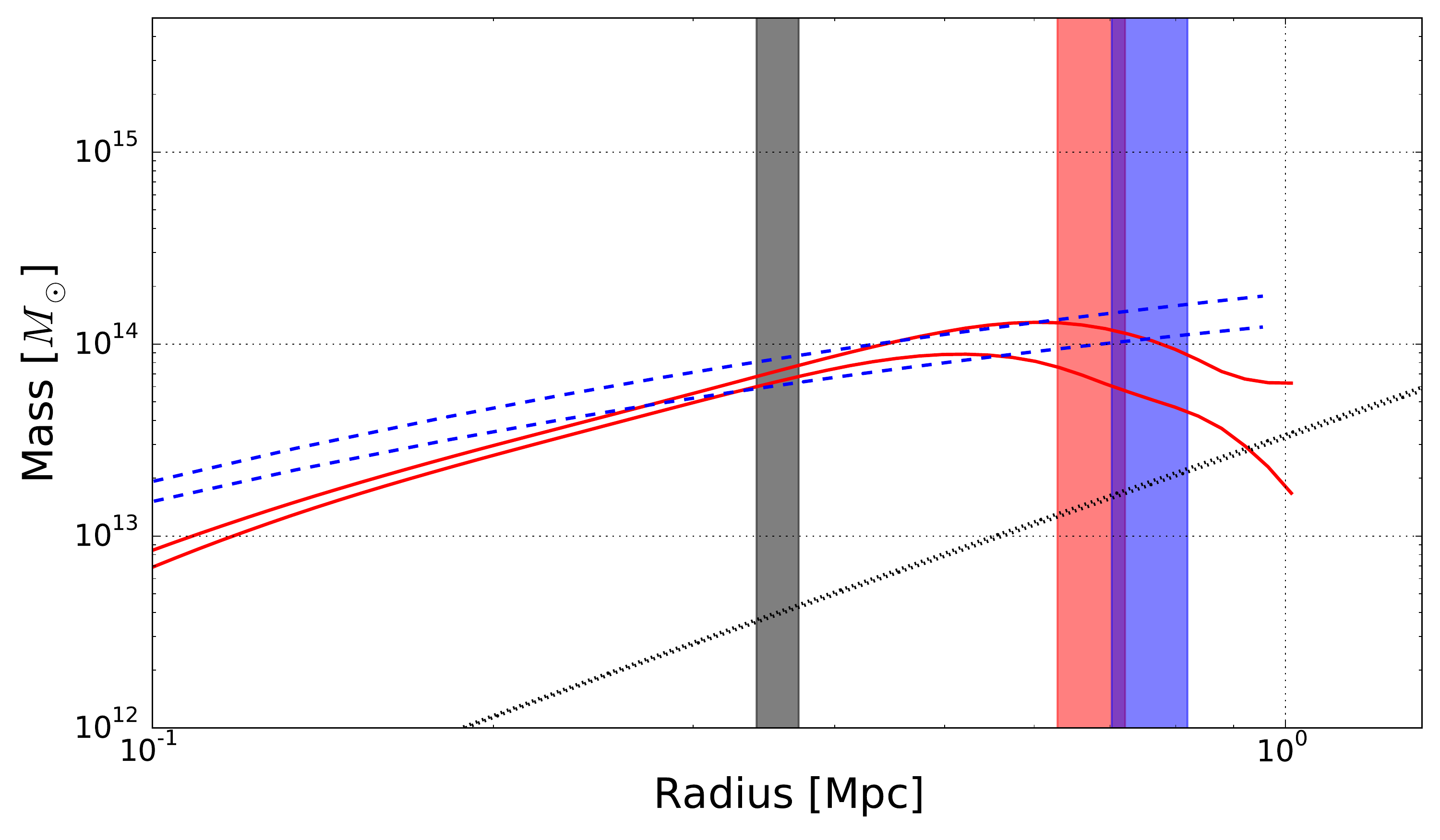}
	\caption{As Fig. \ref{fig:app_2A0335} but for A2634.}
	\label{fig:app_A2634}
\end{figure}
\begin{figure}
	\centering
	\includegraphics[width=0.45\textwidth]{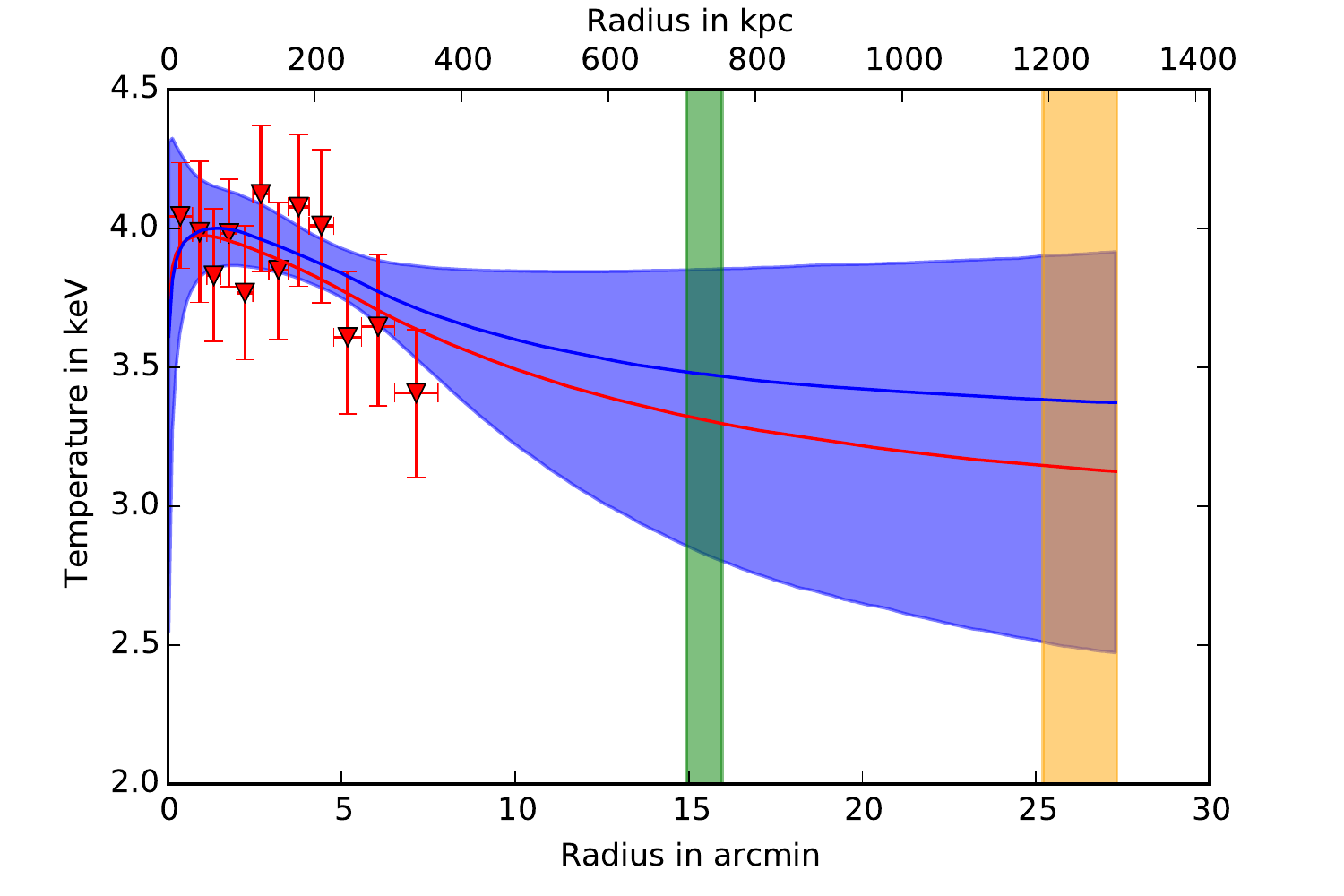}
	\includegraphics[width=0.45\textwidth]{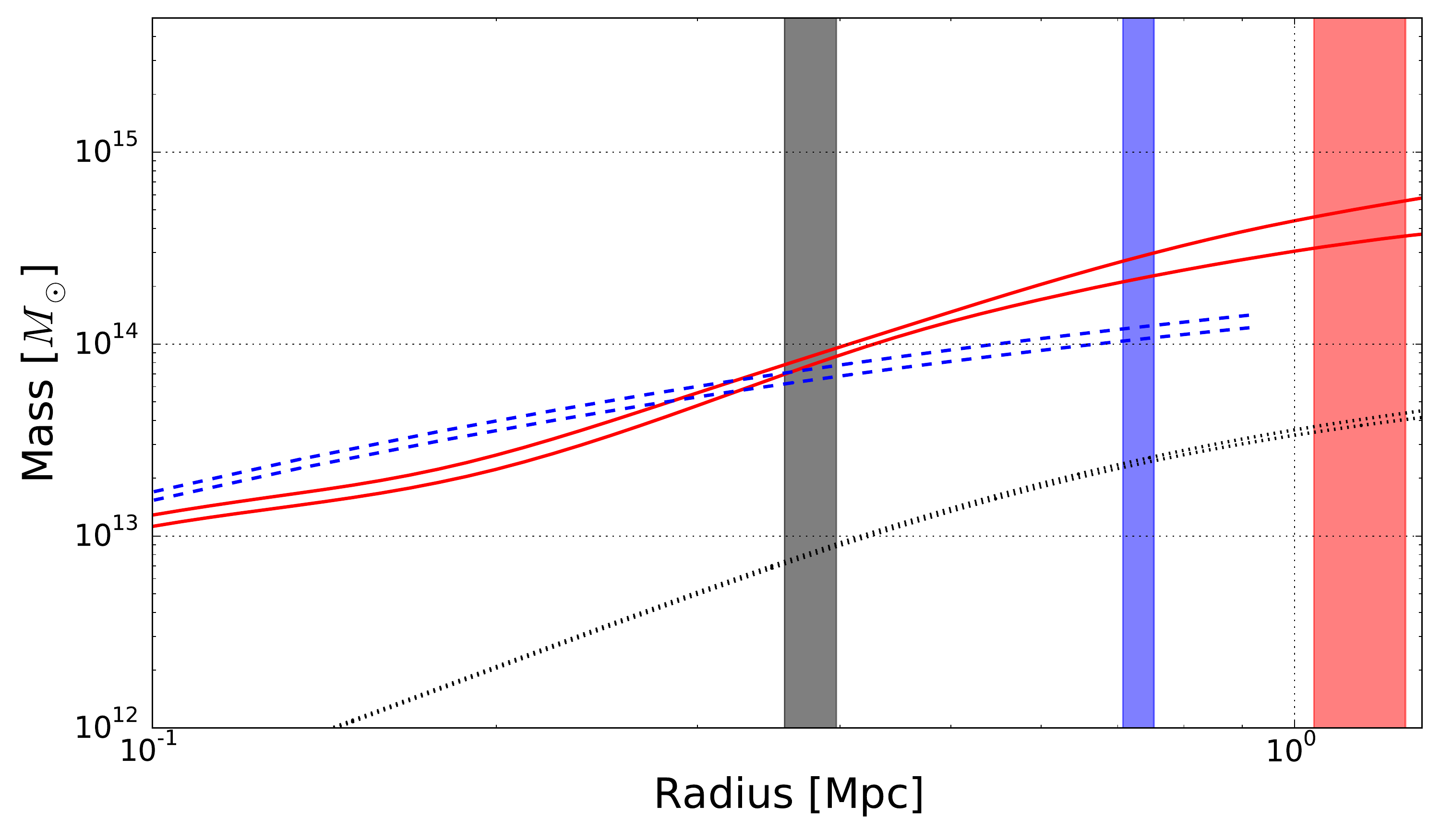}
	\caption{As Fig. \ref{fig:app_2A0335} but for A2657.}
	\label{fig:app_A2657}
\end{figure}
\clearpage
\begin{figure}
	\centering
	\includegraphics[width=0.45\textwidth]{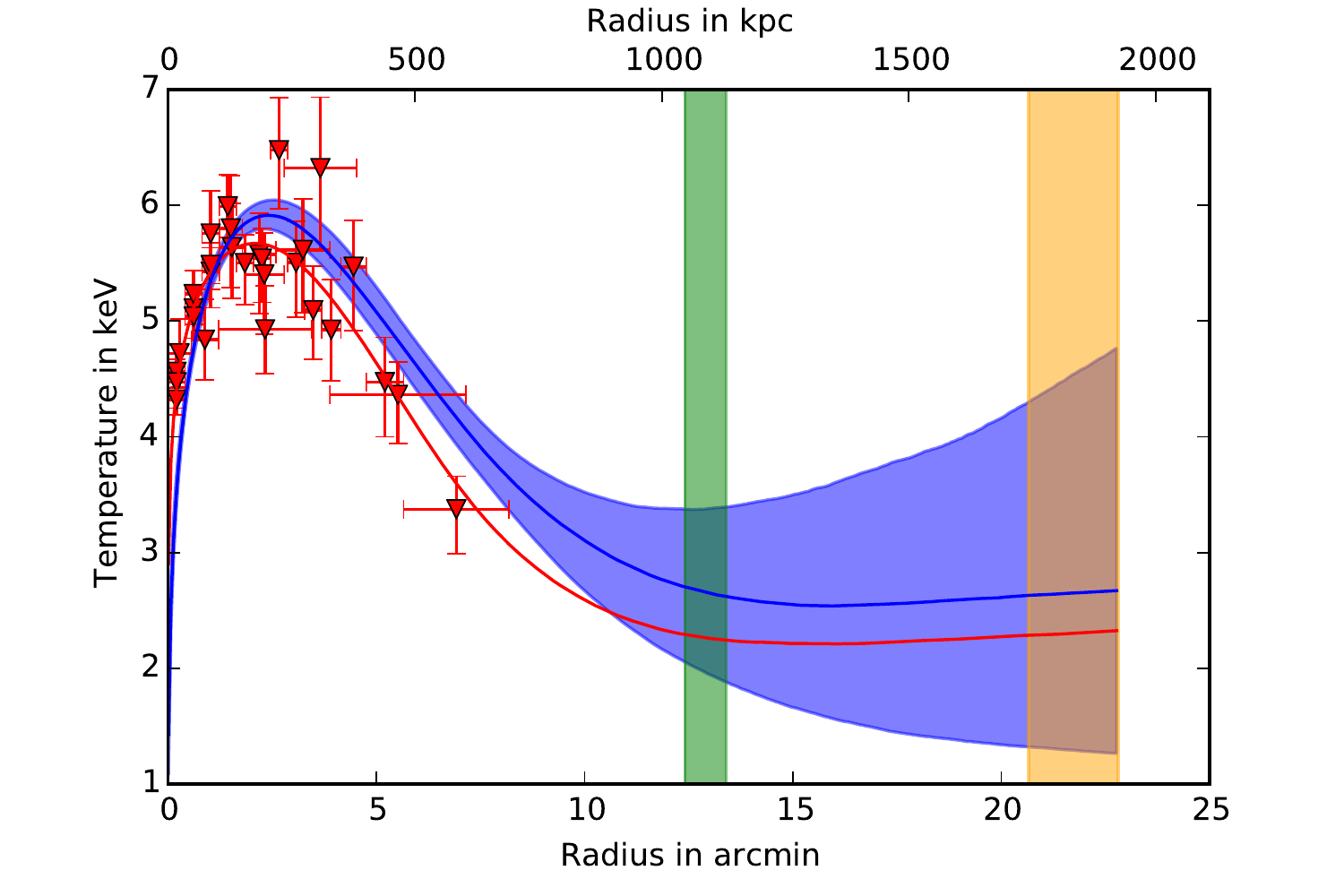}
	\includegraphics[width=0.45\textwidth]{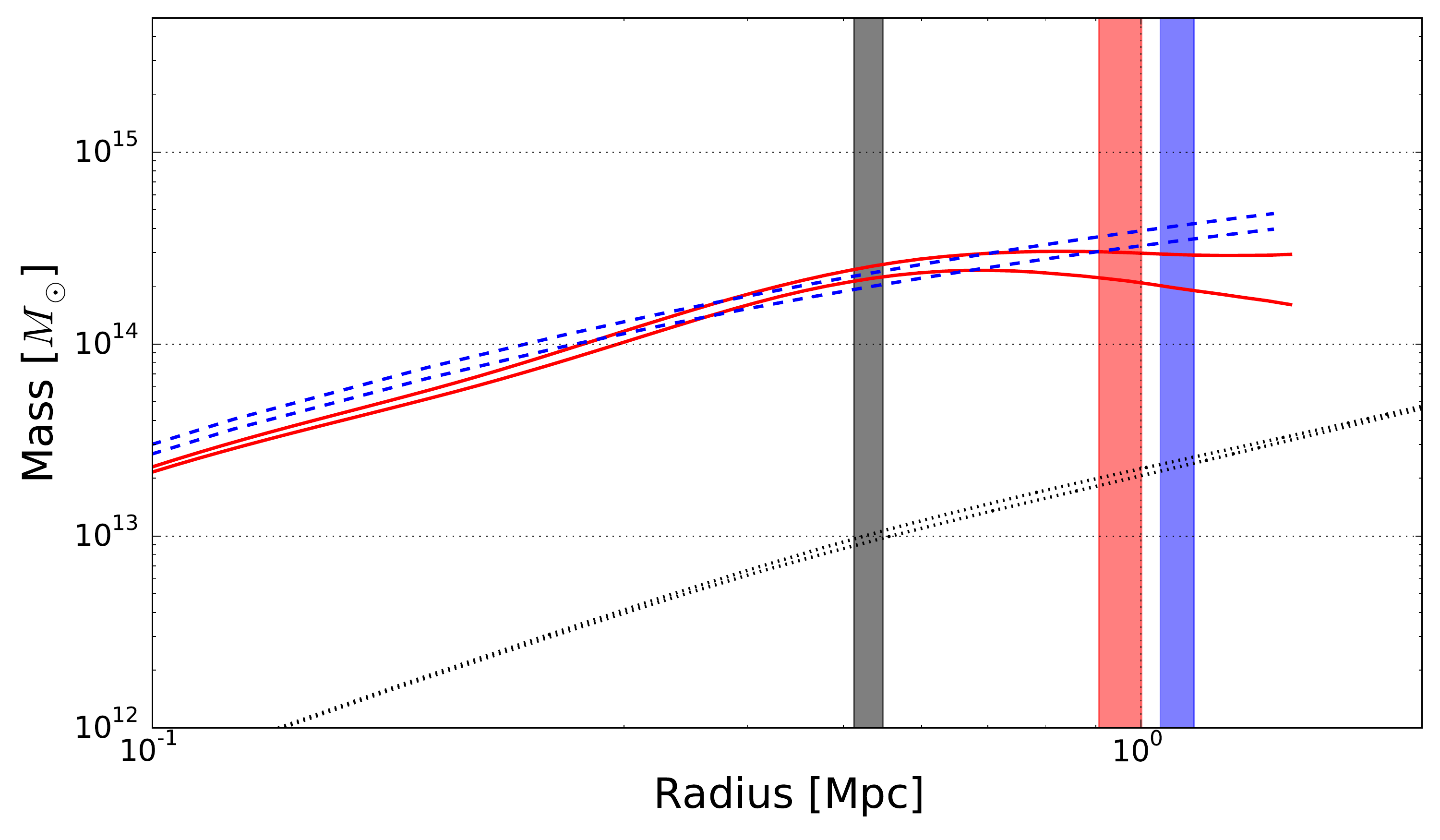}
	\caption{As Fig. \ref{fig:app_2A0335} but for A3112.}
	\label{fig:app_A3112}
\end{figure}
\begin{figure}
	\centering
	\includegraphics[width=0.45\textwidth]{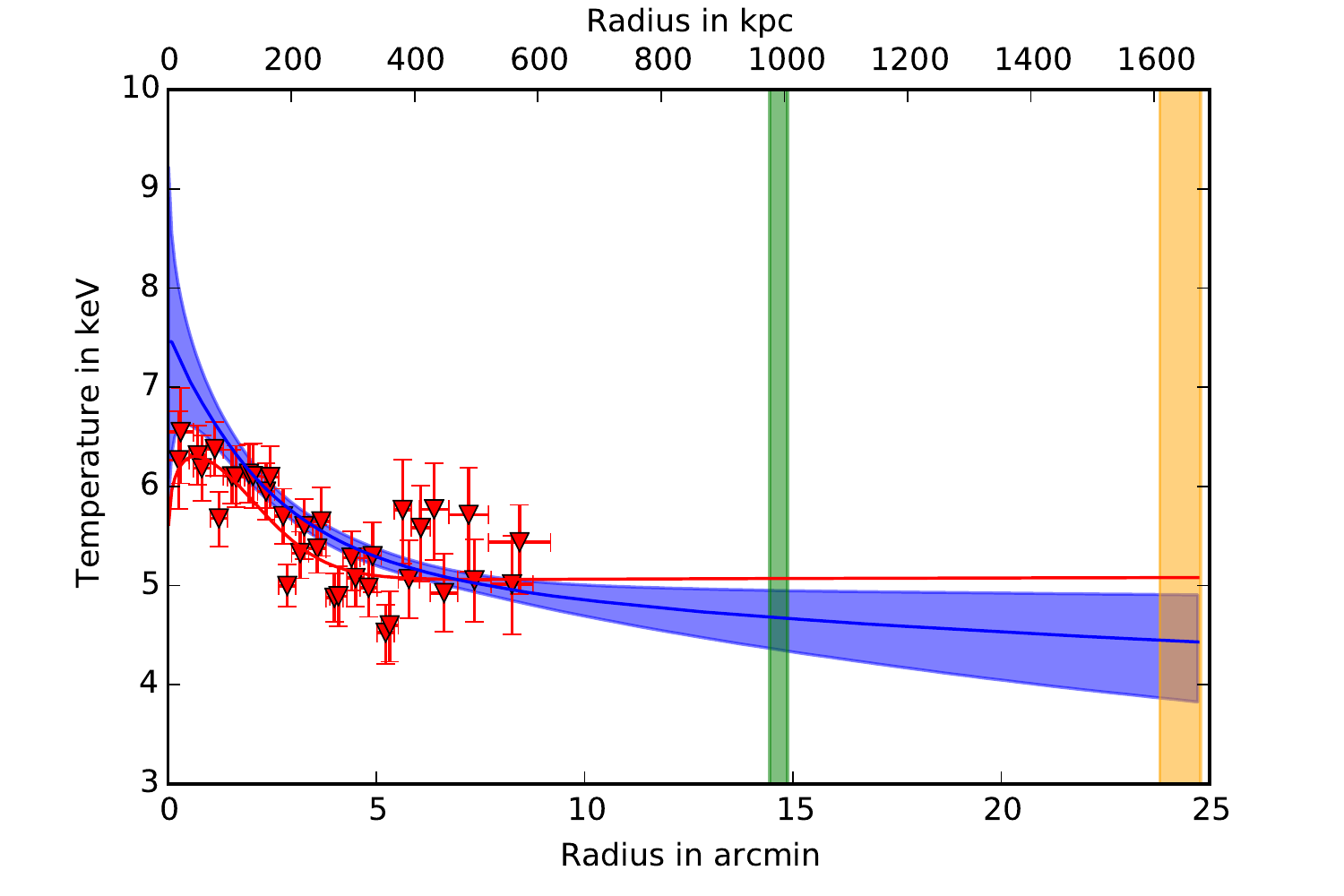}
	\includegraphics[width=0.45\textwidth]{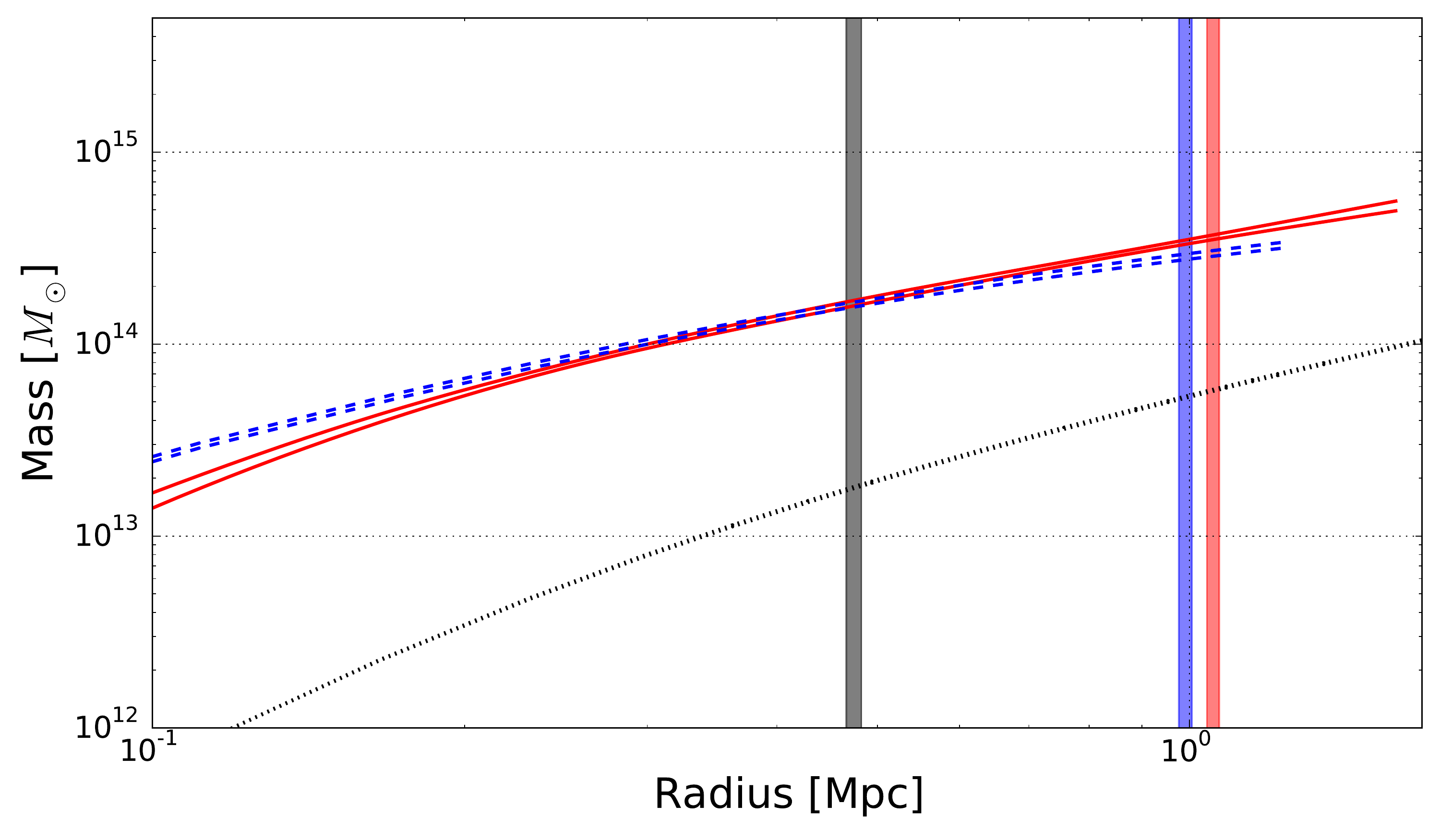}
	\caption{As Fig. \ref{fig:app_2A0335} but for A3158.}
	\label{fig:app_A3158}
\end{figure}
\begin{figure}
	\centering
	\includegraphics[width=0.45\textwidth]{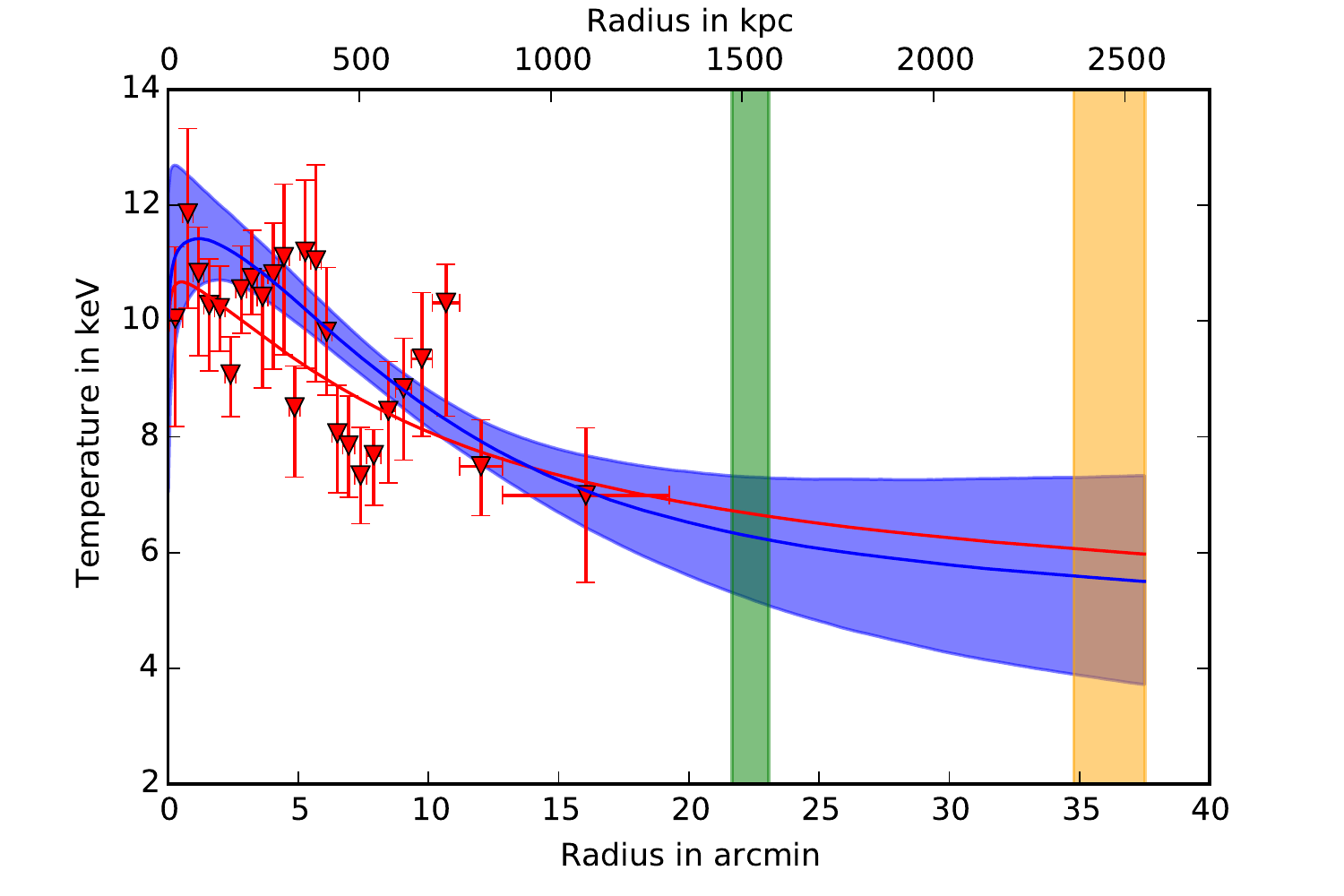}
	\includegraphics[width=0.45\textwidth]{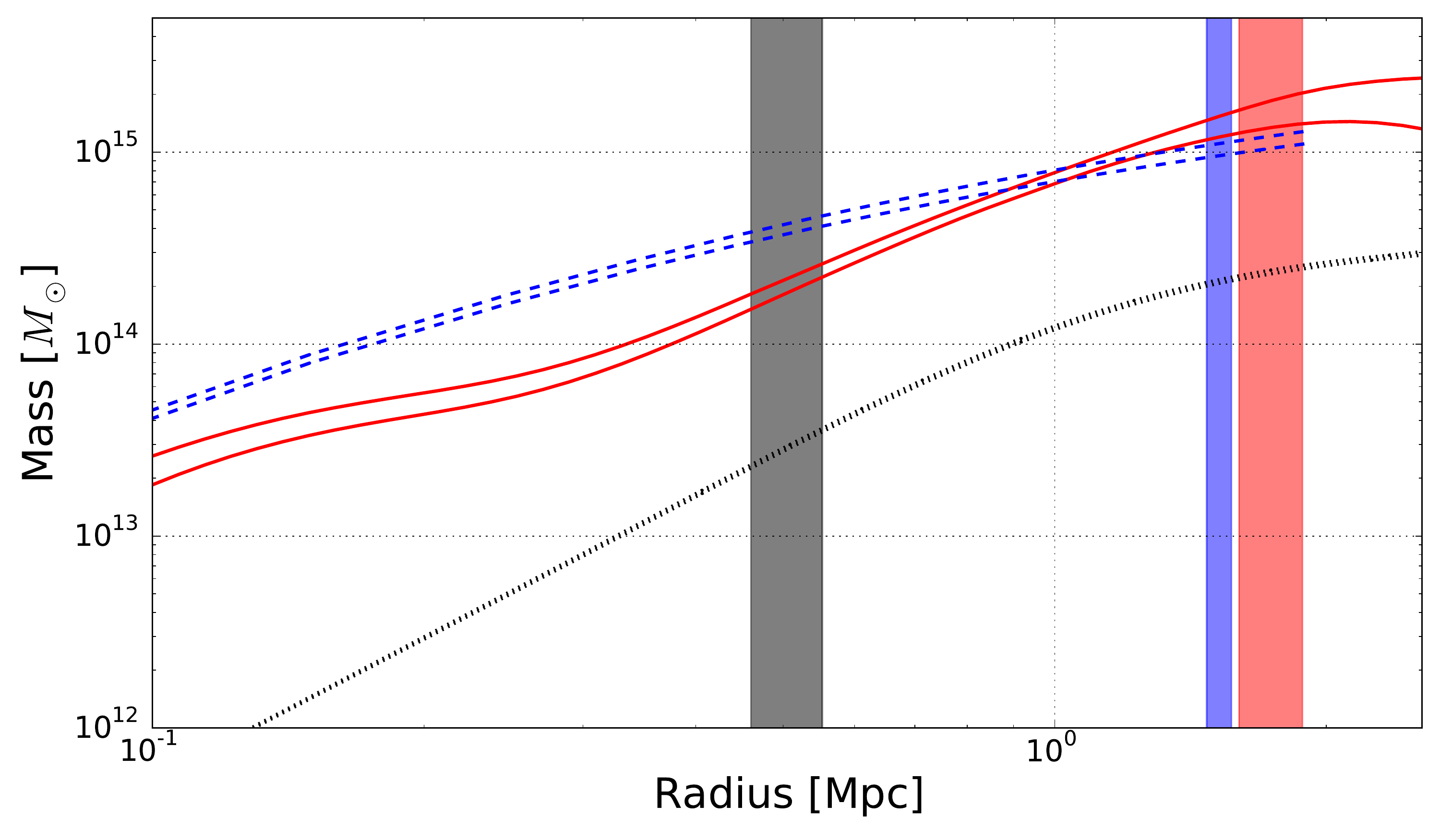}
	\caption{As Fig. \ref{fig:app_2A0335} but for A3266.}
	\label{fig:app_A3266}
\end{figure}
\clearpage
\begin{figure}
	\centering
	\includegraphics[width=0.45\textwidth]{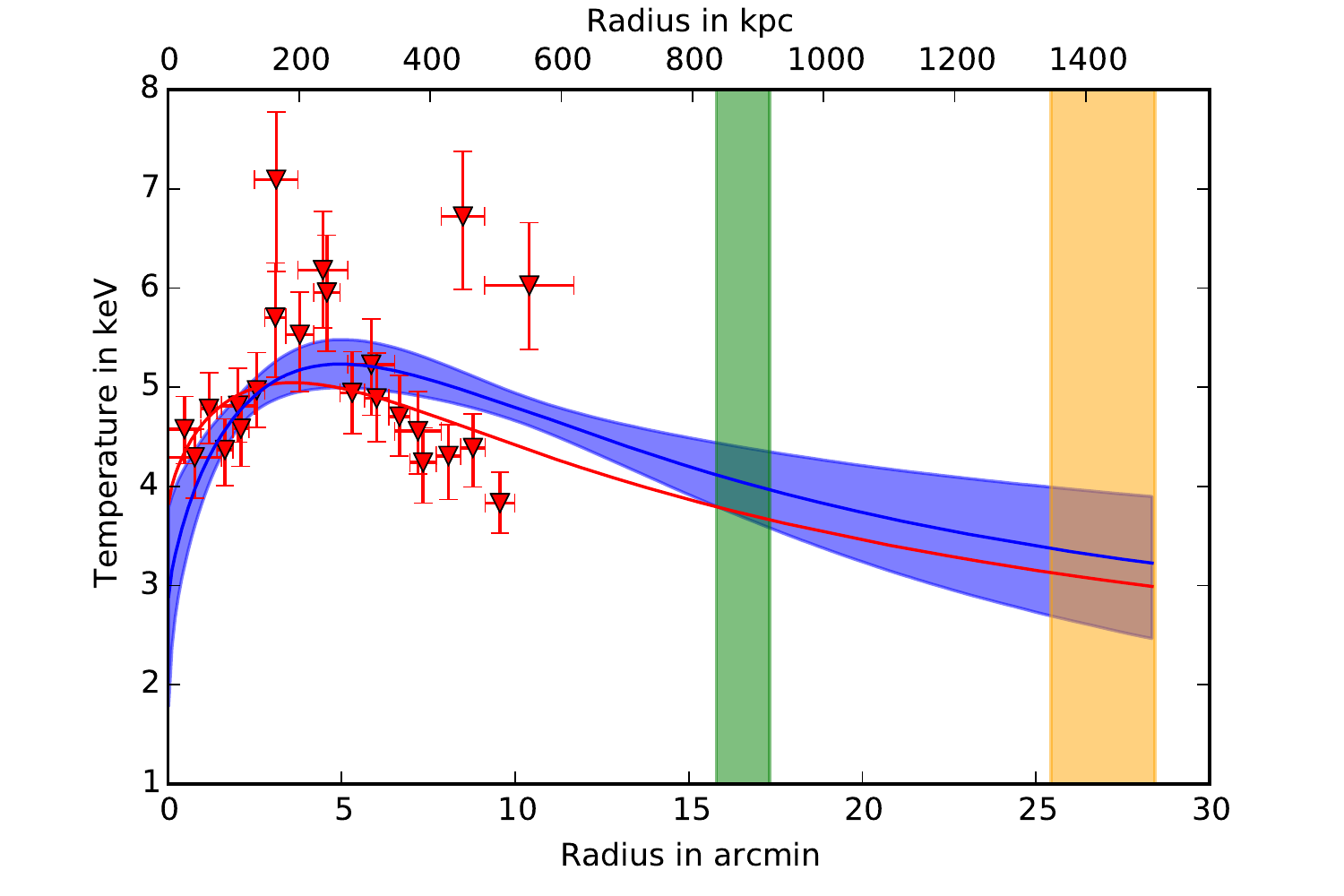}
	\includegraphics[width=0.45\textwidth]{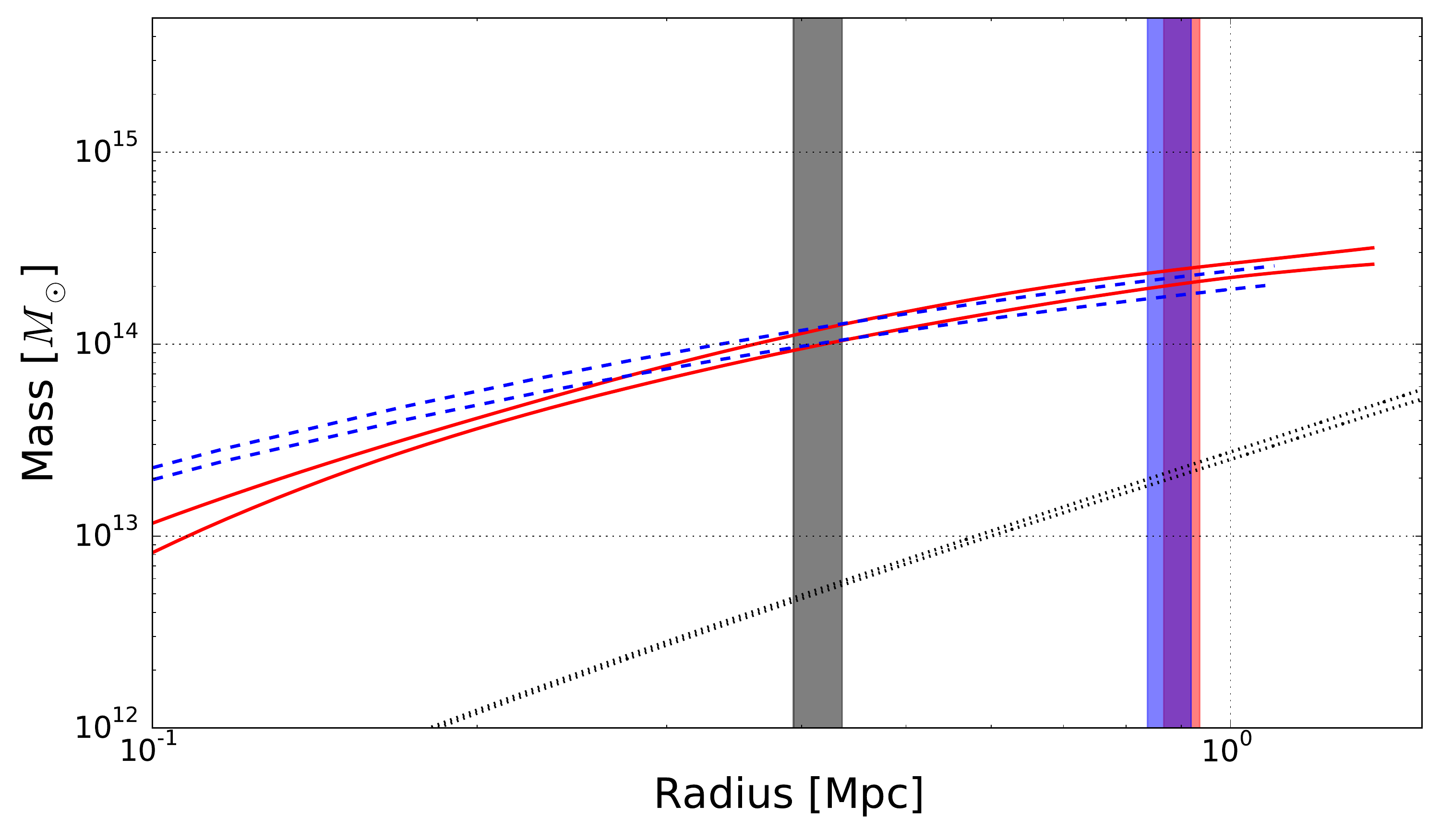}
	\caption{As Fig. \ref{fig:app_2A0335} but for A3376.}
	\label{fig:app_A3376}
\end{figure}
\begin{figure}
	\centering
	\includegraphics[width=0.45\textwidth]{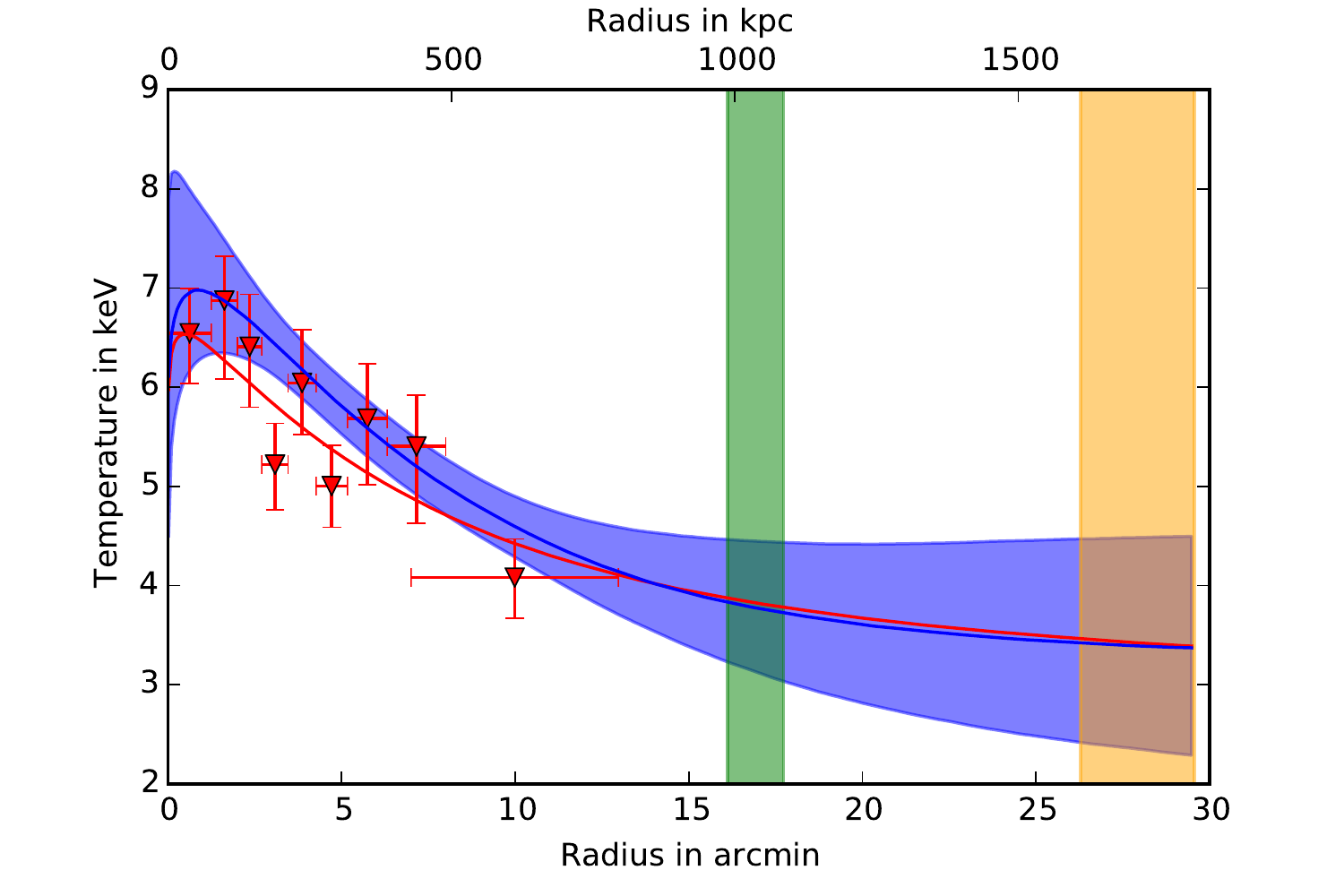}
	\includegraphics[width=0.45\textwidth]{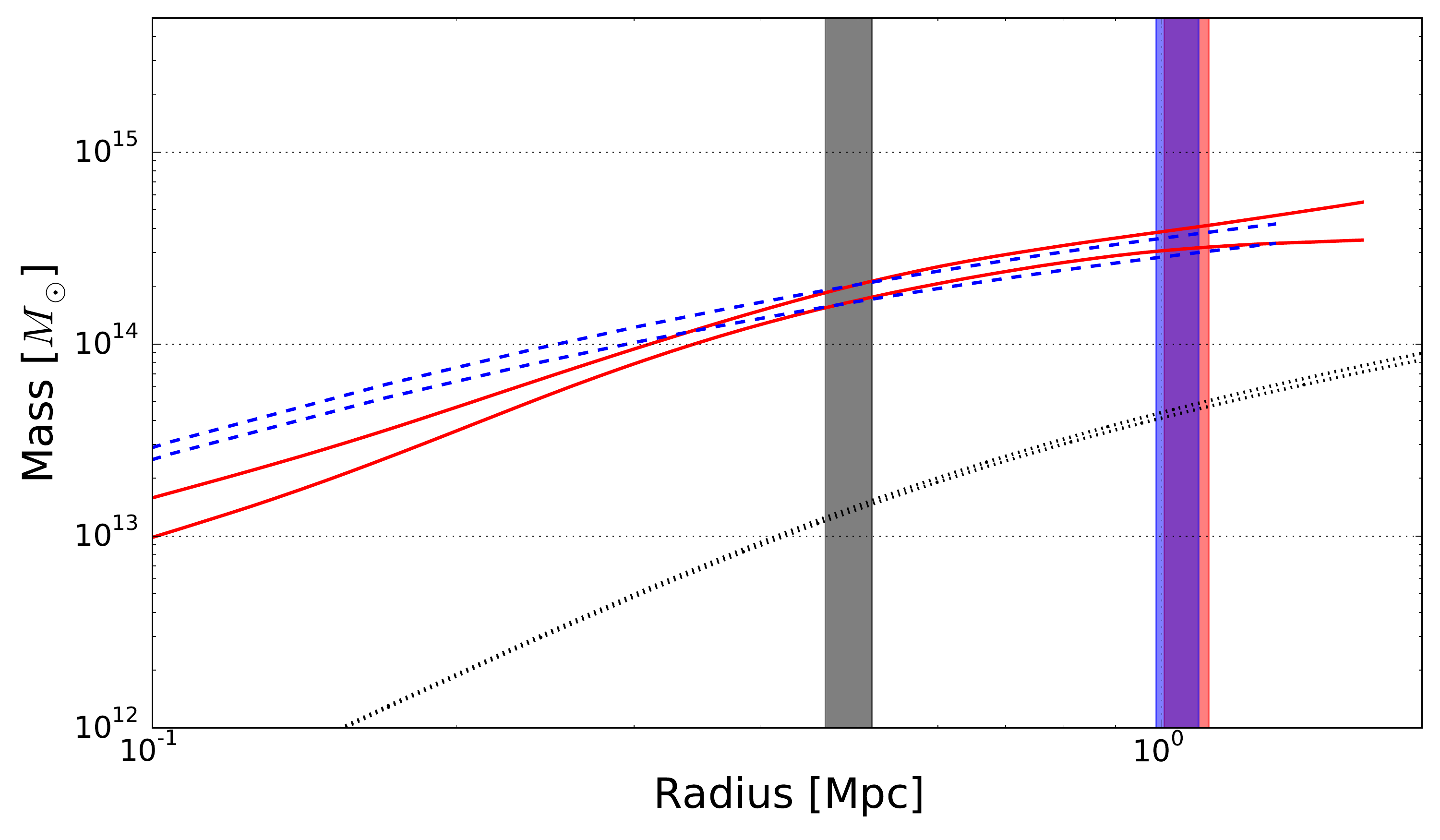}
	\caption{As Fig. \ref{fig:app_2A0335} but for A3391.}
	\label{fig:app_A3391}
\end{figure}
\begin{figure}
	\centering
	\includegraphics[width=0.45\textwidth]{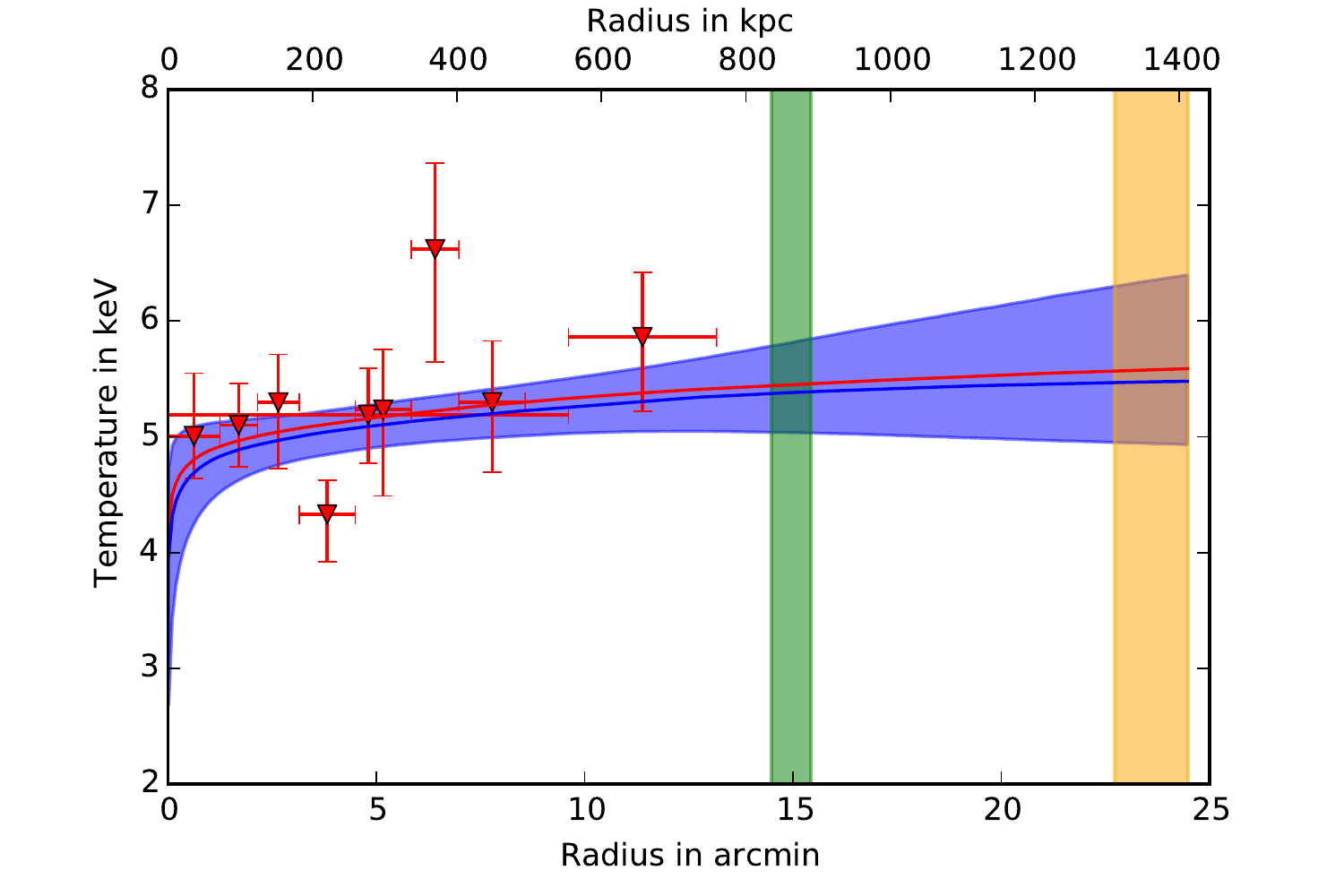}
	\includegraphics[width=0.45\textwidth]{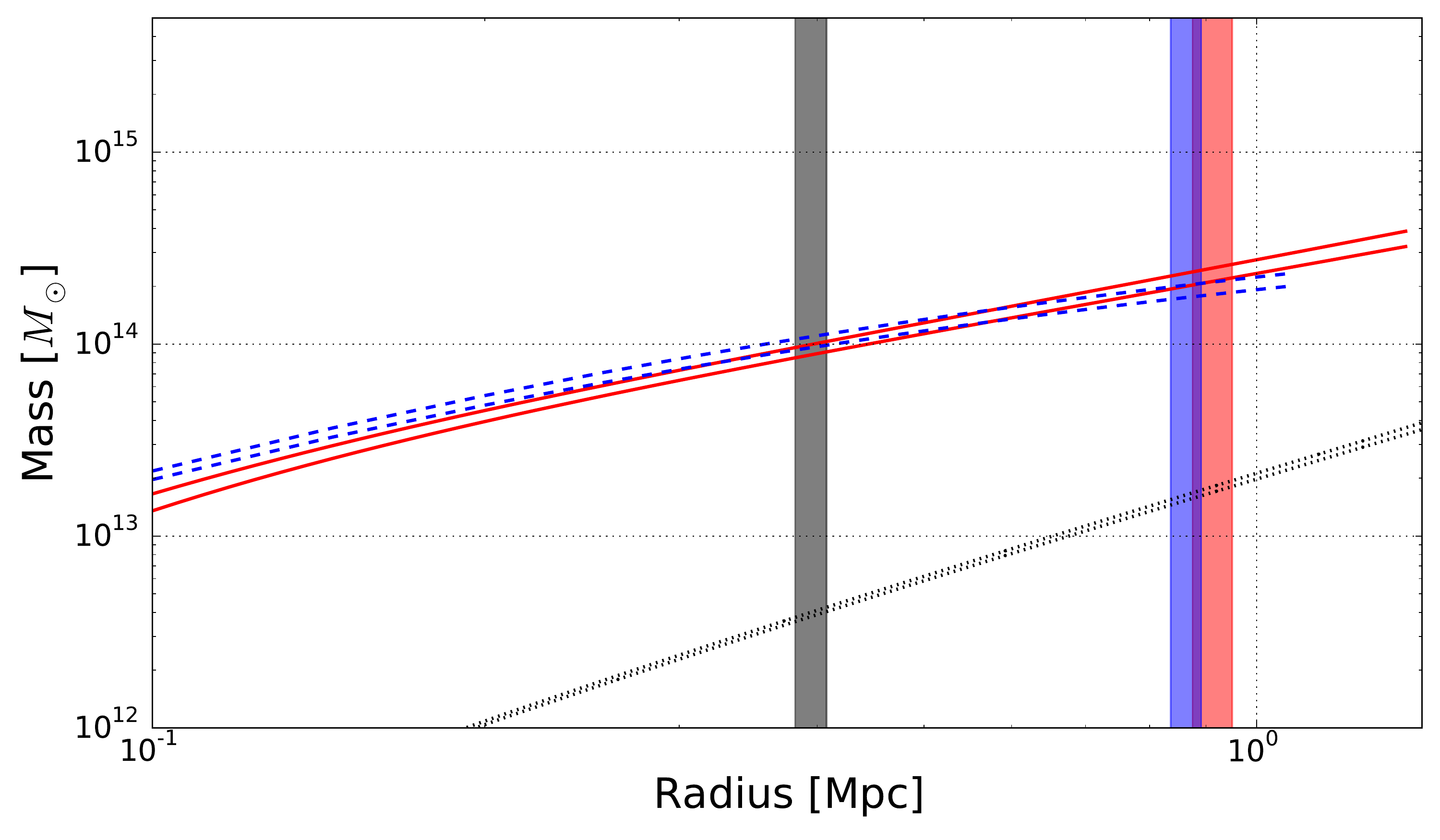}
	\caption{As Fig. \ref{fig:app_2A0335} but for A3395.}
	\label{fig:app_A3395}
\end{figure}
\clearpage
\begin{figure}
	\centering
	\includegraphics[width=0.45\textwidth]{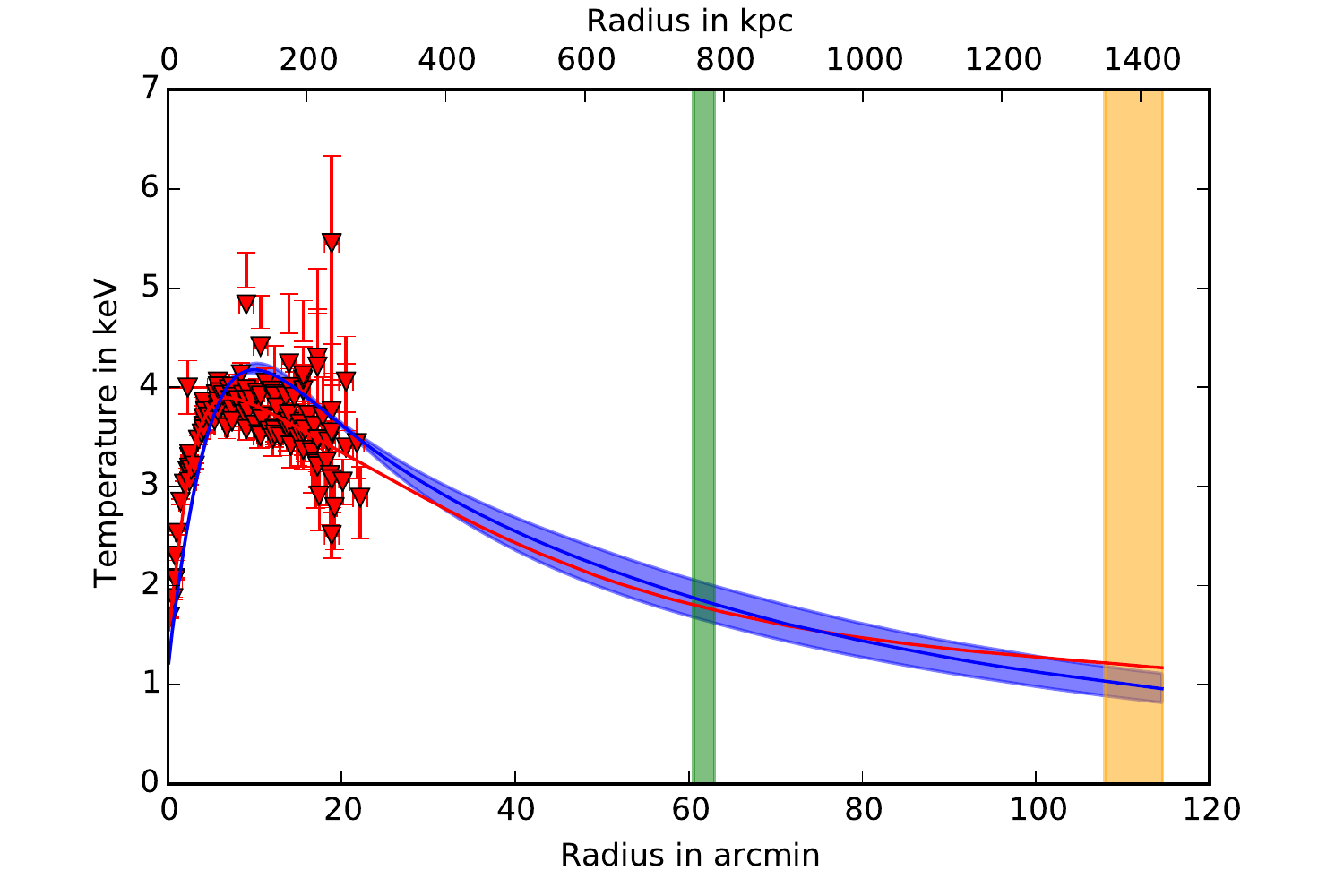}
	\includegraphics[width=0.45\textwidth]{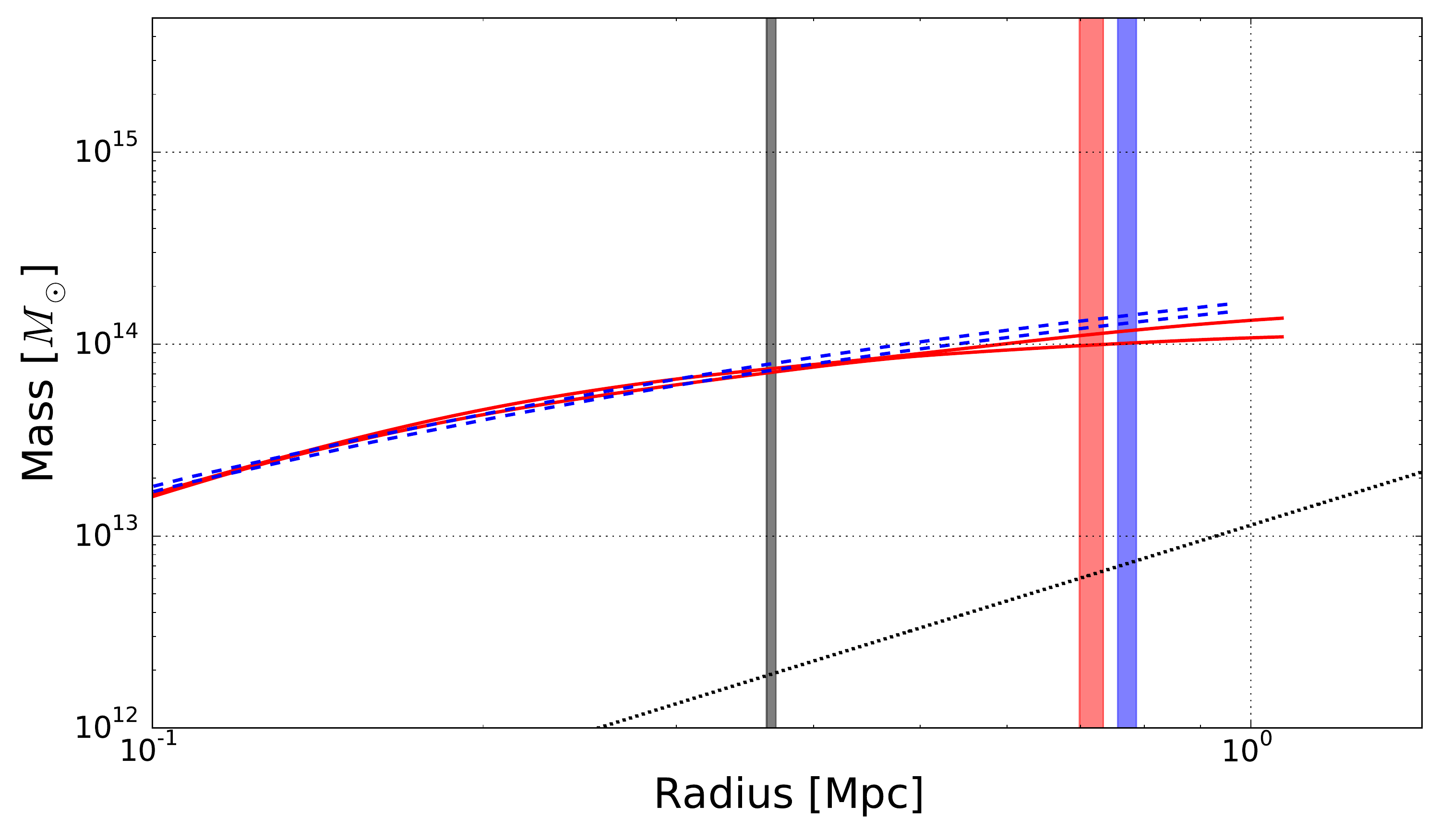}
	\caption{As Fig. \ref{fig:app_2A0335} but for A3526.}
	\label{fig:app_A3526}
\end{figure}
\begin{figure}
	\centering
	\includegraphics[width=0.45\textwidth]{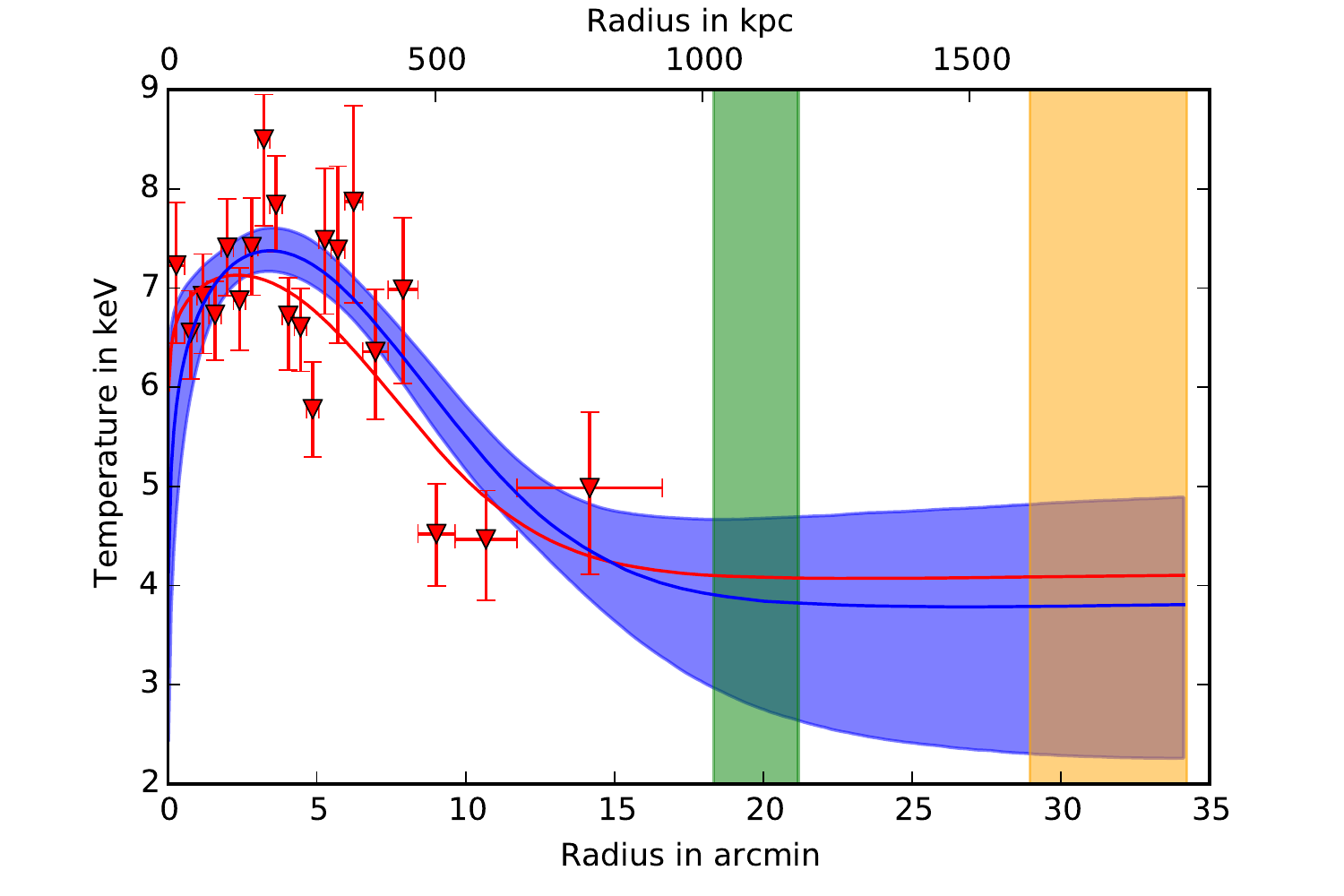}
	\includegraphics[width=0.45\textwidth]{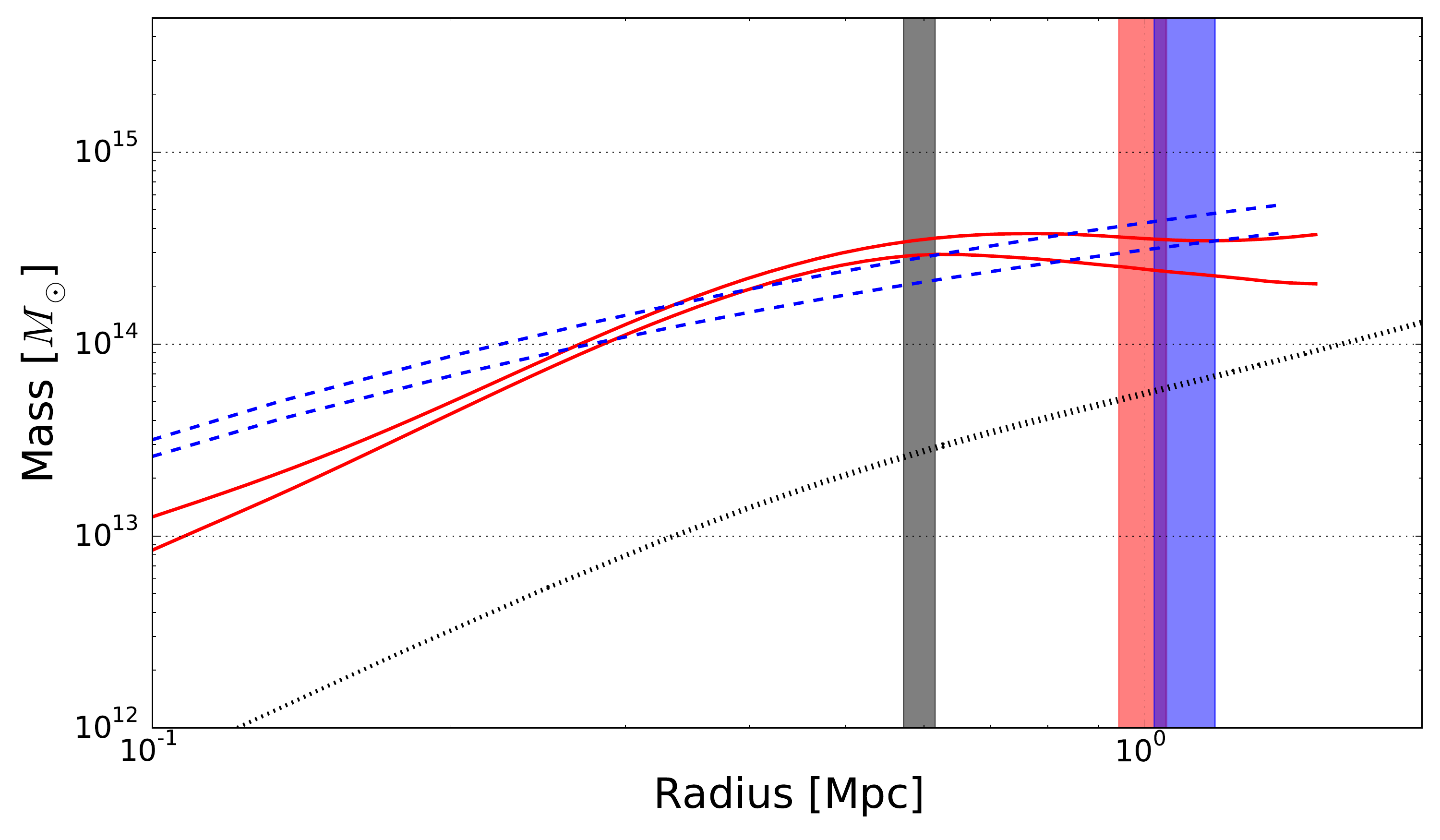}
	\caption{As Fig. \ref{fig:app_2A0335} but for A3558.}
	\label{fig:app_A3558}
\end{figure}
\begin{figure}
	\centering
	\includegraphics[width=0.45\textwidth]{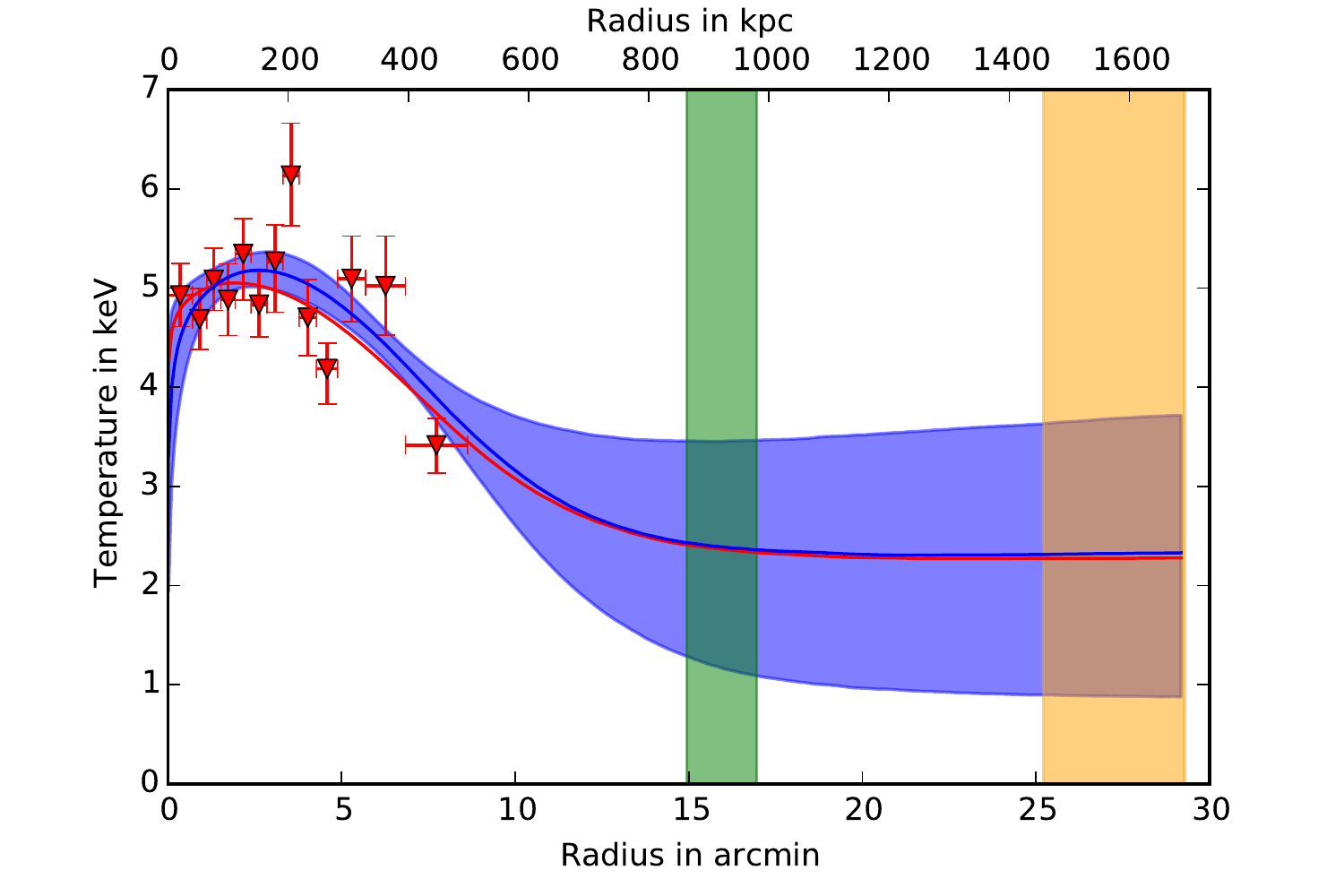}
	\includegraphics[width=0.45\textwidth]{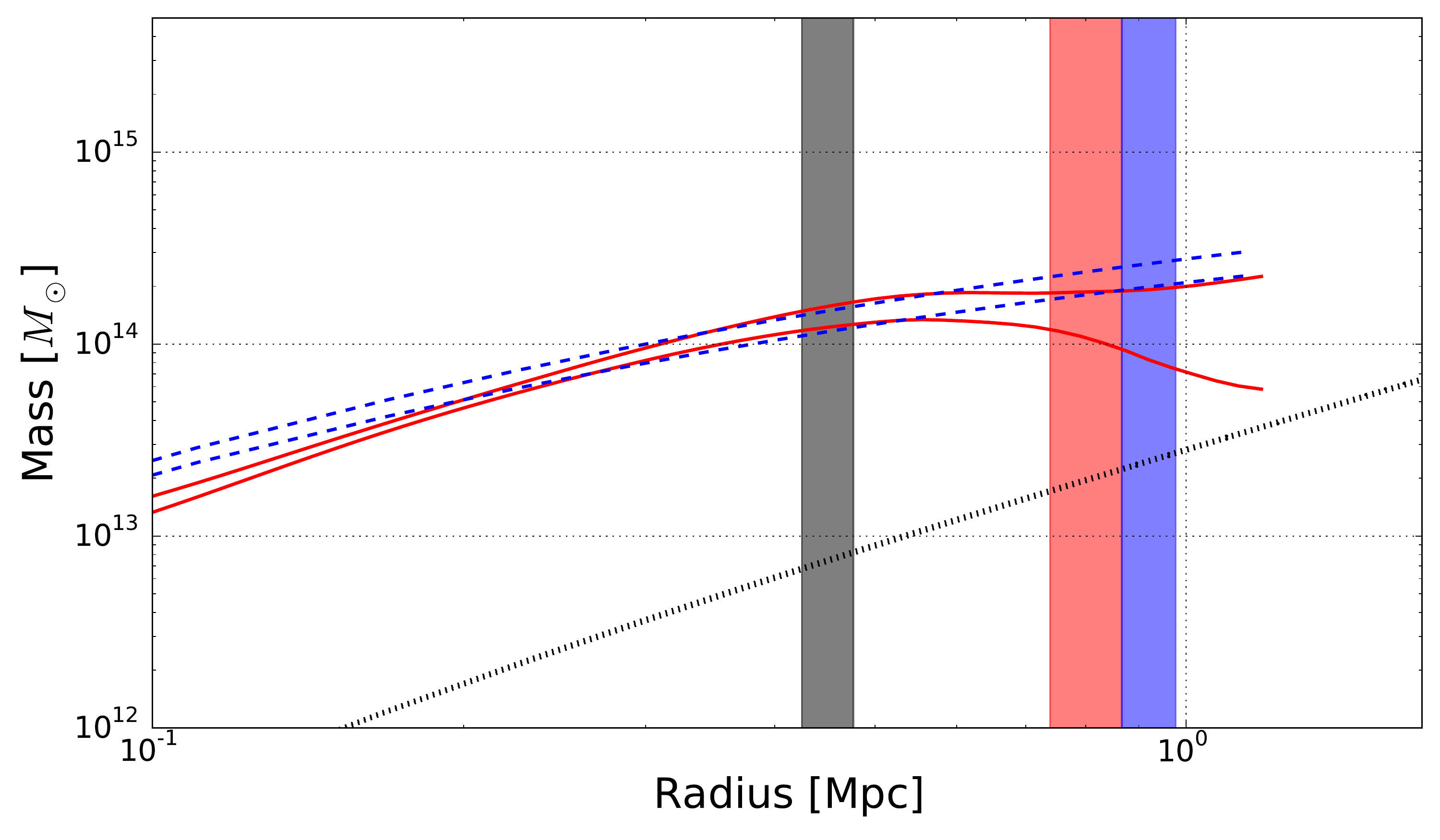}
	\caption{As Fig. \ref{fig:app_2A0335} but for A3562.}
	\label{fig:app_A3562}
\end{figure}
\clearpage
\begin{figure}
	\centering
	\includegraphics[width=0.45\textwidth]{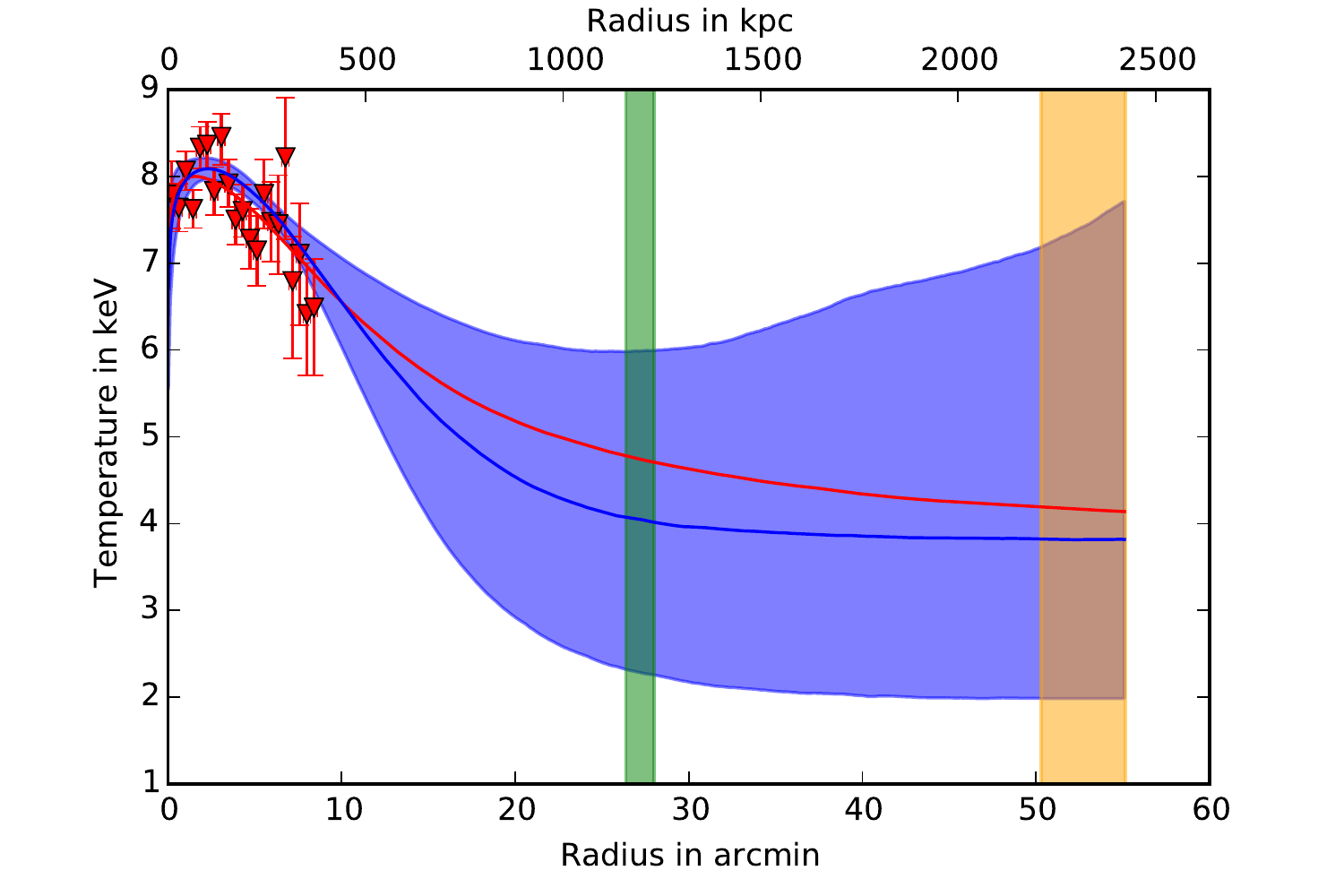}
	\includegraphics[width=0.45\textwidth]{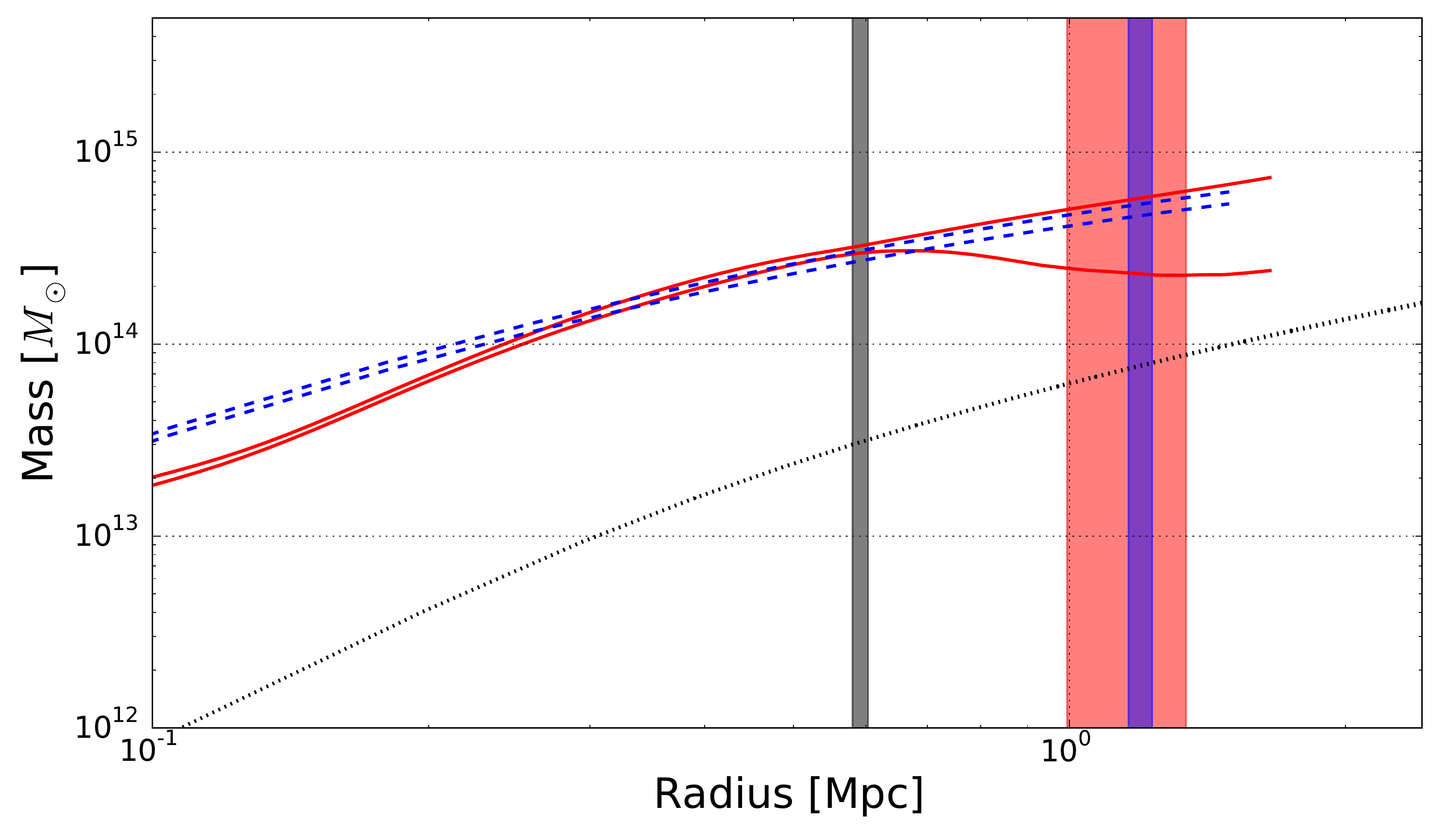}
	\caption{As Fig. \ref{fig:app_2A0335} but for A3571.}
	\label{fig:app_A3571}
\end{figure}
\begin{figure}
	\centering
	\includegraphics[width=0.45\textwidth]{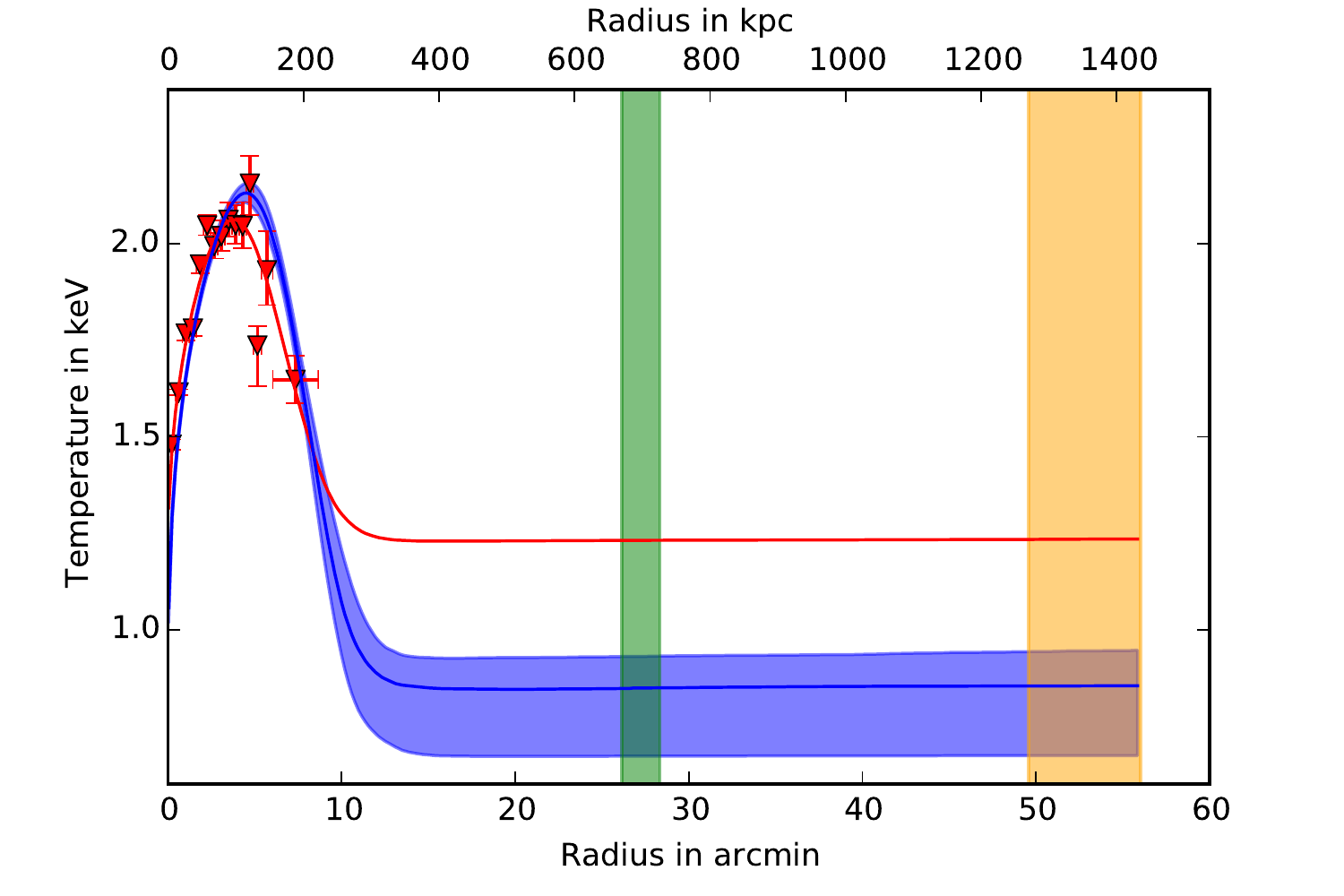}
	\includegraphics[width=0.45\textwidth]{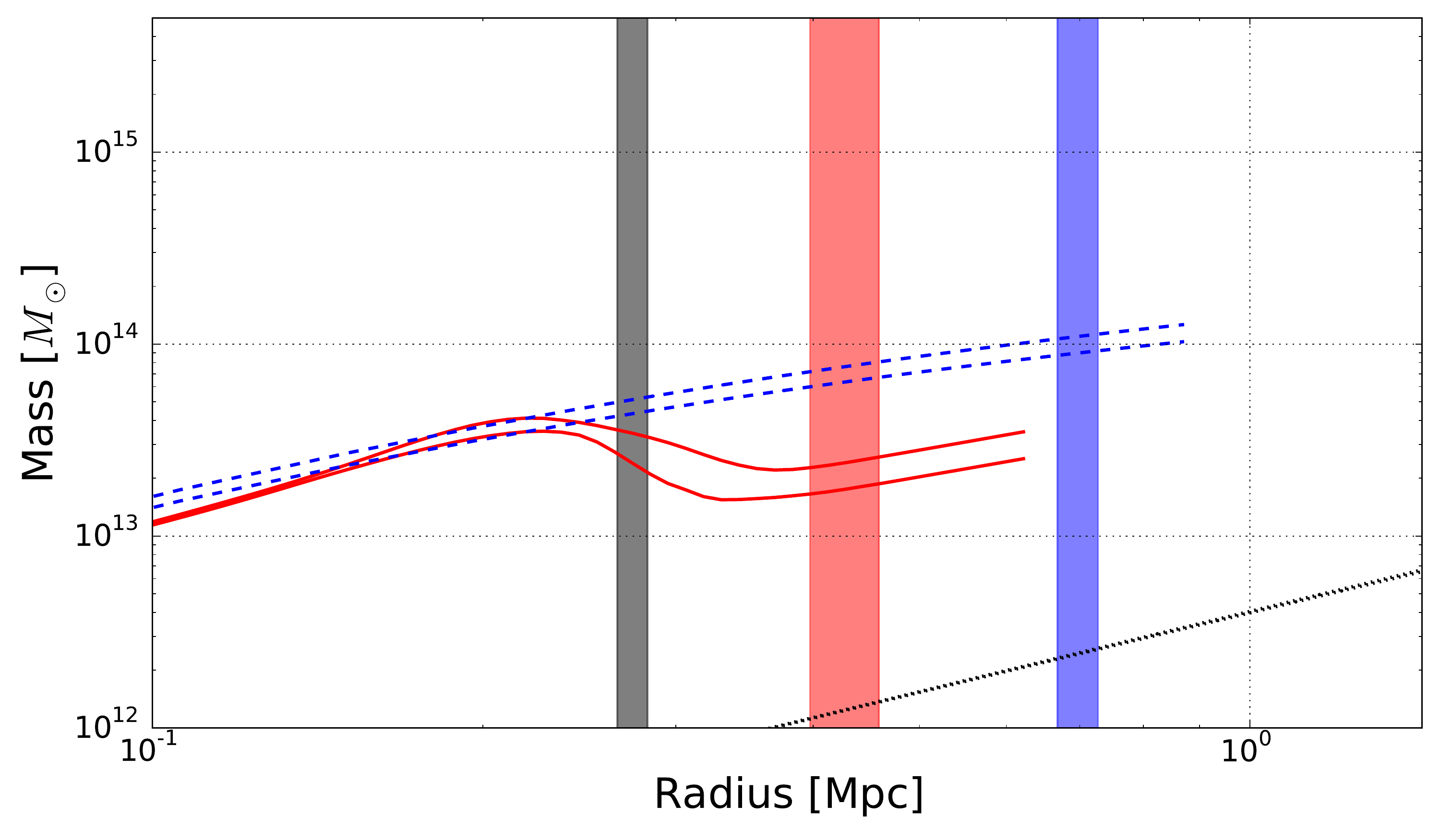}
	\caption{As Fig. \ref{fig:app_2A0335} but for A3581.}
	\label{fig:app_A3581}
\end{figure}
\begin{figure}
	\centering
	\includegraphics[width=0.45\textwidth]{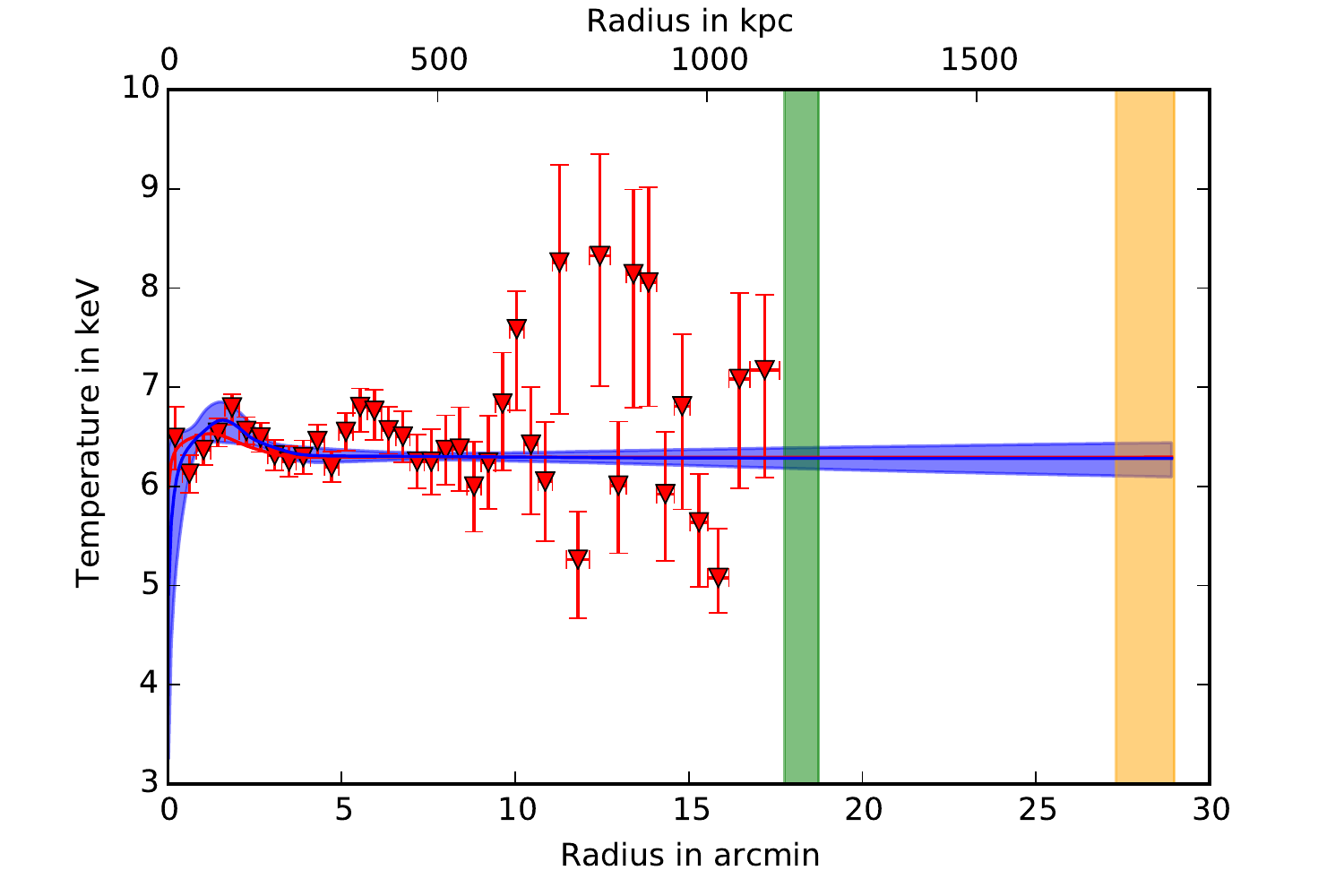}
	\includegraphics[width=0.45\textwidth]{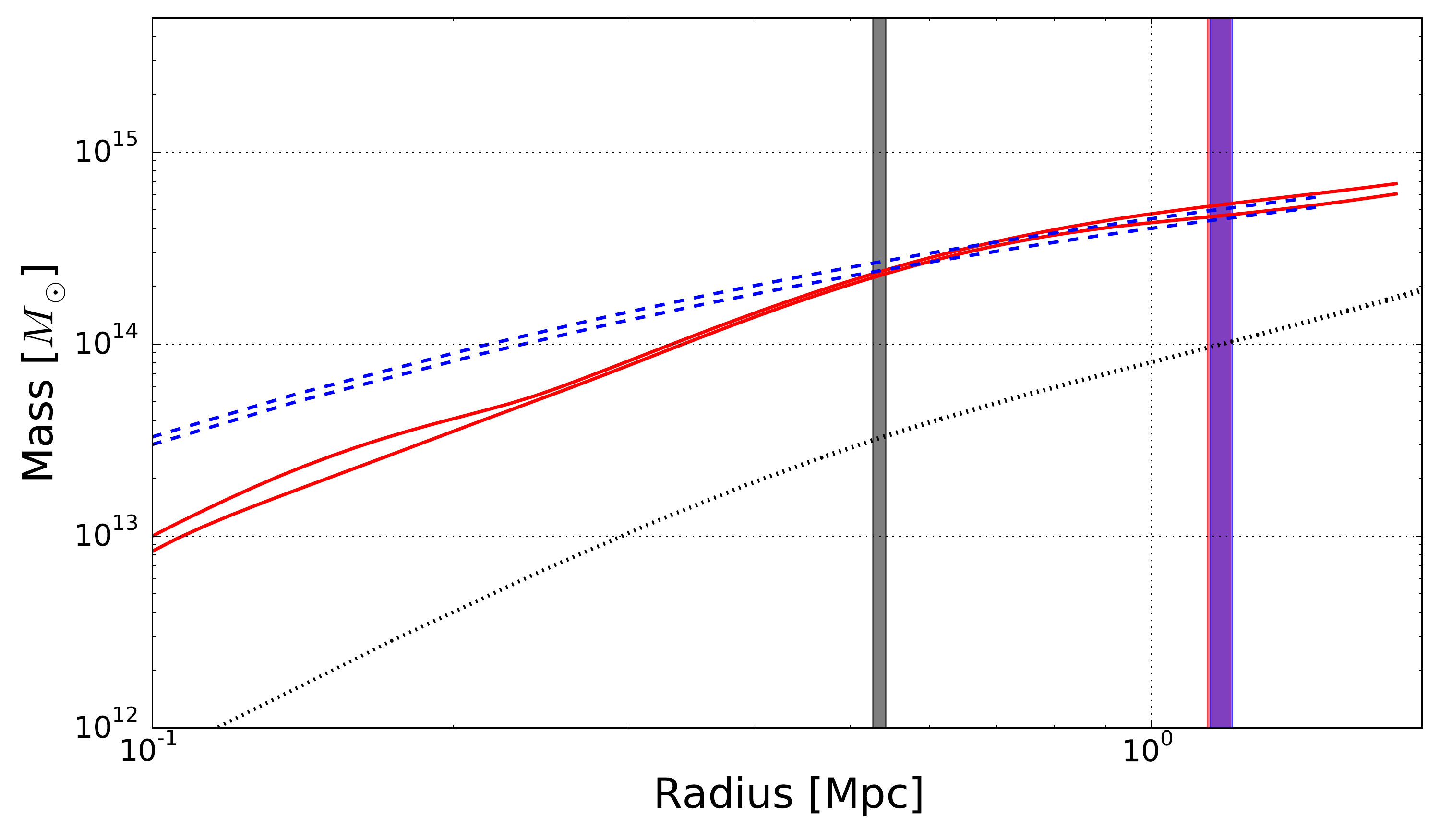}
	\caption{As Fig. \ref{fig:app_2A0335} but for A3667.}
	\label{fig:app_A3667}
\end{figure}
\clearpage
\begin{figure}
	\centering
	\includegraphics[width=0.45\textwidth]{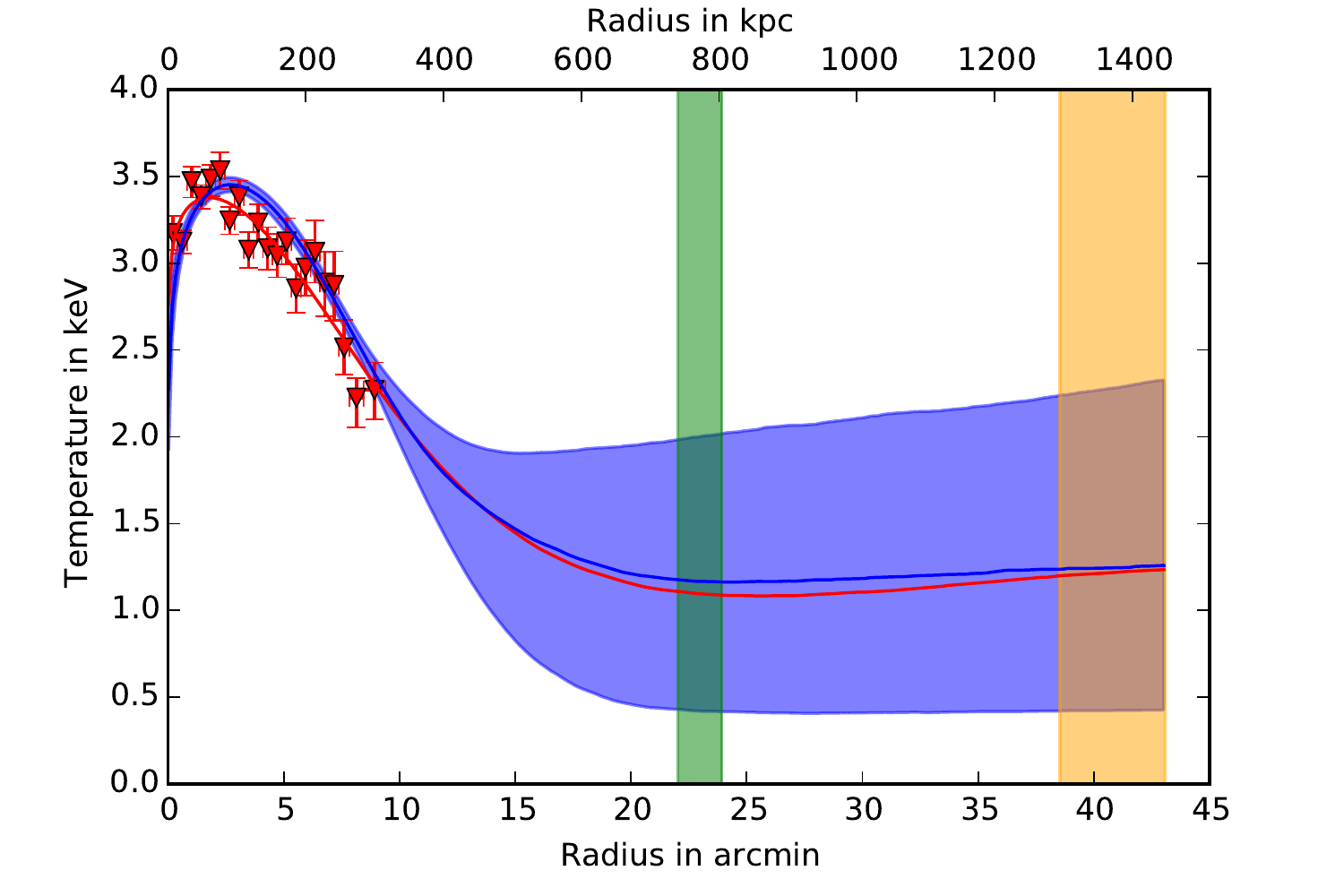}
	\includegraphics[width=0.45\textwidth]{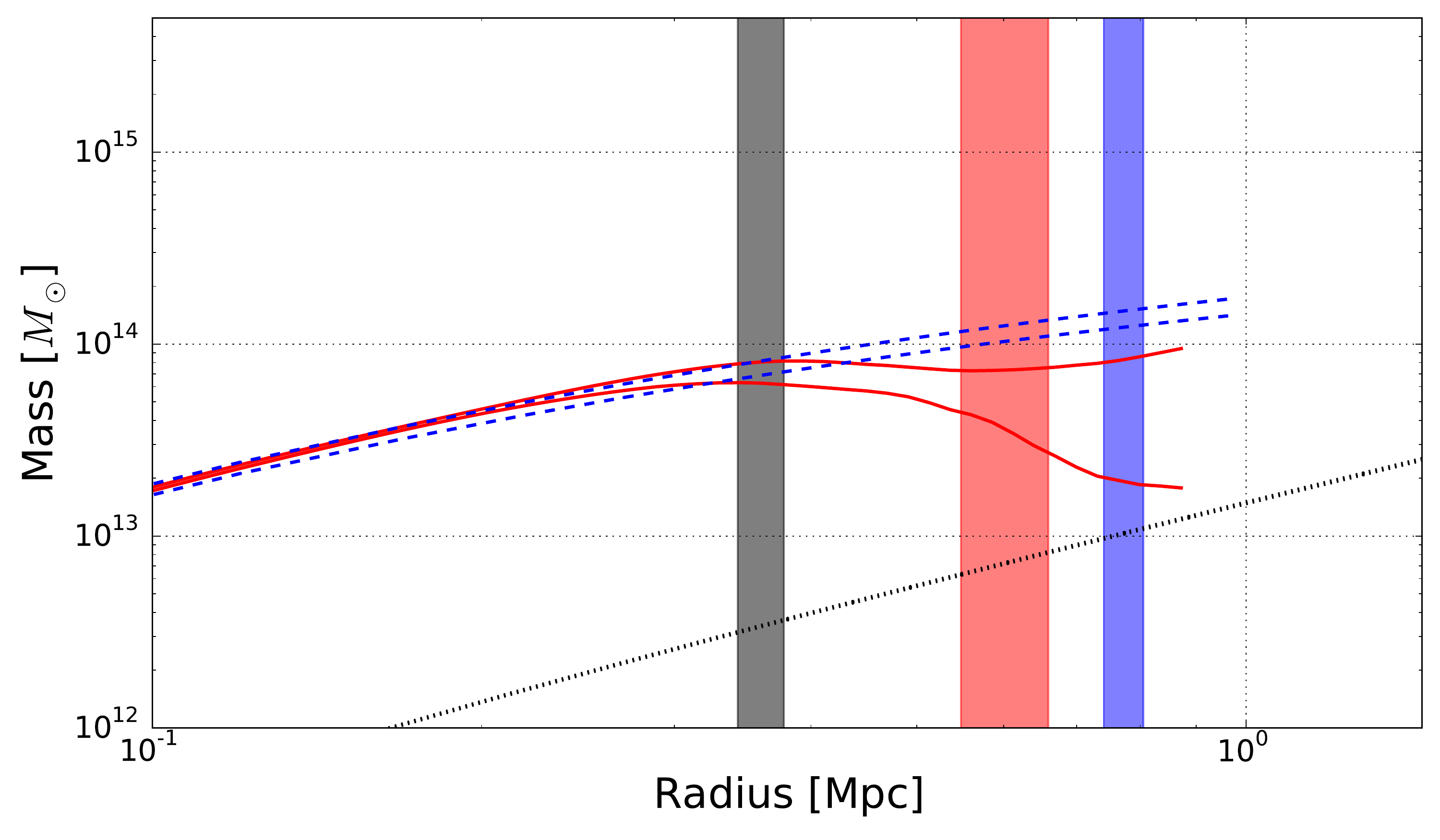}
	\caption{As Fig. \ref{fig:app_2A0335} but for A4038.}
	\label{fig:app_A4038}
\end{figure}
\begin{figure}
	\centering
	\includegraphics[width=0.45\textwidth]{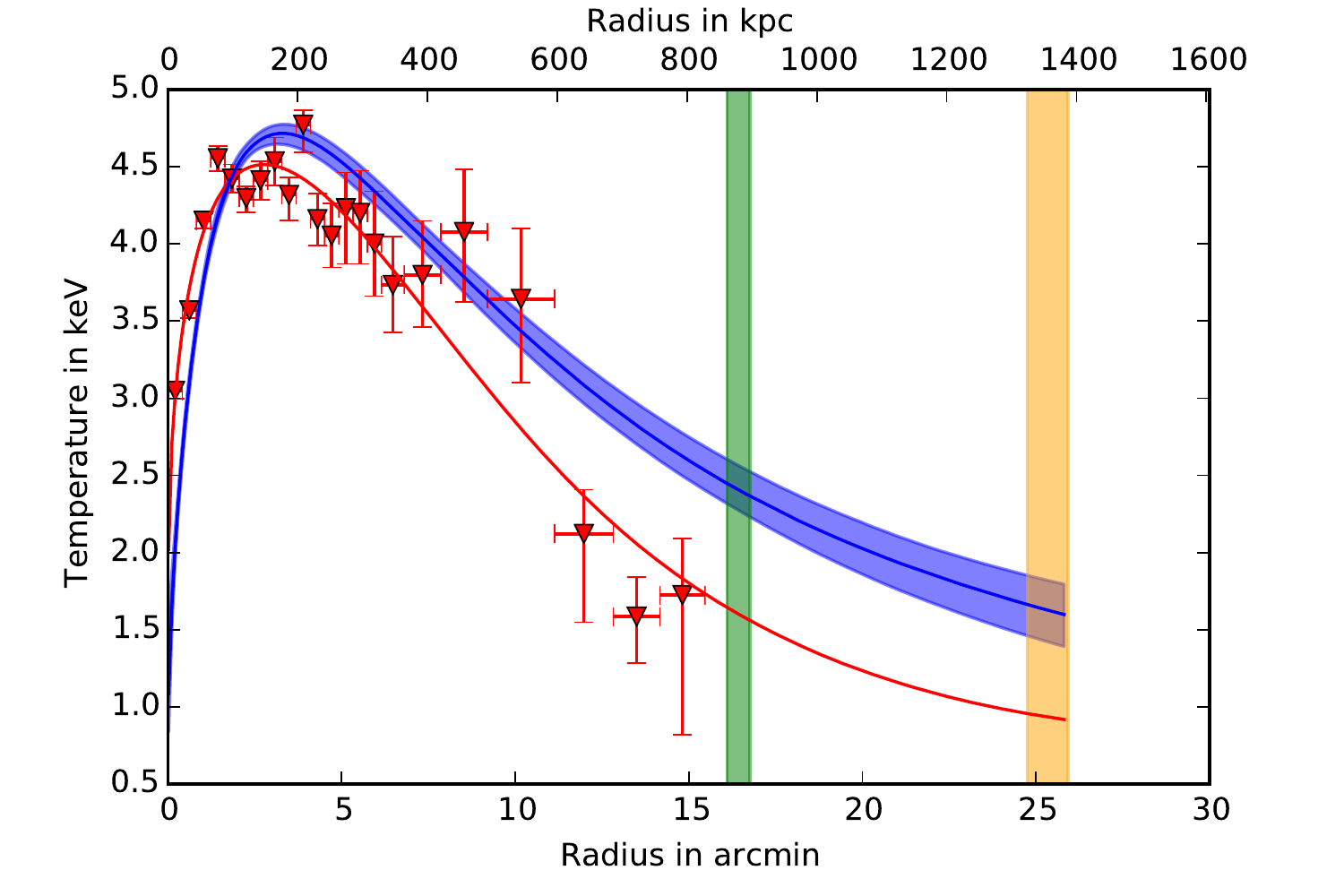}
	\includegraphics[width=0.45\textwidth]{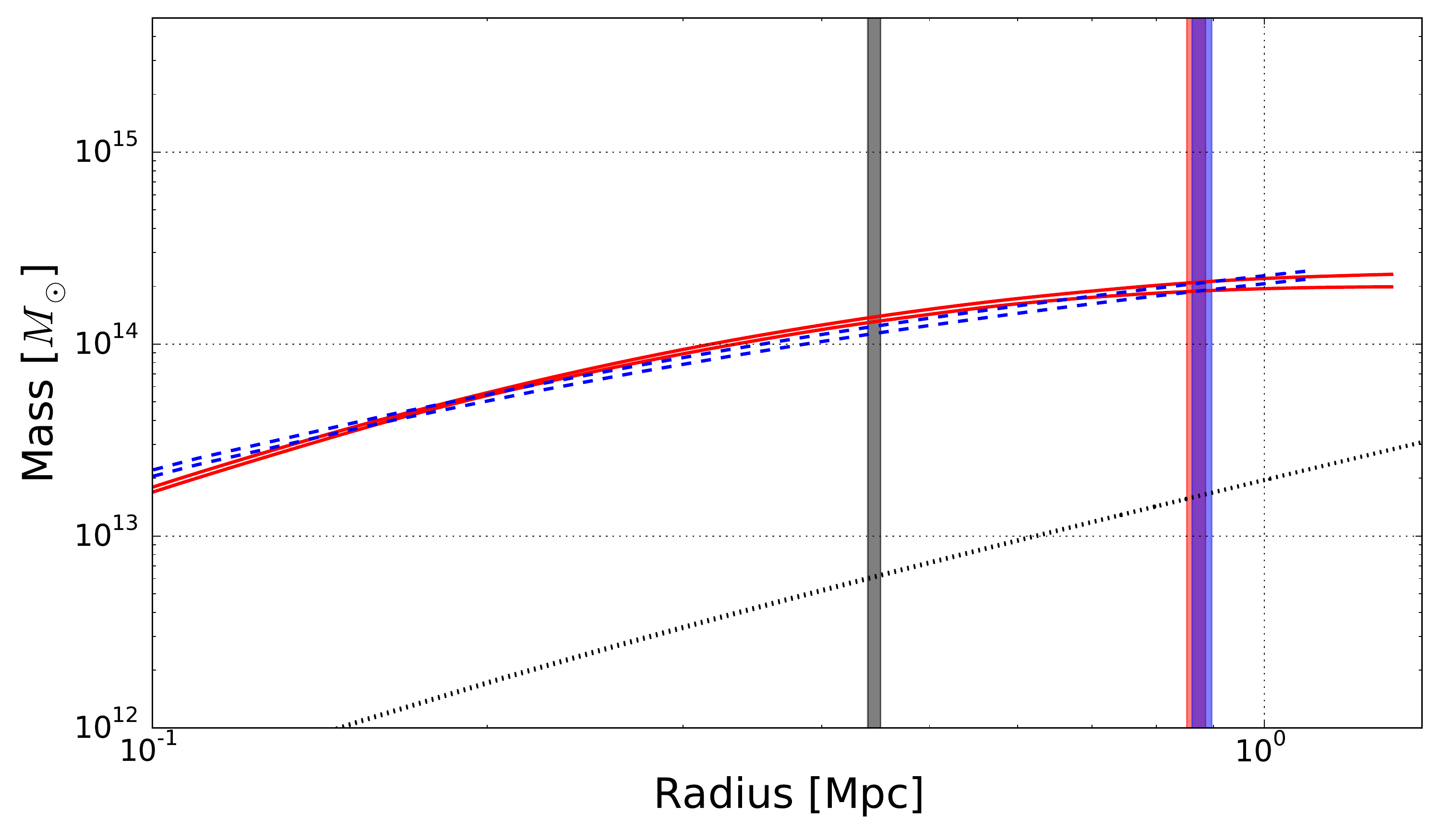}
	\caption{As Fig. \ref{fig:app_2A0335} but for A4059.}
	\label{fig:app_A4059}
\end{figure}
\begin{figure}
	\centering
	\includegraphics[width=0.45\textwidth]{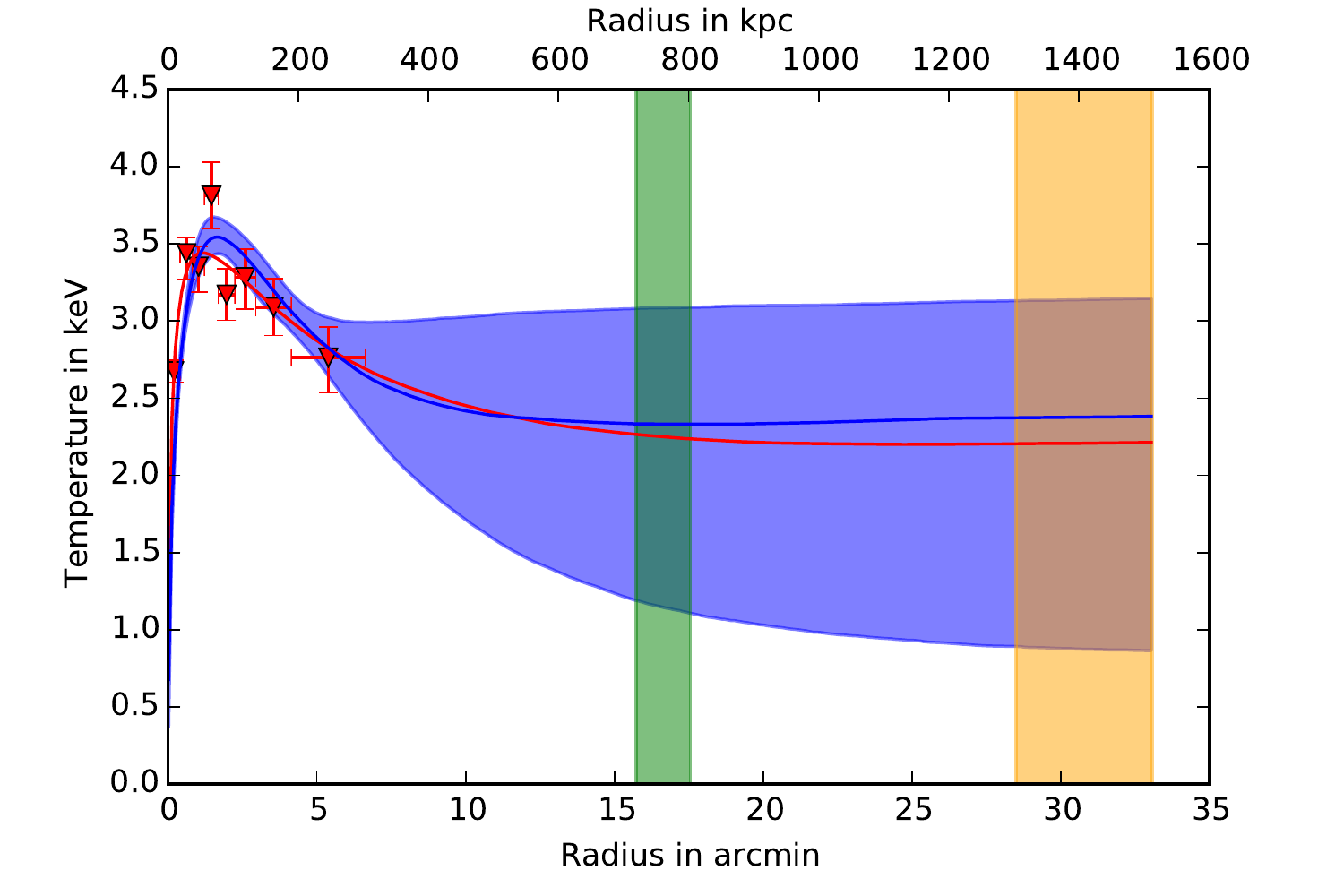}
	\includegraphics[width=0.45\textwidth]{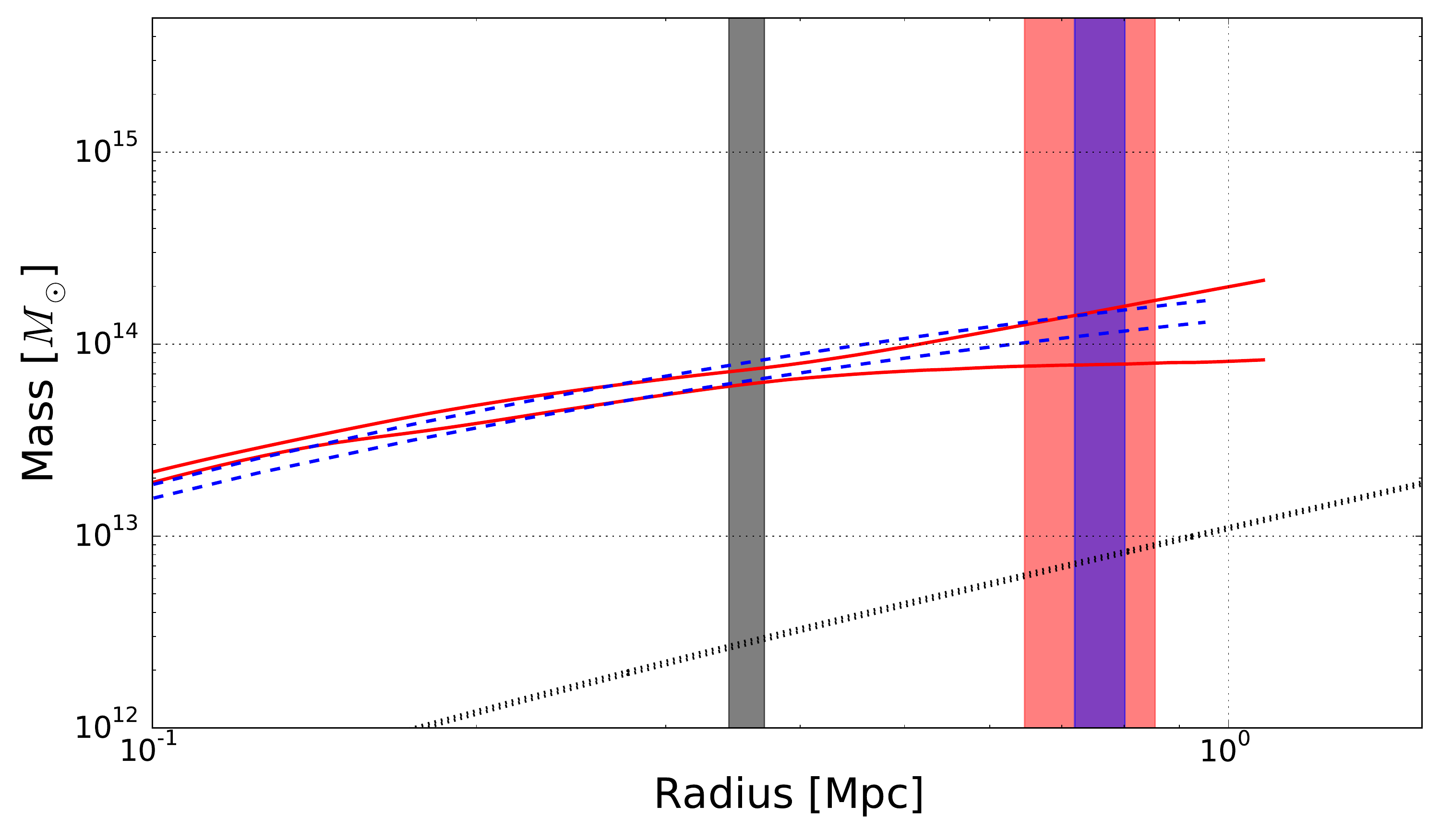}
	\caption{As Fig. \ref{fig:app_2A0335} but for EXO0422.}
	\label{fig:app_EXO0422}
\end{figure}
\clearpage
\begin{figure}
	\centering
	\includegraphics[width=0.45\textwidth]{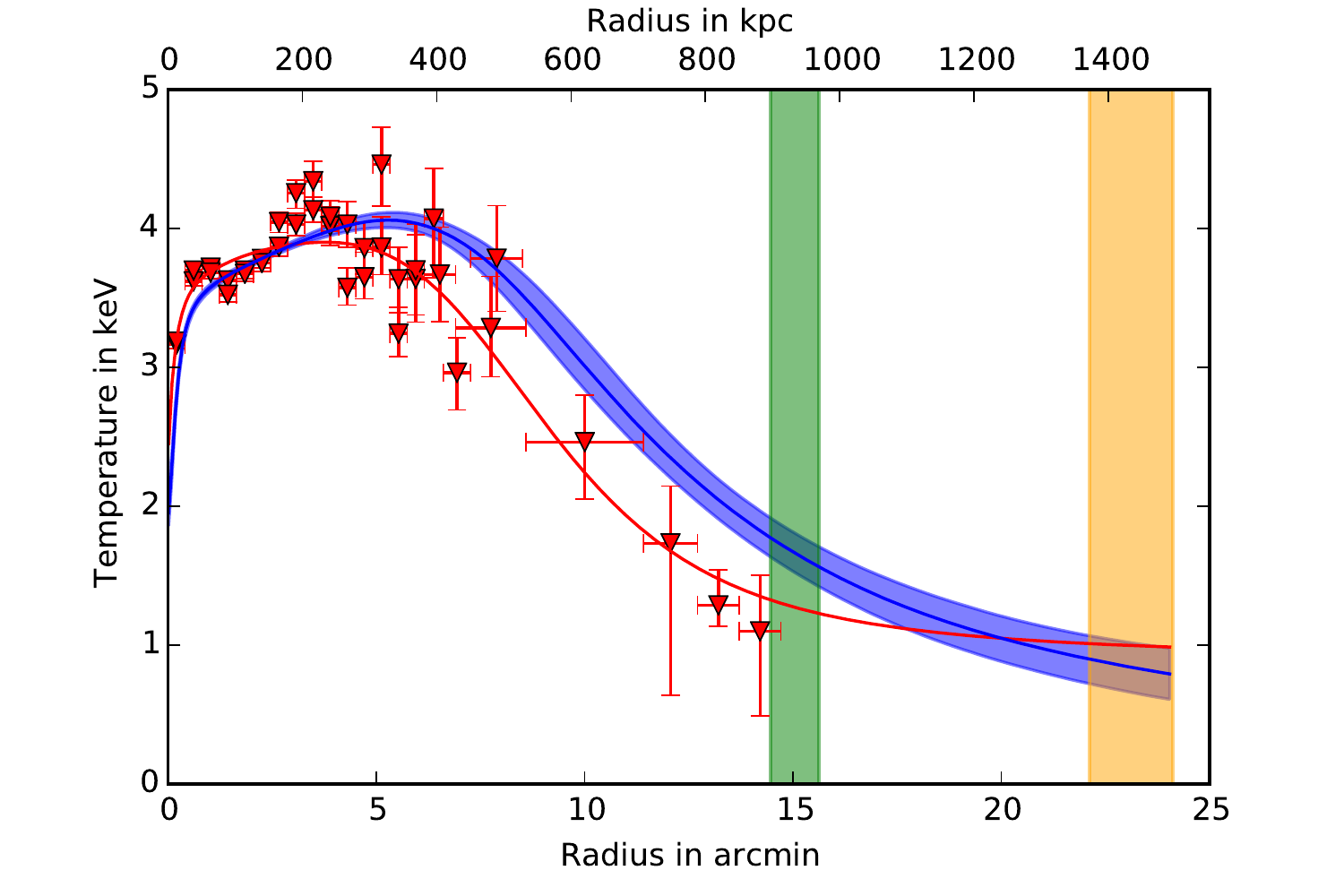}
	\includegraphics[width=0.45\textwidth]{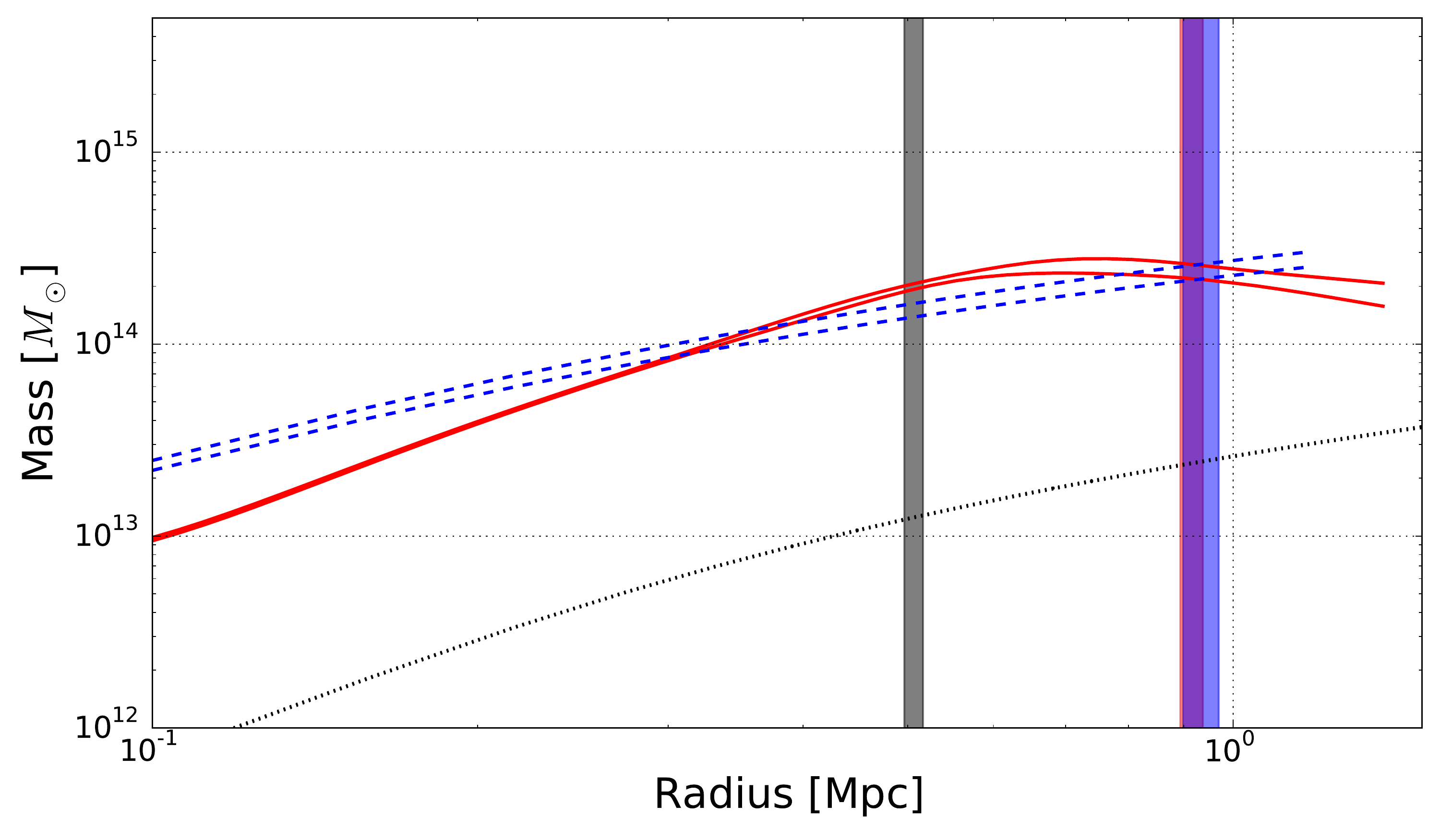}
	\caption{As Fig. \ref{fig:app_2A0335} but for HydraA.}
	\label{fig:app_HydraA}
\end{figure}
\begin{figure}
	\centering
	\includegraphics[width=0.45\textwidth]{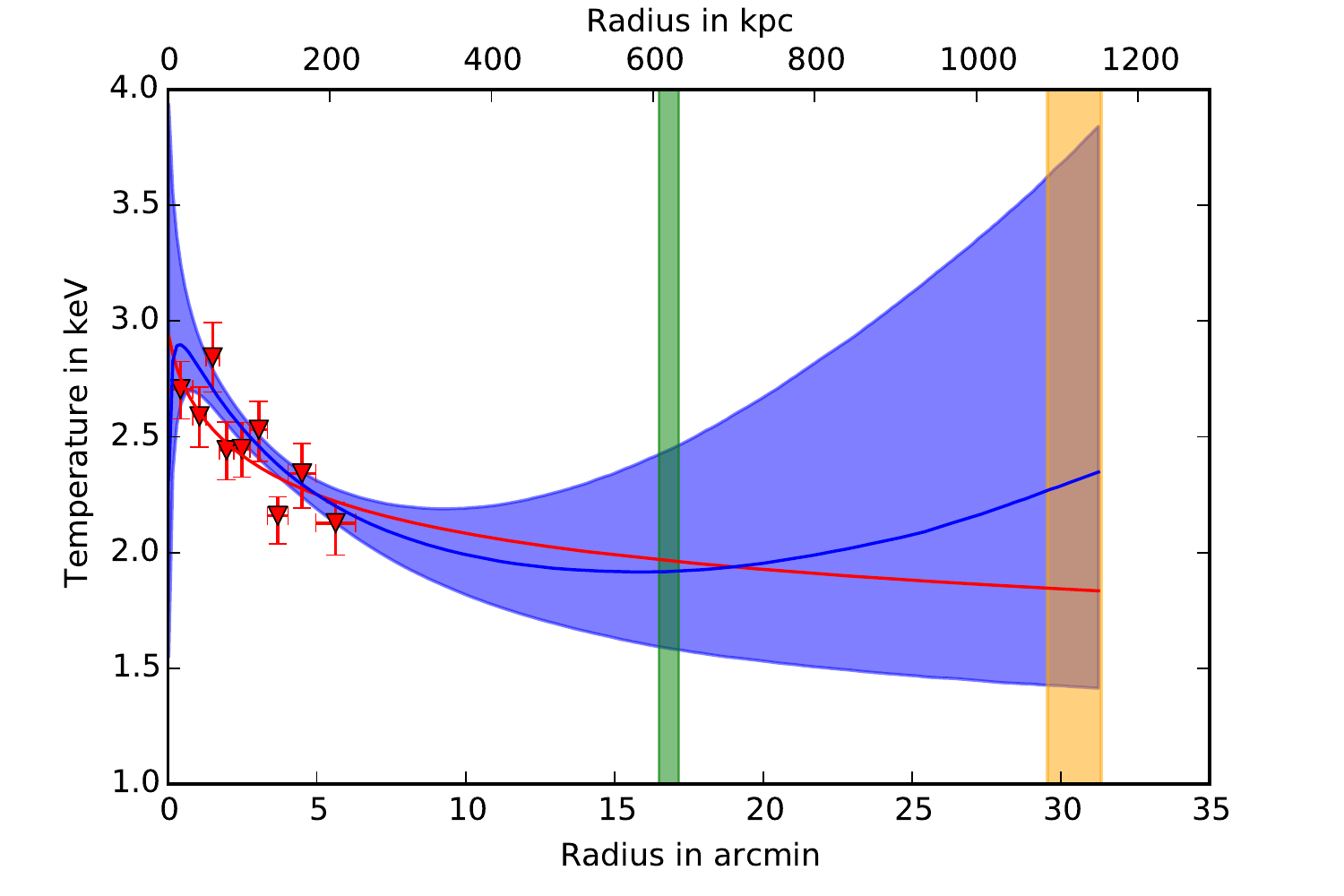}
	\includegraphics[width=0.45\textwidth]{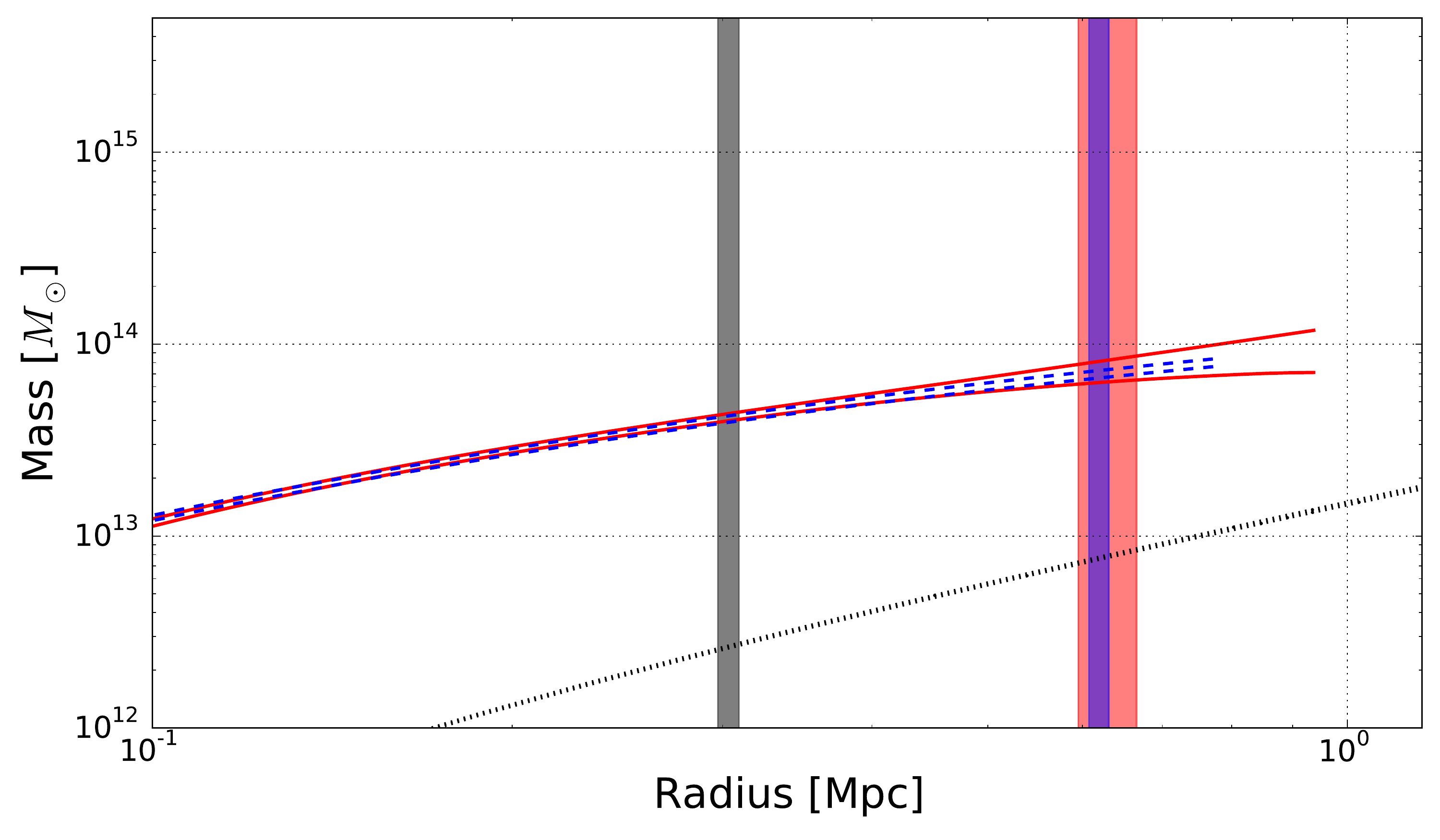}
	\caption{As Fig. \ref{fig:app_2A0335} but for IIIZw54.}
	\label{fig:app_IIIZw54}
\end{figure}
\begin{figure}
	\centering
	\includegraphics[width=0.45\textwidth]{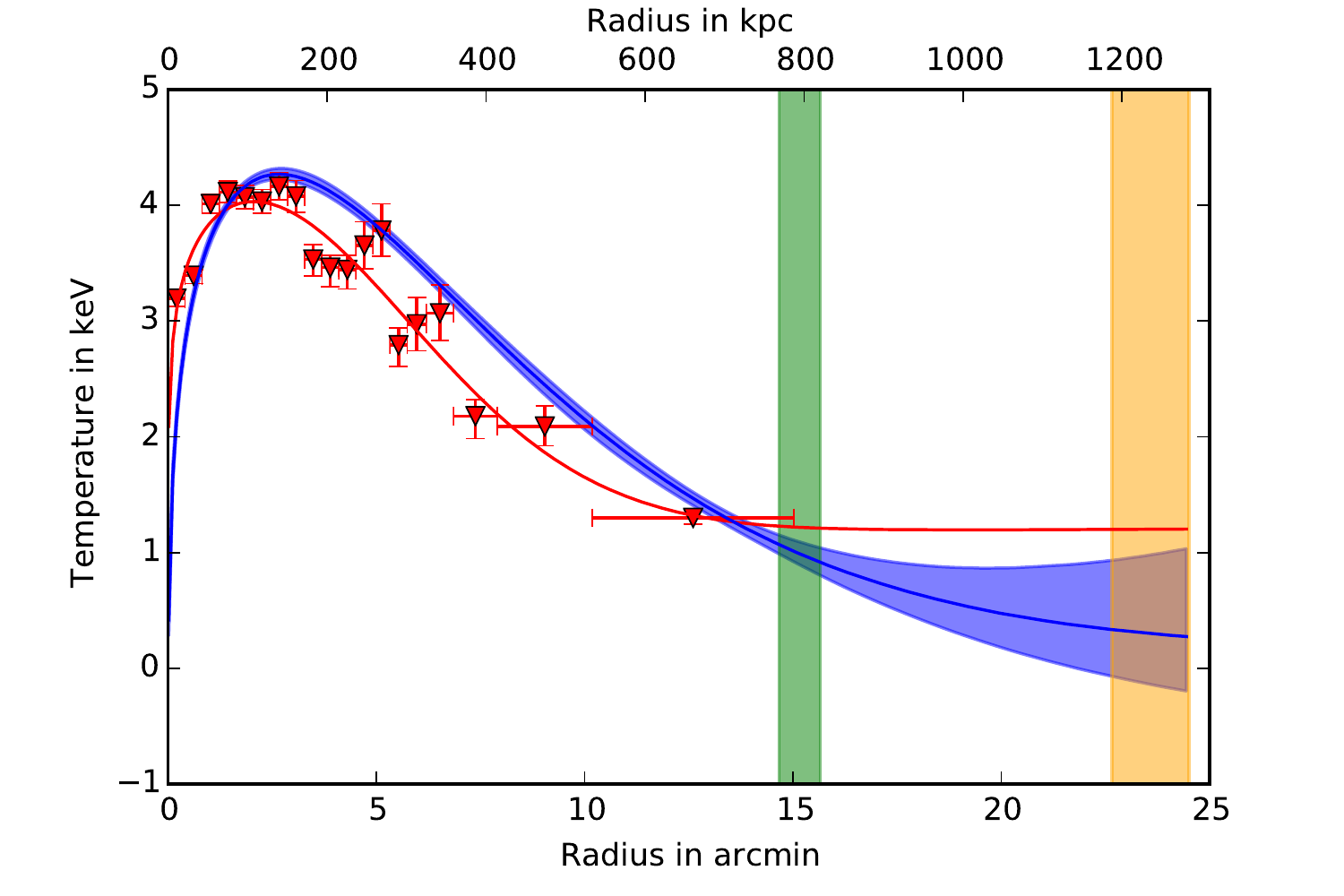}
	\includegraphics[width=0.45\textwidth]{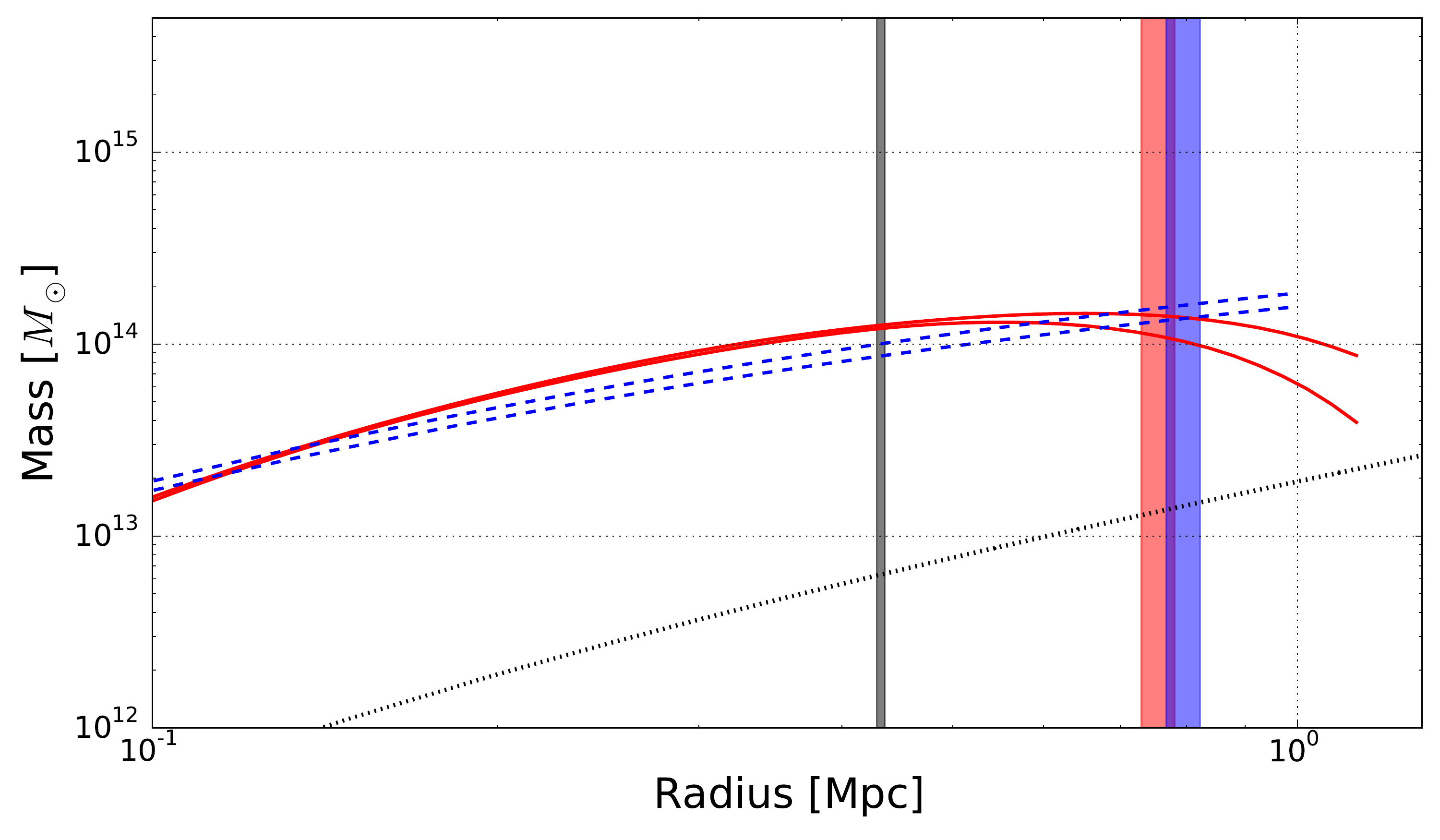}
	\caption{As Fig. \ref{fig:app_2A0335} but for MKW3S.}
	\label{fig:app_MKW3S}
\end{figure}
\clearpage
\begin{figure}
	\centering
	\includegraphics[width=0.45\textwidth]{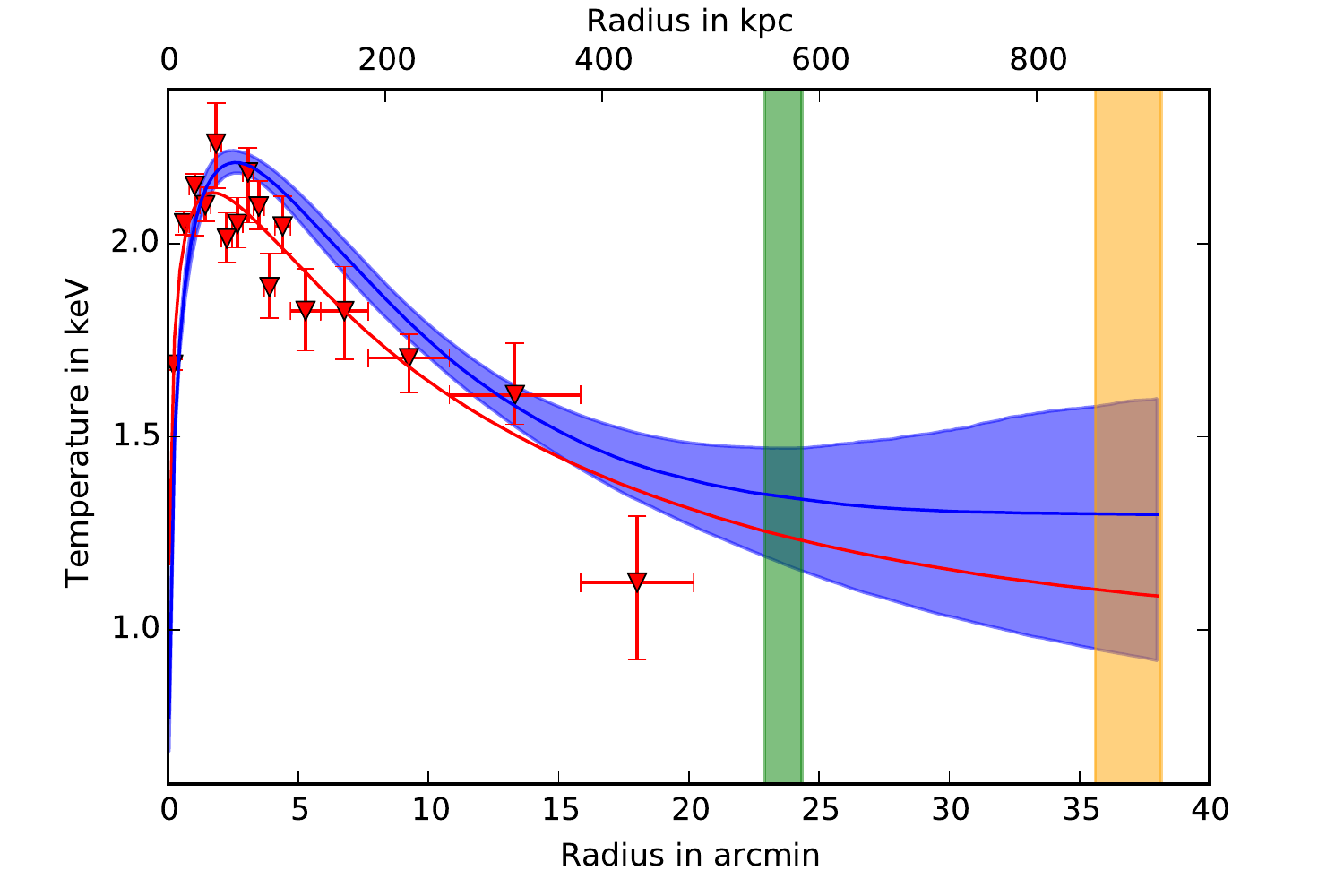}
	\includegraphics[width=0.45\textwidth]{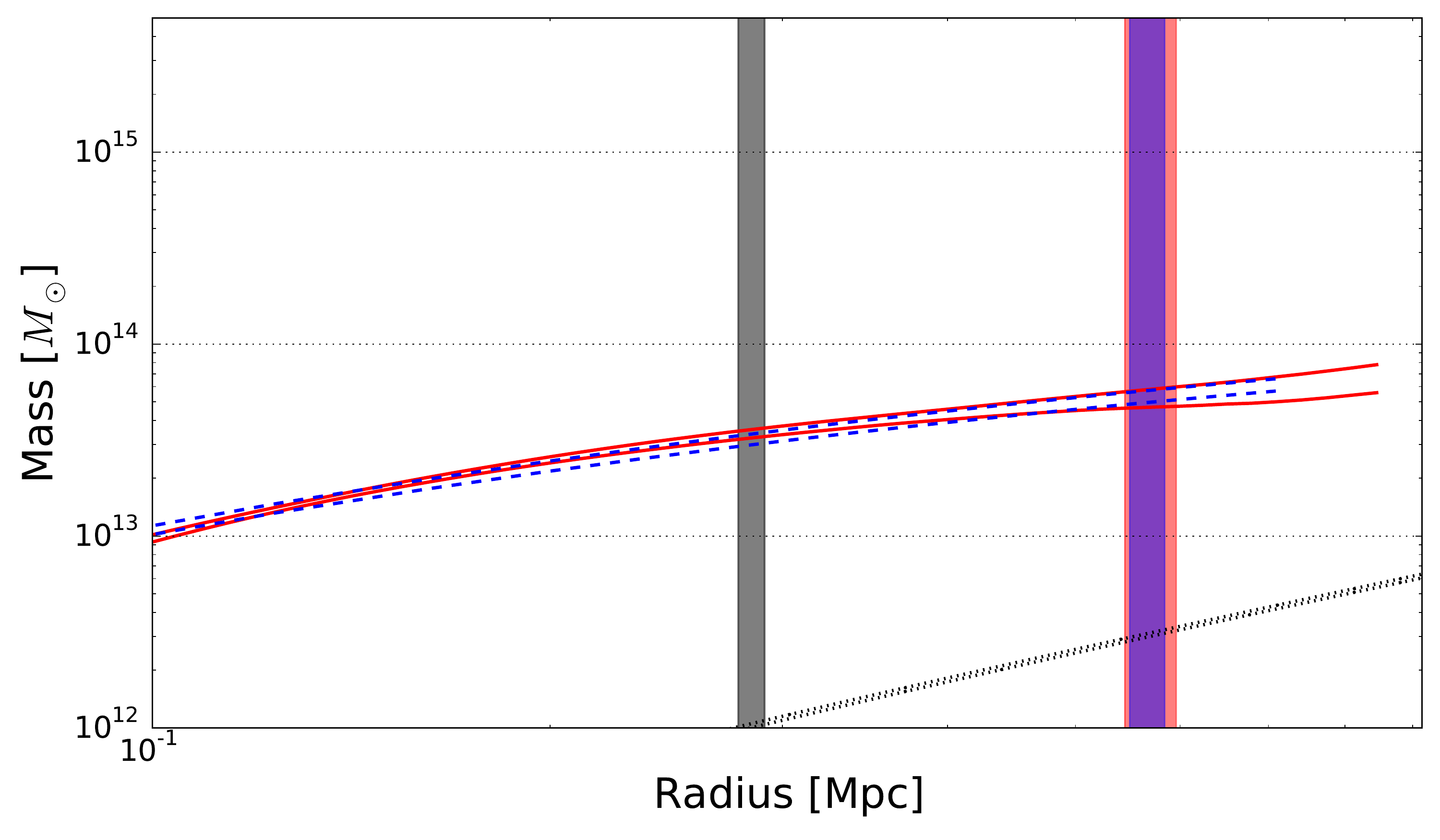}
	\caption{As Fig. \ref{fig:app_2A0335} but for MKW4.}
	\label{fig:app_MKW4}
\end{figure}
\begin{figure}
	\centering
	\includegraphics[width=0.45\textwidth]{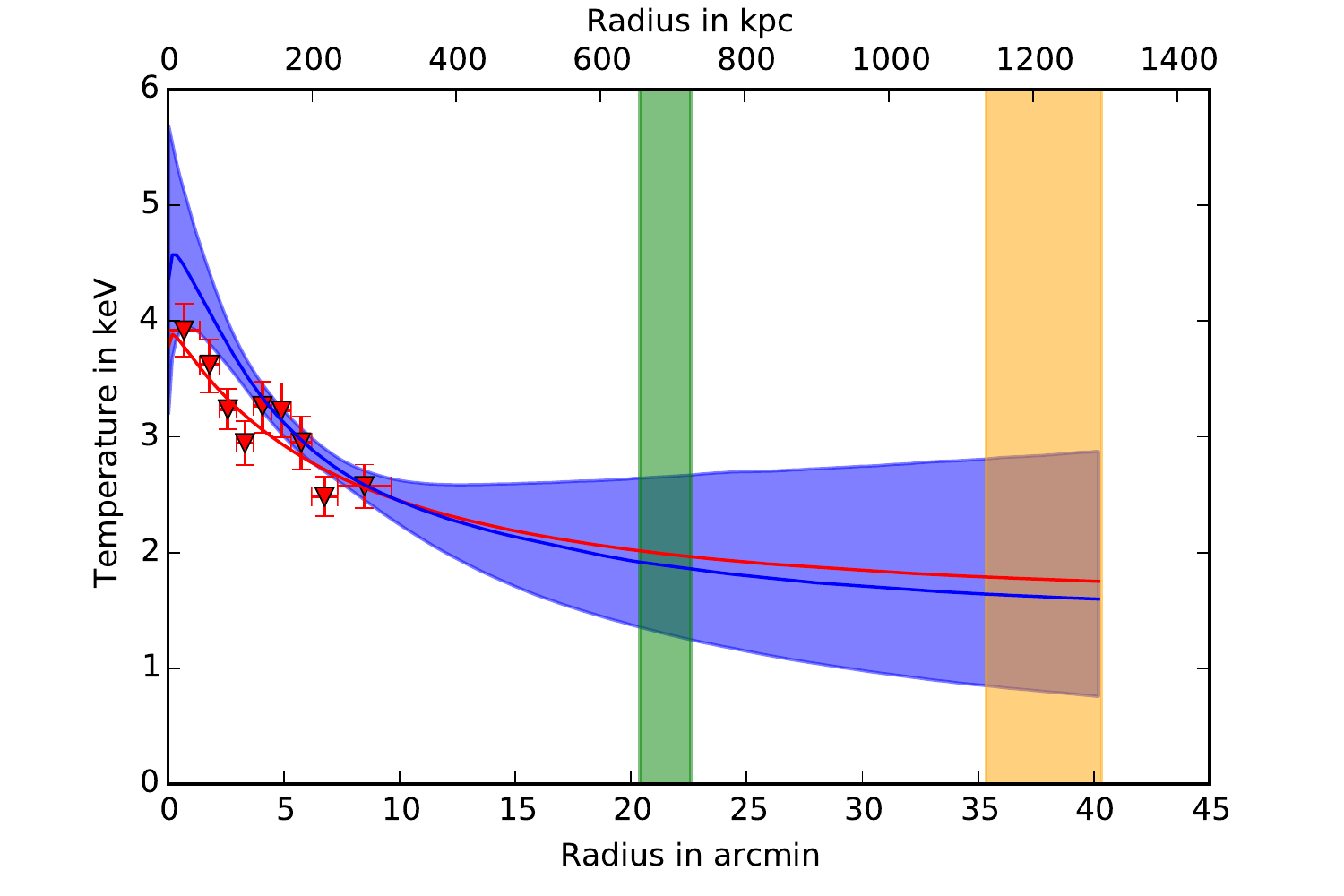}
	\includegraphics[width=0.45\textwidth]{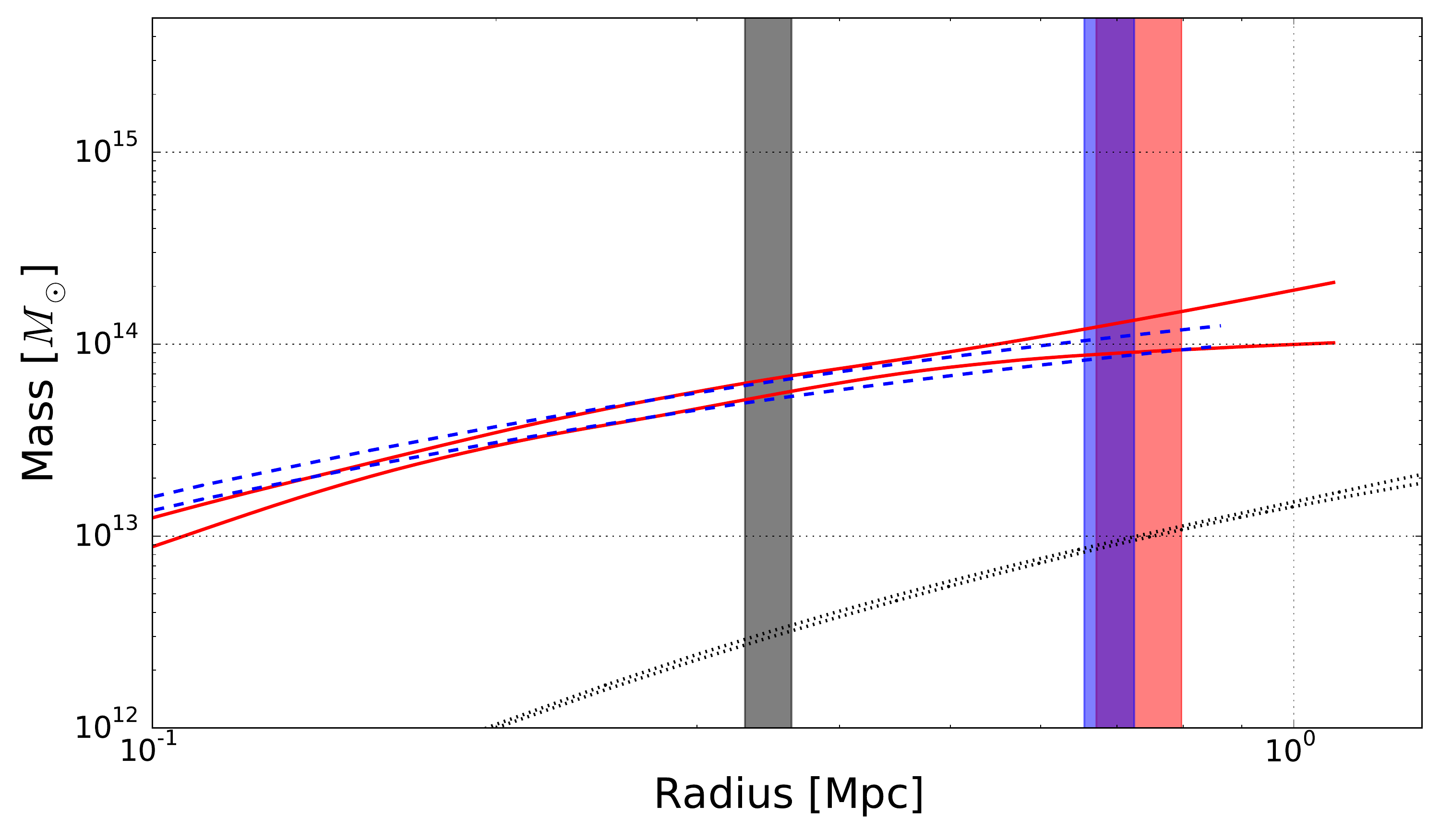}
	\caption{As Fig. \ref{fig:app_2A0335} but for MKW8.}
	\label{fig:app_MKW8}
\end{figure}
\begin{figure}
	\centering
	\includegraphics[width=0.45\textwidth]{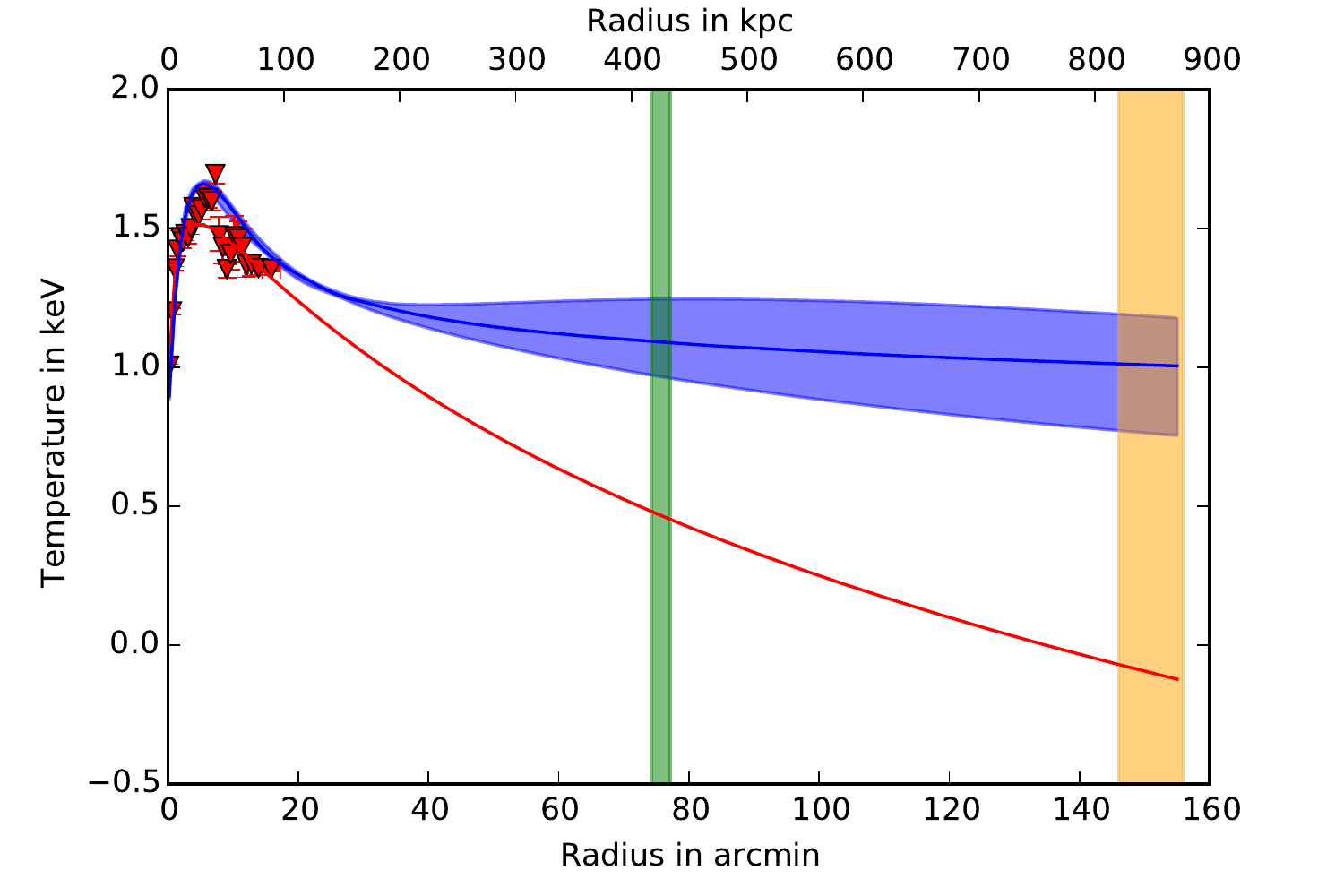}
	\includegraphics[width=0.45\textwidth]{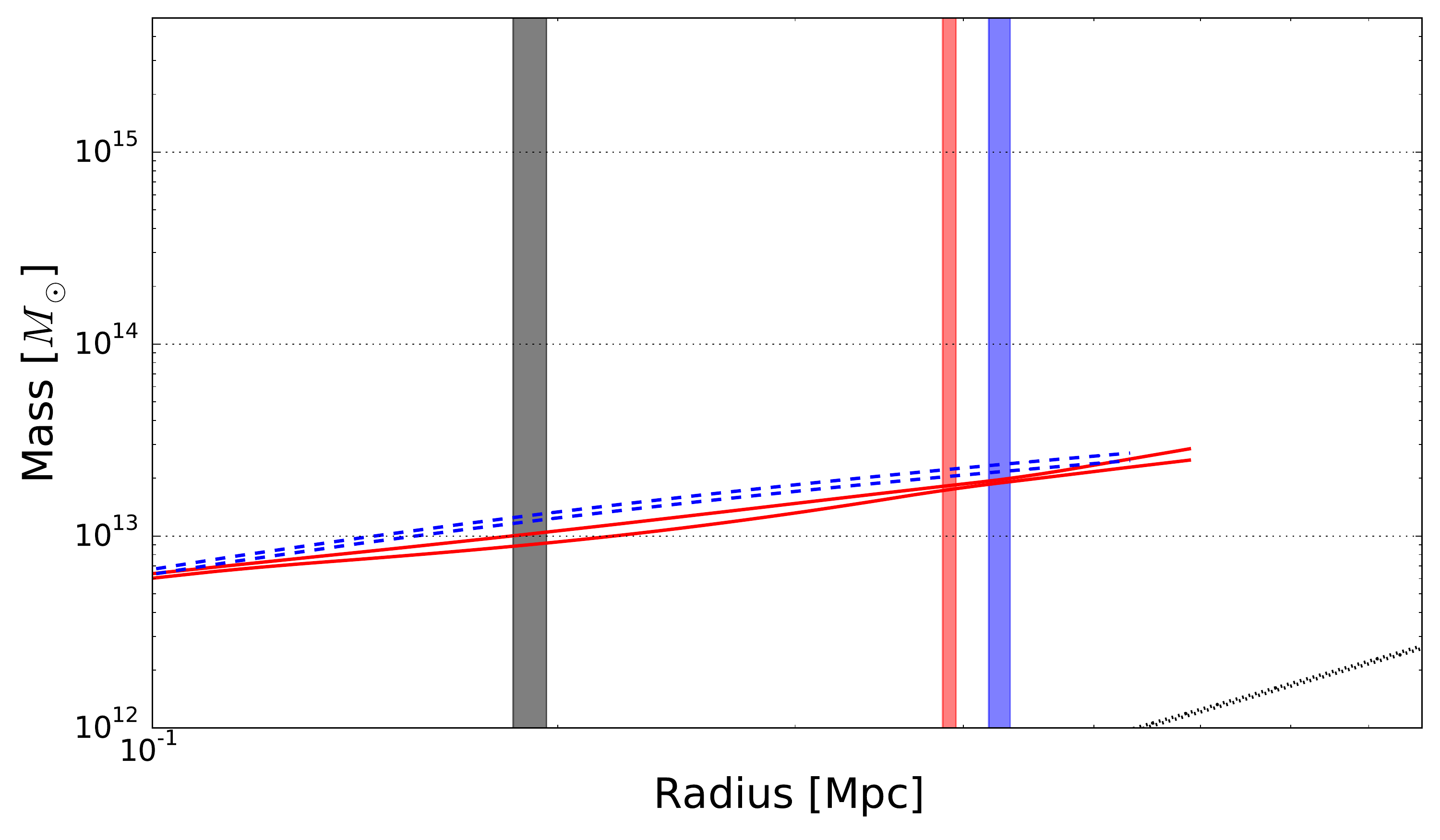}
	\caption{As Fig. \ref{fig:app_2A0335} but for NGC1399.}
	\label{fig:app_NGC1399}
\end{figure}
\clearpage
\begin{figure}
	\centering
	\includegraphics[width=0.45\textwidth]{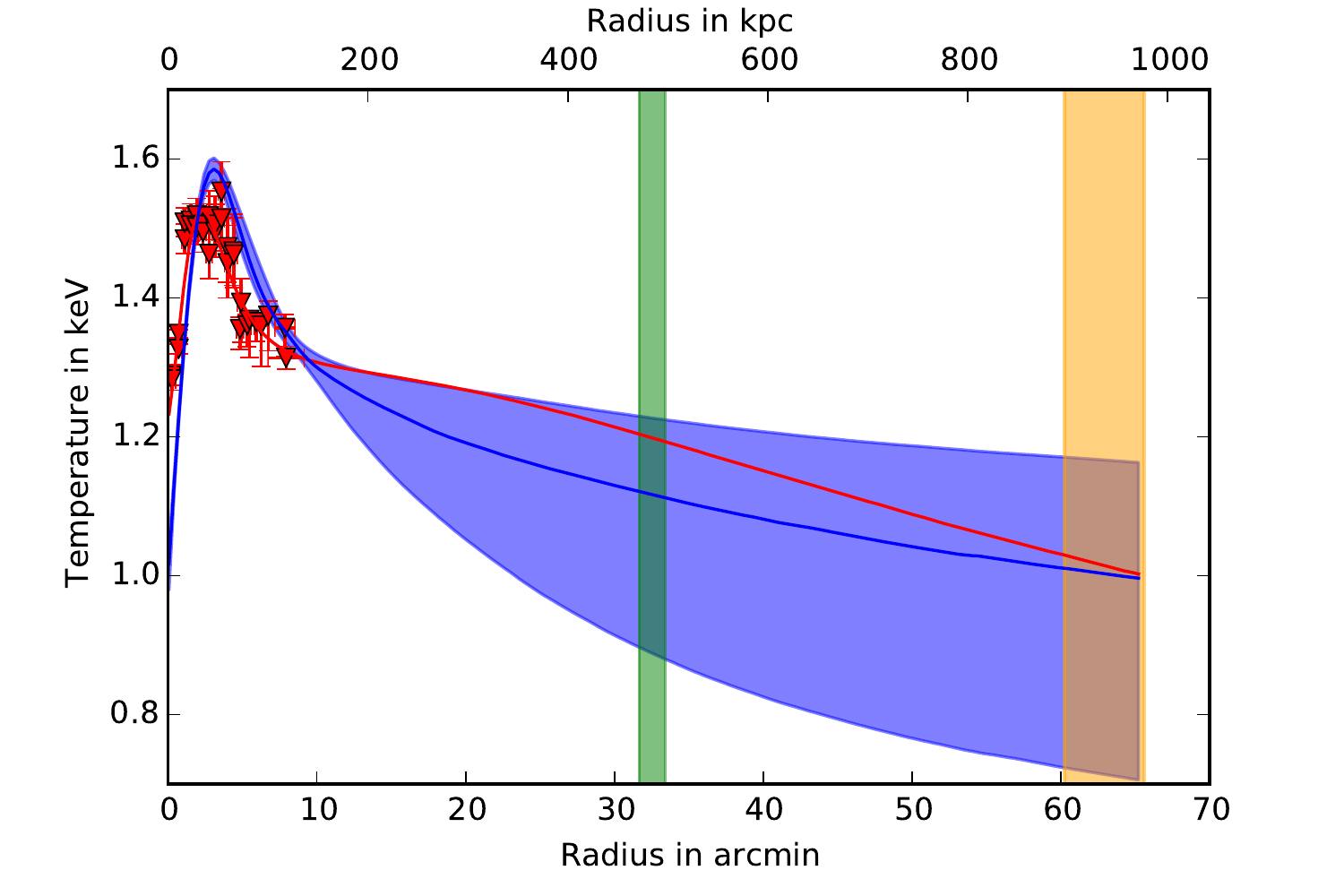}
	\includegraphics[width=0.45\textwidth]{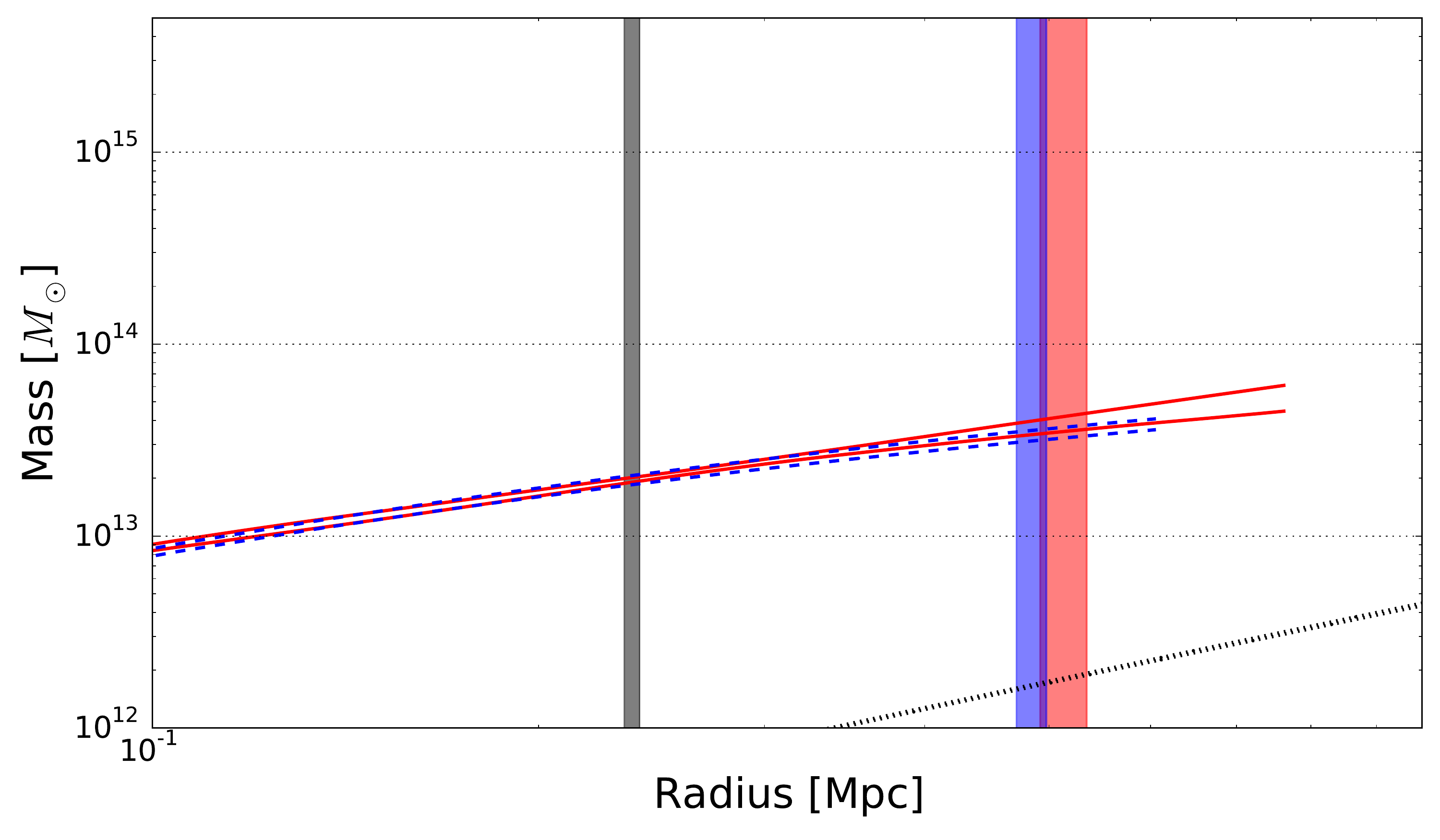}
	\caption{As Fig. \ref{fig:app_2A0335} but for NGC1550.}
	\label{fig:app_NGC1550}
\end{figure}
\begin{figure}
	\centering
	\includegraphics[width=0.45\textwidth]{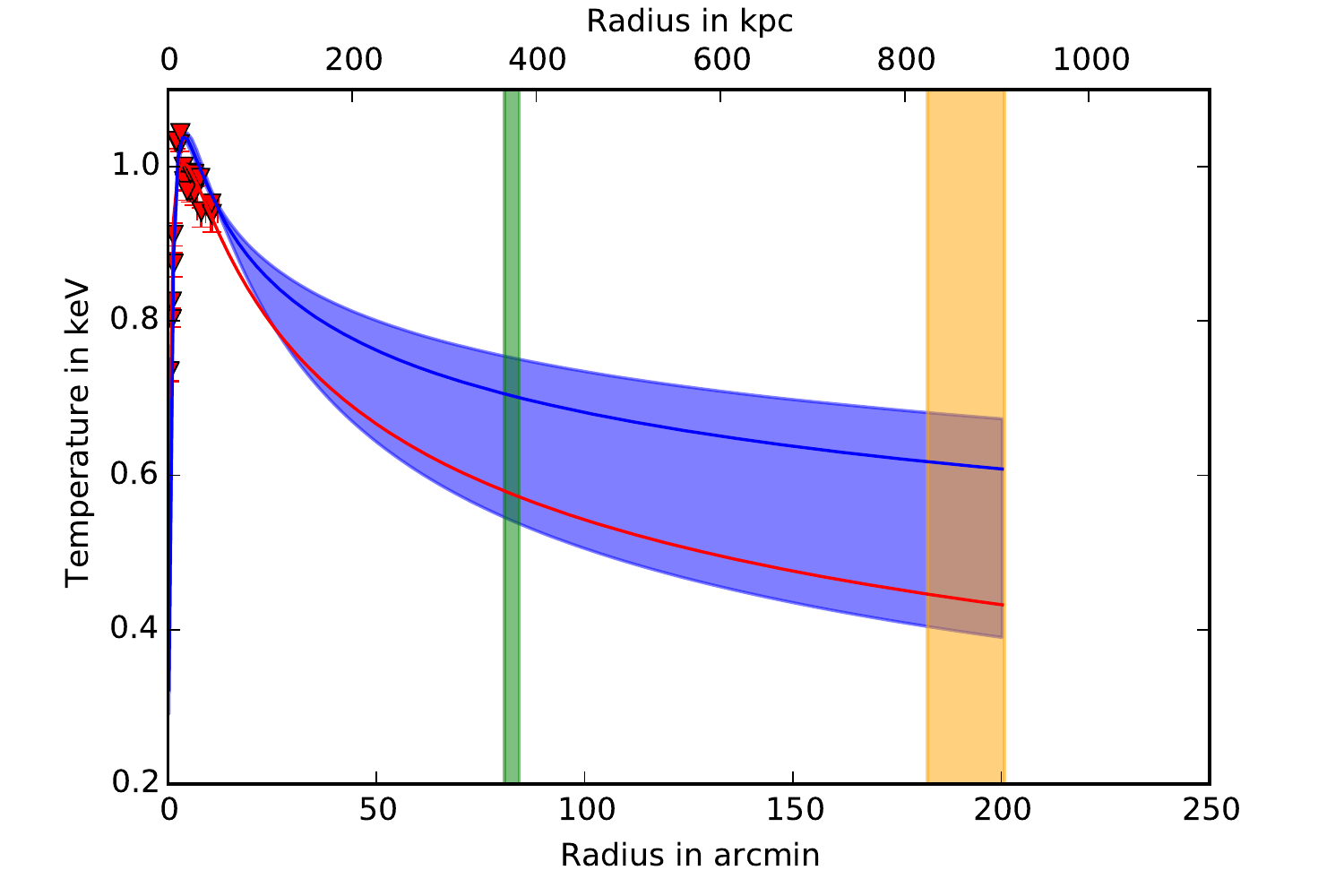}
	\includegraphics[width=0.45\textwidth]{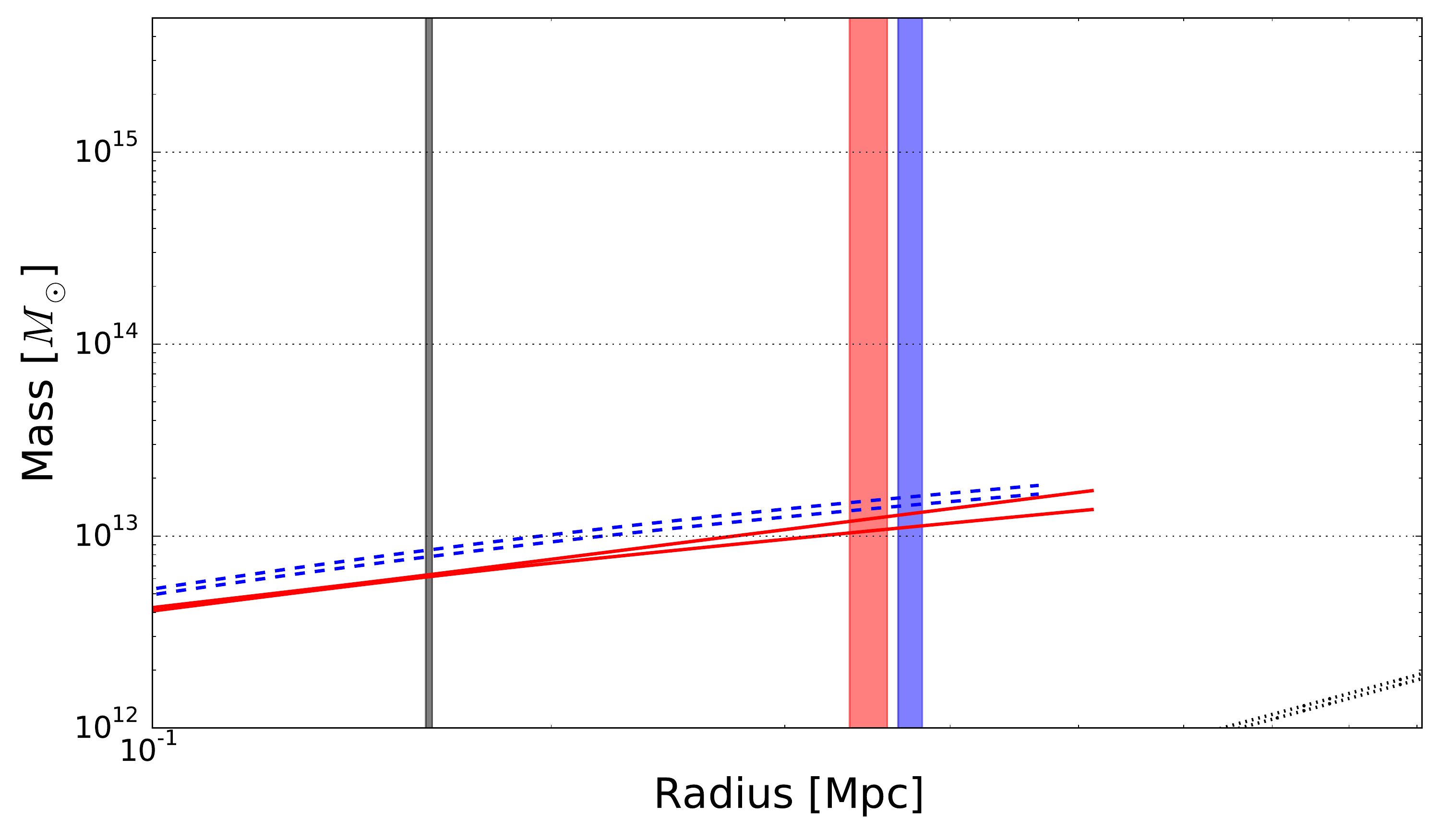}
	\caption{As Fig. \ref{fig:app_2A0335} but for NGC4636.}
	\label{fig:app_NGC4636}
\end{figure}
\begin{figure}
	\centering
	\includegraphics[width=0.45\textwidth]{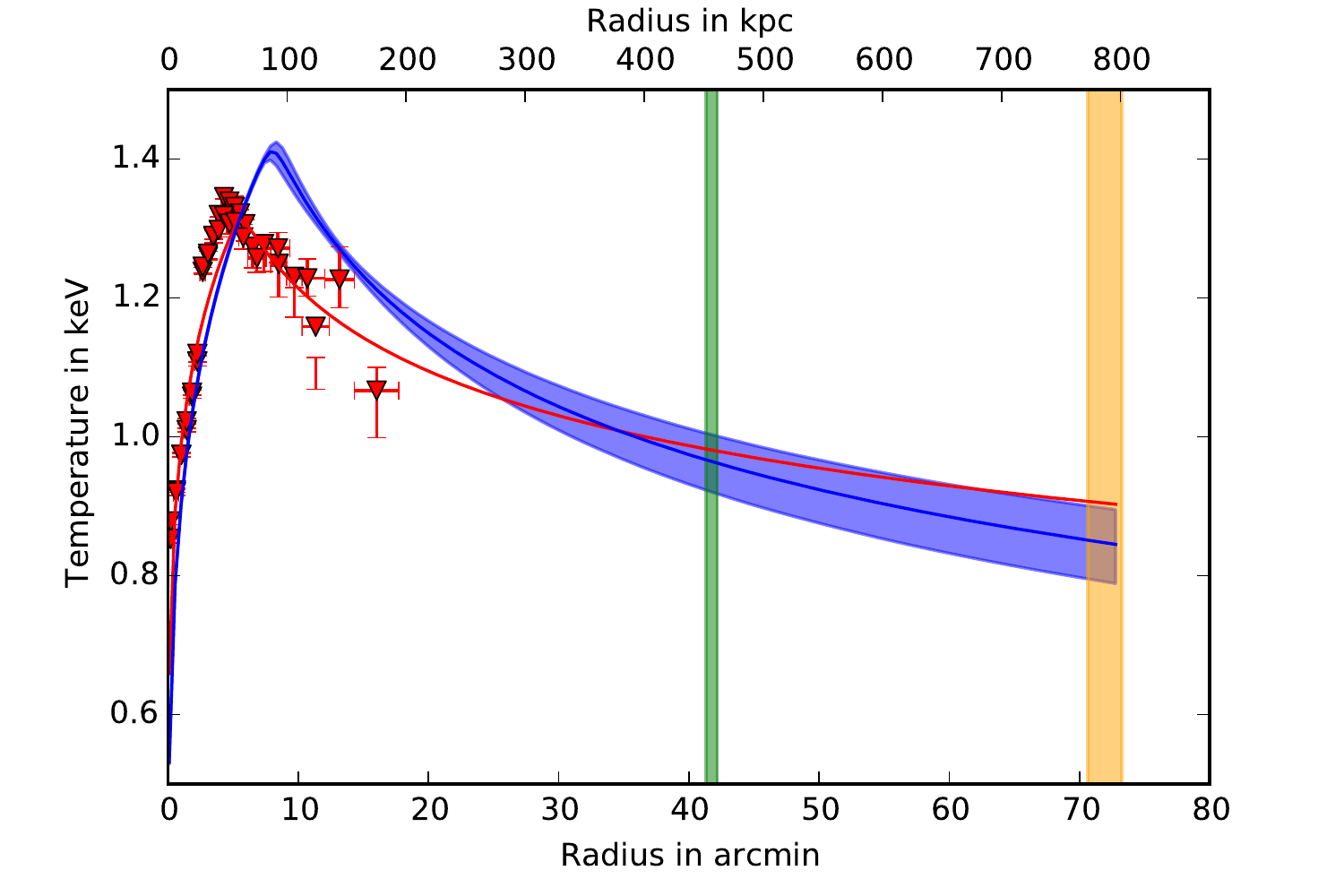}
	\includegraphics[width=0.45\textwidth]{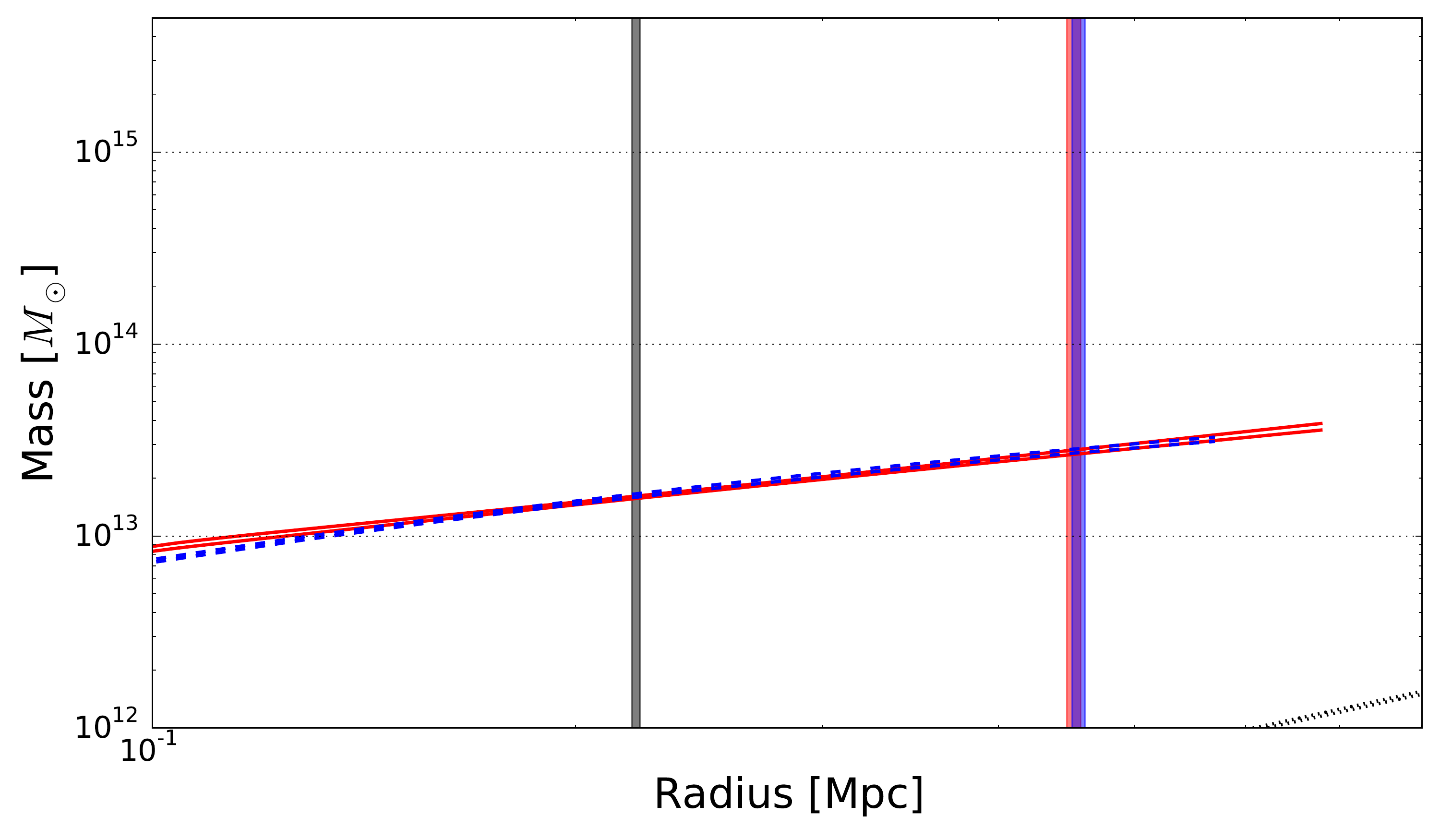}
	\caption{As Fig. \ref{fig:app_2A0335} but for NGC5044.}
	\label{fig:app_NGC5044}
\end{figure}
\clearpage
\begin{figure}
	\centering
	\includegraphics[width=0.45\textwidth]{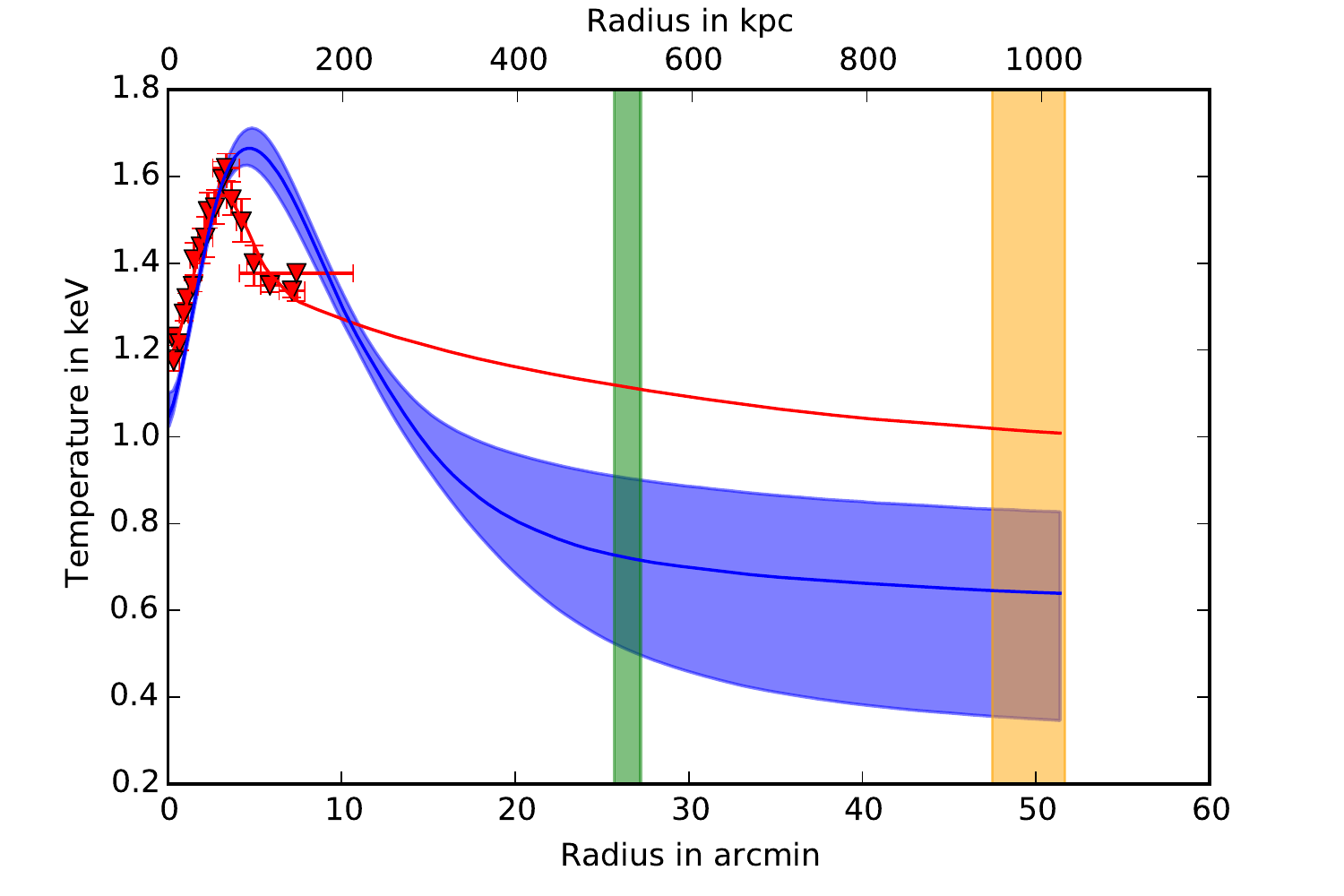}
	\includegraphics[width=0.45\textwidth]{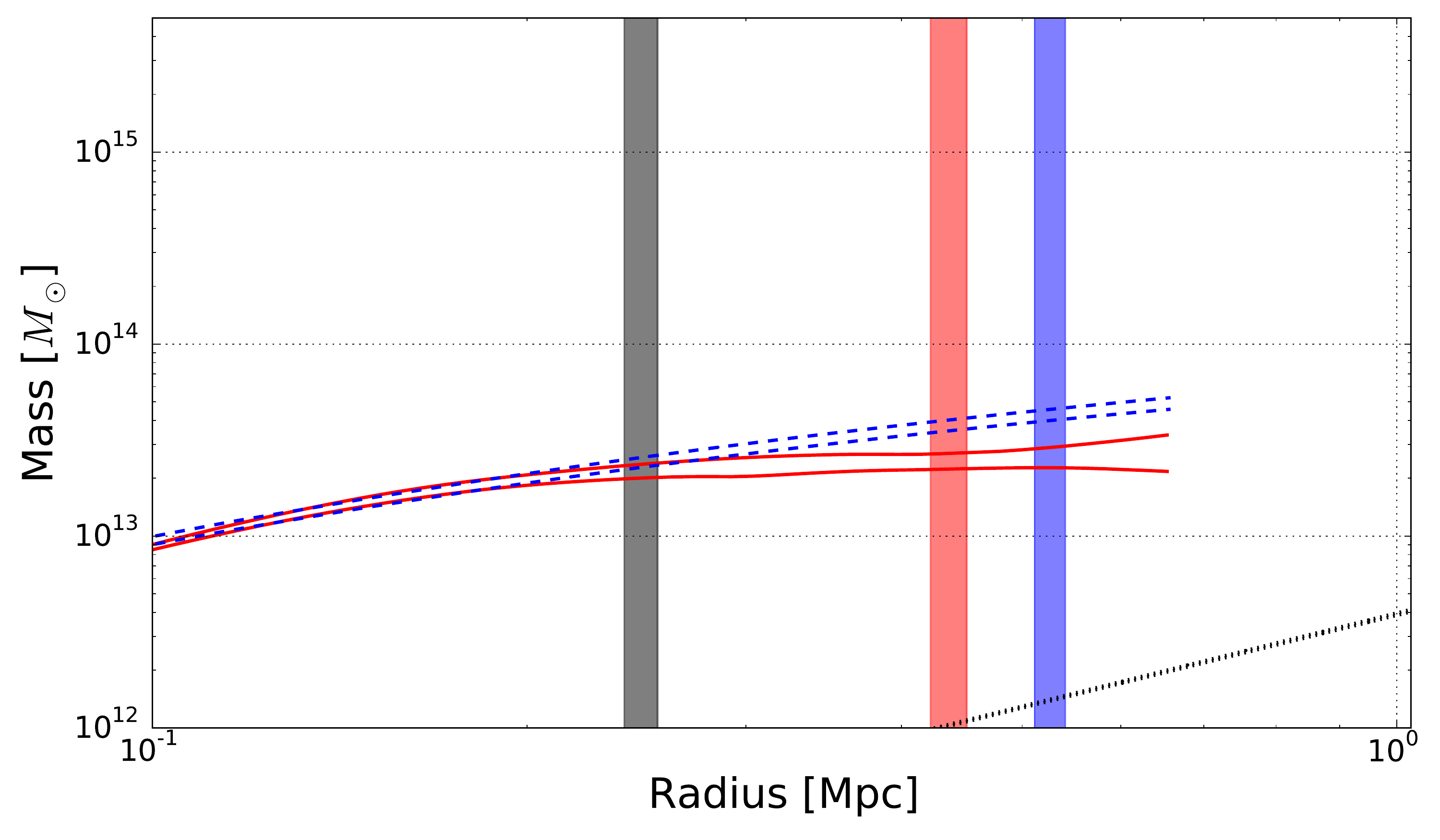}
	\caption{As Fig. \ref{fig:app_2A0335} but for NGC507.}
	\label{fig:app_NGC507}
\end{figure}
\begin{figure}
	\centering
	\includegraphics[width=0.45\textwidth]{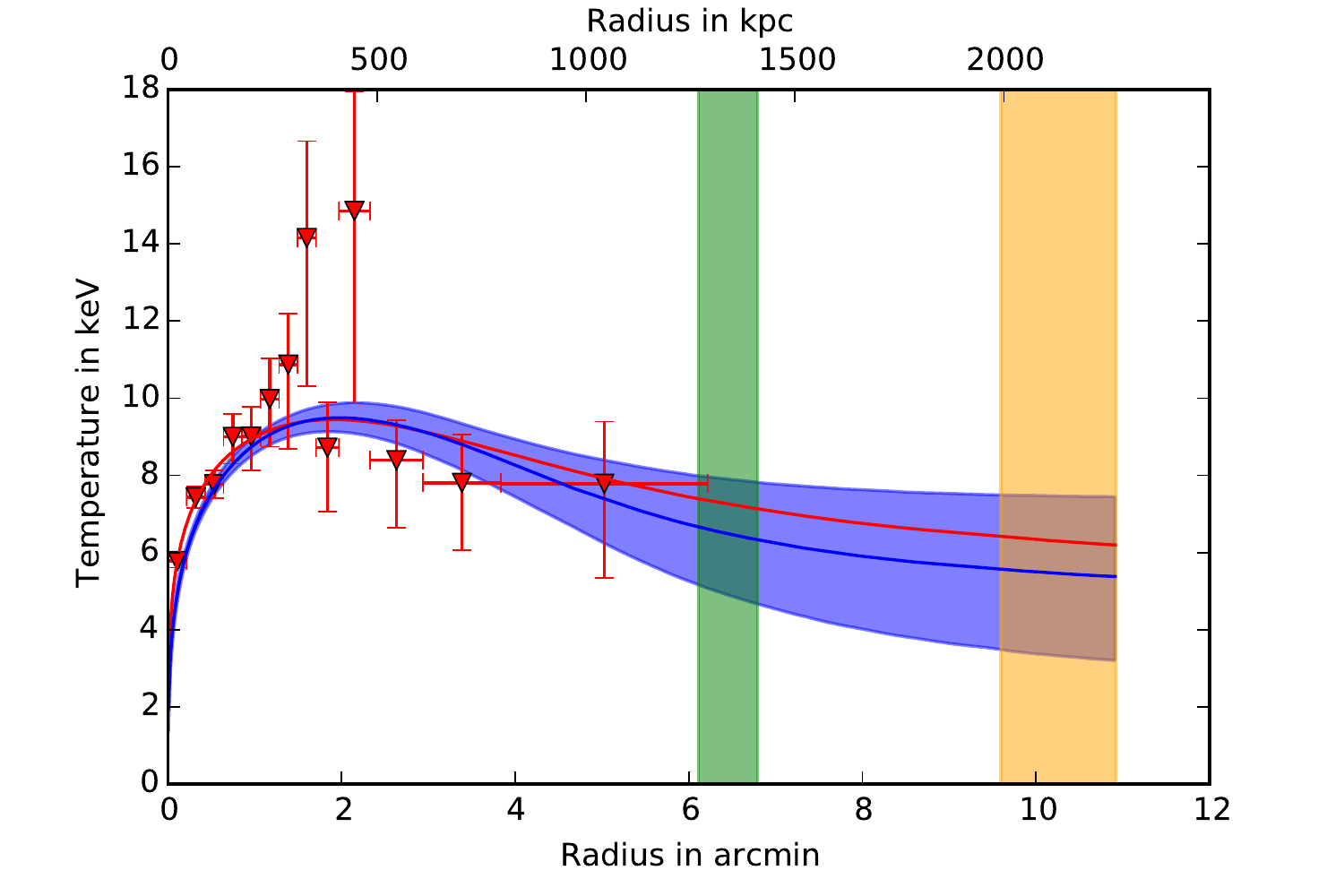}
	\includegraphics[width=0.45\textwidth]{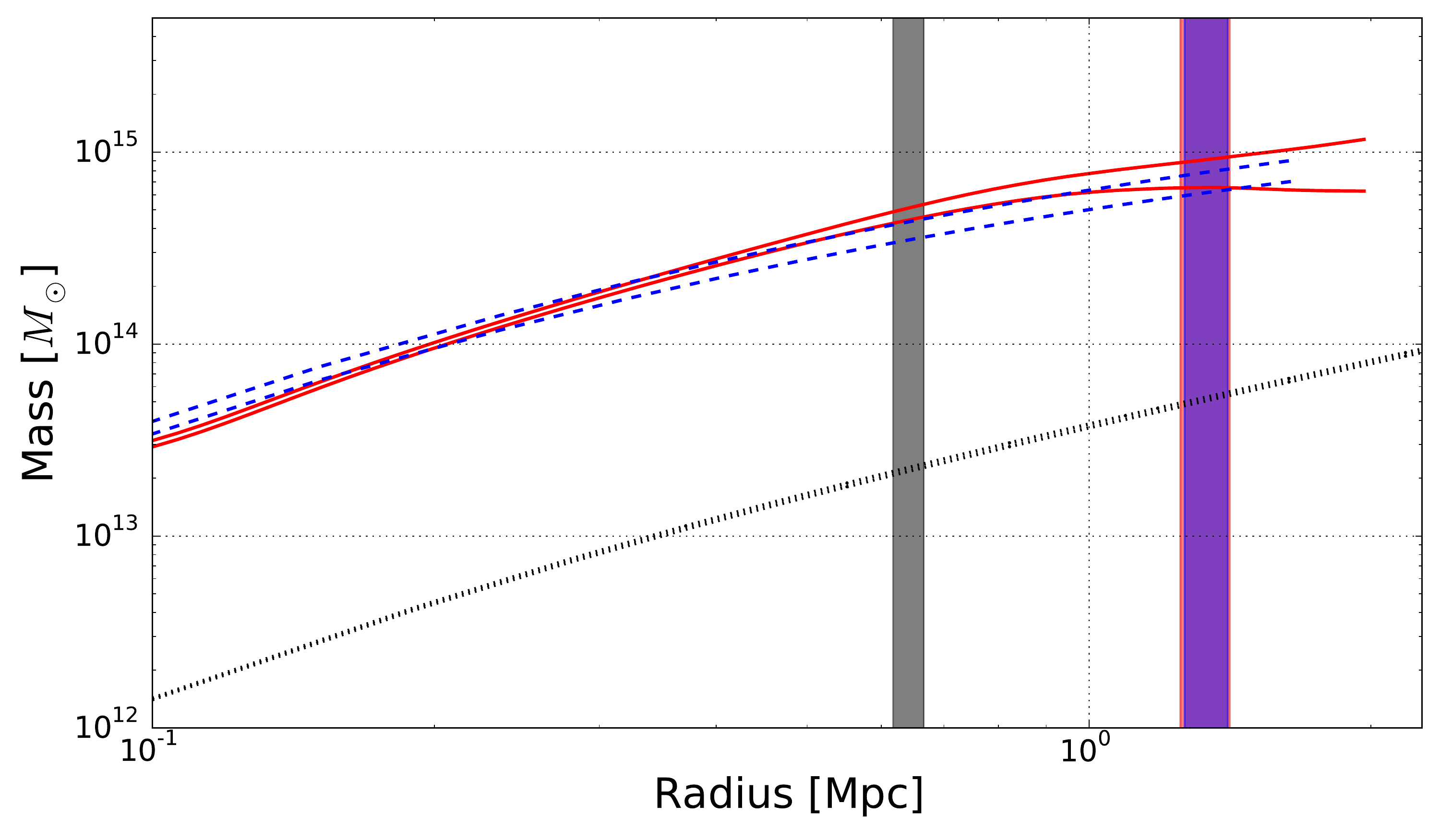}
	\caption{As Fig. \ref{fig:app_2A0335} but for RXCJ1504.}
	\label{fig:app_RXCJ1504}
\end{figure}
\begin{figure}
	\centering
	\includegraphics[width=0.45\textwidth]{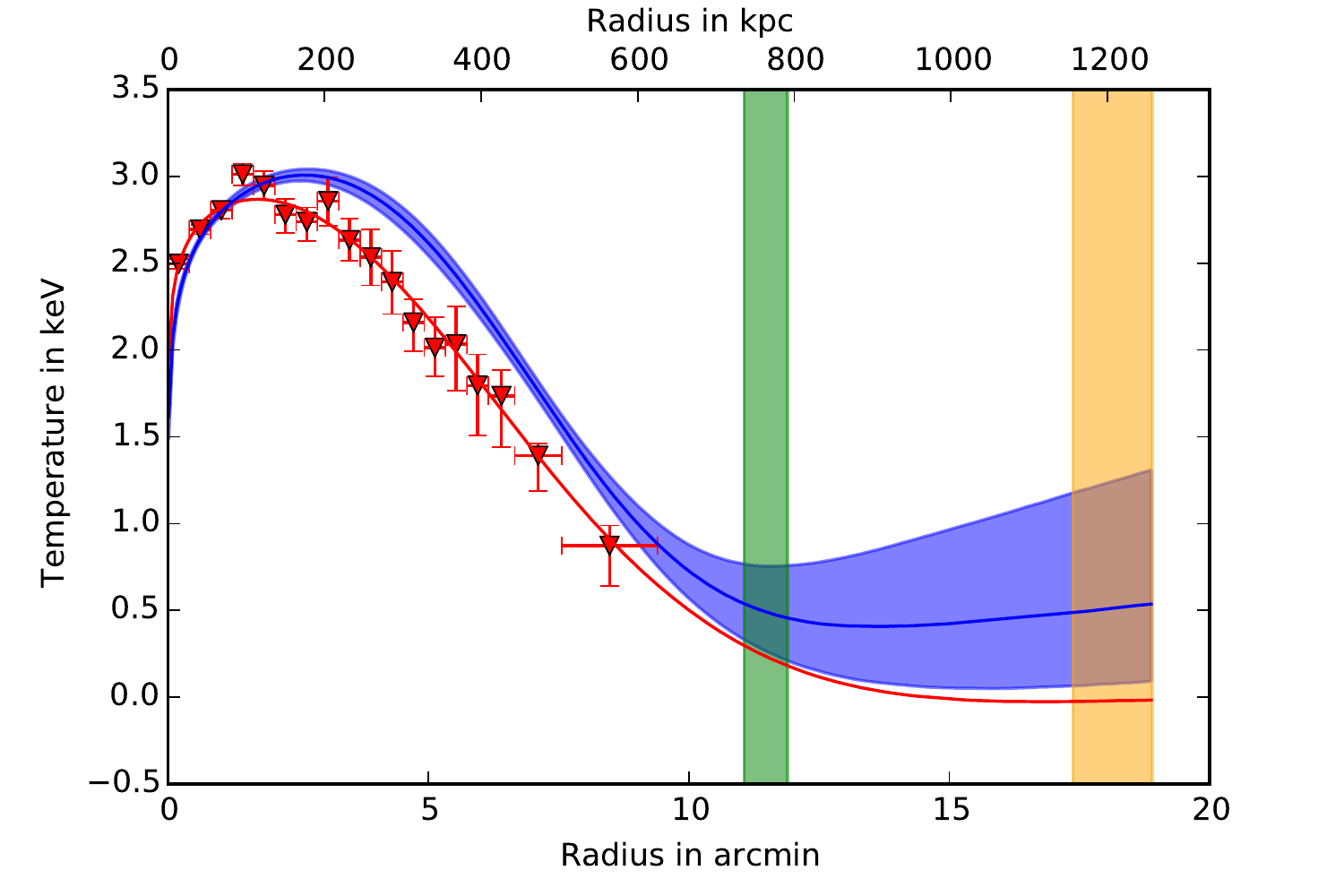}
	\includegraphics[width=0.45\textwidth]{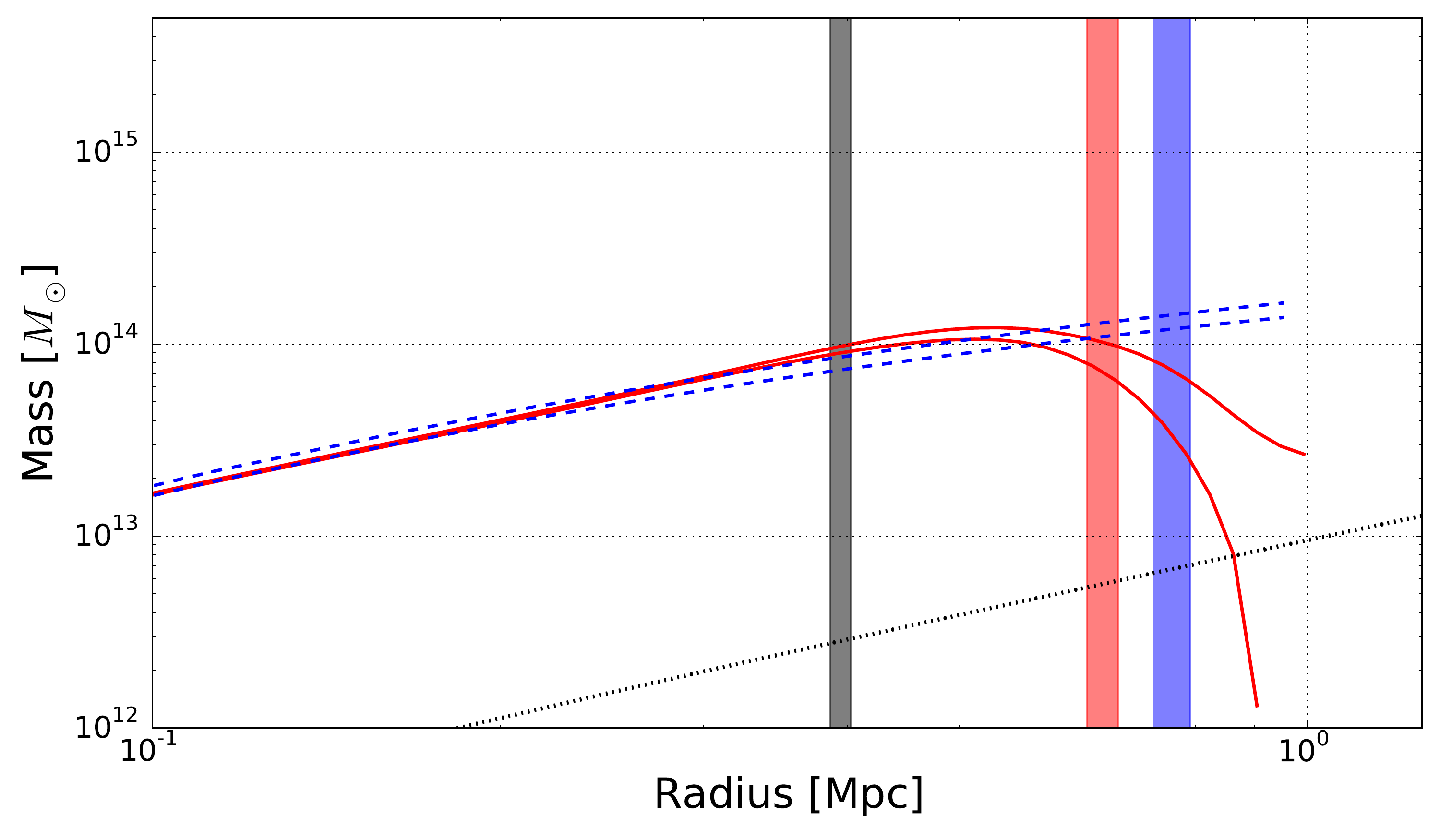}
	\caption{As Fig. \ref{fig:app_2A0335} but for S1101.}
	\label{fig:app_S1101}
\end{figure}
\clearpage
\begin{figure}
	\centering
	\includegraphics[width=0.45\textwidth]{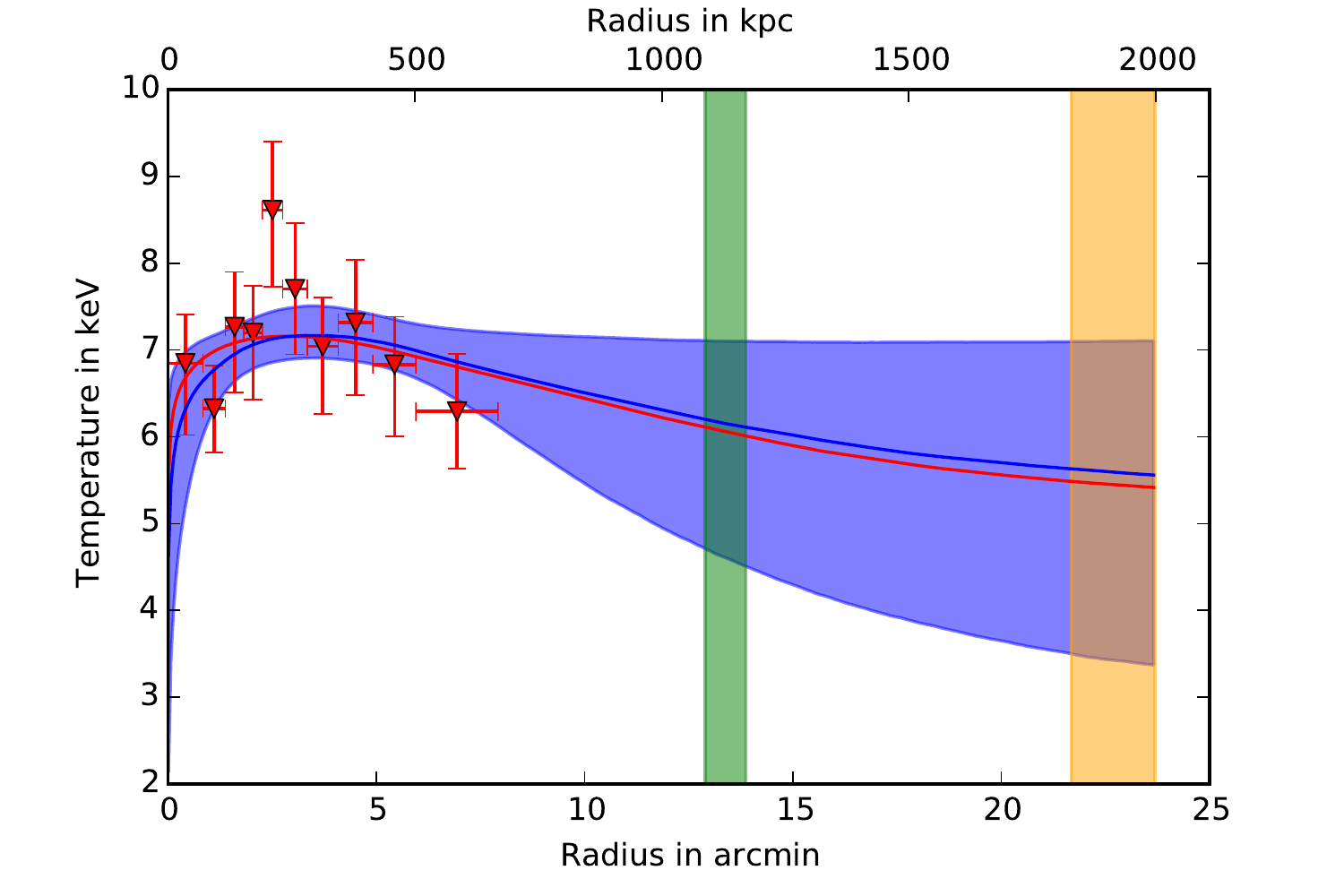}
	\includegraphics[width=0.45\textwidth]{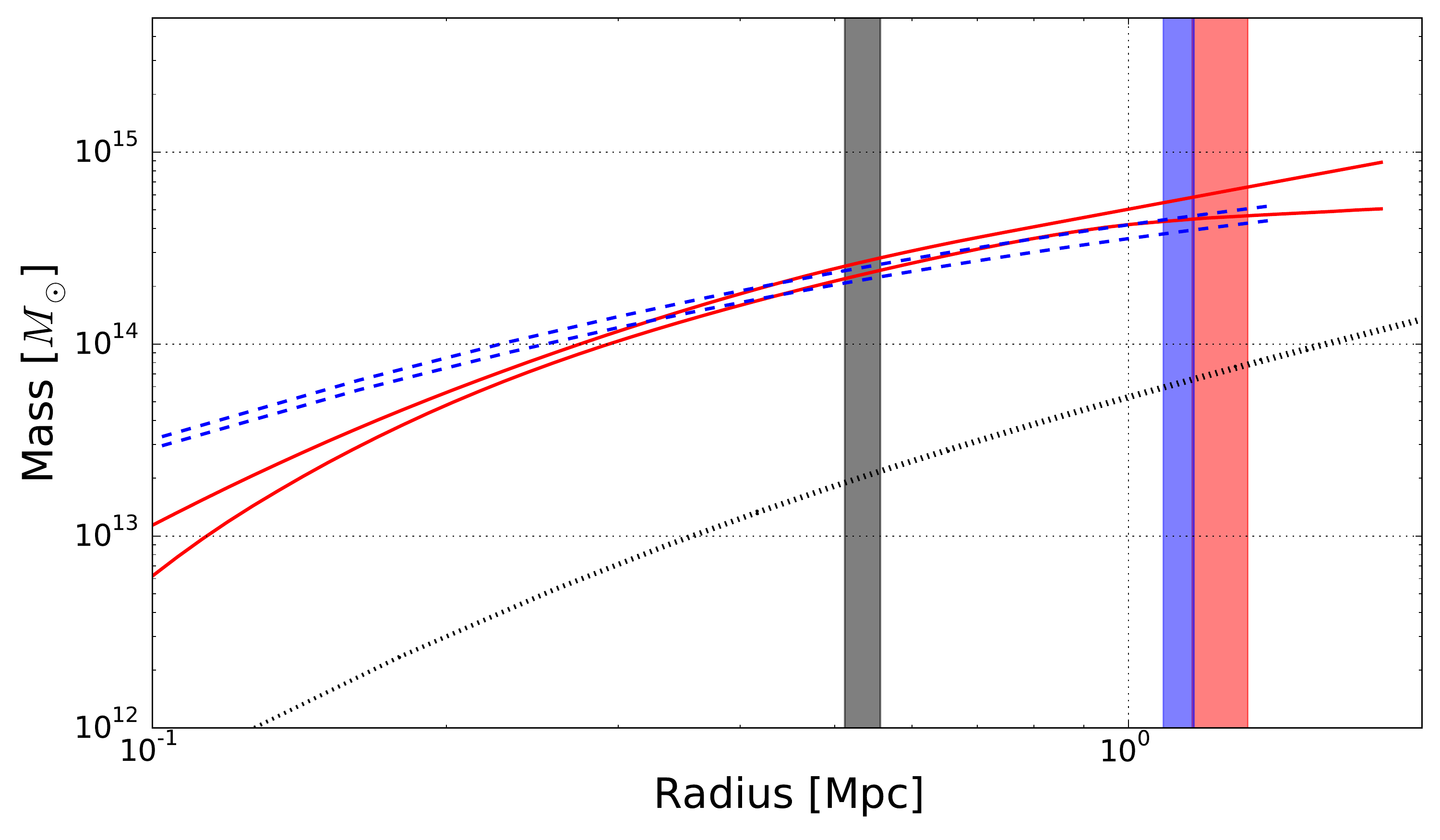}
	\caption{As Fig. \ref{fig:app_2A0335} but for ZwCl1215.}
	\label{fig:app_ZwCl1215}
\end{figure}

%%%%%%%%%%%%%%%%%%%%%%%%%%%%%%%%%%%%%%%%%%%%%%%%%%

% Don't change these lines
\bsp	% typesetting comment
\label{lastpage}
\end{document}